\documentclass[graybox,envcountchap]{svmono}
\usepackage{textcomp,gensymb}
\usepackage{mathptmx}
\usepackage{helvet}
\usepackage{courier}
\usepackage{type1cm}
\usepackage{makeidx} 
\usepackage{graphicx} 
\usepackage{multicol} 
\usepackage[bottom]{footmisc} 
\makeindex 
\usepackage[authoryear,round]{natbib}

\usepackage{authblk}

\begin{document}

\author[1]{Seymur Cahangirov}
\author[2,3]{Hasan Sahin}
\author[4]{Guy Le Lay}
\author[5,6]{Angel Rubio}
\affil[1]{UNAM - National Nanotechnology Research Center, Bilkent University, 06800 Ankara, Turkey\\
email: seymur@unam.bilkent.edu.tr\\
webpage: http://unam.bilkent.edu.tr/~seymur}
\affil[2]{Department of Photonics, Izmir Institute of Technology, 35430 Izmir, Turkey}
\affil[3]{Department of Physics, University of Antwerp, Groenenborgerlaan 171, B-2020 Antwerp, Belgium\\
email: hasan.sahin@uantwerpen.be\\
webpage: http://cmt.ua.ac.be/hsahin}
\affil[4]{Aix Marseille Universit\' e, CNRS, PIIM UMR 7345, 13397, Marseille, France\\
email: guy.lelay@univ-amu.fr}
\affil[5]{Max Planck Institute for the Structure and Dynamics of Matter and Center for Free-Electron Laser Science  Luruper Chaussee 149, 22761 Hamburg, Germany\\
email: angel.rubio@mpsd.mpg.de\\
webpage: http://www.mpsd.mpg.de/113438/theod}
\affil[6]{Nano-Bio Spectroscopy Group and ETSF,  Dpto. Fisica de Materiales, Universidad del Pa\' is Vasco, CFM CSIC-UPV/EHU-MPC and DIPC, 20018 San Sebasti\' an, Spain\\
email: angel.rubio@ehu.es\\
webpage: http://nano-bio.ehu.es/}
\renewcommand\Authands{ and }

\title{Introduction to the physics of Silicene and other 2D materials}
\subtitle{Lecture Notes in Physics}
\maketitle

\include{dedic}
%
%

\extrachap{Acknowledgements}

We would like to thank our collaborators including Salim Ciraci, Francois Peeters, Paola De Padova, Patrick Vogt, Maria Carmen Asensio, Maria Davila, Wenhui Duan, Shou-Cheng Zhang, Lede Xian, Peizhe Tang and many others.

Seymur Cahangirov acknowledges financial support from the Marie Curie grant FP7-PEOPLE-2013-IEF project No. 628876 and the Scientific and Technological Research Council of Turkey (TUBITAK) under the project number 115F388. Hasan Sahin is supported by the FWO Pegasus Long Marie Curie Fellowship. Angel Rubio acknowledges financial support from the European Research Council (ERC-2015-AdG-694097), Spanish grant (FIS2013-46159-C3-1-P), Grupos Consolidados (IT578-13), and AFOSR Grant No. FA2386-15-1-0006 AOARD 144088, H2020-NMP-2014 project MOSTOPHOS, GA no. SEP-210187476 and COST Action MP1306 (EUSpec).
\tableofcontents

\chapter{A Brief History of Silicene} \label{History}

\abstract*{Research on silicene shows a fast and steady growth that has increased our tool-box of novel 2D materials with exceptional potential applications in materials science. Especially after the experimental synthesis of silicene on substrates in 2012 it has attracted substantial interest from both theoretical and experimental communities. Every day, new people from various disciplines join this rapidly growing field. The aim of this book is to serve as a fast entry to the field to these newcomers and as a long-living reference to the growing community. To achieve this goal, the book is designed to emphasize the most crucial developments from both theoretical and experimental point of view since the starting of the silicene field back in 1994 with the first theoretical paper proposing the structure of silicene. We provide the general concepts and ideas such that the book is accessible to everybody from graduate students to senior researchers and we refer the reader interested in the detail to the relevant literature. We now start with a brief history of silicene where we highlight, in the chronological order, the important works that shaped our understanding of silicene.}

Research on silicene shows a fast and steady growth that has increased our tool-box of novel 2D materials with exceptional potential applications in materials science. Especially after the experimental synthesis of silicene on substrates in 2012 has attracted substantial interest from both theoretical and experimental community. Every day, new people from various disciplines join this rapidly growing field. The aim of this book is to serve as a fast entry to the field these newcomers and as a long-living reference to the growing community. To achieve this goal, the book is designed to emphasize the most crucial developments from both theoretical and experimental points of view since the start of the silicene field with the first theoretical paper proposing the structure of silicene. We provide the general concepts and ideas such that the book is accessible to everybody from graduate students to senior researchers and we refer the reader interested in the details to the relevant literature. In the next paragraphs, we present a brief history of silicene where we highlight, in the chronological order, the important works that shaped our understanding of silicene.

The atomic and electronic structure of the materials that we now call silicene and germanene was investigated for the first time by Takeda and Shiraishi, a decade before graphene was obtained by exfoliation from the parent graphite crystal \citep{Takeda1}. Using density functional theory (DFT) \citep{Hohenberg1}, they have shown that it is energetically favorable for silicene and germanene to become buckled instead of staying planar, as carbon atoms do in the case of graphene. The band structure of silicene was also reported but there was no emphasis on the linear crossing at the Fermi level, namely the Dirac cone. This visionary paper was ignored for almost a decade for two main reasons. First, there was a common belief that these two-dimensional (2D) materials cannot exist in nature \citep{Peierls1,Landau1,Mermin1}. Second, it was hard to believe that silicon could acquire an \textit{sp$^2$}-like hybridization because it always preferred \textit{sp$^3$} hybridization \citep{Fagan1}.

In 2004, a monolayer of carbon atoms named graphene was isolated from graphite, the layered bulk allotrope of carbon \citep{Novoselov1}. This left no doubt about the stability of 2D materials. However, later it was discovered that graphene possesses long-wavelength mechanical oscillations \citep{Meyer1} which are the manifestation of the fact that pure 2D crystals are indeed thermodynamically unstable  \citep{Peierls1,Landau1,Mermin1}. Graphite was not the only layered material. The monolayer counterparts of other layered materials like hexagonal boron nitride, metal dichalcogenides and metal oxides were also peeled off or synthesized and studied in detail. The layers of these materials are formed by strong covalent bonds while they are kept together by weak van der Walls interactions. The isolation of these layers has opened a possibility of combining these layers in a desired order to create the so-called van der Walls heterostructures that can be engineered to have novel properties \citep{Geim1,Gao1}.

Besides being a stable 2D material with impressive mechanical properties (which is also true for the BN monolayer) like having 1.0 terapascal Young's modulus while being extremely flexible \citep{Lee1}, graphene also attracted huge interest due to the linearly crossing bands at the Fermi level that allow its electrons to behave as if they have no mass \citep{Novoselov2}. The rise of graphene triggered interest in other possible group IV 2D materials like silicene, germanene and stanene (also written stannene, but rarely, from stannum in Latin, or sometimes called tinene) that would have similar electronic structures. However, they lack a layered parent bulk counterpart like graphite. Synthesis of these 2D materials that lacked bulk counterpart has required a more bottom-up approach, like growing them epitaxially on top of a substrate.

\begin{figure}[t]
\begin{center}
\includegraphics[width=9.5cm]{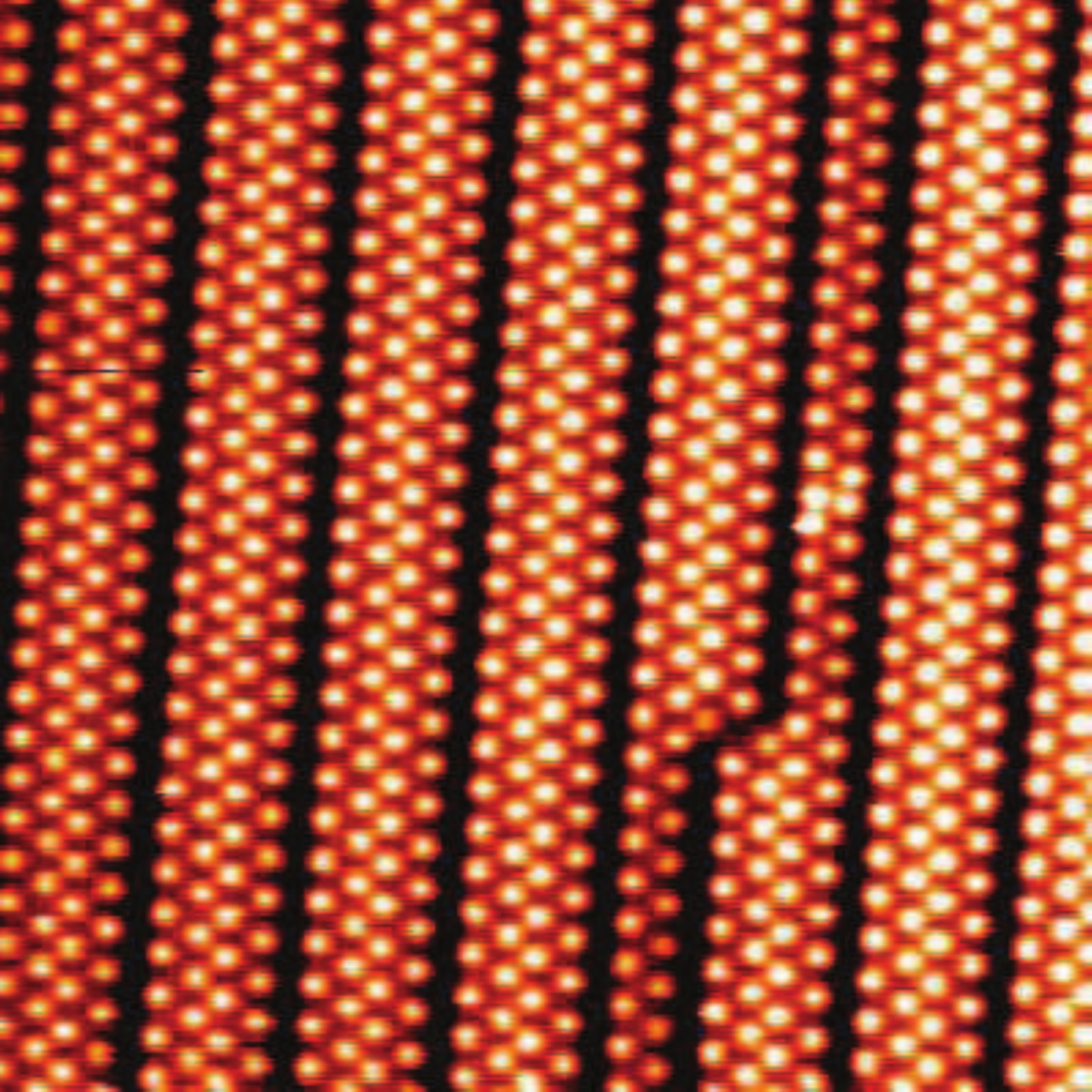}
\caption{15$\times$15 nm$^2$ STM image recorded at 5K (V = $\pm$0.05 V, I = 200 nA) of the quasi perfect grating with a pitch of just 2 nm, formed by self-assembled symmetry breaking silicene nanoribbons, 1.6 nm in width. Adapted from Fig. 2 of \citep{Ronci1}.}
\label{ribbons}
\end{center}
\end{figure}

In 2005, silicene (the 2D allotrope of silicon) started to appear in several theoretical and experimental studies. Self-aligned one dimensional (1D) structures of silicon (see Fig.~\ref{ribbons}) were synthesized on Ag(110) surface \citep{Leandri1}. Guy Le Lay, gave the kick to this study by noticing novel quantized states in the $sp$ region of silver during the angle-resolved photoemission spectroscopy (ARPES) measurements at the Elettra synchrotron radiation source in Trieste, Italy, and inferred the presence of 1D objects prior to their observation in STM imaging. He took inspiration from his earlier work of the reverse system in which Ag was deposited onto silicon surfaces \citep{Lelay1}. The same year, Durgun \textit{et al.} investigated Si nanotubes which were based on the buckled geometry of silicene reported by Takeda and Shiraishi \citep{Durgun1}.

In 2007, the planar structure of silicene was investigated using tight-binding Hamiltonians \citep{Guzman1}. In this study, it was emphasized that, similar to graphene, the planar structure of silicene would also exhibit linearly crossing bands at the Fermi level. The name "silicene" was used for the first time in this study. In 2009, Leb\'egue and Eriksson investigated the electronic properties of several 2D materials including planar structures of silicene and germanene using first-principles DFT calculations \citep{Lebegue1}. Similar to the tight binding results mentioned above, their calculations showed that planar silicene had linear bands crossing at the Fermi level. However, planar germanene was found to be a poor metal since the conduction band was partially filled around the $\Gamma$-point which shifted the Fermi level below the linear crossing at the $K$-point. This study revived interest in group IV 2D materials beyond graphene but it was unfortunate that the authors adopted the planar structure of silicene and germanene which was shown to be energetically less favorable compared to the buckled one \citep{Takeda1}.

In the meantime, after waiting almost a year in the review process, Cahangirov \textit{et al.} reported a study that answered the question of whether silicon could adopt an $sp^2$-like hybridization to make a stable free-standing 2D structure \citep{Cahangirov1}. They have calculated phonon dispersions for both planar and buckled silicene. Planar silicene clearly had imaginary frequencies indicating structural instability, while all phonon modes were positive for the buckled silicene. This meant that buckled silicene had restoring forces to any possible deformation while planar silicene was unstable. They have also emphasized that although buckling was lowering the symmetry of the system, the remaining hexagonal symmetry of the lattice was enough to preserve the linearly crossing bands at the Fermi level. Furthermore, they have shown that similar to their graphene counterparts, armchair silicene nanoribbons displayed the family behavior in the band gap variation with respect to their widths \citep{Cahangirov2}. This work boosted many theoretical and experimental investigations on silicene and other 2D materials beyond graphene \citep{Sahin1,Xu2}.

On the experimental side, Le Lay \textit{et al.} continued to study deposition of silicon onto Ag substrates \citep{DePadova1,DePadova2,Kara1,LeLay2}. Now their work was clearly inspired by the theoretically predicted stability of silicon structures with $sp^2$-like hybridization and their promising electronic properties. They have also discovered that what they thought to be silicon nanowires were in fact silicene nanoribbons epitaxially grown on Ag(110) surface. Using DFT calculations they were able to reveal the detailed atomic structure of these silicene nanoribbons. Their work continued to build up the idea of a possible formation of monolayer silicene on Ag substrates \citep{Aufray1,DePadova3}. ARPES measurements that they performed along the silicene nanoribbons showed linear bands in the vicinity of the Brillouin zone boundary \citep{DePadova3}. These linear bands were interpreted as $\pi$-states of zigzag silicene nanoribbons \citep{DePadova4}.

A gold rush started to synthesize 2D silicene. In 2010, ignoring Guy Le Lay's reservations, some of his former co-workers in Marseille published a letter on the ``epitaxial growth of a silicene sheet'' on a Ag(111) substrate. This letter was based solely on STM topographs with a highly perfect, nearly flat honeycomb appearance but incredibly short lattice parameter, without a single defect on areas of several hundred nm$^2$! This letter has been a nuisance, creating continuing confusion, although the graphene-like honeycomb appearance was proved to be just an illusion: a mere mirage originating from a contrast reversal in STM imaging of a bare Ag(111) surface \citep{Lelay3}. All this perfectly illustrated the famous quote of Luigi Galvani: ``for it is easy in experimentation to be deceived, and to think one has seen and discovered what one has desired to see and discover''.

Research on 2D materials received another boost in 2010 after the Nobel Prize in Physics was awarded to Geim and Novoselov for their work on graphene. Silicene acted as a role model for the possibility of other 2D materials beyond graphene. While synthesis of silicene was still expected, theoretical works on silicene continued to progress. Several works investigated the atomic and electronic properties of fully hydrogenated silicene, named silicane \citep{Houssa1,Garcia1,Osborn1}. The carbon counterpart, named graphane, had been already synthesized \citep{Elias1} and studied theoretically \citep{Sofo1}.

On the other hand, differences between silicene and graphene started to attract some interest. One such difference was originating from the lowered symmetry in silicene due to buckling. Using ab initio calculations, it has been shown that it is possible to tune the band gap of silicene and germanene by applying an electric field in perpendicular direction to the sheets \citep{Ni1,Drummond1}. They concluded that this property could be used to build field effect transistors that can operate at room temperature. Another important difference between graphene and silicene is the strength of the spin-orbit coupling. In the case of graphene, the spin-orbit coupling is so weak that the quantum spin Hall effect (QSHE) can only occur at unrealistically low temperatures \citep{Kane1,Yao1}. Silicene, on the other hand, was shown to preserve its topologically nontrivial electronic structure despite buckling and to be able to host the QSHE up to 18 K, thanks to the larger spin-orbit coupling in silicon atoms \citep{Liu1,Liu2}. In the case of germanene, the QSHE could occur at even higher temperatures up to 277 K and this grows even further for stanene, the tin based analogue, another stable 2D material of the same family that has been synthesized much recently \citep{Zhu1}.

\begin{figure}[t]
\begin{center}
\includegraphics[width=9.5cm]{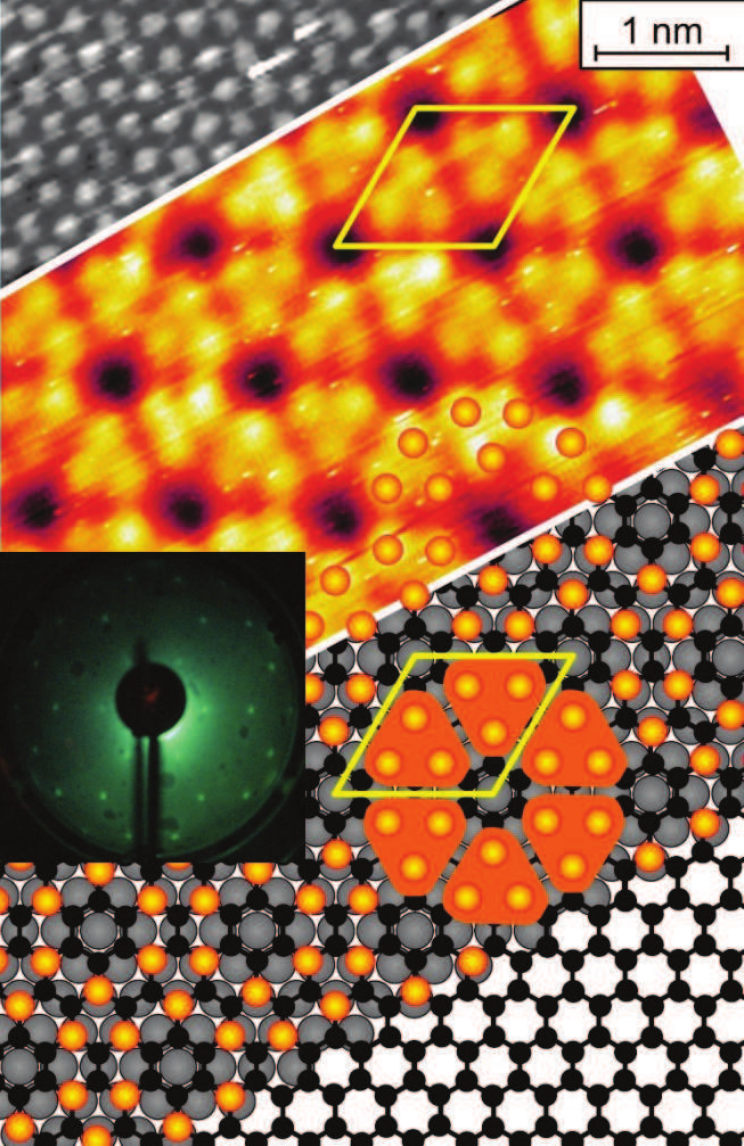}
\caption{The ``flower pattern'' with six bright protrusions per 4$\times$4 unit cell (with respect to Ag(111)) seen in STM images of the 3$\times$3 reconstructed underlying honeycomb silicene monolayer (bottom right corner) and the way it adapts on Ag(111) (top left corner) to form the 3$\times$3/4 $\times$4 (inserted LEED pattern) configuration. Adapted from Fig. 4 of \citep{Vogt1}.}
\label{flower}
\end{center}
\end{figure}

In 2012, the first compelling experimental evidence for the existence of silicene was reported \citep{Vogt1}. In this study, they deposited silicon on top of an Ag(111) substrate instead of Ag(110). As a result, a monolayer of silicene with 3$\times$3 reconstruction was formed, although not directly imaged. The 3$\times$3 supercell of silicene was perfectly matched with the 4$\times$4 supercell of the Ag(111) surface. Looking from a distance, the STM image consisted of bright spots arranged in a honeycomb lattice, but, zooming in, it was clear that actually each bright spot was formed by three protruding Si atoms making up the so called ``flower pattern'', i.e., a  3$\times$3 reconstructed silicene cell, as shown in Fig.~\ref{flower}. Having in mind an underlying, hidden, silicene honeycomb structure, Vogt \textit{et al.} were able reproduce the experimentally observed STM pattern by geometry optimization using DFT calculations. It turns out, as will be seen in more details below, that silicene, on all templates used for growth until now, is always reconstructed: the 1$\times$1 primitive cell has not yet been observed. The reason is that the first silicene layer conforms to the substrate by changing its lattice parameter and buckling with respect to free standing silicene to find commensurate epitaxial structures. Furthermore, Vogt \textit{et al.} did ARPES measurements in which they have detected linear surface bands (i.e. non-dispersing with the perpendicular component of the wave vector) starting at 0.3~eV below the Fermi level and extending down to -3~eV with a Fermi velocity of 1.3$\times$10$^6$ m/s when silicene was present on the surface. No such state was detected on the bare Ag(111) surface in the absence of silicene on top.

Later in the very same year, two more groups reported the experimental realization of silicene using either Ag(111) surface \citep{Feng1} or ZrB$_2$(0001) surface \citep{Fleurence1} as a substrate. Besides the aforementioned 3$\times$3 reconstructed superstructure, Feng \textit{et al.} reported other phases that would emerge when the experimental conditions were varied. In particular, a $\sqrt{3}\times\sqrt{3}$ reconstructed silicene structure with a lattice constant of 0.64~nm was observed. This value is $\sim$5$\%$ lower than the number that one would get by multiplying the 1$\times$1 unitcell lattice constant by $\sqrt{3}$. The STM image of this structure had bright spots arranged in a honeycomb lattice with $\sqrt{3}\times\sqrt{3}$ periodicity. Further deposition of silicon resulted in a second layer of $\sqrt{3}\times\sqrt{3}$ reconstructed silicene grown on top of the first one. This was interpreted as an evidence that $\sqrt{3}\times\sqrt{3}$ structure was intrinsic to silicene and was not dictated by the interaction with the Ag substrate. This claim was further supported by the fact that the shrunk lattice constant of the $\sqrt{3}\times\sqrt{3}$ structure could not be matched by anything on the Ag(111) surface. Although an atomic model for the $\sqrt{3}\times\sqrt{3}$ structure was proposed, it could not explain why the structure was shrinking \citep{Feng1,Chen1}. Later, the same group reported STS measurements on the $\sqrt{3}\times\sqrt{3}$ structure \citep{Chen1}. They have observed wavelike quasiparticle interference patterns near the boundaries of islands formed by the second $\sqrt{3}\times\sqrt{3}$ silicene layer grown on top of the first one. This suggested the intervalley and intravalley scattering of charge carriers. Furthermore, they have shown that the wavelength of the quasiparticle interference was changing linearly with respect to the applied voltage. This was attributed to a Dirac-like behavior of electrons close to the Fermi level and the Fermi velocity calculated from the slope of this line was reported to be (1.2$\pm$0.1)$\times$10$^6$m/s which was in agreement with ARPES results for monolayer silicene \citep{Vogt1} and also with those for multilayer silicene nanoribbons \citep{DePadova5} on Ag(110) surfaces. Note that, here the Fermi velocity was obtained from the empty states, at variance with ARPES measuring the filled states.

Upon experimental realization, silicene became a focus of attention and was dubbed ``graphene's silicon cousin'' \citep{Nature1,Brumfiel1}. It's exotic electronic structure was explored in a further detail \citep{Ezawa1}. In 2013,  $\sqrt{3}\times\sqrt{3}$ reconstructed silicene was synthesized on yet another metallic substrate, this time Ir(111) \citep{Meng1}. Meanwhile, the nature of the linear bands of silicene on Ag(111) observed in experiments became a hot debate \citep{Lin1,Chen2,Arafune1,Chen3,Wang1,Guo1,Cahangirov3,Gori1}. Some part of the community continued to interpret them as Dirac Fermions originating from silicene while others claimed that they were $sp$ bands of bulk Ag. Each side had reasonable arguments to support their view. The measured Fermi velocity and the range in which the linear bands were extending were to large compared to what was expected from silicene. On the other hand, the linear bands disappeared when silicene was not on top of the Ag substrate. Although this debate is still not fully resolved, there is a growing consensus that the linear bands are originating from the hybridization between silicene and Ag states. Indeed, to put the exotic electronic structure of silicene in use one needs to synthesize it on semiconducting substrates. This seems to remain as a challenge for experimentalists in years to come. Yet, this most probably will be achieved, provided a good template is used. A first indication is the growth of 2D Si nanosheets with local hexagonal structure on a MoS$_2$ surface \citep{Chiappe1}.

Meanwhile, De Padova \textit{et al.} continued studying multilayer silicene \citep{DePadova5,DePadova6,Vogt2}. They interpreted the $\sqrt{3}\times\sqrt{3}$ reconstructed structure as the bilayer silicene, since it would usually appear as islands on top of the 3$\times$3 phase \citep{DePadova6}. Using STM they found the $\sqrt{3}\times\sqrt{3}$ surface to be $\sim$2~\AA~above the 3$\times$3 surface. It was shown that $\sqrt{3}\times\sqrt{3}$ reconstruction was preserved when further layers were grown \citep{Vogt2}. The distance between consecutive $\sqrt{3}\times\sqrt{3}$ surfaces was measured to be $\sim$3~\AA. Interestingly, they have found that the multilayer silicene was metallic. 

Although, the $\sqrt{3}\times\sqrt{3}$ reconstruction with shrunk lattice constant and honeycomb STM appearance was observed in many experiments its atomic structure was still a mystery. In 2014, Cahangirov \textit{et al.} proposed a model that could explain these experimental observations \citep{Cahangirov4}. They have shown that formation of the 3$\times$3 structure on Ag(111) substrate was energetically favorable, but sending more Si atoms on top of this already formed 3$\times$3 structure would transform it to a $\sqrt{3}\times\sqrt{3}$ structure through formation of dumbbell units. The dumbbell units were formed spontaneously whereby a new-coming Si atom would push one of the silicene atoms down and make the same connections itself on the top \citep{Kaltsas1,Ozcelik1,Marutheeswaran1}. It is energetically favorable for dumbbell units to arrange themselves in a $\sqrt{3}\times\sqrt{3}$ honeycomb lattice and to shrink the lattice constant down to 0.64~nm which is exactly what was measured in experiments. Later, it was shown that upon further deposition of Si atoms new layers with $\sqrt{3}\times\sqrt{3}$ reconstruction would form ultimately leading to a bulk structure that they named layered dumbbell silicite \citep{Cahangirov5}.

After many experimental advances in silicene research, a new goal was set: to synthesize germanene and stanene. However, it was not possible to synthesize germanene onto Ag(111) because Ge tends to make an ordered Ag$_2$Ge surface alloy. Instead, D\' avila \textit{et al.} deposited germanium on top of the Au(111) surface \citep{Davila1}. Although they observed several coexisting phases, a $\sqrt{3}\times\sqrt{3}$ germanene cell matched with a $\sqrt{7}\times\sqrt{7}$ Au(111) one was the best fitting structure to their experimental data and DFT calculations. More work needs to be done to fully characterize the growth and structure of germanene on Au(111) surface. Yet, the progress is very rapid and just at the beginning of 2015, Derivaz \textit{et al.} managed to synthesize at only 80$\degree$C a unique, 2$\times$2 reconstructed, germanene phase on the aluminum (111) surface \citep{Derivaz1}. Much recently, stanene was synthesized by molecular beam epitaxy on Bi$_2$Te$_3$(111) substrate \citep{Zhu1}. The atomic and electronic characterization of stanene was made by STM and ARPES measurements that showed good agreement with DFT calculations.

In 2015, while this book was still in development, the first silicene field-effect transistor was fabricated and was shown to operate at room temperature \citep{Tao1}. To achieve this, they have first synthesized silicene on top of a thin silver film on mica and covered it in situ by a thin alumina layer. Then they delaminated the silver film, flipped the whole structure upside down, and etched some part of the silver until silicene was reached. This resulted in a silicene strip with two Ag pads on top that acted as metal electrodes. This study attracted huge interest and was praised as a ``proof of principle'' for usage of silicene in electronic devices \citep{Peplow1,Lelay4}. Clearly, on the occasion of the 50$^{th}$ anniversary of Moore's law, it is a hallmark in the quest for 2D materials for electronic applications \citep{Schwierz1}.
\chapter{Freestanding Silicene} \label{Freestanding}

\abstract*{Each chapter should be preceded by an abstract (10--15 lines long) that summarizes the content.}

Obtaining a freestanding 2D graphene flake is relatively easy because it has a naturally occurring 3D layered parent material, graphite, made up of graphene layers weakly bound to each other by van der Waals interaction. In fact, graphite is energetically more favorable than diamond (one the most stable and hard materials on Earth) that is the $sp^3$ hybridized allotrope of carbon. To prepare freestanding graphene, it is enough to come up with a smart procedure for isolating the weakly bound layers of graphite. The same is also true for other layered materials like hexagonal boron nitride, black phosphorus, metal dichalcogenides and oxides. Silicene, on the other hand, doesn't have a naturally occurring 3D parent material since silicon atoms prefer $sp^3$ hybridization over $sp^2$ hybridization. This makes the synthesis of freestanding silicene very hard, if not impossible. However, it is possible to epitaxially grow silicene on metal substrates and make use of its intrinsic properties by transferring it to an insulating substrate \citep{Tao1}. In this Chapter, we focus on intrinsic properties of freestanding silicene in the absence of the metallic substrate.

\section{Reconstructions of bulk silicon surfaces}

Synthesis of 2D crystal structures can be achieved through various experimental techniques such as chemical vapor deposition, micromechanical cleavage, liquid exfoliation and dry exfoliation. Whether it is grown on a surface or cleaved from bulk, the atomic structure of a natural 2D material resembles one of its crystallographic surfaces and planes. Therefore, to apprehend the formation and unique features of silicene, understanding the properties of various surfaces of bulk silicon is essential.

A surface can be described as the truncated form of the bulk material. In case of a crystal, the surface region is defined as the few outermost layers of the material. For covalently bonded crystals, when the truncation takes place, the presence of dangling bonds at the surface, that contain less than two spin-paired electrons, is inevitable at the surface. Due to the presence of such dangling bonds the structural, electronic and magnetic properties at the surface may become significantly different from those of the bulk material.

\begin{figure}[t]
\includegraphics[width=11.5cm]{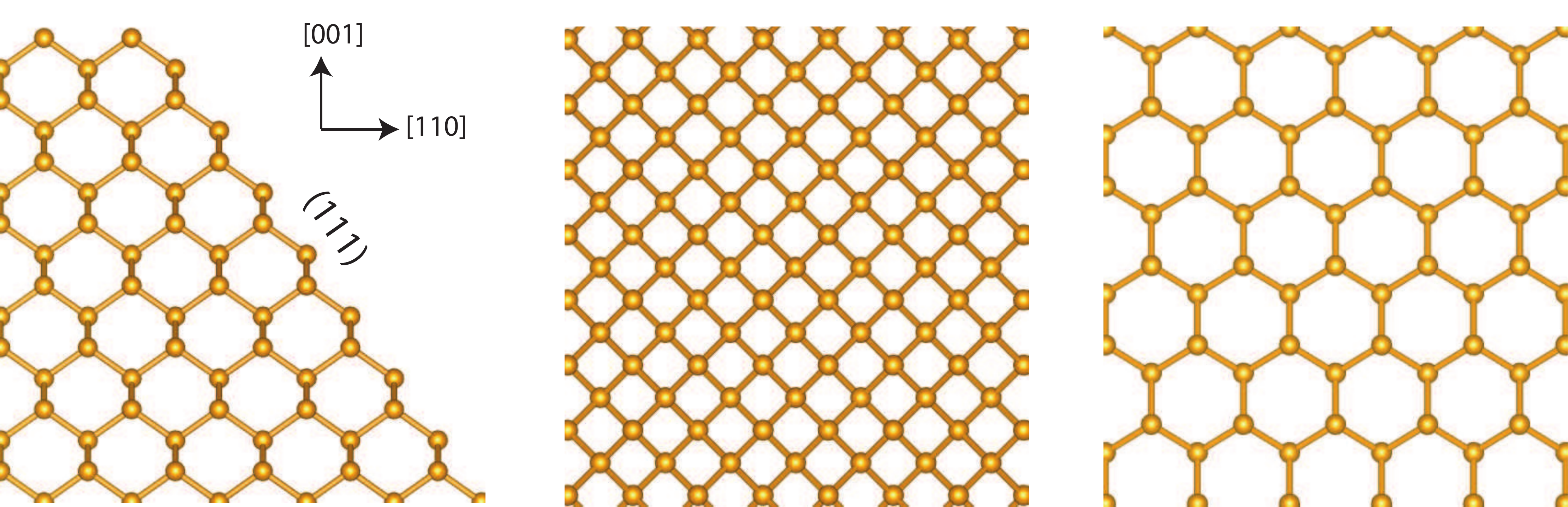}
\caption{\label{hs1} (Left panel) View of unreconstructed surfaces of silicon 
through the (1\={1}0) plane. Atomic arrangements at (110) and (111) surfaces are 
shown in the mid and right panels, respectively.}
\end{figure}

The space lattice of bulk silicon, which has a diamond structure, is face-centered cubic. In this structural arrangement, where each Si atom is tetrahedrally bonded to neighboring atoms, some crystallographic planes display quite different structural properties. Noteworthy, as seen in Fig.~\ref{hs1}, especially the buckled hexagonal symmetry of the Si(111) surface exactly resembles the monolayer honeycomb lattice structure of silicene. The formation of a Si(111) surface is energetically the most favorable one. Since the Si(111) surface has the lowest density of dangling bonds, it has the lowest surface energy. Therefore, since formation of such a surface is favored over the other types of surfaces, the natural cleavage plane and growth direction of bulk Si can be expected to be [111] direction. 

When a surface is formed, firstly surface atoms tend to change their atomic position inward or outward slightly a phenomenon called \textit{surface relaxation}.  Furthermore, the presence of dangling bonds at the surface results in an unstable atomic order and these atoms tend to change their position to find the lowest energy configuration. Such a structural transformation driven by the dangling hybrids is called as \textit{surface reconstruction}. As proposed by Jahn and Teller, if there is a degeneracy in the ground state of a molecular structure its atoms will re-arrange toward a lower symmetry \citep{jt}. Such a structural-degeneracy-driven deformation is called a Jahn-Teller distortion. Similarly, upon the relaxation and reconstruction mechanisms, surface structure spontaneously releases its surface energy (or surface tension) by reducing the number of dangling bonds. 

\begin{figure}[t]
\includegraphics[width=11.5cm]{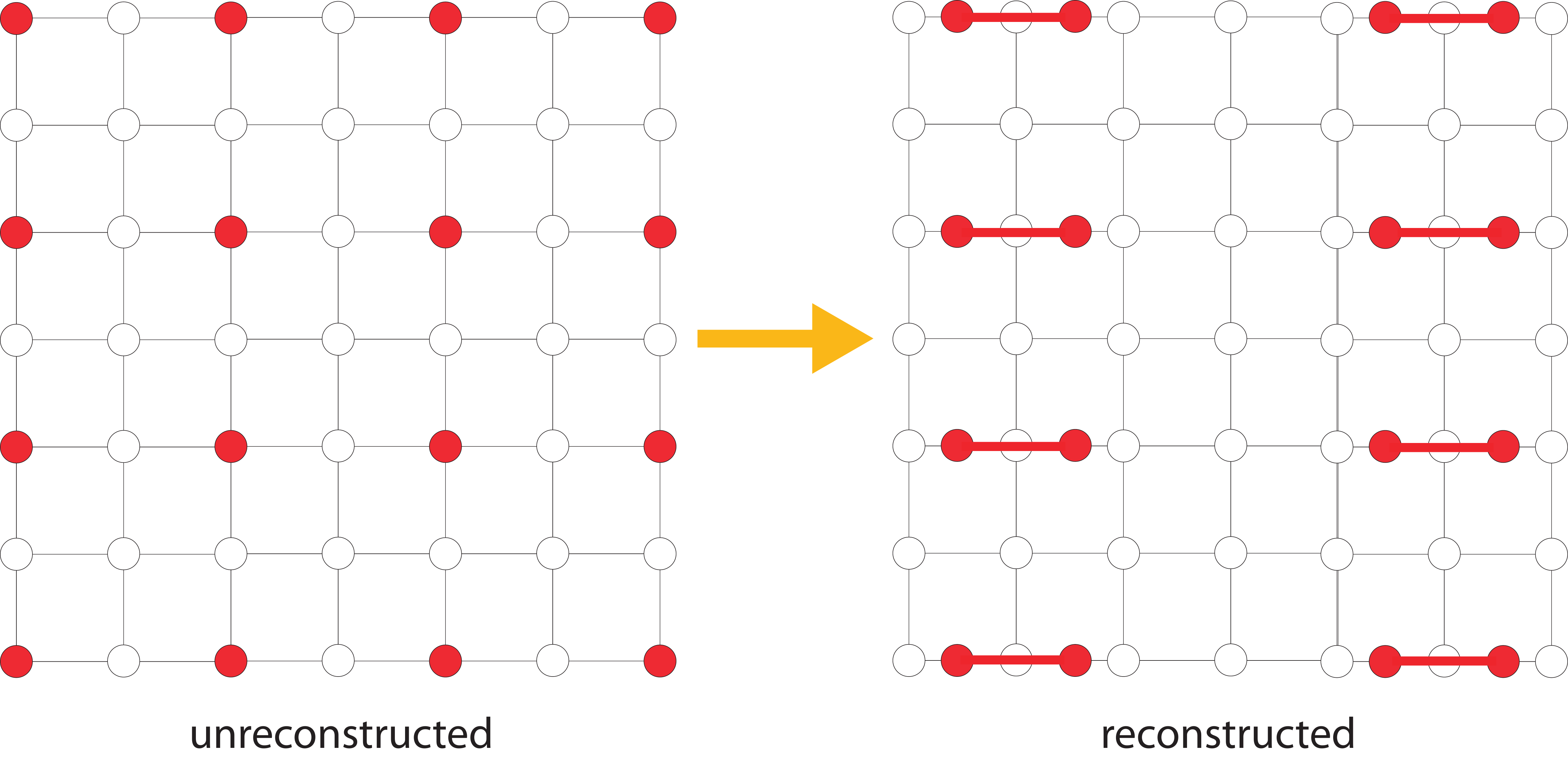}
\includegraphics[width=11.5cm]{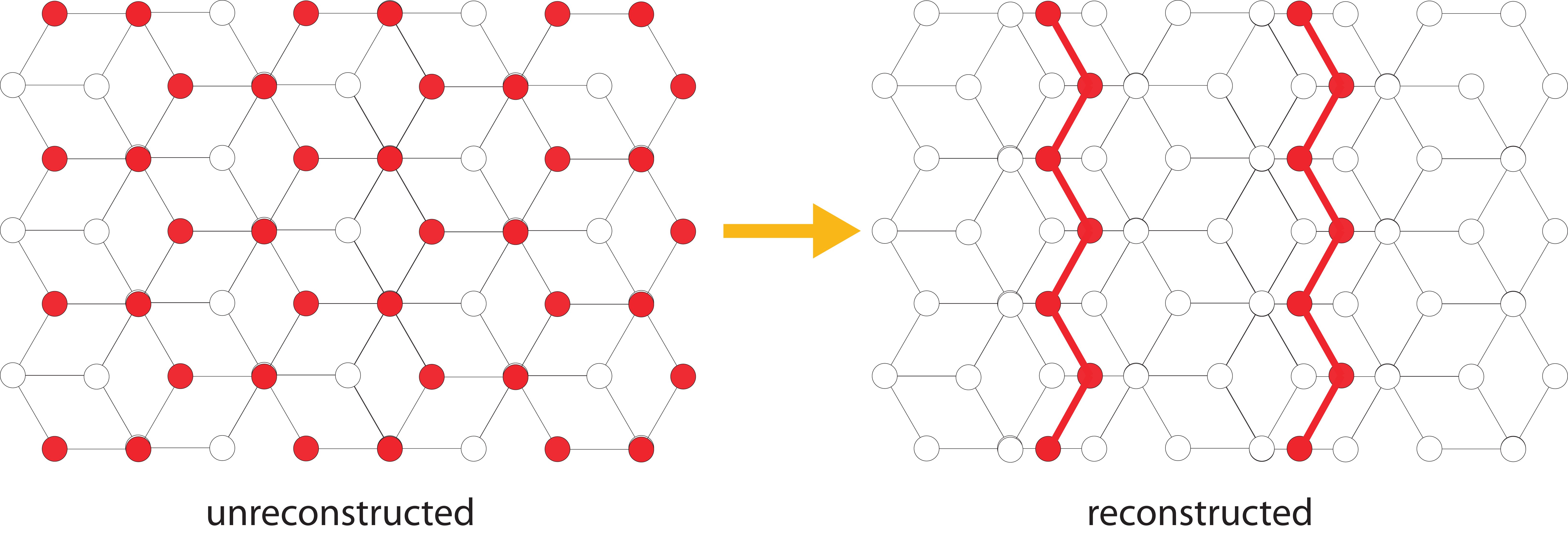}
\caption{\label{2x1} Top panels: top views for the Si(100) unreconstructed 
and (2$\times$1) reconstructed surfaces. Uppermost surface Si atoms are colored red. 
Lower panels: top views for the unreconstructed Si(111) surface and its (2$\times$1) reconstruction. Atoms at uppermost two layers are represented by red balls.}
\end{figure}

Since silicon devices are mostly grown on Si(100) substrates, revisiting its characteristic properties is of importance for silicene studies. At the (100) silicon plane each atom is bonded to two underlying atoms and therefore, one can expect creation of two dangling hybrids upon the surface creation. Studies have shown that while the resulting surface has a 2$\times$1 periodicity at room temperature (see Fig.~\ref{2x1}), thanks to the diversity of arrangements of dangling hybrids the formation of this surface may take place in different ways \citep{ph,po,kr}.

The Si(111) surface also form a 2$\times$1 reconstruction upon in situ cleavage under ultra-high vacuum \citep{land}. In this configuration $\pi$-bonded chains of locally $sp^{2}$-bonded atoms create one-dimensional Si chains at the surface, as initially proposed by Pandley \citep{Pandley1}. Since Si(111)-(2$\times$1) structure is irreversibly transformed into a 7$\times$7 reconstruction at $\sim$~700 K, followed by a reversible  7$\times$7~$\leftrightarrow$~1$\times$1 transition to a 1$\times$1 structure at $\sim$~1100 K, it can be considered as a metastable phase of the surface. 

The 7$\times$7 reconstruction of the Si (111) surface, composed of 49 Si(111) primitive cells, is much more complex. Although it was first observed very early in a Low Energy Electron Diffraction (LEED) experiment \citep{sch59} due to the insufficiency of experimental characterization tools the atomic structure of the surface was an intriguing open question for a long term. The determination of the complicated surface topology of the Si(111)-(7$\times$7) surface was performed only in 1985 by Takayanagi et al., surprisingly by transmission electron diffraction using thinned Si(111) sample in an UHV microscope \citep{Takayanagi1}. 

\begin{figure}[t]
\includegraphics[width=11.5cm]{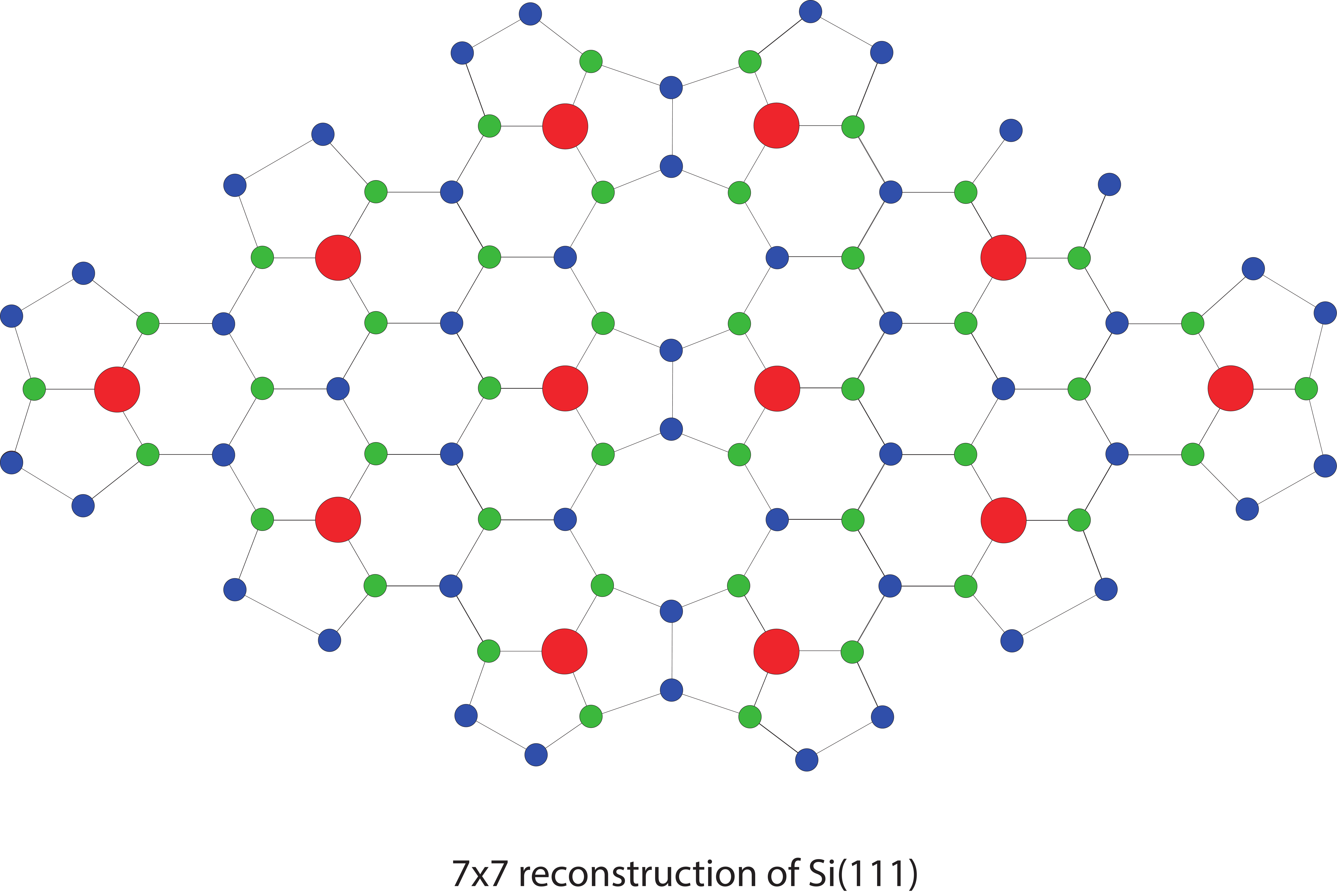}
\caption{\label{7x7} Top view for the (7$\times$7) reconstruction of the Si(111) 
surface. Adatoms, first-layer-atoms and second layer atoms are shown by red, green 
and blue balls, respectively.}
\end{figure}

\begin{figure}[t]
\includegraphics[width=11.5cm]{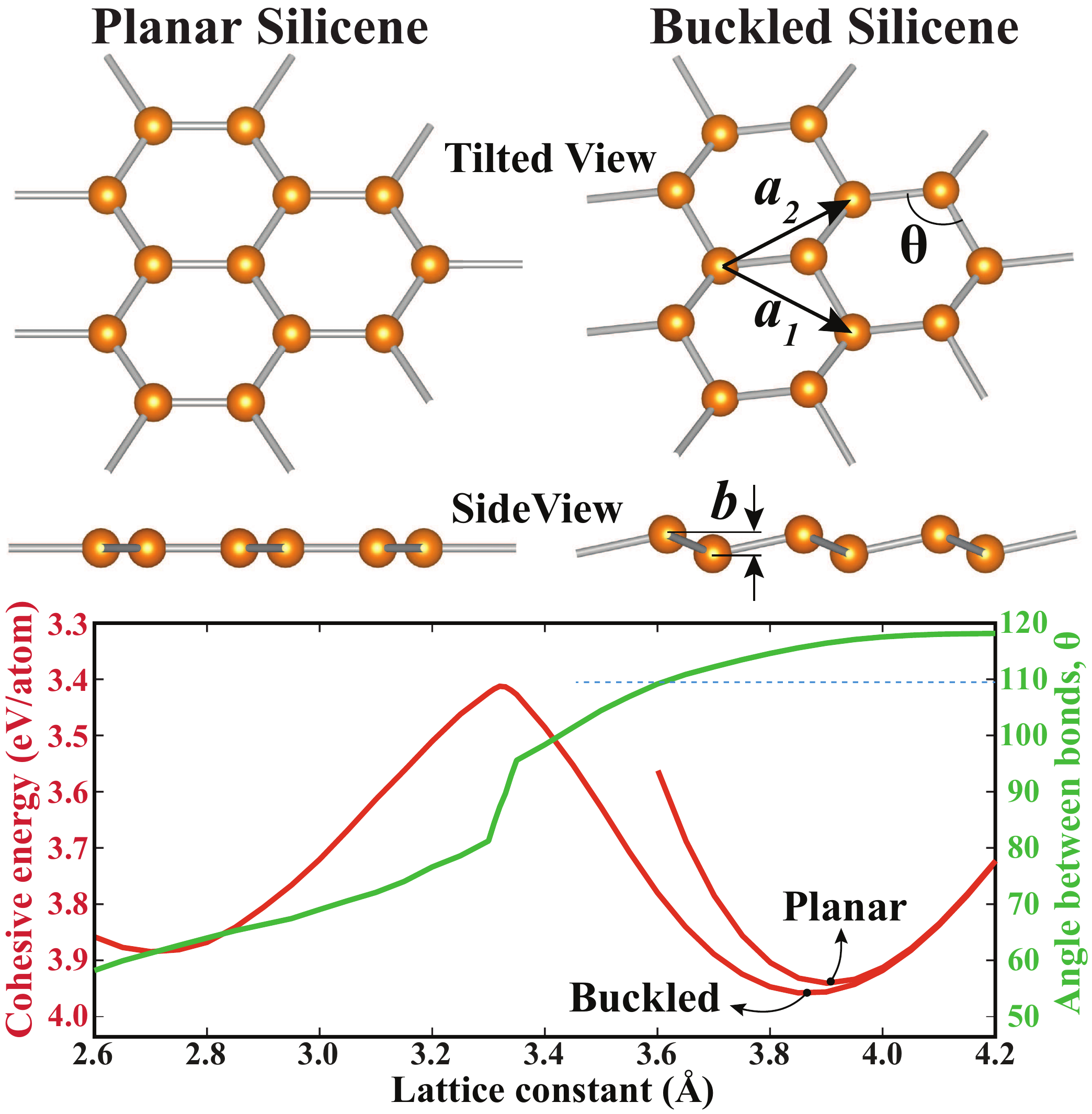}
\caption{The top panel shows ball and stick models of planar and buckled silicene from tilted and side views. The bottom panel shows the variations of the cohesive energy (red solid line) and bond angle (green solid line) with respect to the lattice constant. Red and blue curves correspond to planar and buckled silicene, respectively. The thin dashed red line represents the ideal bond angle of $sp^3$ hybridization.}
\label{relaxation}
\end{figure}

As shown in Fig.~\ref{7x7}, the 7$\times$7 reconstruction takes place, according to the so-called \textit{dimer adatom stacking-fault model}, by (i) formation of hole sites at each cell corner, (ii) formation of pentagons, which are linked by the dimers at the second layer, in the middle of the 7$\times$7 unitcell and (iii) adsorption of 12 adatoms on this surface. Upon this reconstruction, the number of surface dangling bonds is reduced from 49 to 19 (12 for the adatoms, six for the so-called rest atoms and one for the atom in the center of the corner hole) per unitcell. Therefore, the energy of the unreconstructed Si(111) surface is significantly lowered through the 7$\times$7 reconstruction.

These early studies clearly show that silicon, besides being the key material in semiconductor device technology possesses easy-to-reconstruct complex surfaces, which offer a rich playground for variety of electronic and physical properties.

\section{Atomic structure and stability of freestanding silicene} \label{atomic}

If one searches for the silicon counterpart of graphene, a natural start would be to replace C atoms with Si and find the optimum lattice constant. A standard density functional theory (DFT) calculation \citep{Kresse1} performed using the projector-augmented wave (PAW) method \citep{Blochl1} and the PBE exchange-correlation functional \citep{Perdew1} results in a planar silicene with a lattice constant of 3.90~\AA. However, this optimum value is found only because one puts a constraint by confining Si atoms in a single plane. Once this constraint is removed, silicene goes to a more energetically favorable state that has a lattice constant of 3.87~\AA~and a slight buckling of 0.45~\AA~(see Fig.~\ref{relaxation}). There are two main reasons why this buckling occurs in silicene and not in graphene. First, Si atoms prefer $sp^3$ hybridization over $sp^2$ which pulls bond angles from 120$\degree$ towards 109.5$\degree$. Second, the distance between Si atoms in silicene becomes rather large to maintain a strong $\pi$-bonding. Despite this, the hybridization of silicene is closer to $sp^2$ than to $sp^3$. This can be quantified by making use of the fact that in the ideal cases the degree of hybridization, that is $D$ of $sp^D$, is correlated with the ideal bond angles $\theta$ in such a way that $D=-1/cos(\theta)$. This relation holds true for carbon allotropes like atomic chains, graphene and diamond that have $sp$, $sp^2$ and $sp^3$ hybridization with bond angles 180$\degree$, 120$\degree$ and 109.5$\degree$, respectively. In the case of silicene, the bond angle is $\theta$=116.2$\degree$ which gives $D \sim 2.27$.

In Fig.~\ref{relaxation} we present the variations of the cohesive energy and bond angle of silicene as the lattice constant is changed. The energy minima for planar and slightly buckled silicene are clearly seen. As the lattice constant is squeezed by $\sim$0.5~\AA, the system enters to a new regime. This is accompanied with a sudden change in the bond angle. A new energy minimum is found when the lattice constant is squeezed even further. However, the geometry of this new energy minimum is quite different from that of silicene. The bond angles almost approach 60$\degree$; as a consequence, the second neighbor Si atom is almost as close as the first one which means that each Si atom has nine neighbors. It was shown that this unlikely structure is unstable \citep{Cahangirov1}.

The buckled structure of silicene was first proposed by Takeda and Shiraishi in 1994 while the name silicene was coined by Guzm\' an-Verri and Lew Yan Voon in a 2007 publication that investigated the planar structure \citep{Takeda1,Guzman1}. However, silicene remained ignored until it was shown that the buckled structure was indeed thermodynamically stable \citep{Cahangirov1}. This was achieved by calculation of vibrational modes for both planar and buckled silicene. While for planar silicene the phonon dispersion had imaginary frequencies in the buckled case all modes were positive over the whole Brillouin zone, indicating that there was a restoring force for all possible atomic displacements.

\begin{figure}[t]
\includegraphics[width=11.5cm]{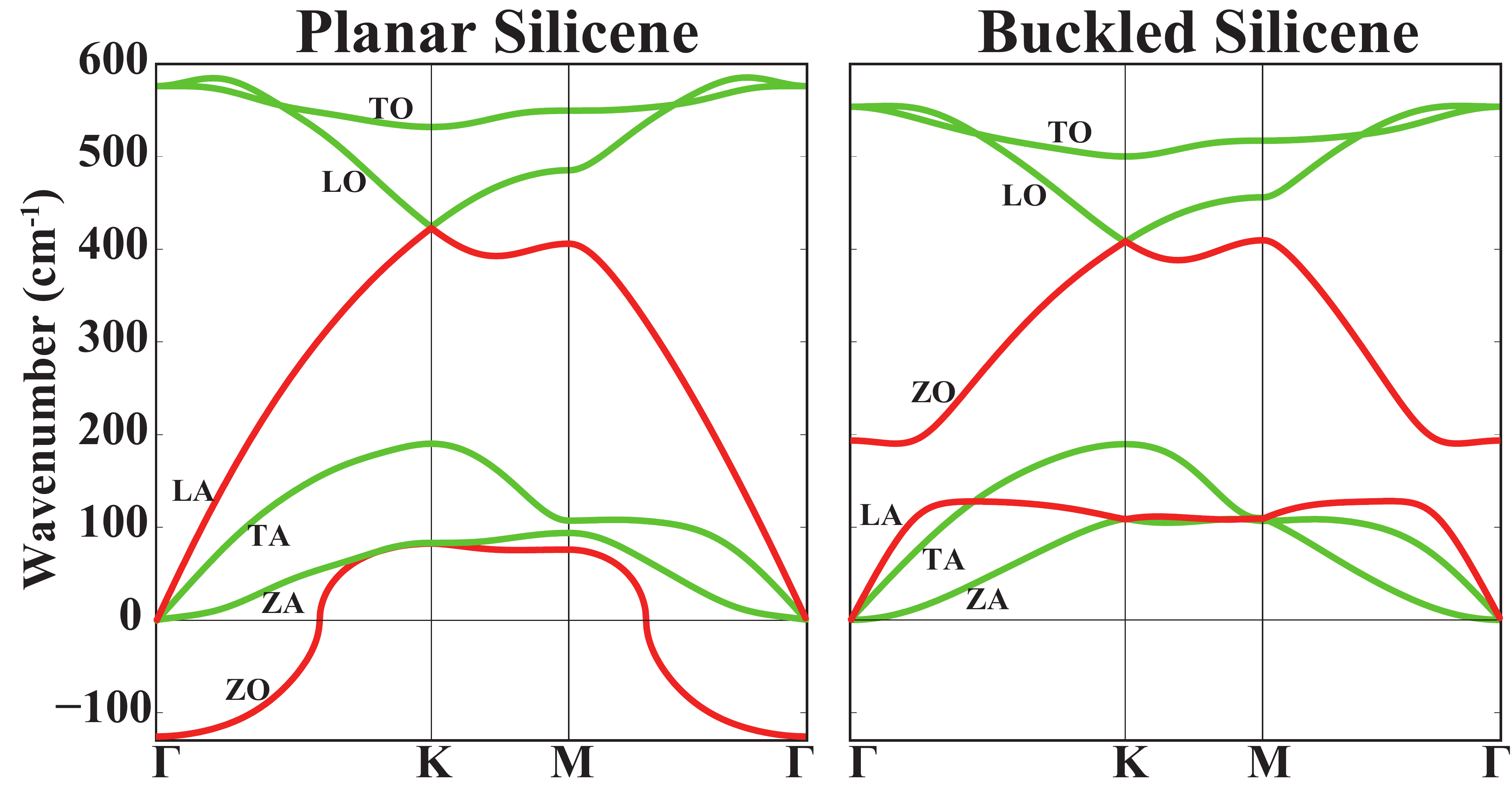}
\caption{Phonons of planar and buckled silicene. The out of plane, transverse and longitudinal acoustic and optical modes are denoted by ZA, TA, LA, ZO, TO and LO, respectively. The LA and ZO modes change significantly upon structural transformation. These modes are shown by red lines to call attention to this change.}
\label{phonons}
\end{figure}

The phonon dispersions for planar and buckled silicene are shown in Fig.~\ref{phonons}. The out of plane optical mode (ZO) of planar silicene has imaginary values at the $\Gamma$ point indicating instability. This mode corresponds to displacement of Si atoms in one sublattice up and atoms in the other sublattice down along the axis passing perpendicular to planar silicene. It means that if one tries to push atoms of planar silicene in this particular way, there will be no restoring force against it. When we go from planar to buckled silicene the LA and ZO modes change significantly while others remain almost the same. Upon buckling, the imaginary part of the ZO mode in planar silicene moves up to 200 cm$^{-1}$ replacing the portion of the LA mode that was connected to the LO mode. Meanwhile, the positive portion of the ZO mode in planar silicene becomes a part of the new LA mode.

Note that the stability analysis presented here ignores many experimentally relevant effects like thermal fluctuations and reactivity. Suspended graphene sheets were shown to have ripples caused by thermal fluctuations that reach 1~nm in height \citep{Meyer1}. A similar behavior is expected for silicene. However, due to its high reactivity, it is almost impossible to synthesize silicene in its freestanding form \citep{Brumfiel1}. Instead, growth of silicene requires the presence of a substrate that can stabilize its two-dimensional structure that is not the ground state of silicon. The sticky nature of silicene is also the reason why the layered allotrope of silicon, like graphite, doesn't exist in nature. All this is excellently expressed in the ``silicene interlude'' written by the 1981 Chemistry Nobel Prize Laureate Roald Hoffmann. He explains that in silicon chemistry, single $\sigma$-bonds are favored over double $\pi$-bonds and the latter need to be protected sterically in order to be isolated \citep{Hoffmann1}. For this reason, freestanding silicene will latch to any molecular dirt in its environment. Fully hydrogenated silicene or silicane, on the other hand, should be possible to synthesize in the freestanding form.

\section{Electronic structure of freestanding silicene} \label{electronic}

\begin{figure}[t]
\includegraphics[width=11.5cm]{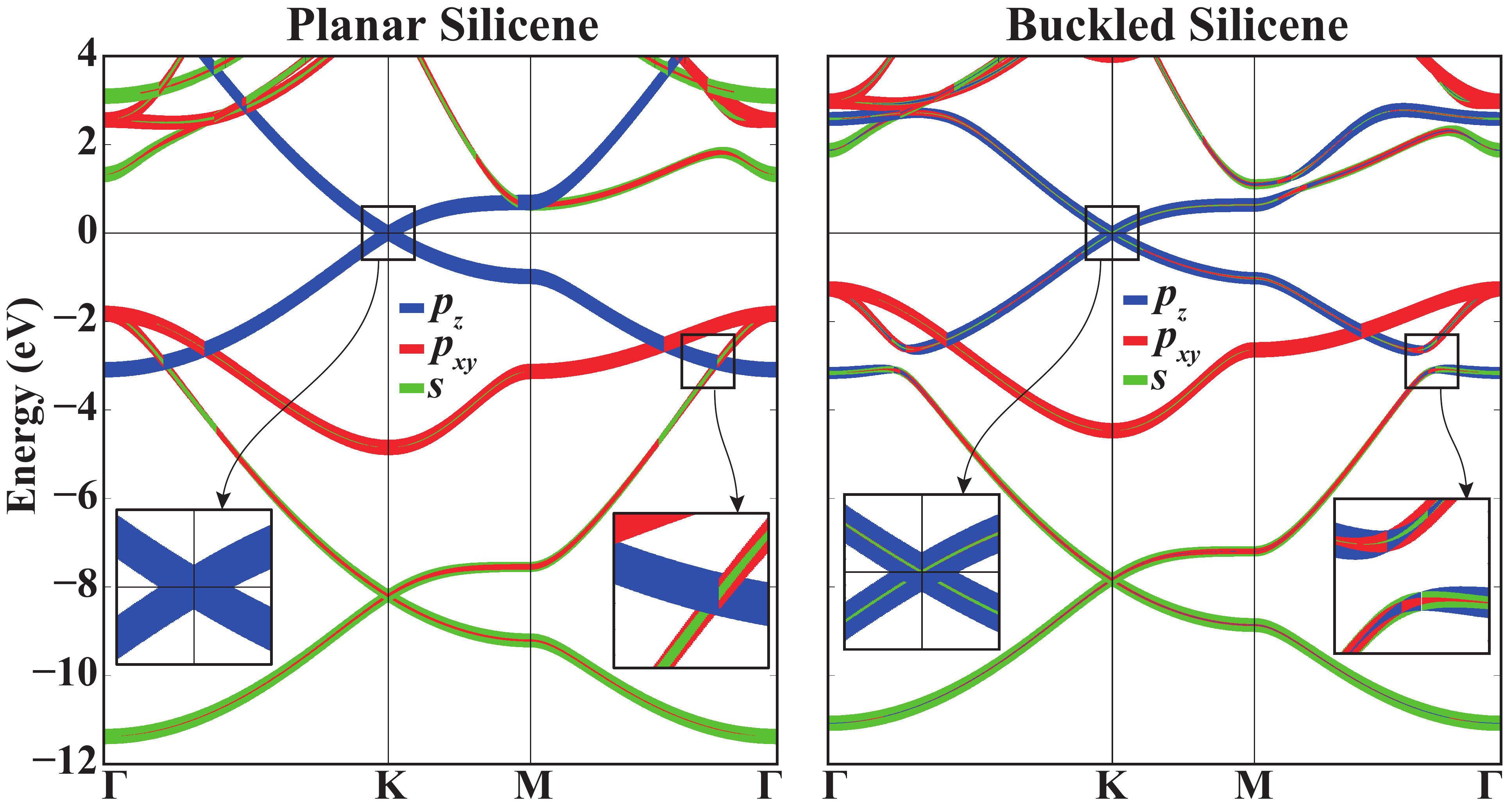}
\caption{The electronic structures of planar and buckled silicene. The orbital character of each state is shown by colors. The width of each line is proportional to the contribution from the orbital denoted by the color (blue, red, and green for $p_z$, $p_{xy}$ and $s$ orbitals, respectively) of that line. The orbital with lower contribution is plotted on top of the one that has higher contribution, which makes all contributions visible. Some parts of the bands are shown in more detail in the insets.}
\label{bands}
\end{figure}

Although freestanding silicene is buckled, the hybridization in Si atoms remains closer to $sp^2$ rather than becoming $sp^3$. In Fig.~\ref{bands} we compare the electronic structure of planar and buckled silicene. When silicene is planar, the $p_z$ orbitals do not mix with others. This is clearly seen in the linearly crossing bands at the K point that are solely composed of $p_z$ orbitals. However, upon buckling, these states get a minor but noticeable contribution from both $s$ and $p_{xy}$ orbitals. Similarly, $p_z$ bands cross $s$ and $p_{xy}$ bands near the $\Gamma$ point in planar silicene while in buckled silicene this crossing splits and bands rehybridize. The electrons of buckled silicene behave as massless Dirac fermions due to the linearly crossing bands at the Fermi level. The Fermi velocity of these bands is around 0.53$\times$10$^6$ m/s when the LDA or PBE functionals are employed and increases to 0.68$\times$10$^6$ m/s when a hybrid functional like HSE06 is used \citep{Drummond1}. This velocity is close to that of freestanding graphene which is experimentally measured to be around 1$\times$10$^6$ m/s \citep{Novoselov2}.

Furthermore, silicene also has properties that are not found in graphene. Due to the buckled structure it is possible to tune the electronic properties of silicene by applying an electric field in the perpendicular direction \citep{Drummond1}. The slope of the band gap with respect to the applied perpendicular electric field was found to be $\sim$~0.07~e\AA. This value is eight times higher when it is estimated using first-order perturbation theory that does not include screening which suggests that silicene has strong sublattice polarizability. Owing to the larger spin-orbit coupling in Si compared to C, silicene has topologically non-trivial states with a spin-orbit gap of 1.5 meV \citep{Liu1,Drummond1}. The topological insulator state of silicene protects a gapless spectrum of edge states. When the perpendicular electric field reaches the critical value of 20 mV/\AA, silicene makes transition from the topological to a band insulator state \citep{Drummond1}. These states are more pronounced in germanene and stanene as will be discussed in Chapter~\ref{Germanene}. Depending on the applied exchange field and perpendicular electric field, a broad range of phases emerge in silicene that include a quantum anomalous Hall insulator, a valley polarized metal, a marginal valley polarized metal, a quantum spin Hall insulator and a band insulator \citep{Ezawa1}. 

\section{Hydrogenation: Silicane and Germanane}

Silicene not only provides an alternative monolayer to graphene but also prepares a ground for derivation of novel silicon-based functionalized materials. Rapidly growing experimental and theoretical research on graphene derivatives have shown that production of high quality functionalized crystals, each with different electronic and magnetic properties, can be realized in many different ways (such as hydrogenation, fluorination, chlorination, oxygenation, defect patterning, strain application and adatom/molecule decoration). Thanks to the advances in functionalization techniques, novel silicene derivatives, which are promising to be integrated into a widespread collection of nanoscale optoelectronic devices, have emerged. Here we review how silicene can be functionalized and what are the characteristic properties of silicene-based functionalized materials.

Hydrogenation process is a chemical reaction of hydrogen molecule with a surface, molecule or an element. While most of the hydrogenation processes need presence of a metallic catalyst, non-catalytic hydrogenation can only take place at high temperatures. Alkenes are hydrocarbons with at least one double bond.  However, reaction of the carbon-carbon double bond in an Alkene with hydrogen leads to formation of an Alkane. Alkanes are hydrocarbons with only single bonds between the carbon atoms. Following the terminology of chemistry one can name the conversion from graphene to fully hydrogenated graphene as conversion from graphene to graphane. As firstly predicted by Sofo \textit{et al.},  Graphane, a $sp^3$ hybridized hexagonal network of graphene hydrogenated on both sides of the plane in an alternating manner, is stable and its binding energy is comparable to other hydrocarbons such as benzene, cyclohexane, and polyethylene \citep{Sofo1}. Graphane, with formula CH, has been reported as an insulator with nonmagnetic ground state.  Following experimental efforts by Elias \textit{et al.} graphane was synthesized by cold hydrogen plasma treatment of graphene \citep{Elias1}. The characteristic insulating nature of graphane (with about 5~eV bandgap) and its possible use in nanodevice technologies have been well-documented in a review \citep{sah15}.

It was shown that, the optical properties of graphane are dominated by the localized charge-transfer excitations governed by the enhanced electron correlations in a 2D dielectric medium \citep{Cudazzo1}. The strong electron-hole interaction in graphane leads to the appearance of small radius bound excitons which open the path towards the possibility of excitonic Bose-Einstein condensation that might be observed experimentally. This can also be true for silicon and germanium counterparts of graphane discussed below.

\begin{figure}[t]
\includegraphics[width=11.5cm]{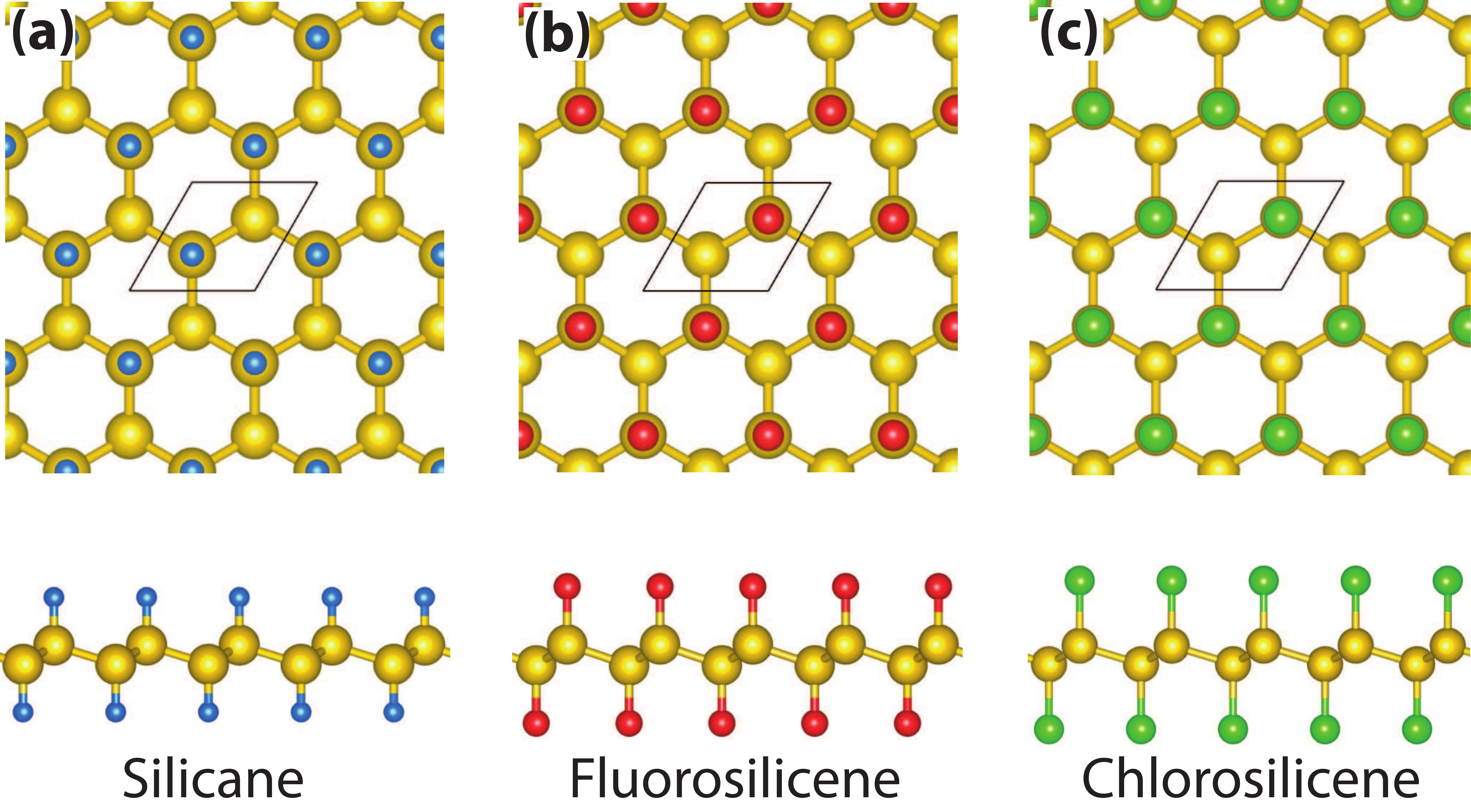}
\caption{\label{der1} Top and side views of the atomic structures of fully (a) hydrogenated, (b) fluorinated and (c) chlorinated silicene. Hydrogen, fluorine and chlorine atoms are represented by blue, red and green colors, respectively.}
\end{figure}

It is known that similar to graphene structural analogue of benzene can also be formed by silicon atoms \citep{Barton1}. However, differing from benzene rings silicon rings are stabilized in a chair-shaped atomic structure \citep{Abersfelder1}. Therefore, the buckled geometry of monolayer silicene provides an ideal ground for hydrogen treatment of its surfaces. By adopting the terminology used for graphene, a one-by-one hydrogenation process turns silicene into silicane. In order to avoid the confusion between the names 'graphene' and 'graphane', generally graphane's name is written as "graphAne" in the literature. Similarly, one can make use of the same definition to name silicAne and germanAne, the fully-hydrogenated derivatives of silicene and germanene.

\begin{figure}[t]
\includegraphics[width=11.5cm]{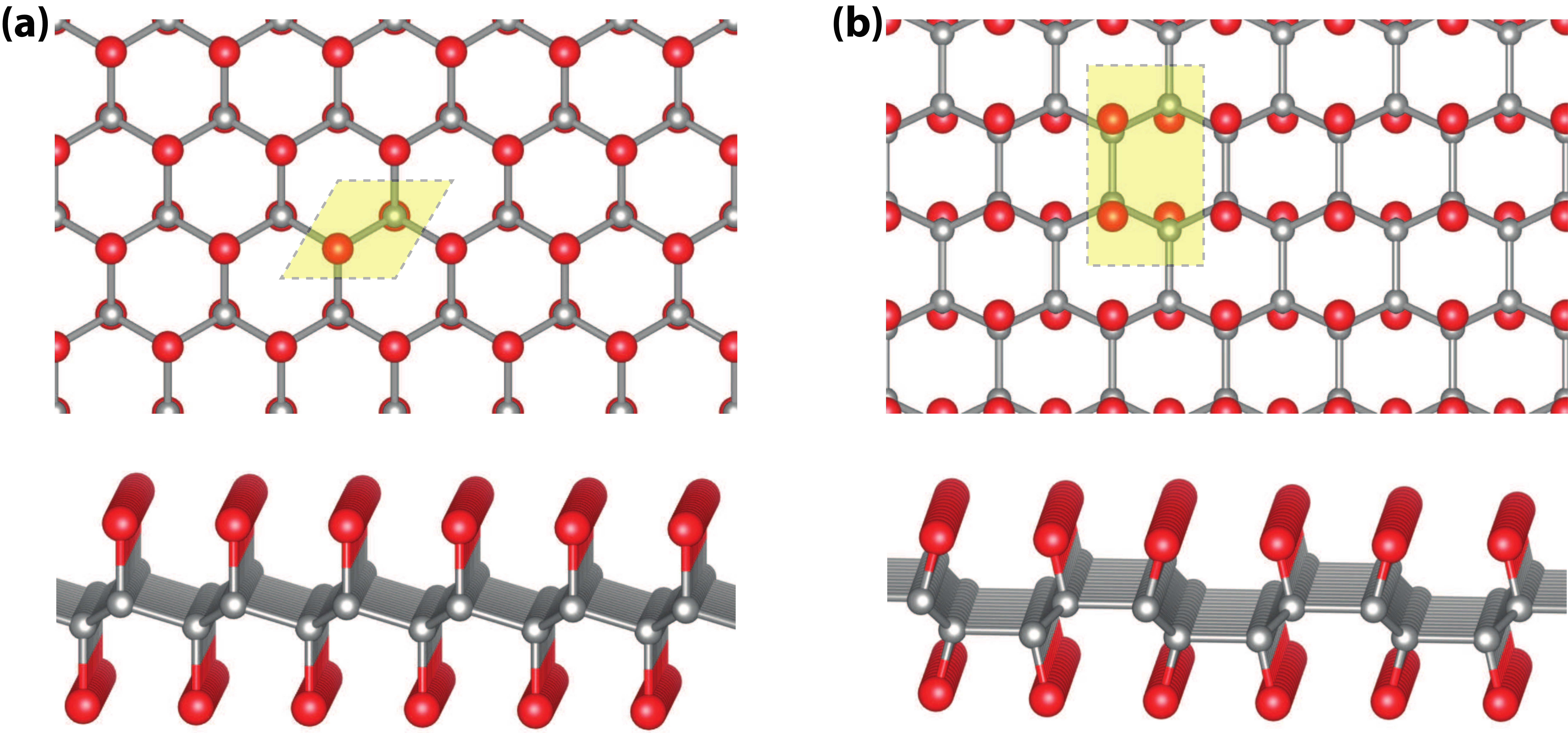}
\caption{\label{der2} Atomic structures of (a) chair and (b) boat conformers of fully hydrogenated silicene. Primitive unitcells of the structures are highlighted by yellow color.}
\end{figure}

As shown in Fig. \ref{der2}(a), silicane which is analogous to graphane has a buckled honeycomb structure with a single hydrogen atom attached to each Si site on either sides of the 2D crystal. Voon \textit{et al.}, for the first time, studied the structural and electronic  properties of hydrogenated silicene (and germanene counterparts) using the ab-initio DFT methodology \citep{Voo10}. It was reported that the buckling of silicene increases from 0.44 to 0.72~\AA~  upon full hydrogenation. Their structural analysis on silicane and germanane revealed that the chair configuration is energetically favored over the boat structure. Voon \textit{et al.} also reported that full hydrogenation opens a large bandgap in the silicene's electronic structure, typically, in the 3-4~eV range for silicane (with LDA, the approximated value is only 2~eV). In their analysis, where the spin-orbit interaction was ignored, it was shown that semimetallic silicene (germanene) turns into an indirect (a direct) bandgap semiconductor. Tight-binding parameters for silicane and germanane were also derived \citep{Zolyomi1}.

\begin{figure}[t]
\includegraphics[width=11.5cm]{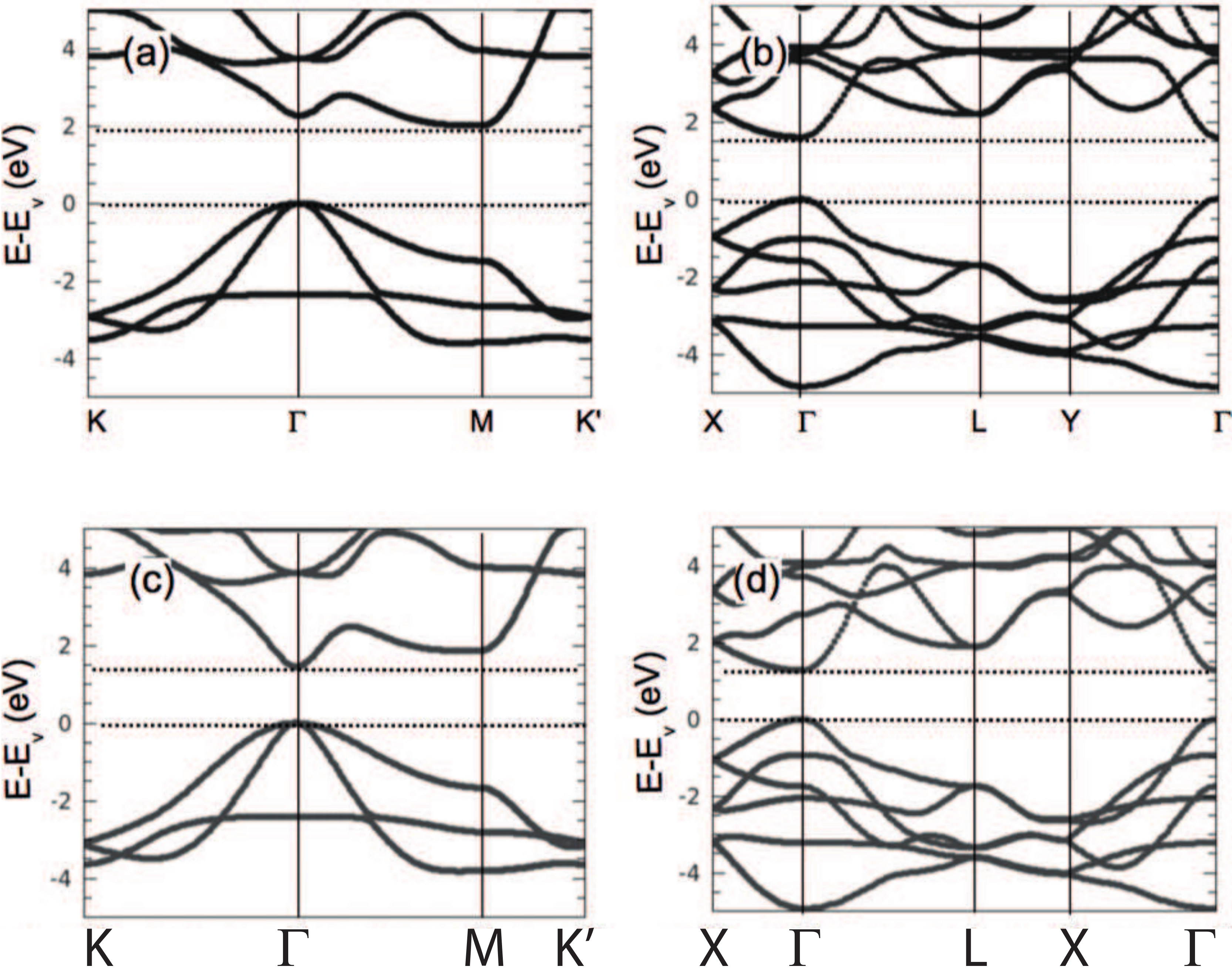}
\caption{\label{der3} Energy band structures, calculated using the LDA functional, for (a) chairlike silicane, (b) boatlike silicane, (c) chairlike germanane, and (d) boatlike germanane. The reference zero energy level corresponds to the top of the valence band. Adapted from \citep{Houssa1}.}
\end{figure}

Similarly, Houssa \textit{et al.} performed first principles calculations based on DFT, within LDA, HSE and GW, to investigate the electronic properties of silicane and germanane. They predicted that two different configurations (chairlike and boatlike) are equally stable, and, depending on the synthesis procedure, that both configurations can be formed. While the chair configuration of silicane is an indirect semiconductor, the boat structure is found to be a direct bandgap material in all approaches (LDA, HSE and GW). It was also shown that chairlike germanane is a direct gap semiconductor with an energy bandgap of 3.5 eV, which makes it useful for optical applications in the blue/violet range \citep{Houssa1}.

Following these preliminary studies many groups have shown that silicane forms a chairlike structure in its ground state. Although, the presence of an indirect electronic bandgap is one of the main disadvantage of silicane, its employ in optoelectronic device applications still can be realized by making use of hydrogenated bilayer silicene. As shown by Huang \textit{et al.}, in contrast with single layer silicene and silicane, hydrogenated bilayer silicene is optically active and a promising material for the optical applications \citep{Hua14}.  It was also concluded that depending on hydrogen concentrations double-sided SiH$_x$ structures can have direct (or quasidirect) band gaps within the range of RGB colors, which shows that these double-sided SiH$_x$ structures can be used as light emitters for white LEDs. 

The substrate plays an important role in determining the structural and electronic properties of silicene-based materials (see Chapter~\ref{Substrate}). In a theoretical study which takes into account the effect of the substrate it was shown that the coverage and arrangement of the absorbed hydrogen atoms on silicene changed significantly the electronic properties, such as the direct/indirect band gaps or metallic/semiconducting features \citep{Zha12}. They also reported that half-hydrogenated silicene with chair-like structure had ferromagnetic character.

One of the main drawback of silicene and silicane studies is that synthesis of their free standing monolayer crystals have not been achieved so far. However, Qiu \textit{et al.} demonstrated the possibility of ordered and reversible hydrogenation of silicene on an Ag(111) substrate. In their study, by combining ab initio DFT calculations and scanning tunneling microscopy measurements \citep{Qiu1}, it was found that  hydrogenation of silicene at room temperature results in formation of perfectly ordered $\gamma$-(3$\times$3) superstructure. Moreover, when the reconstructed hydrogenated silicene sheet is heated up to 450 K, dehydrogenation takes place and the surface restores to its initial form which is pristine silicene layer. Such a reversible reaction of silicene sheets with hydrogen allows tunable crossover between semiconducting and metallic phases.

The in-plane thermal conductivity of graphene, as a consequence of quantum confinement in 2D, is the highest among well-known materials. Therefore, one can expect high thermal transport and specific heat also in silicene and related materials. Classical molecular dynamics calculations of graphene supported silicene revealed that the thermal transport behavior of such hetero-interfaces can be dramatically tuned by hydrogenation \citep{liu14}. It was found that by changing the hydrogen coverage the thermal conductivity can be controllably manipulated and maximized up to five times larger than that of pristine silicene/graphene hetero-interface.

Ferromagnetism in monolayer crystal structures is essential for nanoscale spintronics device applications. Hydrogenation is also an efficient way of inducing ferromagnetism in silicene. It was demonstrated that half-hydrogenation breaks the extended $\pi$-bonding network of silicene, leaving the electrons in the unsaturated Si atoms localized and unpaired, and thus it exhibits ferromagnetic semiconducting behavior with a band gap of 0.95 eV. The long-range ferromagnetic coupling between Si atoms was also predicted by Zhang and Yan, with a Curie temperature of about 300 K \citep{Zha12b}.

Full hydrogenation of germanene, the first synthesis of germanane, was successfully achieved by Bianco \textit{et al.} in 2013. The hydrogenated germanene (with the chemical formula GeH) is obtained from the topochemical deintercalation of CaGe2 \citep{bia13}. The obtained GeH monolayers GeHs are thermally stable up to 75~$\degree$C, but above this temperature amorphization and dehydrogenation begin to occur. These authors also demonstrated that easy transfer of these single crystals can be achieved through mechanical exfoliation onto SiO$_2$/Si surfaces. The theoretically predicted bandgap of 1.53~eV and the high electron mobility which is five times higher than for bare germanene of germanane show the potential use in optoelectronic device applications.

\section{Oxygenation}

Due to its technological importance, the interaction of graphitic materials with oxygen has always been one of the main focus of materials science. It has been reported by many groups that synthesis of new products can be achieved by either a chemical reaction or by gaseous diffusion of oxygen through the material. However, in contrast to single crystals of halogenated graphenes,formation of a well-ordered oxidized monolayer graphene has never been achieved so far. Instead, researchers synthesized grapheneoxides (GOs) which is composed of a graphene layer surrounded by many hydrocarbons and chemical impurities.

\begin{figure}[t]
\includegraphics[width=11.5cm]{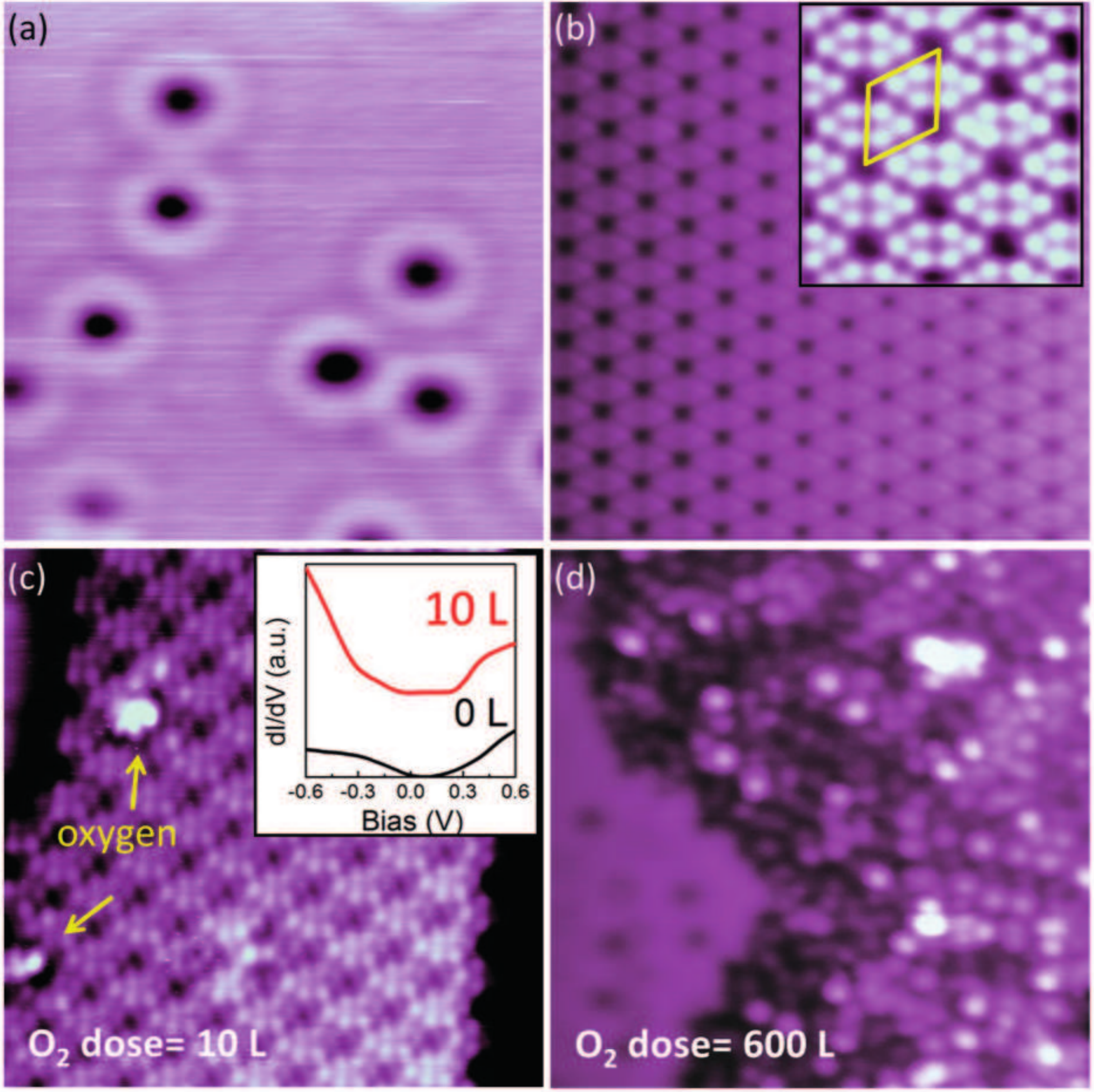}
\caption{\label{oxy} (a) STM topographical image of clean Ag(111) substrate, (b) 3$\times$3 silicene on 4$\times$4 Ag(111), (c) STM image of silicene layer oxidized by an oxygen dose of 10 L. O adatoms prefer to reside at bridge sites. The inset contains STS spectra of silicene and silicene oxide samples, indicating that there is gap opening due to oxidation. (d) STM image of the 3$\times$3 silicene sheet oxidized under 600 L O$_2$. The bare Ag(111) surface can be seen at the bottom left of (d). Adapted from \citep{Xu14}.}
\end{figure}

It was demonstrated that oxidation leads to dramatic changes in the structural and electronic properties of silicene. While the binding energy of silicene on Ag surface is around 0.7 eV, Si-O bonds have binding energy of 4-12 eV \citep{Xu14,Zhang2}. As shown in Fig.\ref{oxy}, oxygen adatoms prefer to form Si-O-Si bonds at bridge sites in the 3$\times$3 silicene surface without resulting in any deformation on the Ag substrate. ARPES measurements revealed that oxidation of silicene breaks the Si-Ag hybridization and that the Ag(111) Shockley surface state can be revived. It was also reported that after the oxidation process, the semimetallic behavior disappears and the silicene oxide exhibits a disordered structure with a semiconductor-like character. Similarly, by performing low-temperature scanning tunneling microscopy and in situ Raman spectroscopy measurements it was shown that depending on the adsorption configurations and amounts of oxygen adatoms on the silicene surface different buckled structures with different bandgaps can be obtained \citep{Du14}. In addition, very recently the microscopic mechanism of the reaction of O$_2$ with silicene was studied by first-principles molecular dynamics simulations. It was found that oxidation of the surface is accelerated by the synergetic effect of molecular O$_2$ dissociation and subsequent local structural relaxations \citep{mor15}.

Partial and full oxidation of silicene has been studied in detail \citep{Wang2,Ozcelik3}. Depending on the amount and type of oxidizing agents, the honeycomb structure of silicene may be preserved, distorted or destroyed. The properties of the resulting materials also vary significantly from metals and semimetals to semiconductors and insulators \citep{Wang2}. The 2D honeycomb counterpart of $\alpha$-quartz was also shown to be stable \citep{Ozcelik3}. The geometry of this structure is composed of reentrant Si-O bonds which makes its Poisson ratio negative. When pulled in a certain direction, such auxetic materials enlarge (instead of shrinking) in the perpendicular direction. 2D $\alpha$-quartz also has a high piezoelectric coefficient and its nanoribbons show metallic or semiconducting behavior depending on their chirality \citep{Ozcelik3}.

\section{Interaction with halogens}

Halogenation is the general term used for reaction of an alkene structure with a halogen atom. When an alkene interacts with a halogen atom, reaction takes place at the carbon-carbon double bond. Regarding monolayer crystal structures, the first example of full halogenation was achieved using XeF$_2$ leading to a fully fluorinated structure, which is called fluorographene, is a high-quality insulator (resistivity $>$10$^{12}$ ohm) with an optical gap of 3~eV \citep{nai10}. It inherits the mechanical strength of graphene, exhibiting a Young's modulus of 100 N m$^{-1}$ and sustaining strains of 15 percent. Fluorographene is inert and stable up to 400 $\degree$C even in air, similar to Teflon. Moreover, even if bare graphene does not interact with molecular Chlorine, experimental realization of chlorination of graphene can be achieved by using a photochlorination technique \citep{li11}.

Since silicene electronically and structurally shares some of the unique properties of graphene, one may expect similar properties upon halogenation of silicene. Recently, the geometry, electronic structure and  mechanical properties of halogenated silicene SiX (X = F, Cl, Br and I) in various conformers were studied using first-principles calculations within the DFT \citep{Zha15b}. Chair conformation of fluorinated and chlorinated silicenes are alike Fig.~\ref{der2}(b) and (c). It was shown that halogenated silicene has enhanced stability compared to silicene, a moderate and tunable direct gap, and small carrier masses. The element- and conformer-dependence of the energy gap could be well understood by the variance of buckling and a bond energy perturbation theory based on orbital hybridization.

\begin{figure}[t]
\includegraphics[width=11.5cm]{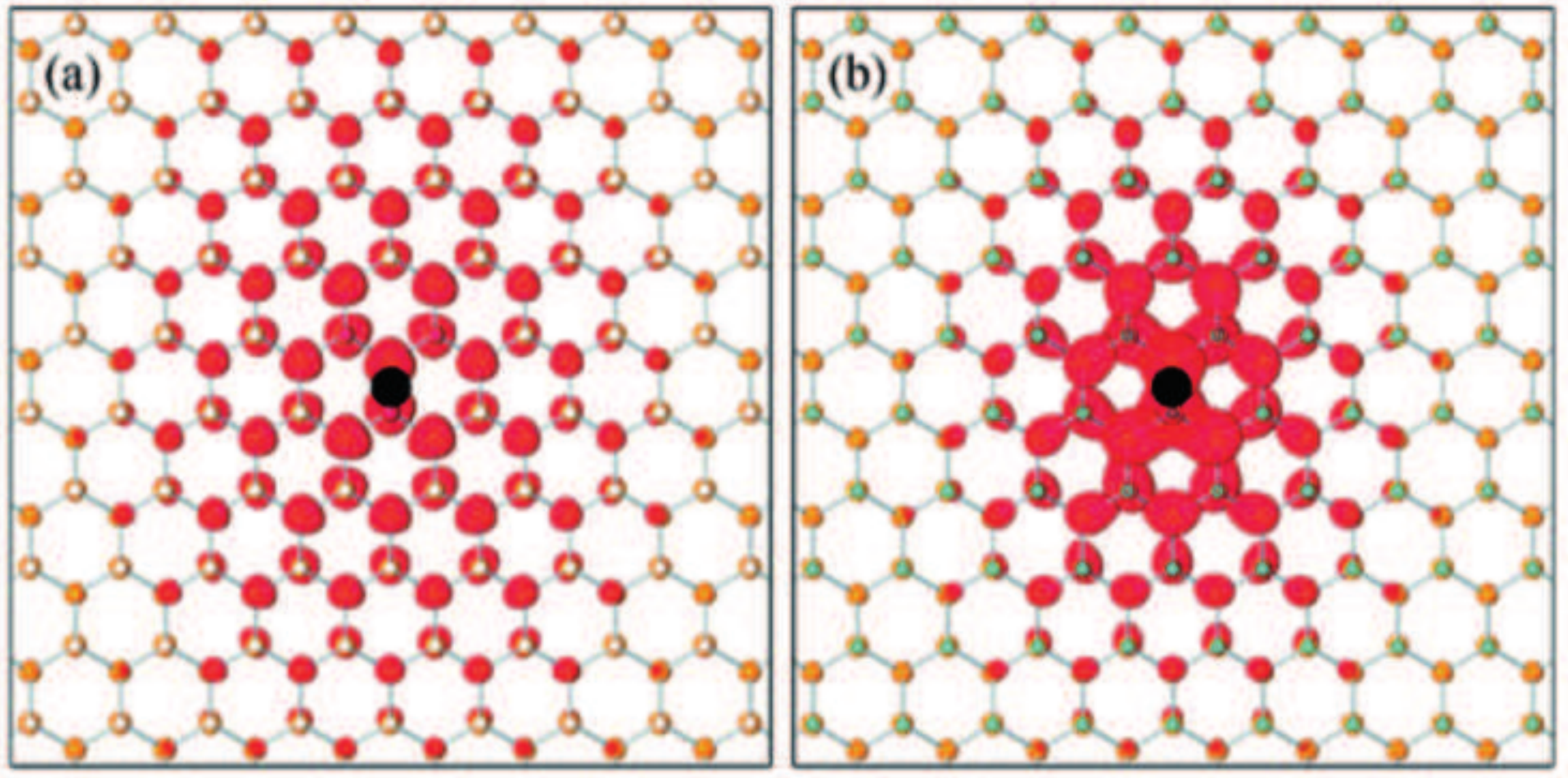}
\caption{\label{der4} Electron probability distribution for the first bound excitonic state of (a) silicane and (b) fluorosilicene. Adapted from \citep{Wei13}.}
\end{figure}

Many-body effects in fluorosilicene and silicane, as well as armchair silicene nanoribbons (ASiNRs) were studied using Green's function perturbation theory \citep{Wei13}. It was found that in addition to the remarkable self-energy effects, the optical absorption properties of silicane and fluorosilicene are dominated by strong excitonic effects with formation of bound excitons with considerable binding energies.

Moreover, the structural, electronic and magnetic properties of half-fluorinated silicene sheets were investigated using first-principles simulation in the framework of DFT \citep{Wan15}. They reported that half-fluorinated silicene sheets with zigzag, boat-like or chair-like configurations were confirmed as dynamically stable based on phonon calculations. Upon the adsorption of fluorine, direct energy band gaps opened in both zigzag and boat-like conformations. Moreover, the half-fluorinated silicene with chair-like configuration showed an antiferromagnetic behavior that mainly stems from fluorine-free Si atoms.

Recently, two-dimensional 2D topological insulators (TIs) in functionalized germanenes (GeX, X = H, F, Cl, Br, or I) were proposed \citep{Si14}. It was found that GeI is a 2D TI with a bulk gap of about 0.3 eV, while GeH, GeF, GeCl, and GeBr can be transformed into TIs with sizable gaps under achievable tensile strains. They reported that the coupling of the $p_{xy}$ orbitals of Ge and heavy halogens in forming the $\sigma$ orbitals also plays a key role in the further enlargement of the gaps in halogenated germanenes.

\section{Functionalization of silicene} \label{functionalization}
Functionalization of silicene by various adatoms has been studied extensively \citep{Sahin2,Silvek1,Ozcelik1}. The interaction of silicene with metal adatoms is quite strong compared to graphene \citep{Sahin2}. Alkali metals like Li, Na, and K, adsorb to the hollow sites without distorting the lattice. Upon significant charge transfer, silicene becomes metallic. Adsorption of alkaline-earth metals like Be, Mg and Ca, on the other hand, turns silicene into a narrow gap semiconductor. Silicene becomes half-metallic when it is functionalized by transition metal adatoms like Ti and Cr \citep{Sahin2}.

\begin{figure}[t]\label{ad}
\includegraphics[width=11.5cm]{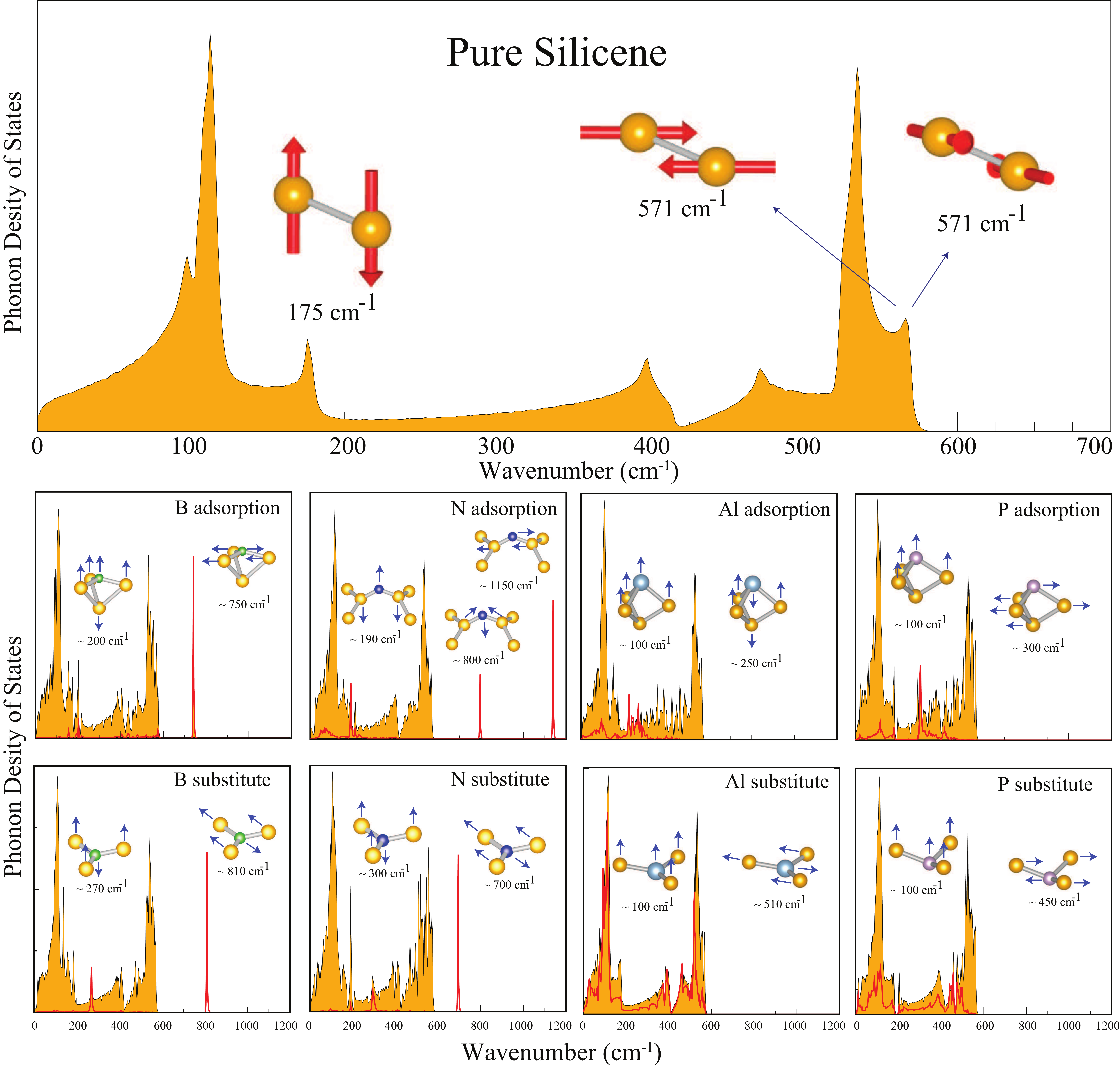}
\caption{ Phonon density of states for adsorption and substitution cases of B, N, Al and P on silicene. The total DOS of the structures are shown by the filled area. The projected DOS belonging to foreign atoms are represented by red lines. Gamma point vibrational motions are also depicted.}
\end{figure}

Adsorption of Al and P, that are group III and V elements from the Si row, result in dumbbell structures discussed below \citep{Silvek1}. Adsorption of B also makes a dumbbell structure, while a N adatom prefers the bridge site of silicene. Both adsorption and substitution of these group III and V elements make silicene metallic. As shown in Fig.~\ref{ad}, adsorption/substitution of first-row elements do not result in a significant change in the phonon DOS. Due to the coupling of B and N adsorbates/substituents to the acoustical phonon branch of pristine silicene, several sharp peaks appear between 200 and 300 cm$^{-1}$. Obviously, among the two adsorbates nitrogen atoms are more likely to mix with silicene’s acoustic phonons. In addition to these, high-frequency adsorbate/substituent induced modes appear between 700 and 1200 cm$^{-1}$. As shown in the Fig.~\ref{ad}, these high energy modes correspond to in-plane bond-stretching motion of the adatom and the neighboring silicon atoms. However, since the second row elements Al and P have similar atomic weights with Si, their adsorption/substitution modifies the vibrational spectra of silicene in a different way. As seen from Fig.~\ref{ad}, the characteristic behavior of second row elements is the absence of high-frequency bond stretching modes. It is also seen that while substituent B and N atoms do not couple with optical modes of silicene, Al and P atoms entirely contribute to both acoustic and optical modes. While the presence of first row elements B and N can be monitored by Raman spectroscopy measurements, second-row elements Al and P have no clear fingerprint in their phonon dispersion.

The presence of Stone-Wales defects, i.e. a pentagon associated to a heptagon, can significantly affect the site preference of adatoms \citep{Sahin3}. A N adatom, for example, prefers to attach to defect-free sites. The energy barrier for the formation of Stone-Wales defects in freestanding silicene was shown to be much smaller compared to that of graphene \citep{Sahin3,Ozcelik2}.

\section{Dumbbell structure}

\begin{figure}[t]
\includegraphics[width=11.5cm]{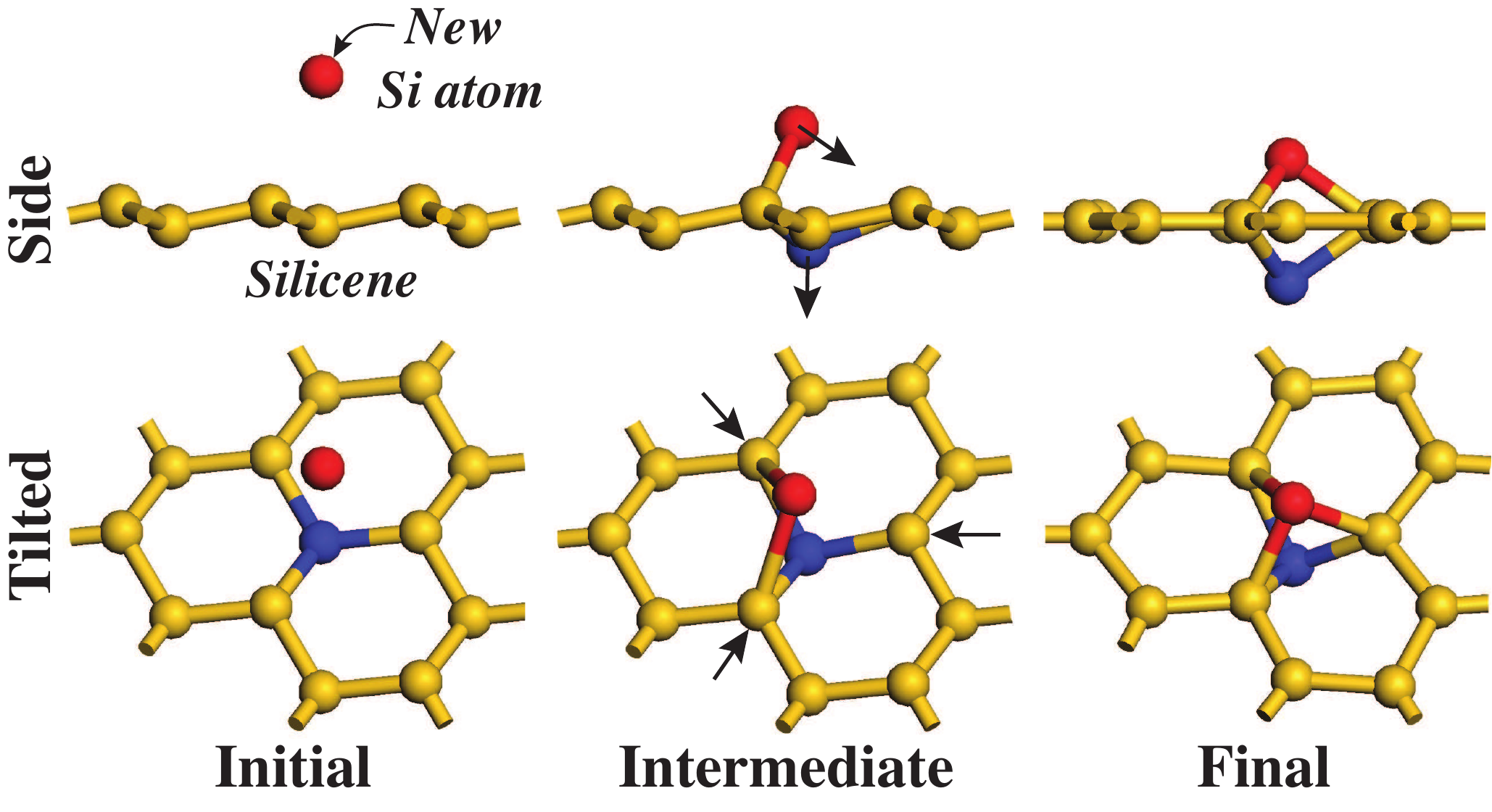}
\caption{Formation of the dumbbell building block units starting from freestanding silicene.}
\label{formation}
\end{figure}

One of the most interesting cases occur when the adatom is also silicon. A Si adatom first attaches to dangling bonds of silicene and then forms bridge bonds with two-second neighbor Si atoms of silicene thereby increasing the coordination number of these Si atoms from three to four. In search for tetrahedral orientation, these four bonds then force the atoms to move towards the directions shown in the middle panel of Fig.~\ref{formation}. As a result, the new Si adatom sits 1.38~\AA~ above the top site of silicene while at the same time pushing down the Si atom just below it by the same amount. These two atoms are connected to other Si atoms by three bonds that are almost perpendicular to each other. The resulting geometry is called the dumbbell (DB) structure \citep{Ozcelik1,Kaltsas1,Cahangirov4}. The DB formation is an exothermic process and occurs spontaneously without need to overcome any kind of barrier. In the case of a C atom adsorbed on graphene, the DB structure does not form because it is energetically less favorable compared to the configuration in which the C adatom is attached to the bridge site of graphene \citep{Ozcelik2}.

\begin{figure}[t]
\includegraphics[width=11.5cm]{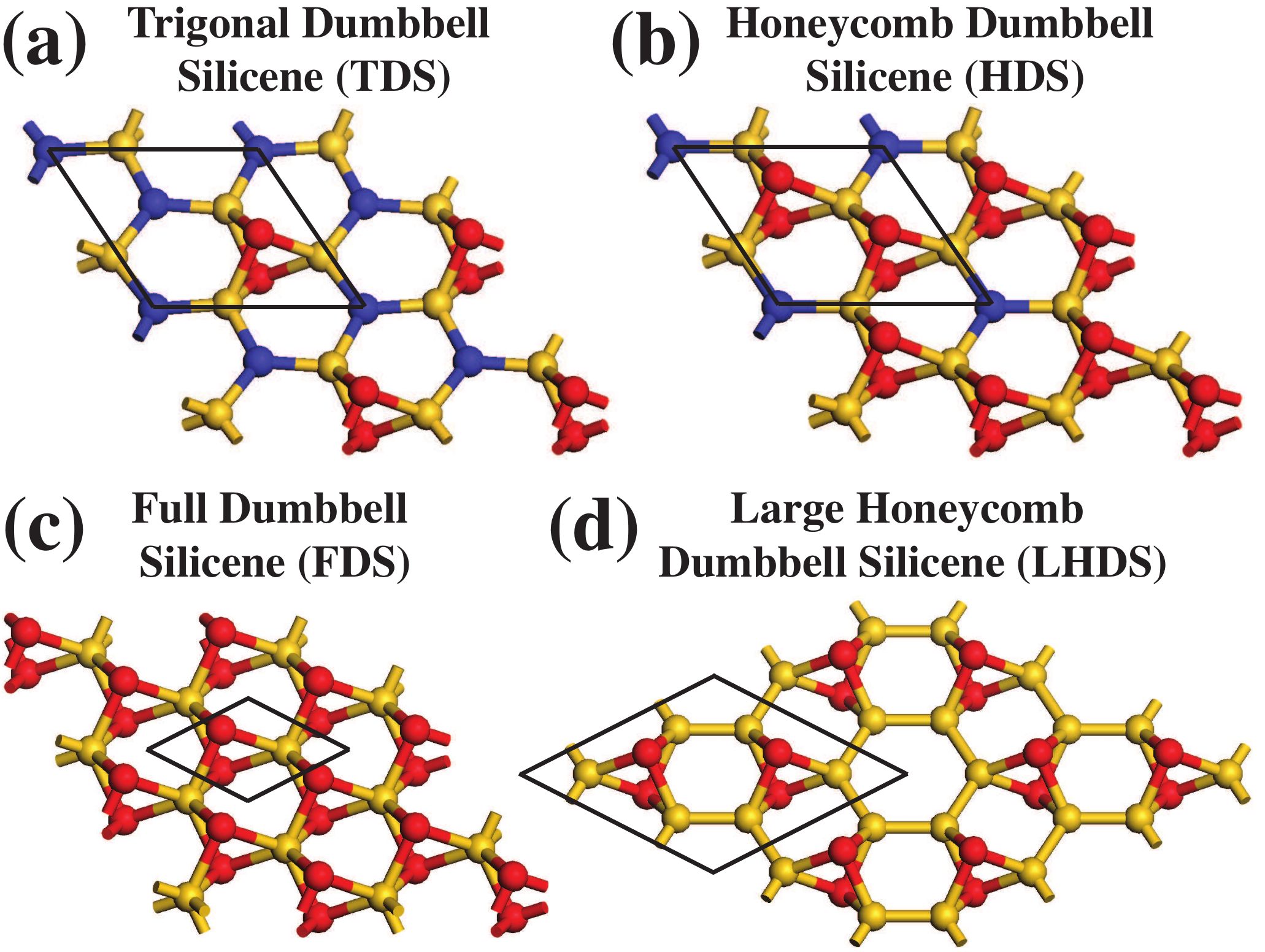}
\caption{Atomic structure of (a) $\sqrt{3} \times \sqrt{3}$ trigonal dumbbell silicene (TDS), (b) $\sqrt{3} \times \sqrt{3}$ honeycomb dumbbell silicene (HDS), (c) $1 \times 1$ full dumbbell silicene (FDS) and (d) $2 \times 2$ large honeycomb dumbbell silicene (LHDS). The unit cells are delineated by solid black lines. Atoms having different environment are represented by balls having different colors.}
\label{dbstructures}
\end{figure}

Calculations show that when a single DB unit is placed in an $n \times n$ unit cell the cohesive energy per Si atom is maximized when $n = \sqrt{3}$ and decreases monotonically for $n \ge 2$ \citep{Cahangirov4}. We refer to the structure having a single DB unit in the $\sqrt{3} \times \sqrt{3}$ unit cell as trigonal dumbbell silicene (TDS) due to the trigonal lattice formed by DB atoms, as shown in Fig.~\ref{dbstructures}(a) \citep{Kaltsas1, Cahangirov4}. As seen in Table~\ref{dbdata}, TDS is energetically more favorable than freestanding silicene \citep{Ozcelik1, Kaltsas1}. Interestingly, the cohesive energy per Si atom is further increased when another DB unit is created in the $\sqrt{3} \times \sqrt{3}$ unit cell of TDS. We refer to this new structure as honeycomb dumbbell silicene (HDS) due to the honeycomb structure formed by two DB units in the $\sqrt{3} \times \sqrt{3}$ unit cell (see Fig.~\ref{dbstructures}((b)). The atomic structure of HDS is crucial to understand the $\sqrt{3} \times \sqrt{3}$ reconstruction that emerges when silicene is epitaxially grown on Ag(111) substrates \citep{Cahangirov4,Feng1,Chen1,Vogt2}. Adding another DB unit in the $\sqrt{3} \times \sqrt{3}$ unit cell of HDS results in a 1$\times$1 structure composed of DB atoms connected by sixfold coordinated Si atoms (see Fig.~\ref{dbstructures}((c)). The cohesive energy of this structure, that we refer to as full dumbbell silicene (FDS), is less than that of TDS and HDS.

\begin{table}[t]
\caption{Cohesive energy and $\sqrt{3} \times \sqrt{3}$ lattice constant of slightly buckled silicene compared with that of dumbbell structures.}
\label{dbdata} 
\begin{tabular}{p{4cm}p{1.4cm}p{1.4cm}p{1.4cm}p{1.4cm}p{1.4cm}}
\hline\noalign{\smallskip}
~ & Silicene & TDS & LHDS & HDS & FDS \\
\noalign{\smallskip}\svhline\noalign{\smallskip}
Cohesive energy (eV/atom) & 3.958 & 4.013 & 4.161 & 4.018 & 3.973 \\
$\sqrt{3} \times \sqrt{3}$ lattice constant (\AA) & 6.69 & 6.52 & 6.43 & 6.38 & 6.23 \\
\noalign{\smallskip}\hline\noalign{\smallskip}
\end{tabular}
\end{table}

\begin{figure}[t]
\includegraphics[width=11.5cm]{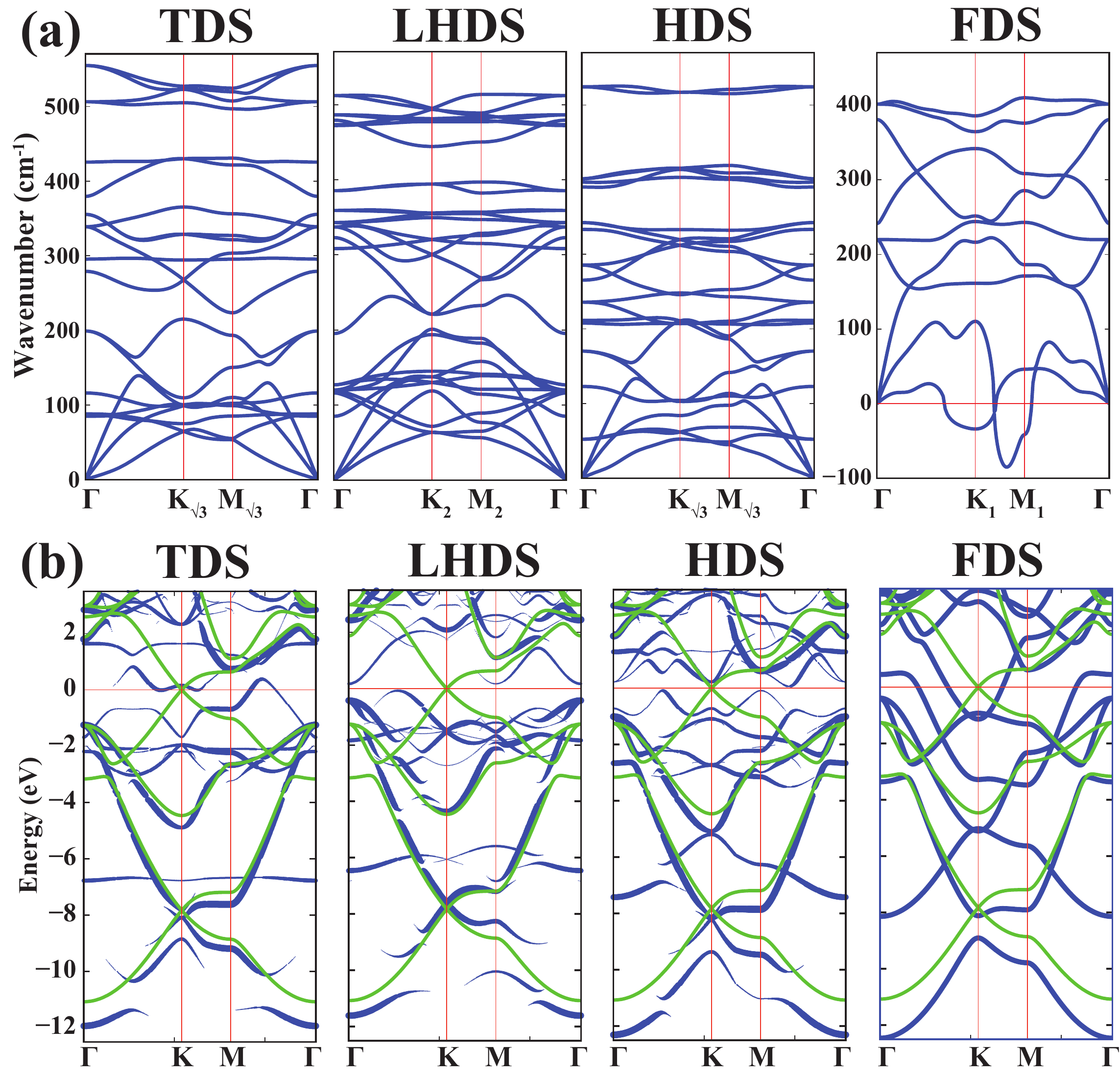}
\caption{(a) The phononic band dispersions of the TDS, LHDS, HDS and FDS structures. The K and M points in the BZs of the 1$\times$1, 2$\times$2 and $\sqrt{3} \times \sqrt{3}$ unit cells are indicated by subscripts. (b) The electronic band dispersions of the TDS, LHDS and HDS structures unfolded to the BZ of the free-standing 1$\times$1 silicene. The thickness of the blue line at a certain point of the Brillouin zone is proportional to the weight of that state in the unfolded band \citep{Allen1}. The bands of the FDS structure do not need to be unfolded because it has 1$\times$1 periodicity. The superimposed electronic band structures of the free-standing 1$\times$1 silicene are shown by green lines.}
\label{dbbands}
\end{figure}

We should emphasize that it is the interplay between two competing effects that makes HDS the most favorable $\sqrt{3} \times \sqrt{3}$ structure. While formation of new DBs and thus new bonds increases the cohesive energy, the increase in the coordination number beyond four decreases it. As seen in Fig.~\ref{dbstructures}, the coordination number of yellow atoms in the TDS structure is four while in HDS it is five. Apparently, the formation of a new DB and hence new bonds compensates the energy required to form the peculiar five-fold coordination. However, it fails to compensate the six-fold coordination of Si atoms forming the middle atomic layer of FDS. This arguments led us to investigate another DB structure that has even larger cohesive energy per atom compared to HDS. This structure has two DB units arranged in a honeycomb lattice in a 2$\times$2 unit cell. Here the packing of DB units is dense compared to TDS but sparse compared to HDS. In this structure, the honeycomb lattice formed by dumbbell units is larger compared to the one formed in HDS, hence we refer to this structure as large honeycomb dumbbell silicene (LHDS). As seen in Fig.~\ref{dbstructures}(d), the maximum coordination of Si atoms in the LHDS is four. Since there are more DB units in LHDS compared to TDS and no hypervalent Si atoms as in HDS, the cohesive energy per atom of freestanding LHDS is higher than both TDS and HDS.

In Fig.~\ref{dbbands}(a) we show the calculated phonon dispersions of the TDS, LHDS, HDS and FDS structures, showing that the frequencies of all the modes are positive over the whole Brillouin zone (BZ) in the TDS, LHDS and HDS, cases while there are imaginary frequencies near the BZ boundary in the FDS one. This means that TDS, LHDS and HDS are thermodynamically stable structures while FDS is unstable. This also implies that the stability of TDS, LHDS and HDS structures does not depend on the substrate and thus these structures can exist in their freestanding configuration.

In Fig.~\ref{dbbands}(b) we present the electronic band dispersions of DB silicene structures. To compare with free-standing silicene we unfold the bands of all structures into the BZ of the 1$\times$1 primitive cell, except that of FDS, which already has this periodicity \citep{Cahangirov4}. The structures are intentionally ranked starting with TDS in which DB units are the most sparse and ending with FDS in which they are the densest. In this way, one can immediately see how the flat band around -7~eV that comes from the weakly-interacting DB units of TDS is gradually turned into the highly-dispersive ($\sim$1.5~eV) band that comes from the strong interaction between DB units that are densely packed in the FDS structure. While this deep band of FDS is easily traced back to TDS the other band of FDS that is originating from the DB units appears much higher and crosses the Fermi level. It is much harder to clearly associate this latter band with its counterparts in TDS, LHDS or HDS. This indicates that in these structures there is a complex interaction between the states originating from the DB units and the $\pi$-states coming from other Si atoms. These results might be used in experiments to identify the formation of DB units.

\begin{figure}[t]
\begin{center}
\includegraphics[width=11cm]{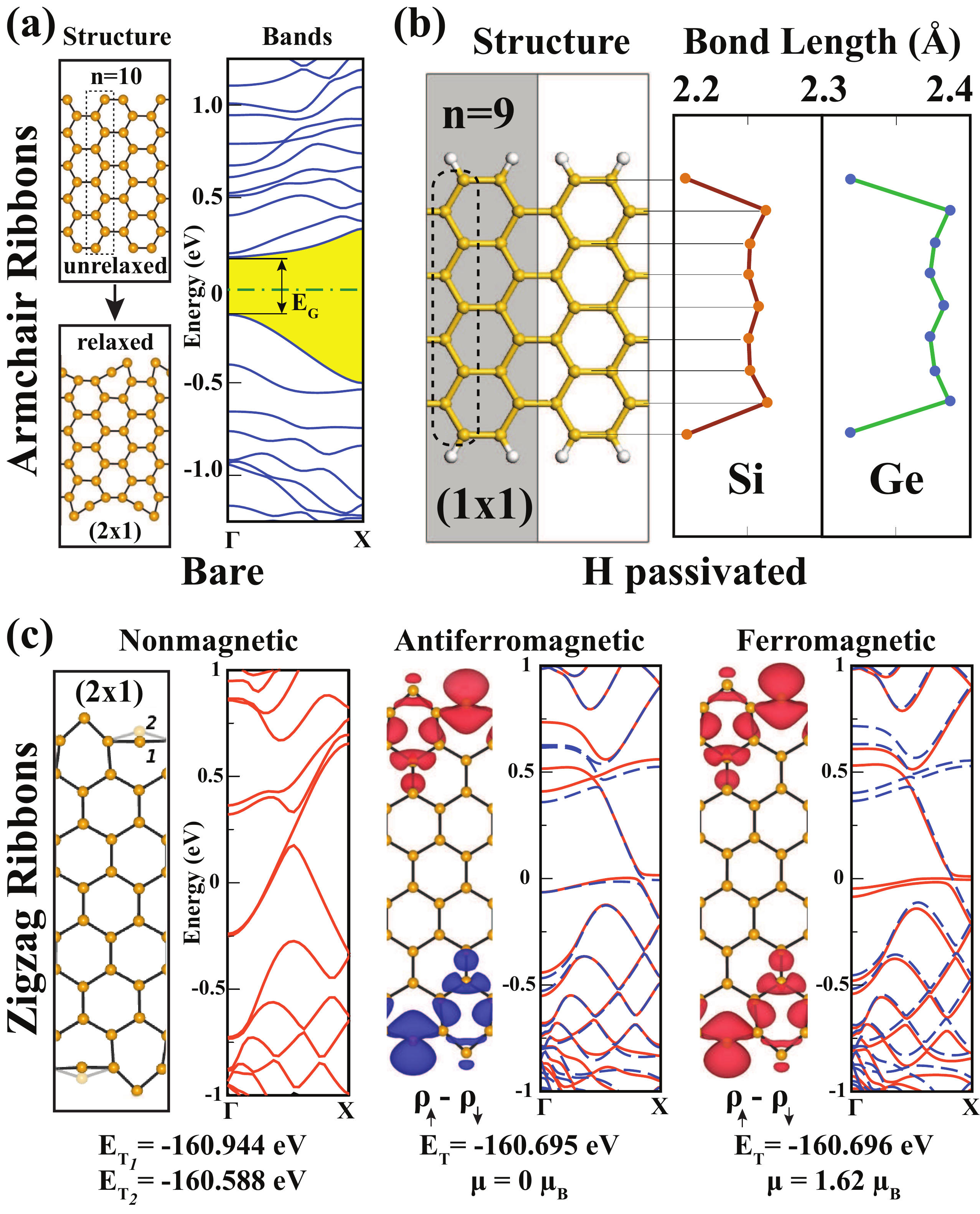}
\end{center}
\caption{(a) Atomic structure of bare armchair silicene nanoribbon (SiNR) with n=10 atoms (which defines the width) in a $1 \times 1$ unitcell. Upon relaxation edge atoms get reconstructed and the unitcell is doubled. Band structure of optimized bare armchair SiNR with n=10 width. (b) Atomic structure and bond length alternation of hydrogen passivated armchair SiNR with width n=9. (c) Atomic and electronic structure of nonmagnetic and magnetic states of $2 \times 1$ reconstructed zigzag silicene nanoribbons. The isosurface of the total charge density difference between the up-spin and the down-spin states are given for the antiferromagnetic and ferromagnetic states \citep{Cahangirov1,Cahangirov2}.}
\label{nanoribbons1}
\end{figure}

\begin{figure}[t]
\begin{center}
\includegraphics[width=11cm]{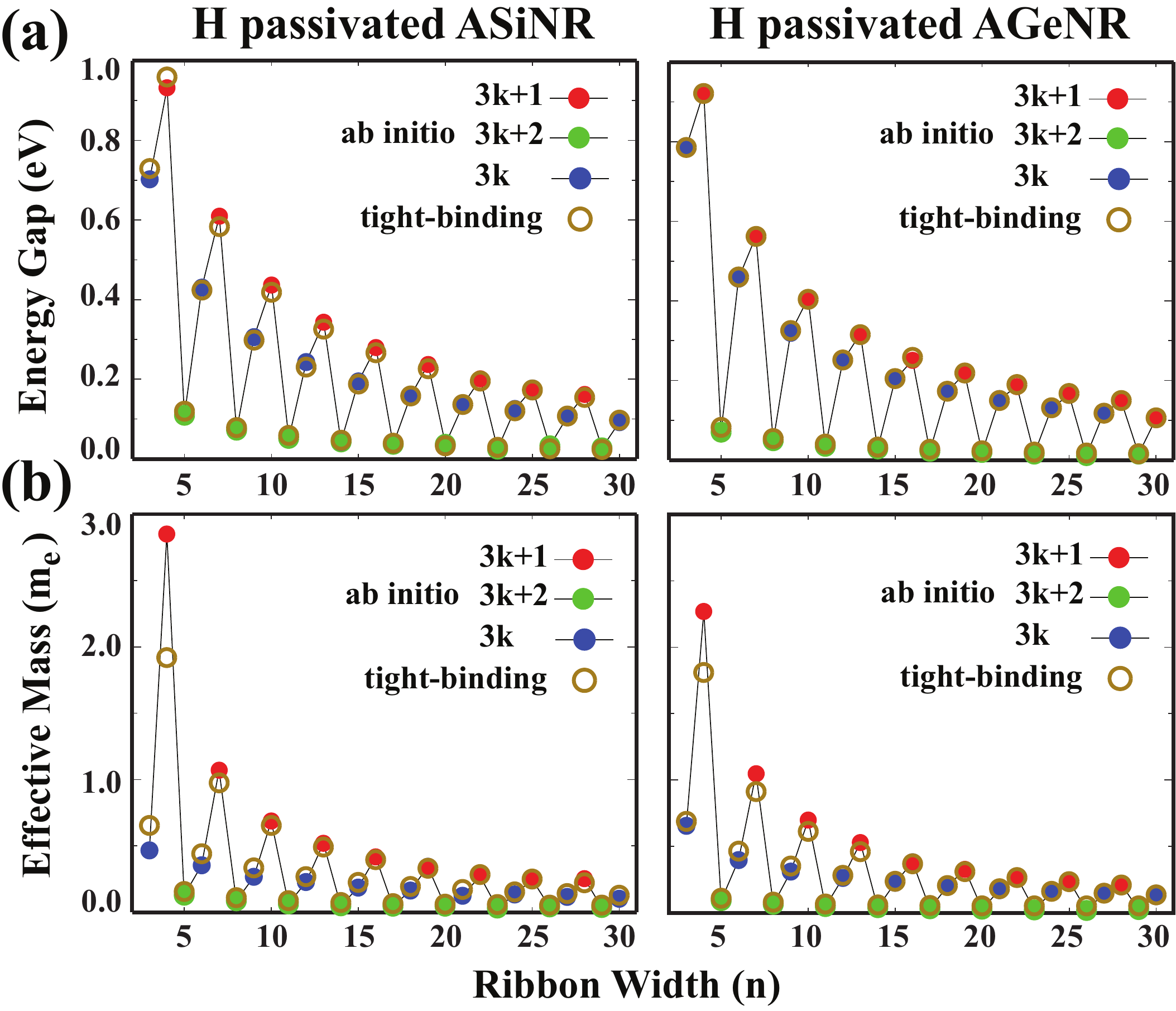}
\end{center}
\caption{The family behavior in (a) the energy gaps and (b) effective masses of hydrogen saturated nanoribbons of silicene and germanene. Adapted from \citep{Cahangirov2}.}
\label{nanoribbons2}
\end{figure}

\section{Nanoribbons} \label{nanoribbons}

The atomic and electronic structures of zigzag and armchair silicene nanoribbons was studied in detail \citep{Cahangirov1,Cahangirov2}. As seen in Fig.~\ref{nanoribbons1}, both zigzag and armchair silicene nanoribbons (SiNR) acquire a 2$\times$1 reconstruction when the edges are not saturated. In Fig.~\ref{nanoribbons1}(c) we compare the total energies of two nonmagnetic and two magnetic structures. The difference between the two nonmagnetic structures is the position of one of the edge atoms denoted by 1 and 2. In the geometry denoted by 1, the corresponding edge atom significantly deviates from the nanoribbon plane thereby gaining enough energy to make this nonmagnetic structure the ground state. This metallic nonmagnetic ground state is in contrast to semiconducting zigzag graphene nanoribbons that have localized states along their edges with opposite spins. One can take advantage of this property to make zigzag graphene nanoribbons half-metallic by applying an in-plane homogeneous electric field \citep{Son1}. It was shown that, zigzag SiNRs can acquire magnetic properties through hydrogen saturation at the edges \citep{Zberecki1}. Furthermore, when doped with N atoms, zigzag SiNRs become half-metallic with opposite spins making up the conduction and valence band edge states that touch each other at the Fermi level \citep{Zheng1}.

Both bare and hydrogen saturated armchair nanoribbons of silicene and germanene are semiconductors and show family behavior in their band gaps with respect to their widths \citep{Cahangirov2}. It means that, if one chooses three armchair nanoribbons having 3$k$-1, 3$k$ and 3$k$+1 atoms per unit cell, where $k$ is a positive integer, then the former will have the lowest gap, the latter will have the largest gap while the middle one will have a gap that is in between. As $k$ is increased, the band gaps of all three families approach to zero, due to the quantum confinement (and the fact that 2D silicene is a semimetal) as illustrated in Fig.~\ref{nanoribbons2}(a). This trend originates from the folding of the band structure of silicene and germanene and the same is also true for the armchair graphene nanoribbons \citep{Son2}. This property was experimentally confirmed in the case of graphene nanoribbons using lithographic processes \citep{Han1}. 

The family behavior in armchair SiNRs can be reproduced using a much simpler two-parameter tight-binding model that also works in the case of armchair graphene nanoribbons \citep{Son2}. Here one starts by noting that the bond length is almost constant throughout the hydrogen passivated armchair nanoribbons of silicene and germanene, except for the sudden decrease at the edges, as illustrated in Fig.~\ref{nanoribbons1}(b). Hence, two hopping parameters are defined; one for the edges and another for the rest of the structure. Since the bond lengths are smaller at the edges, the corresponding hopping is slightly higher. As seen in Fig.~\ref{nanoribbons2}, it is possible to reproduce both energy gaps and the effective masses of the first conduction bands by manually fitting these two parameters. As a result of this fitting, the tight-binding parameters were determined for hydrogen saturated silicene and germanene armchair nanoribbons to be $t_{Si}=1.03~eV$, $t_{Ge}=1.05~eV$, $t_{Si,edge}=1.15~eV$, and $t_{Ge,edge}=1.13~eV$. Interestingly, the ratio of the two tight binding parameters is the same for graphene and silicene.

The confinement of electrons was studied in superlattices formed by armchair SiNRs having different widths \citep{Cahangirov2}. Here computationally feasible structures were calculated using ab-initio techniques and it was verified that the aforementioned tight-binding model works even for such unusual geometries. Then, much larger systems were investigated by the tight-binding model. It was possible to explain the obtained confinement trends using the family behavior and also including the effects due to the interface.

The thermoelectric properties of silicene and of its nanoribbons were also studied in detail \citep{Hu1,Zberecki2,Zberecki3,Yang1,An1,Wierzbicki1}. The thermal conduction of silicene and SiNRs significantly increases with applied tensile strain until reaching a plateau, in contrast to graphene which has lower thermal conductivity when stretched \citep{Hu1}. This was explained by the fact that the out-of-plane flexural phonon modes of silicene are stiffened with applied tensile strain while in the case of graphene they are softened. The in-plane transverse and longitudinal modes, on the other hand, are softened in both silicene and graphene which also explains the plateau observed in silicene at high strain. Doping zigzag SiNRs with Al and P atoms increases the Seebeck coefficient \citep{Zberecki3}. The figures of merit of thermoelectric energy conversion were found to be approaching and in some cases passing 1 for the narrow SiNRs \citep{Yang1}. Finally, single spin negative differential resistance and strong spin Seebeck effect can be achieved by engineering defects on zigzag SiNRs \citep{An1}.
\chapter{Silicene on Ag Substrate} \label{Substrate}

\abstract*{Each chapter should be preceded by an abstract (10--15 lines long) that summarizes the content.}

The isolation of graphene sheets from its parent crystal graphite has given the kick to experimental research on its prototypical 2D elemental cousin, silicene \citep{Brumfiel1}. Unlike graphene, silicene lacks a layered parent material from which it could be derived by exfoliation, as mentioned in Chapter~\ref{Freestanding}. Hence, the efforts of making the silicene dream a reality were focused on epitaxial growth of silicene on substrates. The first synthesis of epitaxial silicene on silver (111) \citep{Vogt1,Lin2} and zirconium diboride templates \citep{Fleurence1} and next on an  iridium (111) surface \citep{Meng1}, has boosted research on other elemental group IV graphene-like materials, namely, germanene and stanene \citep{Matthes1,Xu1}. The boom is motivated by several new possibilities envisaged for future electronics, typically because of the anticipated very high mobilities for silicene and germanene \citep{Ye1}, as well as potential optical applications \citep{Matthes1}. It is also fuelled by their predicted robust 2D topological insulator characters \citep{Liu1,Ezawa1} and potential high temperature superconductor character \citep{Chen4,Zhang1}. One of the most promising candidates as a substrate is Ag because from the studies of the reverse system, where Ag atoms were deposited on silicon substrate, it was known that Ag and silicon make sharp interfaces without making silicide compounds \citep{Lelay1}. Indeed, studies on synthesis and characterization of silicene is mainly focused on using Ag(111) as substrates and hence we think it is important to understand this particular system. In this Chapter, we present the experimental and theoretical studies investigating the atomic and electronic structure of silicene on Ag substrates.

\begin{figure}[t]
\includegraphics[width=11.5cm]{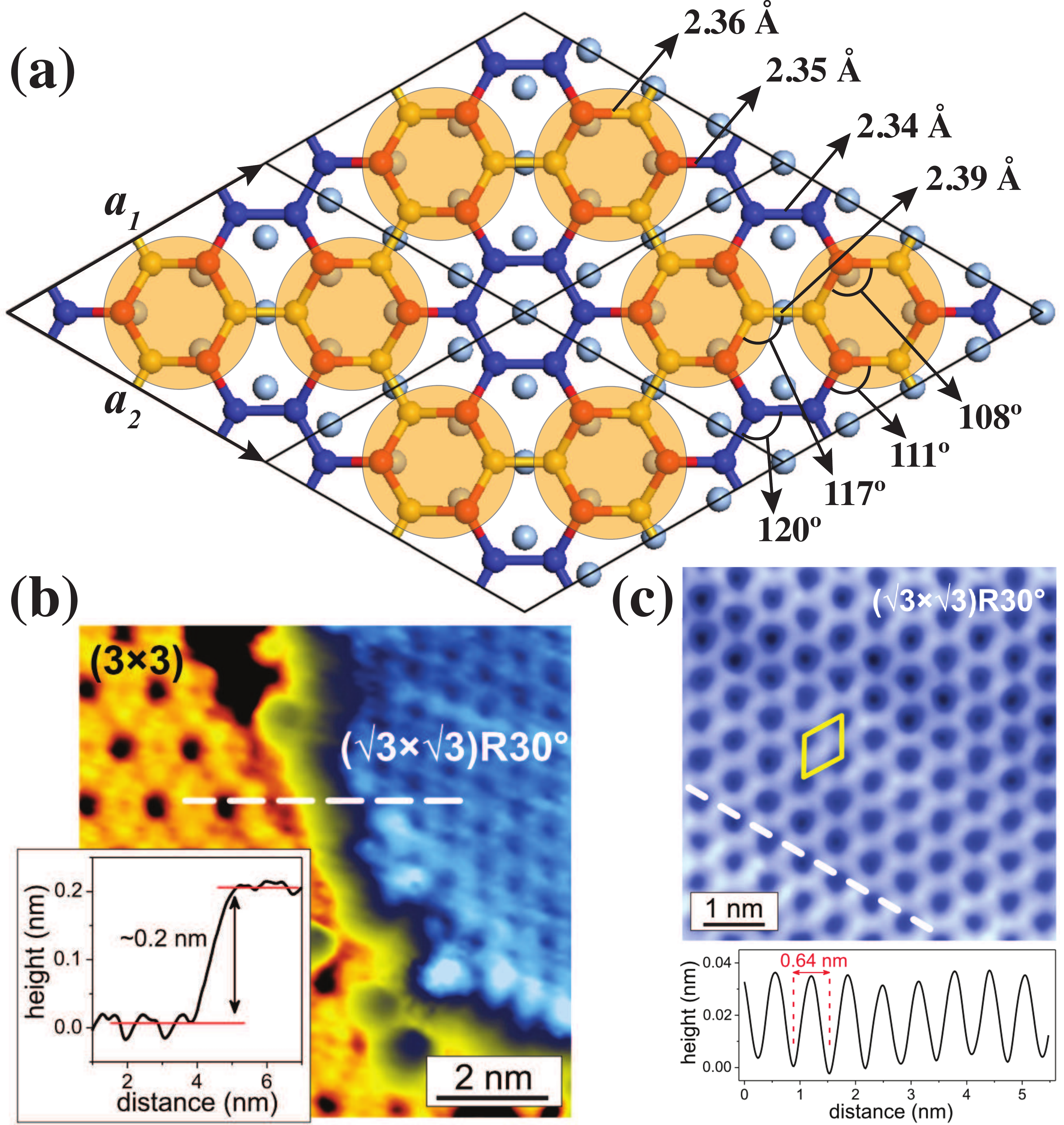}
\caption{(a) Top view of 3$\times$3 silicene matched with a 4$\times$4 Ag(111) surface supercell. Only the topmost atomic layer of Ag(111) is shown and represented by light blue balls. Red, yellow and blue balls represent Si atoms that are positioned near the top, hollow and the bridge sites of the Ag(111) surface. Red Si atoms that are positioned on top of Ag atoms protrude up. Yellow Si atoms are mainly interacting with the Ag atoms that sit under the bonds connecting them. When the STM resolution is not high enough, the three protruding red Si atoms are seen as a single dot represented by transparent orange circles drawn around them. These circles form large hexagons that have 3$\times$3 periodicity. (b) STM image and line profile scanning from 3$\times$3 to $\sqrt{3}\times\sqrt{3}$ silicene on Ag(111). (c) STM image and line profile of $\sqrt{3}\times\sqrt{3}$ silicene on Ag(111). Adapted from \citep{Vogt2}}
\label{monostructure}
\end{figure}

\section{Experimental evidence} \label{substrate}

Silicene was synthesized for the first time on Ag (111) substrates \citep{Vogt1}. Their STM measurements revealed the ``flower pattern'' (see Fig~\ref{flower}) originating from protruding atoms of 3$\times$3 reconstructed silicene matching the 4$\times$4 supercell of the Ag (111) surface as shown in Fig~\ref{monostructure}. This structure was also confirmed by DFT calculations where the geometry optimization starting from unreconstructed 3$\times$3 supercell of silicene on top of 4$\times$4 supercell of Ag (111) slabs resulted in the reconstructed silicene that reproduced the STM image observed in experiments. Each of these three protrusions seen in STM image make a group of Si atoms that belong to the same silicene sublattice. However, there are two such groups in every 3$\times$3 unitcell that belong to the different sublattices of silicene. This becomes evident when H is deposited on 3$\times$3 silicene which results in a highly asymmetric STM image due to the fact that H atoms prefer to bind only to one of the two sublattices. The atomic structure of 3$\times$3 phase of silicene was also confirmed by reflection high energy positron diffraction (RHEPD) as well as low energy electron diffraction (LEED) experiments \citep{Fukaya1,Kawahara1}. Angle resolved photoemmision spectroscopy (ARPES) measurements of the 3$\times$3 silicene phase revealed a linear band starting 0.3~eV below the Fermi level and extending all the way down to -3~eV with a slope of $\sim$~1.3$\times$10$^6$m/s \citep{Vogt1}. This linear band was not detected when silicene was absent. However, the extension and slope of the linear band was too high to be attributed solely to silicene. This created a debate on the origin of these bands that is discussed below.

The archetype single layer silicene, which is the 3$\times$3 phase having a unique orientation, results from a delicate balance between the impinging Si flux (yielding, typically, completion in about 30 minutes), the surface diffusion of the deposited Si atoms on the bare and silicene covered areas and the competing in-diffusion toward the sub-surface. The growth is driven by these kinetic processes, which, actually, gives a very narrow substrate temperature window of about 200-220$\degree$C \citep{Vogt1,Lin2}. Here silicene forms a highly ordered structure, which can cover 95$\%$ of the crystal surface \citep{Fukaya1}, because of the exact correspondence between 3 silicene basis vectors and 4 nearest neighbor Ag-Ag distances. The ``flower pattern'' observed both in STM and non-contact AFM imaging results from the puckered Si atoms sitting nearly on top of Ag atoms, giving a total corrugation of $\sim$ 0.07~nm in the silicene sheet \citep{Vogt1,Resta1}.

Already from $\sim$250$\degree$C a new 2D phase of silicene develops, co-existing with domains of the  3$\times$3~/~4$\times$4 phase. Since control of the substrate temperature is not easy in this temperature regime where most pyrometers are inoperative and where thermocouples, depending on their locations, generally give improper values, in many cases, mixed 3$\times$3~/~4$\times$4 and $\sqrt{7}\times\sqrt{7}$~/~$\sqrt{13}\times\sqrt{13}$ domains are simultaneously observed \citep{Lin2}. In the latter case, four rotated domains, imposed by symmetry are present, since the $\sqrt{7}\times\sqrt{7}$ silicene domains are rotated by $\pm$~19.1$\degree$ with respect to each of the two $\sqrt{13}\times\sqrt{13}$ Ag super cells, which are themselves rotated by $\pm$~13.9$\degree$ with respect to the main [-110] and alike directions of the Ag(111) surface. These domains are buckled, differently from the 3$\times$3~/~4$\times$4 case and are accordingly slightly expanded, while remaining commensurate, to accommodate a 4~$\%$ reduction in Si coverage ratio (from $\theta_{Si}$ = 1.125 for the 3$\times$3~/~4$\times$4 phase to $\theta_{Si}$ = 1.077 for this new one), signaling a self-healing process of the silicene mesh, while some of its atoms have disappeared below the surface. These four domains have been imaged simultaneously on the same STM topograph \citep{Resta1}.

Above about 250$\degree$C, the in-diffusion process is even more active, and, depending on the incident Si flux, a $\sqrt{3}\times\sqrt{3}$ first layer phase may occur, as emphasized by Chen \textit{et al.}, although this has not been reproduced, to the best of our knowledge, by other groups \citep{Chen1}. Instead, a pseudo ``$2\sqrt{3}\times2\sqrt{3}$'' phase is generally obtained \citep{Jamgotchian1,Jamgotchian2,Cinquanta1}; here, the quotation marks indicate its inherent highly disordered and non-periodic nature resulting from the penetration of Si atoms below the surface \citep{Liu4}. Typically, Auger and photoelectron spectroscopy measurements reveal its sudden death, to end, in a dynamic fating process at 300$\degree$C, on the one hand, in multilayer islands through a dewetting mechanism \citep{Acun1,Moras1,Mannix1}, and, on the other hand, in the formation of an alloy \citep{Rahman1}, most probably buried below the surface since the Si LVV signal vanishes in Auger Electron Spectroscopy measurements \citep{Liu4}. Finally, we note in passing that a highly perfect, ``($2\sqrt{3}\times2\sqrt{3}R$)30$\degree$ silicene phase'', prematurely claimed to have been grown essentially without any defect between 220$\degree$C and 250$\degree$C \citep{Lalmi1,Enriquez1}, has been proven to be just a delusive phase, i.e., a contrast reversed appearance of the bare Ag(111) surface \citep{Lelay3}. At this stage it is worth stressing that both 3$\times$3~/~4$\times$4 and $\sqrt{7}\times\sqrt{7}$~/~$\sqrt{13}\times\sqrt{13}$ first layer silicene phases as well as the defected, high temperature ``$2\sqrt{3}\times2\sqrt{3}$'' phase, can be encapsulated in situ with an Al$_2$0$_3$ ultra-thin capping layer. This capping layer, which preserves their integrity, allows for ex situ Raman studies \citep{Cinquanta1}. It has further permitted the fabrication of the first bottom gate field effect transistor based on a silicene channel \citep{Tao1}, a clear breakthrough toward device applications \citep{Lelay4}.

Unlike the other phases mentioned above, the $\sqrt{3}\times\sqrt{3}$ reconstructed phase of silicene that was observed quite frequently in experiments is not matched by any lattice vector of the Ag (111) substrate \citep{Feng1, Chen1, Vogt2}. The $\sqrt{3}\times\sqrt{3}$ reconstructed silicene was first reported by Feng \textit{et al.} They measured the lattice constant to be 0.64~nm which is $\sim$~5~$\%$ less compared to ideal silicene while the STM image was composed of bright triangular spots arranged in a $\sqrt{3}\times\sqrt{3}$ honeycomb lattice \citep{Feng1}. They have also shown that the same STM image persists in the second layer which hinted that the $\sqrt{3}\times\sqrt{3}$ structure was intrinsic and not formed due to the matching with Ag substrate. Later the $\sqrt{3}\times\sqrt{3}$ reconstruction was also observed in multilayer silicene grown on Ag (111) substrates \citep{Vogt2}. The distance between consecutive layers was measured to be around 3.0-3.1~\AA. Several models were proposed to describe the origin and the atomic structure of the $\sqrt{3}\times\sqrt{3}$ reconstruction. These models are also discussed below.

There are other experimental results worth mentioning. First, this is the ordered adsorption of seven H atoms per unit cell upon in situ exposure of the 3$\times$3~/~4$\times$4 phase at room temperature to atomic hydrogen \citep{Qiu1}. This could be a first step toward silicane synthesis; it proves, in any case, the controllable functionalization of silicene, and, possibly opens a route for tuning magnetic properties \citep{Zheng2}. In this sense, one notices that six among these seven H atoms sit on the same triangular silicene sublattice, while the seventh lies on the other one. Second, it is the observation of large hexagonal patterns with no long range order but remarkable vortices in Low Temperature STM images of triangular $\sqrt{7}\times\sqrt{7}$~/~$\sqrt{13}\times\sqrt{13}$ domains \citep{Liu5}. We believe that this vortex structure originates due to the stress imposed on silicene upon cooling. Silicene is very likely to have a negative thermal expansion coefficient unlike the silver substrate, exalted by the fact that bulk silicon has such a behavior as well as graphene with the same topology. When cooling, silicene, which is expanded in this phase with respect to the prototype 3$\times$3~/~4$\times$4 phase, feels too much stress and relaxes it by making vortex-like structures, at variance with the 3$\times$3~/~4$\times$4 phase which is unchanged down to $\sim$800 mK.

\section{Growth mechanism} \label{growthmech}

The subsequent growth of $\sqrt{3}\times\sqrt{3}$ reconstructed silicene after formation of 3$\times$3 silicene was studied by DFT calculations \citep{Cahangirov4}. Here we summarize results of that study. As Si atoms are deposited on Ag(111) surface they search for the optimum structure that minimizes the energy. In the absence of the Ag substrate this optimum structure is the cubic diamond structure that has a cohesive energy of 4.598 eV/atom according to DFT calculations. However, in the presence of the Ag substrate a monolayer of silicene that has primarily 3$\times$3 reconstruction is formed. Here, the 3$\times$3 supercell of silicene is matched with the 4$\times$4 supercell of the Ag(111) surface, as shown in the Fig.~\ref{monostructure}(a). If we remove the Ag substrate and freeze the Si atoms of the 3$\times$3 reconstructed silicene to calculate its cohesive energy, it turns out to be 3.850 eV/atom, which is 108 meV lower than the buckled freestanding silicene. In fact, if we start from the freestanding 3$\times$3 reconstructed silicene and relax the structure, it goes to the buckled structure, meaning that it is not even a local energy minimum in the absence of the Ag substrate. However, the cohesive energy of 3$\times$3 structure surpasses that of the cubic diamond structure when it is placed on the Ag(111) surface, as seen in the Table~\ref{monodata}. This strong interaction between silicene and the Ag substrate explains the growth of monolayer silicene instead of clustering of Si atoms into bulk structures.

\begin{table}[t]
\caption{Cohesive energies per Si atom and per unit area are given for the 3$\times$3 reconstructed silicene, TDS, LHDS, and HDS structures on the Ag(111) surface.}
\label{monodata} 
\begin{tabular}{p{5cm}p{1.5cm}p{1.5cm}p{1.5cm}p{1.5cm}}
\hline\noalign{\smallskip}
~ & 3$\times$3 & TDS & LHDS & HDS\\
\noalign{\smallskip}\svhline\noalign{\smallskip}
Cohesive energy per atom (eV/atom) & 4.877 & 4.663 & 4.483 & 4.471 \\
Cohesive energy per area (eV/\AA$^2$) & 0.759 & 0.887 & 0.938 & 1.014\\
\noalign{\smallskip}\hline\noalign{\smallskip}
\end{tabular}
\end{table}

One of the proposed models to explain the growth of the $\sqrt{3}\times\sqrt{3}$ structure is based on the dumbbell structures described in Chapter~\ref{Freestanding}. Note that the 3$\times$3 silicene matches the 4$\times$4 Ag(111) supercell, while the DB structures cannot be matched because their lattice constant is squeezed as the density of DB units is increased as seen in Table~\ref{dbdata}. To include the effect of Ag, the 4$\times$4 Ag (111) slab composed of five layers is first squeezed to match the lattice of the 3$\times$3 supercell of the DB structures and then the system is optimized by keeping the Ag atoms fixed. Then the energy of the squeezed Ag substrate in the absence of Si atoms is calculated. The energy difference between these two systems gives the cohesive energies of the DB structures. As seen in Table~\ref{monodata}, the cohesive energy per Si atom is maximized in the 3$\times$3 silicene while the cohesive energy per area is maximized for the $\sqrt{3}\times\sqrt{3}$ HDS structure. According to the model, when Si atoms are first deposited on Ag (111) substrate, they form the 3$\times$3 reconstructed silicene that has the highest cohesive energy per Si atom, as seen in Table~\ref{monodata}. At first, the DB units that spontaneously form on 3$\times$3 silicene diffuse and annihilate at the edges and contribute to the growth of even larger 3$\times$3 regions. Once 3$\times$3 silicene reaches sufficiently large area, the DB units compete to form the most energetic structure in a given area covered by 3$\times$3 silicene. To achieve the highest cohesive energy per area, the DB units arrange themselves to form the $\sqrt{3}\times\sqrt{3}$ HDS structure.

\begin{figure}[t]
\includegraphics[width=11.5cm]{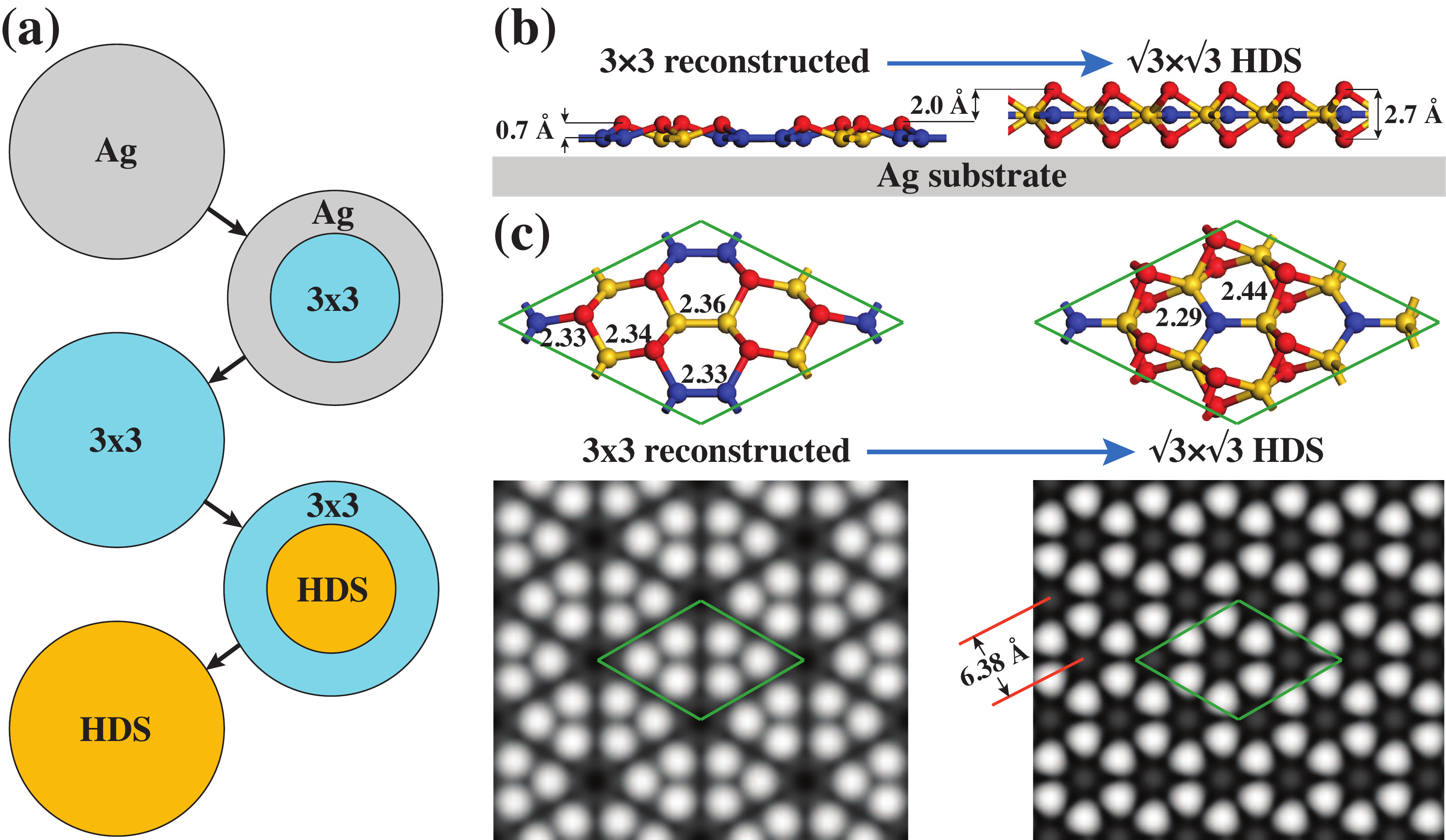}
\caption{(a) Growth sequence of the 3$\times$3 reconstructed silicene and of the HDS structure on the Ag(111) substrate. (b) Schematic depiction of the structural transformation from the 3$\times$3 to the $\sqrt{3} \times \sqrt{3}$ reconstruction from the side view. (c) Tilted view of the atomic structures and calculated STM images. Green lozenges represent the 3$\times$3 supercell.}
\label{monogrowth}
\end{figure}

The schematic sketch of this growth model is presented in Fig.~\ref{monogrowth}(a). This picture is in accordance with experiments in which the $\sqrt{3}\times\sqrt{3}$ structure usually appears as islands on top of the first 3$\times$3 silicene layer. As seen in Fig.~\ref{monogrowth}(b) and (d), the model also excellently reproduces the distance between $\sqrt{3}\times\sqrt{3}$ and 3$\times$3 surfaces measured to be $\sim$~2~\AA~\citep{Vogt2}. Furthermore, as shown in Fig.~\ref{monogrowth}(c) and (e), the simulated STM image of HDS has the same honeycomb pattern as the one obtained in experiments while the $\sqrt{3}\times\sqrt{3}$ lattice constant of HDS that is calculated to be 6.38~\AA, excellently matches the measured value that is $\sim$~6.4~\AA~\citep{Feng1,Chen1,Vogt2}. This model can also be extended to explain multilayers that have $\sqrt{3}\times\sqrt{3}$ reconstruction as discussed in Chapter~\ref{Multilayer} \citep{Cahangirov5}.

There are also other proposed models that try to explain the $\sqrt{3}\times\sqrt{3}$ reconstruction. One of them proposes that if monolayer silicene is squeezed enough, then the $\sqrt{3}\times\sqrt{3}$ honeycomb reconstruction becomes energetically more favorable than the ideal 1$\times$1 buckling \citep{Chen1}. However, this happens if the lattice constant is squeezed down to $\sim$~6.3~\AA~and also there is no physical reason for the system to remain in this high energy state. Another model is based on the well-studied Si(111)-Ag$\sqrt{3}\times\sqrt{3}$ system \citep{Ding1,Shirai1}. This model also produces a $\sqrt{3}\times\sqrt{3}$ honeycomb pattern in STM measurements. In this case the bright spots are originating from Ag atoms on top of Si(111). However, it is not clear how the 3$\times$3 structure is transformed into $\sqrt{3}\times\sqrt{3}$ structure with Ag atoms on top and also why the lattice is compressed. Further discussion of this model is presented in Chapter~\ref{Multilayer}. Yet another model suggests that the $\sqrt{3}\times\sqrt{3}$ honeycomb STM image is a result of the atomic scale flip-flop motion at the surface of bilayer Si(111) structure formed on top of the 3$\times$3 structure \citep{Guo2}. Here the authors suggest that there are three possible configurations and the system is alternating between two of them. Since each state produces a trigonal STM pattern, the combination of two of them should produce the expected honeycomb pattern. However, there is no clear reason why the system should choose to alternate only between two states. Furthermore, this model does not explain the 5~\% lattice contraction observed in experiments \citep{Feng1,Chen1,Vogt2}.

\section{The nature of the linear bands} \label{dirac}

There is no doubt that the linear bands are one of the most intriguing features of silicene. The linear dispersion was observed in 3$\times$3 silicene by ARPES measurements and in $\sqrt{3} \times \sqrt{3}$ silicene by analyzing the quasi-particle interference pattern captured by STM measurements \citep{Vogt1,Chen1}. Both measurements reported very high Fermi velocity of about $1.2-1.3 \times 10^6 $m/s. The linear bands were also reported for the multilayer silicene and its ribbons \citep{DePadova5, DePadova8}.

\begin{figure}
\includegraphics[width=11.5cm]{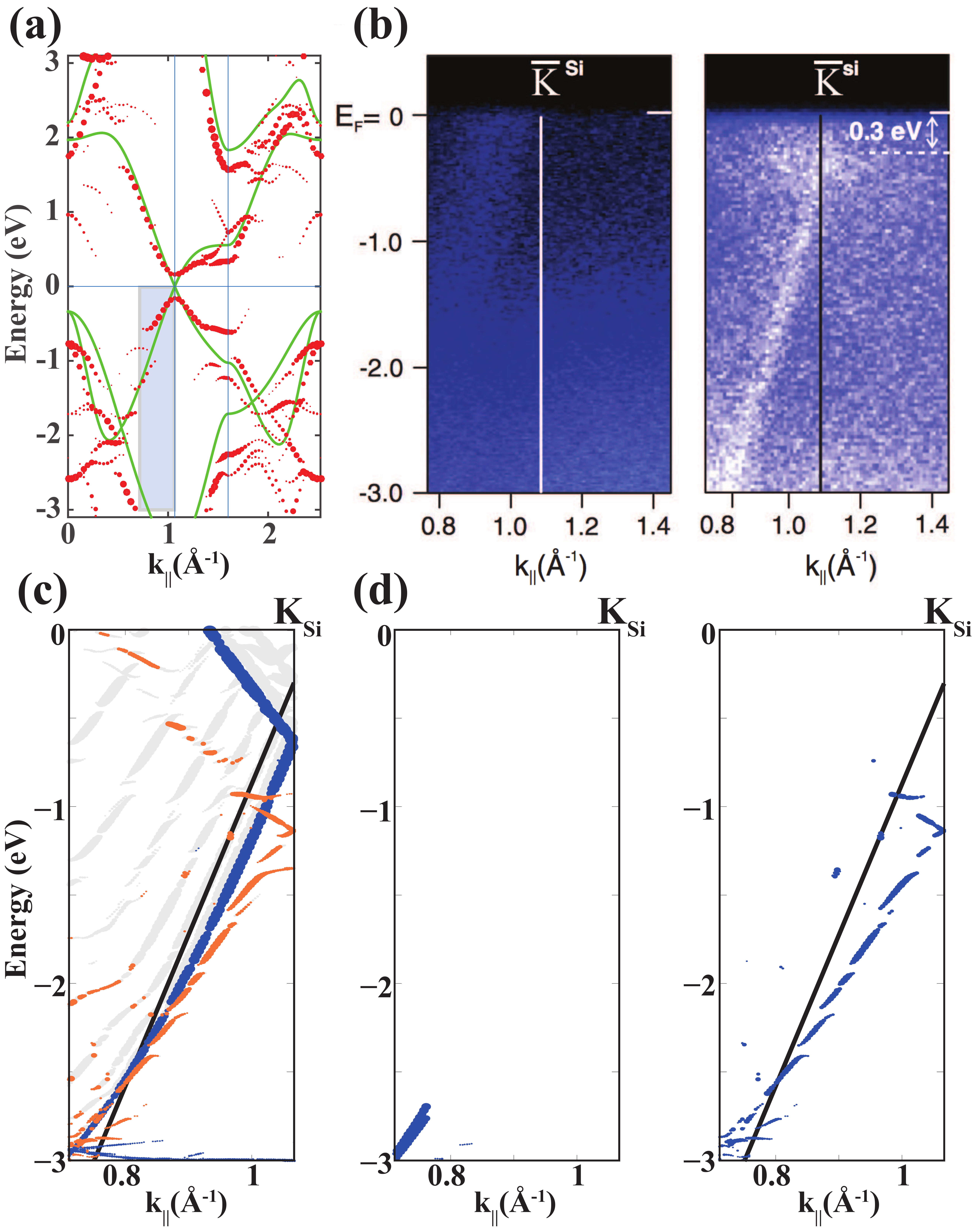}
\caption{(a) Band structure of reconstructed 3$\times$3 silicene (red dots) in the absence of Ag substrate unfolded to 1$\times$1 Brillouin zone of silicene. The dots radii correspond to the weight of the unfolded state. The band structure of ideally buckled silicene is shown by green lines. (b) ARPES data around the K point of 1$\times$1 silicene in the absence (left panel) and presence (right panel) of silicene on Ag(111) substrate (adapted from \citep{Vogt1}). (c) Band structure of 3$\times$3 silicene a 4$\times$4 Ag(111) 11 layer slab in the experimentally relevant range shown by the shaded region in (a). The states contributed by 3 Ag layers in the middle of the slab represent the bulk Ag states and are shown by blue lines. Orange lines are contributed by silicene and 3 Ag layers underneath. The black line represents the experimentally observed linear band. (d) States contributed by 3 Ag layers underneath silicene in the absence (left panel) and presence (right panel) of silicene. Adapted from \citep{Cahangirov3}.}
\label{monoband}
\end{figure}

The 3$\times$3 reconstruction of silicene forms due to the interaction with the Ag substrate and breaks the symmetry needed to preserve the linearly crossing bands at the Fermi level. This is clearly seen in the band structure of the 3$\times$3 reconstructed silicene isolated from the substrate presented in Fig.~\ref{monoband}(a). Here the band structure of the 3$\times$3 reconstructed silicene is unfolded into the Brillouin zone of the ideal 1$\times$1 silicene. Upon reconstruction, the linearly crossing bands are destroyed and instead there is a 0.3~eV gap at the K point. The scanning tunneling spectroscopy (STS) measurement performed under high magnetic field applied in perpendicular direction to the 3$\times$3 silicene on Ag substrate have shown that the peaks corresponding to the Landau levels corresponding to the presence of the Dirac fermions were absent while they were present in the highly oriented pyrolytic graphite samples \citep{Lin1}. This experimental result was supported by DFT calculations to conclude that Dirac fermions of ideal silicene were destroyed due to the symmetry breaking and hybridization with the Ag substrate.

To interpret the ARPES experiments, Cahangirov \textit{et al.} calculated the electronic structure of 3$\times$3 silicene placed on top of 11 layers of 4$\times$4 Ag substrate \citep{Cahangirov3}. Fig.~\ref{monoband}(c) shows the detailed band structure of the silicene/Ag system in the window where the experiments were performed (see Fig.~\ref{monoband}(b)). The blue curve corresponds to the bulk Ag $sp$-band that should not be detected by ARPES which is sensitive to the surface states. Fig.~\ref{monoband}(d) and (e) show the states that have significant contribution from surface Ag states when silicene is absent and present, respectively. Here one can choose the threshold in such a way that the linear band disappears when silicene is absent and appears when it is present thereby mimicking the situation observed in ARPES experiments. This analysis suggests that the linear bands are caused by hybridization between silicene and Ag.  The perpendicular momentum dependence of the electronic states was calculated by a k-projection technique and was used to calculate the contribution of silicene and Ag to the surface band created by hybridization \citep{Chen5}. This study provided further quantitative agreement with experiments while confirming the hybridized nature of the experimentally observed linear bands. The surface band created by the hybridization between silicene and Ag was detected and distinguished from the faintly visible Ag $sp$-bands in the ARPES measurements \citep{Tsoutsou1}. There are many other investigations that have reached to conclusion that the linear bands are due to hybridization between silicene and Ag \citep{Guo1,Wang1}.

The search for Dirac cones is not limited to the silicene on Ag substrates. ARPES measurements performed on calcium disilicide (CaSi$_2$) revealed a massless Dirac cone located at 2~eV below the Fermi level \citep{Noguchi1}. CaSi$_2$ can be considered as buckled silicene sandwiched between the planar atomic planes of Ca. The energy shift in the electronic states of silicene is due to significant charge transfer between Si and Ca atoms. First-principles calculations of the CaSi$_2$ structure revealed that there is, in fact, also a momentum shift in the Dirac cone away from the highly symmetric K point \citep{Dutta1}. This is due to the symmetry breaking between the sublattice atoms of silicene and consequent asymmetric interlayer hopping. The shift in the momentum space is also accompanied with a small energy gap opening between the linearly crossing bands.

The Ag substrate plays a crucial role in growth of silicene, as seen from the previous section. This wouldn't be possible if the interaction between silicene and Ag was too weak. A first-principles study of the electronic charge density between silicene and the Ag(111) substrate has concluded that bonds between Si and Ag atoms don't have covalent character \citep{Stephan1}. However, the strong hybridization with Ag seems to interfere with the delicate electronic structure of silicene, as mentioned above. To avoid this, one has to develop techniques to transfer silicene to less interacting and insulating substrates. Tao \textit{et al.} have taken an important step in this direction \citep{Tao1}. They first grew silicene on Ag and encapsulated it with alumina. Then they flipped the system upside down and etched Ag on silicene, just leaving two Ag pads that they used as metal contacts. In this way they have demonstrated that silicene can operate as an ambipolar field-effect transistor at room temperature.
\chapter{Multilayer Silicene} \label{Multilayer}

\abstract*{Each chapter should be preceded by an abstract (10--15 lines long) that summarizes the content.}

Silicon does not have a naturally occurring layered allotrope like graphite. However, it is possible to grow monolayer silicene on substrates, as we have seen in Chapter~\ref{Substrate}. Extending this idea further, one may wonder whether it is possible to synthesize layered silicon structures by continuing the growth started as a monolayer silicene. In this Chapter we discuss the experimental and theoretical works that are based on this idea of multilayer silicene growth.

\section{Experimental evidence} \label{multilayers}

When Si deposition is prolonged beyond the formation of the first layer 3$\times$3~/~4$\times$4 or $\sqrt{7}\times\sqrt{7}$~/~$\sqrt{13}\times\sqrt{13}$ phases and in the same conditions, growth of multi-layer silicene, which possesses a unique $\sqrt{3}\times\sqrt{3}$R(30$\degree$) (in short $\sqrt{3}\times\sqrt{3}$) structure is obtained. Such films grow in successive terraces, each showing this unique reconstruction. If growth occurs on the prototype 3$\times$3~/~4$\times$4 phase, one gets a single orientation of these terraces. Instead, if growth occurs on the initial $\sqrt{7}\times\sqrt{7}$~/~$\sqrt{13}\times\sqrt{13}$ first layer silicene phase, rotated terraces are obtained; the rotation angles are directly related to those of the first layer domains \citep{Salomon1}. In angle-resolved photoemission (ARPES) measurements, these films possess a Dirac cone at the centre of the Brillouin zone due to back folding of the $\sqrt{3}\times\sqrt{3}$ silicene superstructure with a Fermi velocity about half that of the free standing graphene \citep{DePadova6,DePadova8}. Vogt \textit{et al.} have studied terraces with up to five layers of $\sqrt{3}\times\sqrt{3}$ silicene. They have shown that the height difference between adjacent terraces is $\sim$~3.1~\AA~\citep{Vogt2}. Using in situ a four probe scanning tunneling microscope, a sheet resistance analogous to that of thin films of graphite in nano-grains was determined. De Padova \textit{et al.} took this even further and synthesized few tens monolayers of silicene with $\sqrt{3}\times\sqrt{3}$ reconstruction. Remarkably, these multilayer silicene films survive after exposure in ambient air for a day at least, because just the very top layers are oxidized; the film underneath remains intact, as directly revealed via a graphite-like Raman signature \citep{DePadova7}. Feng \textit{et al.} have investigated bilayers of $\sqrt{3}\times\sqrt{3}$ silicene \citep{Feng1}. They have measured the quasi-particle interference of electrons in the first layer due to the scattering from the islands formed by the second layer grown on top \citep{Chen1}. A linear dispersion with high Fermi velocity was derived from these interference patterns.

\section{Atomic structure} \label{growth}

Although multilayer silicene was grown in many experiments mentioned above its atomic structure has been a subject of debate. The experiments report a 5~$\%$ contracted $\sqrt{3}\times\sqrt{3}$ structure that has a honeycomb appearance in  STM imaging with a $\sim$~3.1~\AA~distance between its layers. However, there is still no structural model that explains all these observations. Here we discuss the proposed models and point out their shortcomings.

\begin{figure}[t]
\includegraphics[width=11.5cm]{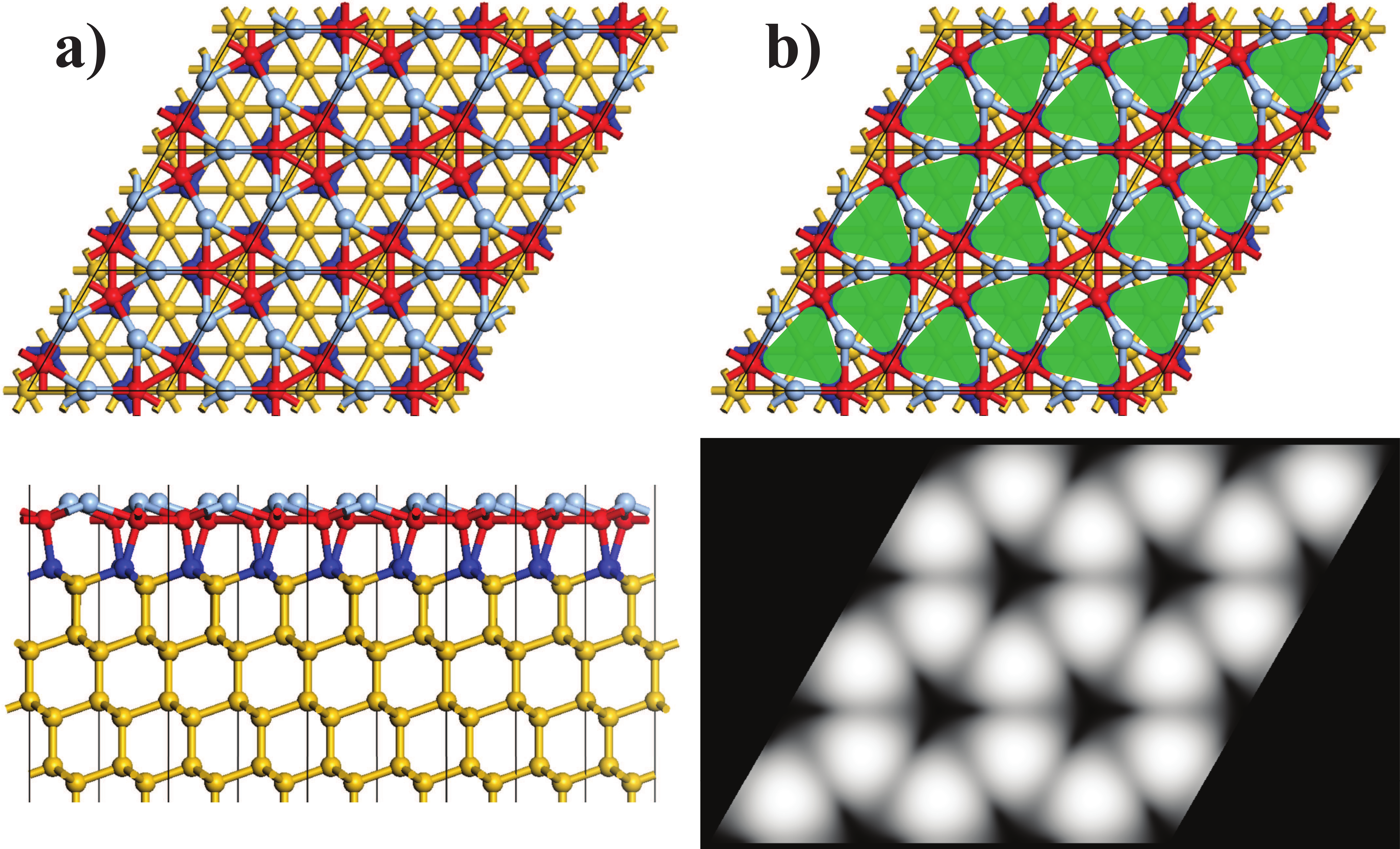}
\caption{(a) Ball and stick representation of the honeycomb-chain trimer (HCT) model \citep{Vlieg1}. The top and side views are presented in the top and bottom panels, respectively. The yellow balls are Si atoms sitting in almost ideal positions of the Si(111) substrate. The topmost layer of Si atoms and the layer below them are represented by the red and the blue balls, respectively. The Ag atoms represented by the light blue balls form the topmost atomic layer of the system by attaching to Si atoms (red balls) below them. (b) Schematic and calculated STM image of HCT structure are presented in top and bottom panels, respectively. The bright triangular spots that are observed in the STM image are represented by green triangles superimposed on the top view of the ball and stick model.}
\label{hct}
\end{figure}

The interlayer separation of multilayer silicene is very close but measurably different from that of Si(111). This inspired models of multilayer silicene that has bulk silicon-like interior with a modified surface structure. One such model is based on the Si(111)-Ag$\sqrt{3}\times\sqrt{3}$ system also mentioned in Chapter~\ref{Substrate}. The various surfaces obtained by the deposition of Ag on Si(111) substrate were studied intensively in the 1980s \citep{Lelay5,Loenen1,Vlieg1,Ding1}. One of the most favorable surfaces that were observed in experiments was the so called honeycomb-chain trimer (HCT) structure \citep{Vlieg1}. As seen in Fig.~\ref{hct}, the HCT model has a $\sqrt{3}\times\sqrt{3}$ honeycomb STM pattern that resembles the one observed in multilayer silicene experiments. Furthermore, the interlayer separation is close to the one obtained in experiments since the bulk region is basically Si(111). The HCT structure makes transition to the so-called inequivalent triangle (IET) structure at low temperatures. This transition could explain the spontaneous symmetry breaking observed in $\sqrt{3}\times\sqrt{3}$ silicene at low temperatures \citep{Chen2,Shirai1}. Finally, it was argued that the slope of the linear portion of the S$_1$ band formed by the Si(111)-Ag$\sqrt{3}\times\sqrt{3}$ surface is comparable to that of the linear bands observed in $\sqrt{3}\times\sqrt{3}$ silicene experiments \citep{Sato1,Chen1,Shirai1}. However, the HCT model does not account for the contraction of the lattice constant observed in experiments. Another model inspired by the bulk silicon structure is the tristable Si(111) bilayer grown on Ag substrate \citep{Guo2}. As mentioned in Chapter~\ref{Substrate}, the flip-flop motion that is suggested to give rise to the honeycomb STM topographs is not supported by convincing arguments. Also it does not explain the lattice contraction and can not be extended to multilayers.

\section{Silicites} \label{silicites}

Here we present a possible growth model of multilayer silicene that produces structures which are in a good agreement with experiments \citep{Cahangirov5}. A realistic growth simulation is really hard to do because one needs to take into account many experimental parameters. It is especially hard to run a molecular dynamics simulation long enough for the atoms to explore the whole energy landscape. Instead, we present structural relaxations accompanied with educated guesses.

\begin{figure}[t]
\includegraphics[width=11.5cm]{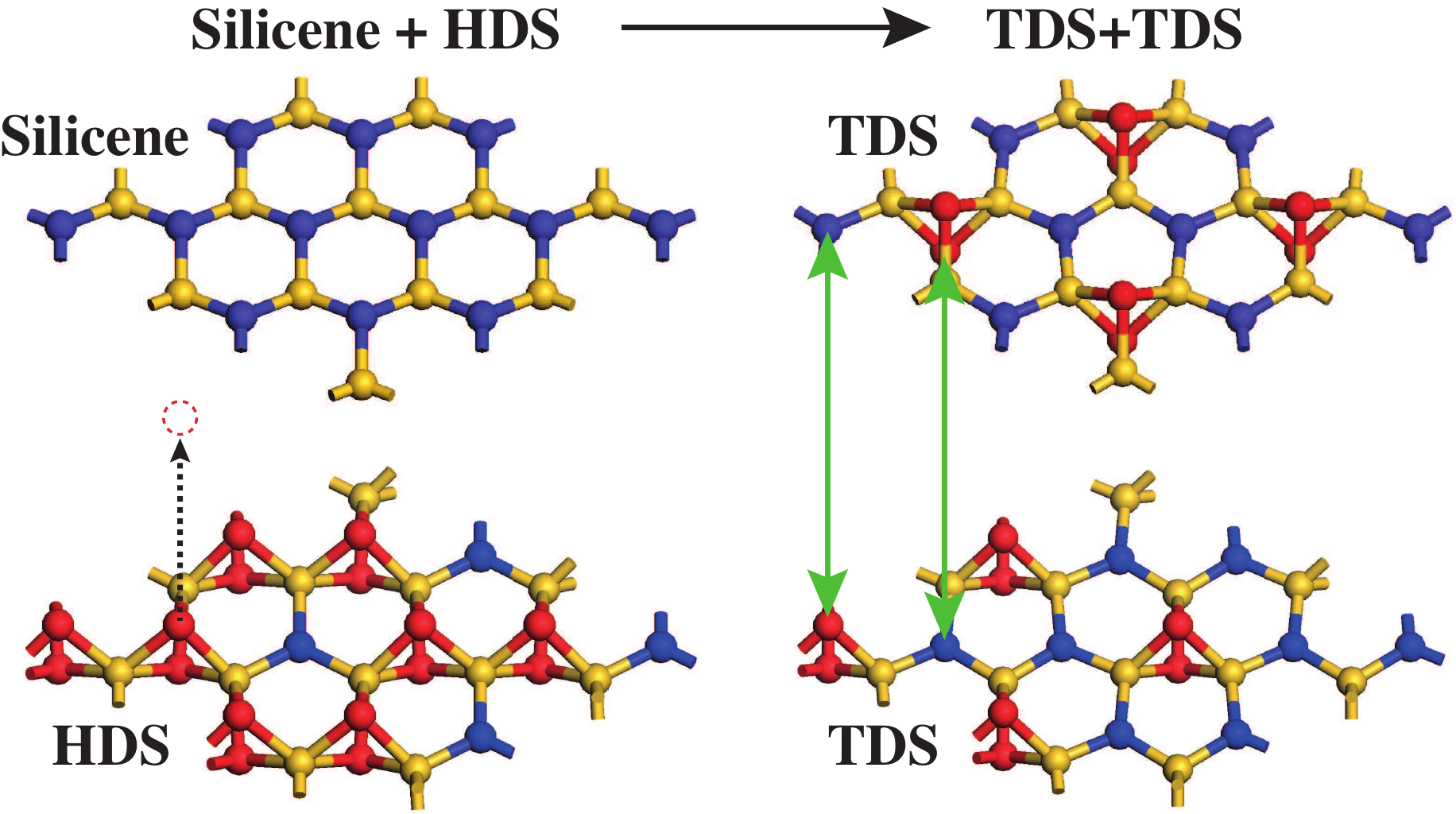}
\caption{A possible growth mechanism of multilayer silicene or silicites. When silicene is put on top of HDS, one of the dumbbell atoms transfer to the silicene layer, as shown by dashed black arrow. As a result, two TDS layers are formed, that connect to each other through covalent bonding between atoms shown by green arrows.}
\label{multigrowth}
\end{figure}

A silicene monolayer is first placed on top of the already formed HDS structure, as seen in Fig~\ref{multigrowth}(a). Upon relaxation of this system, one of the dumbbell atoms in HDS transfers to the silicene layer forming a dumbbell unit there. As a result, HDS loses one dumbbell unit and becomes TDS, while silicene sheet gains one dumbbell unit and also becomes TDS. The two TDS layers become connected to each other by covalent bonds. However, the number and strength of these vertical covalent bonds are less compared to the ones formed between two (111) planes of cubic diamond silicon (cdSi).

\begin{figure}[t]
\includegraphics[width=11.5cm]{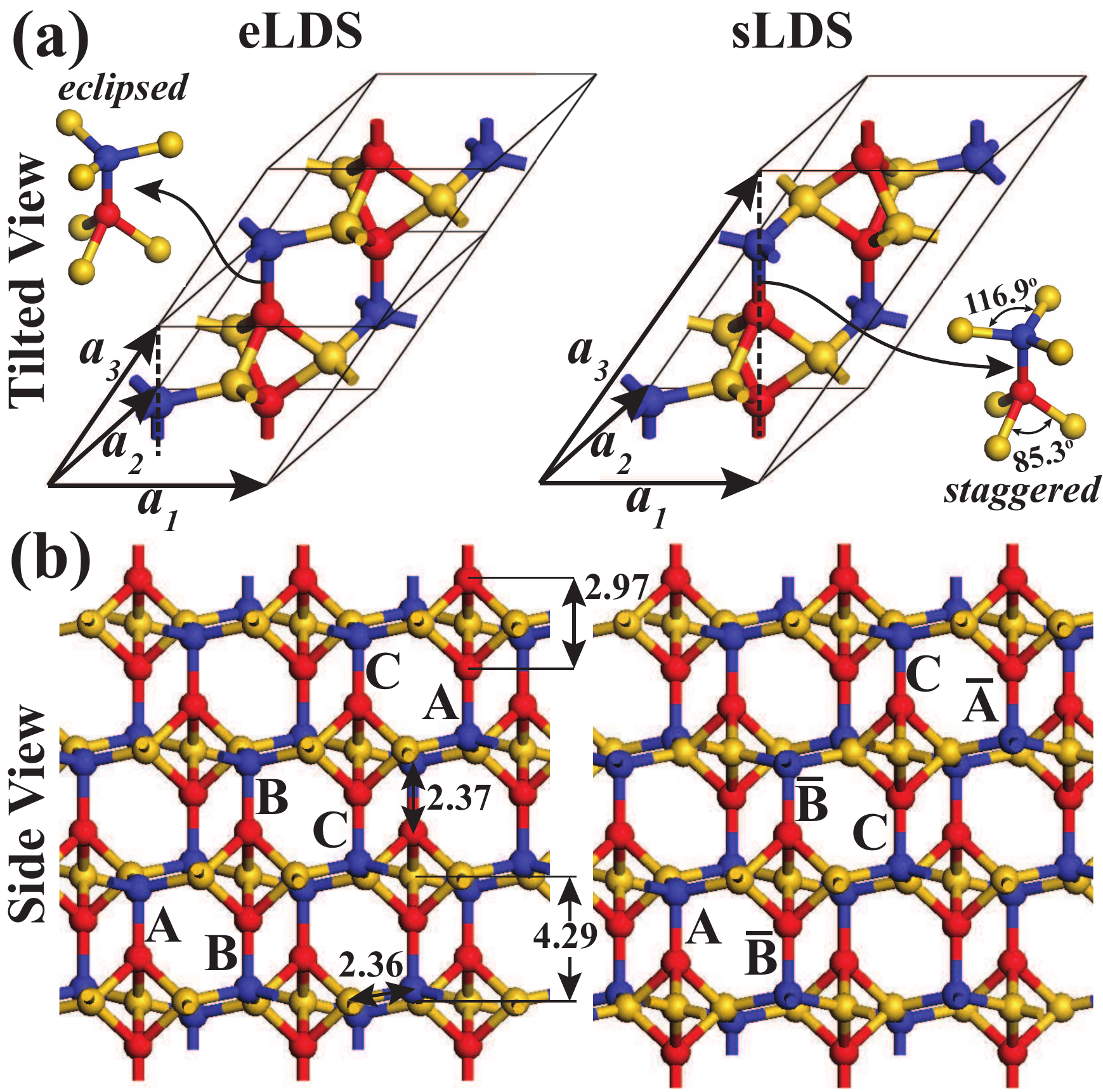}
\caption{(a) The double unit cell of eclipsed layered dumbbell silicite (eLDS) including $N$=7 Si atoms per unit cell and single unit cell of staggered layered dumbbell silicite (sLDS) including $N$=14 Si atoms per unit cell. (b) Side view showing the ABCABC... stacking of eLDS and the A$\bar{B}$C$\bar{A}$B$\bar{C}$A... stacking of sLDS. The bond lengths are given in Angstr\" om \citep{Cahangirov5}.}
\label{multistructure}
\end{figure}

If we continue depositing Si atoms onto the bilayer TDS system, the TDS layer on top will first transform to HDS. This HDS layer will follow the same faith as the original HDS structure, transforming itself to TDS by donating one dumbbell to create another TDS on top, which in turn will transform to yet another HDS layer. This process will continue producing multiple TDS layers connected to each other with an HDS layer on the very top. This is in agreement with experiments that continue to see the $\sqrt{3} \times \sqrt{3}$ honeycomb pattern in the STM measurements performed on multilayer silicene. It is possible to stack TDS layers in eclipsed or staggered fashion, as shown in Fig~\ref{multistructure}(a). The resulting bulk structures are named eclipsed (eLDS) and staggered (sLDS) layered dumbbell silicite, accordingly. All atoms in both eLDS and sLDS structure are fourfold coordinated. However, the covalent bonds significantly deviate from the ideal tetrahedral bonding angle of 109$\degree$. In the eLDS structure the TDS layers are just shifted by one third of their 2D unitcell vectors and stacked on top of each other. As shown in Fig~\ref{multistructure}(a), stacking follows ABCABC... and so on. It cannot be ABABAB... because in that case blue atoms would be connected to dumbbell atoms from both sides, which would unfavorably increase their coordination from four to five. The stacking of the sLDS structure is similar, but the layers are staggered with respect to each other. This is represented by a bar on top of the staggered layers.

Both eLDS and sLDS are open structures similar to the structure of cubic silicon. In fact, the mass densities of eLDS (2.10 g/cm$^3$) and sLDS (2.11 g/cm$^3$) are smaller than that of cdSi (2.28 g/cm$^3$). The interlayer distance in both eLDS and sLDS structures are around 4.3~\AA. This is in contrast to experiments that find the interlayer distance to be 3.1~\AA. Further work needs to be done to resolve this disagreement between theory and experiment. Due to the covalent bonds connecting LDS layers, the interlayer interaction is not like the weak van der Waals interaction found in graphite or MoS$_2$. However, these covalent bonds are sparse compared to those found between Si(111) layers in cubic silicon. The calculated cohesive energies are 4.42~eV and 4.43~eV per atom for eLDS and sLDS, respectively which is very close to that of cdSi (4.60~eV).

\begin{figure}[t]
\includegraphics[width=11.5cm]{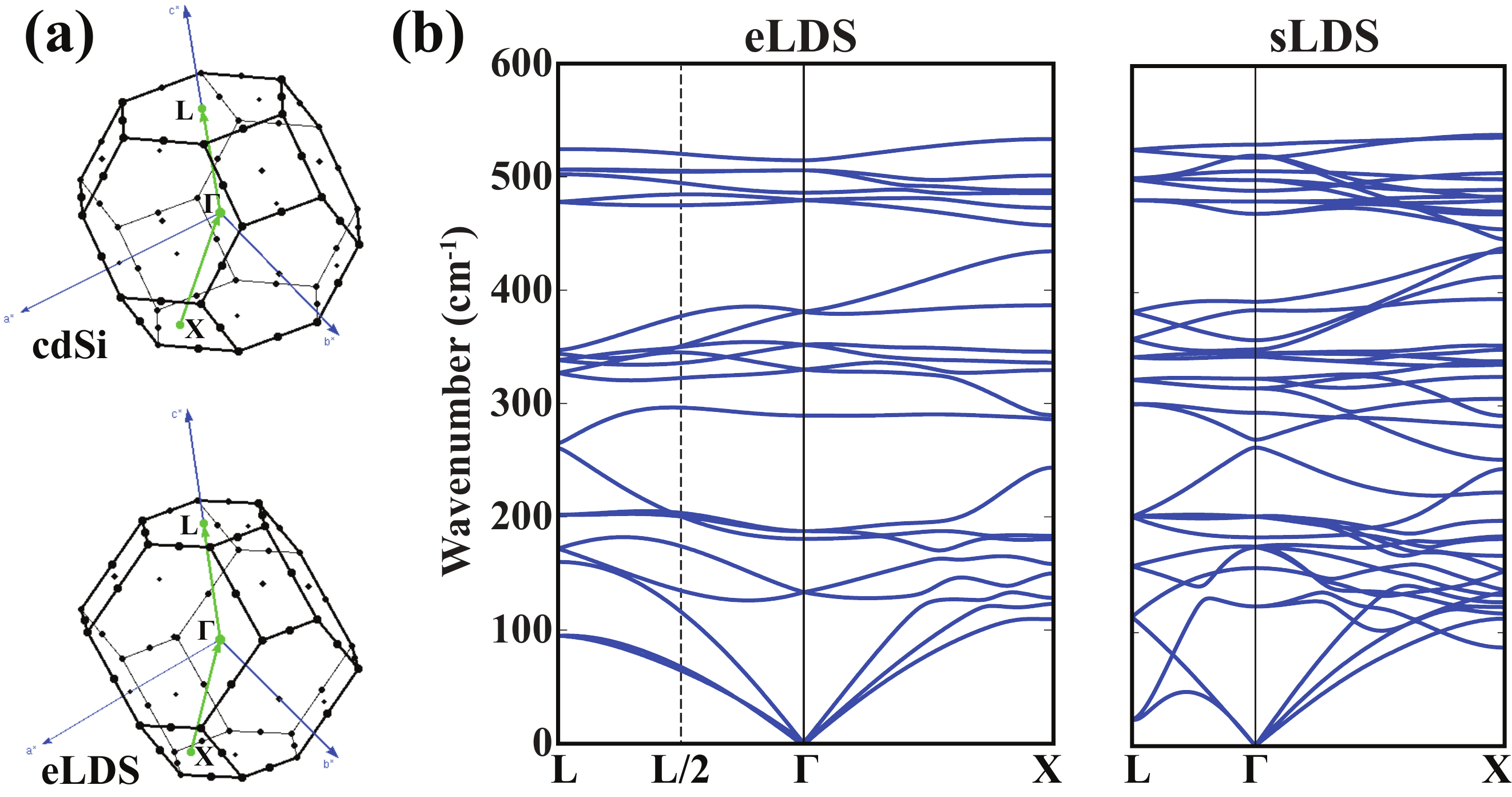}
\caption{(a) Brillouin Zones of cdSi and eLDS with relevant symmetry directions. (b) Phonon bands of eLDS and sLDS. (c) Phonon bands of iLDS.}
\label{multiphonons}
\end{figure}

The calculated vibrational frequencies of eLDS and sLDS phases are all found to be positive. The absence of negative frequencies is taken as an evidence that these layered phases are stable. The phonon bands of these structures presented in Fig.~\ref{multiphonons} disclose interesting dimensionality effects. Specific optical branches are flat and the lower lying branches overlap with the acoustical branches. One of the acoustical branches of the sLDS structure dips in near the point L, indicating phonon softening. We have also performed molecular dynamics simulations where (3$\times$3$\times$4) supercells of eLDS and (3$\times$3$\times$2) supercells of sLDS were kept at 1000 K for 4 ps. No structural deformation was observed in the course of these simulations, which corroborates the stability of these materials.

\begin{figure}[t]
\includegraphics[width=11.5cm]{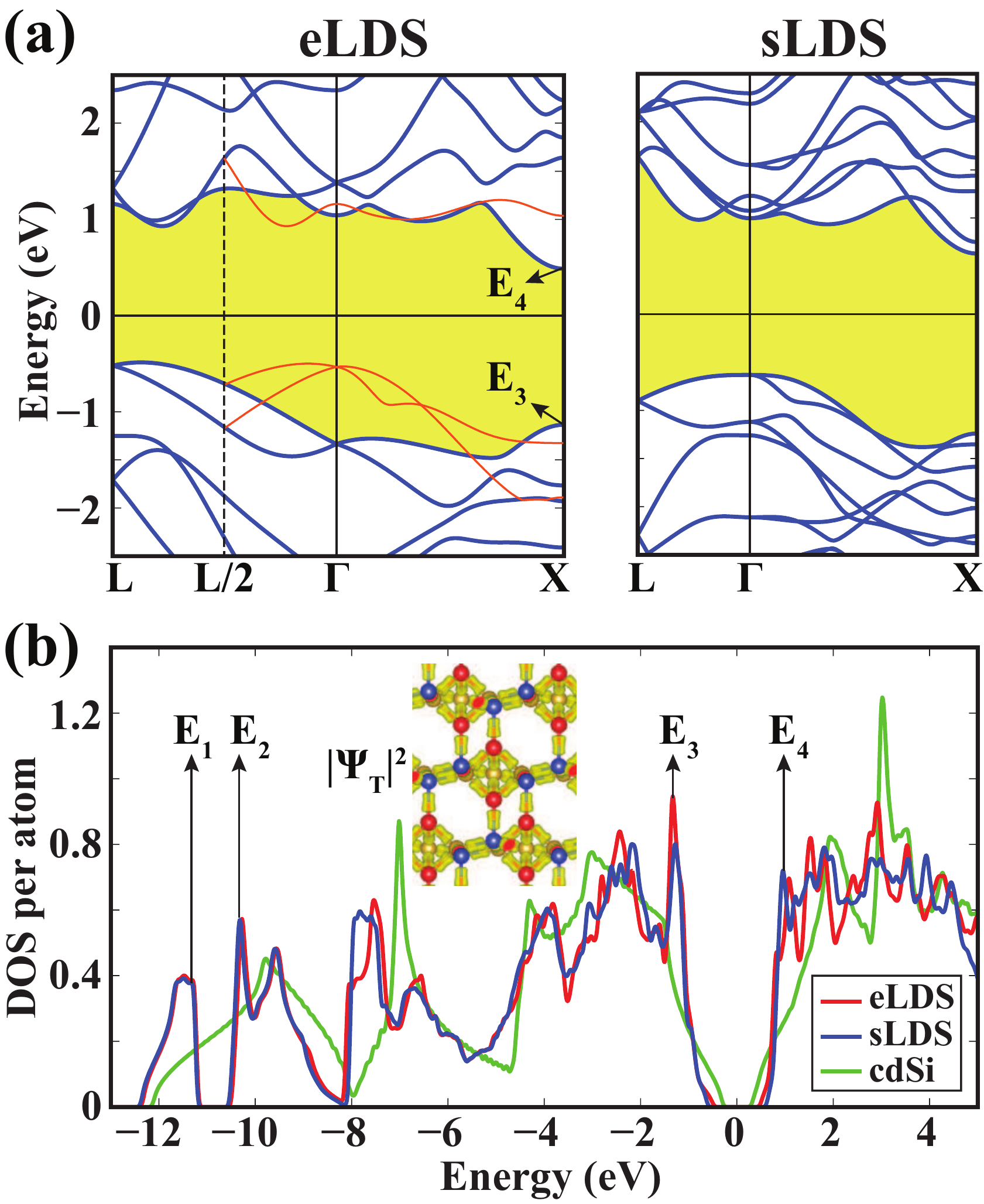}
\caption{(a) Energy band structure of eLDS and sLDS. Zero of energy is set to the the Fermi level. Bands of eLDS folded by doubling the unit cell along $a_{3}$ are shown by red lines. (b) Normalized densities of states (DOS) of eLDS, sLDS and cdSi. The isosurfaces of the total charge density shown by inset confirm the layered nature.}
\label{multibands}
\end{figure}

The layered character of eLDS and sLDS can be conveniently substantiated by investigating the in-plane and out of plane Young modulus and by comparing them with those of cdSi. The perpendicular Young's moduli of eLDS and sLDS are calculated as $Y_{\perp}$=79.6 GPa and 76.4 GPa, respectively,  while the Young's modulus of cdSi along [111] direction is 176.0 GPa and hence more than twice the value of LDS phases. In contrast, the in-plane Young's modulus calculated within  TDS layers of eLDS and sLDS are relatively higher, and are 176.3 GPa and 161.9 GPa, respectively. These values are comparable with the Young's modulus of cdSi calculated in the (111) plane, which is 200 GPa. The dramatic differences between the Young's modulus of LDS structures and that of cdSi calculated in the direction perpendicular to the layers confirm the layered nature of eLDS and sLDS phases.

\begin{figure}[t]
\includegraphics[width=11.5cm]{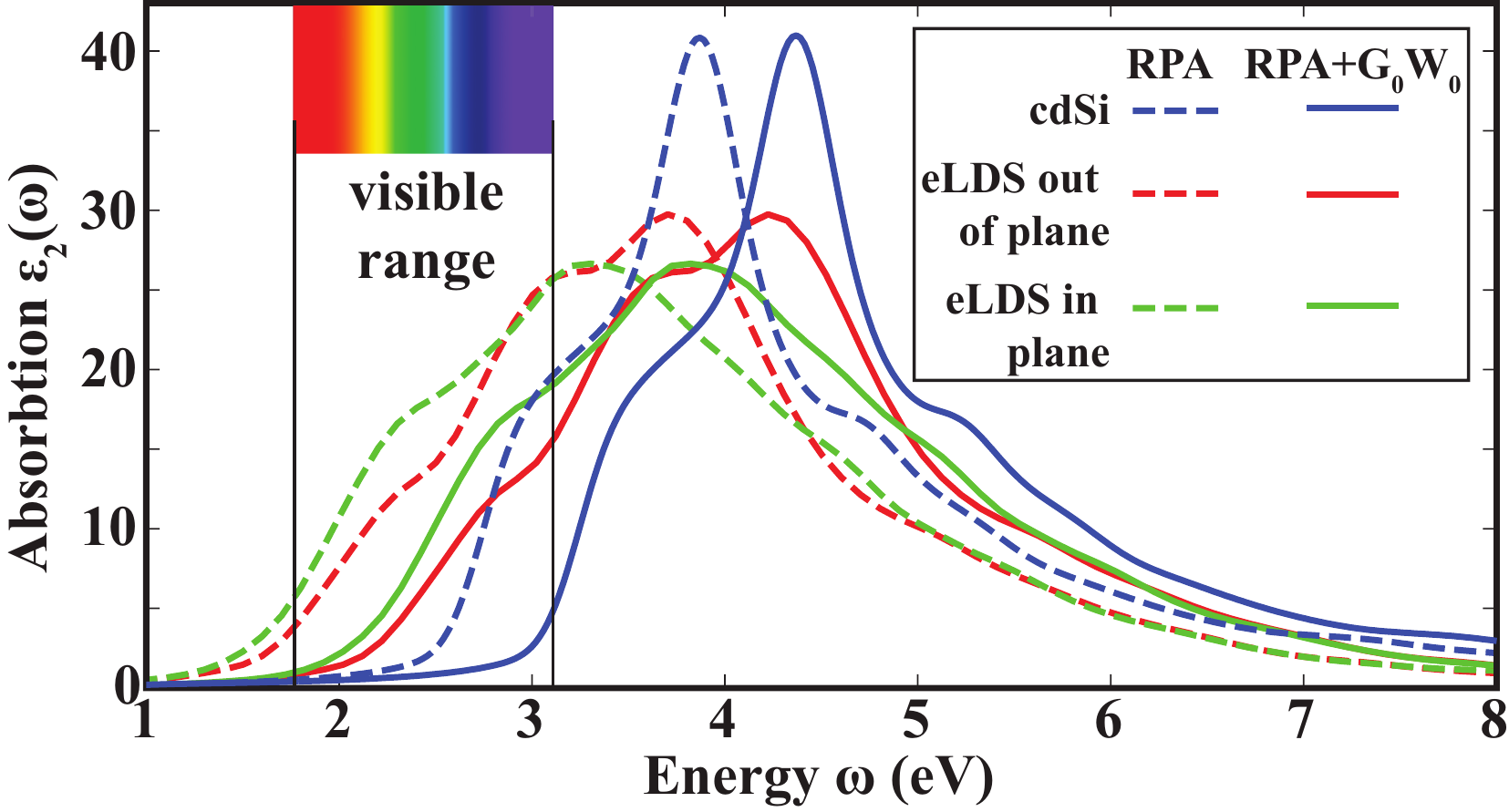}
\caption{The calculated Kohn-Sham and G$_0$W$_0$ RPA optical absorption spectra for eLDS and cdSi.}
\label{absorbtion}
\end{figure}

The electronic structures of eLDS and sLDS phases have indirect band gaps, which are wider than that of cdSi, as shown in Fig.~\ref{multibands}(a). The calculated indirect (direct) band gaps of eLDS and sLDS are 0.98 (1.43) eV and 1.26 (1.65) eV, respectively. The indirect band gap of cdSi is 0.62 eV at the DFT-PBE level while it is increased to 1.12 eV upon including many-body self-energy corrections at the G$_0$W$_0$ level \citep{Hedin1,Shishkin1}. With G$_0$W$_0$ correction the indirect band gap of eLDS increased to 1.52 eV. Indirect (direct) band gaps of eLDS and sLDS calculated by HSE06 hybrid functional are 1.92 eV (2.37 eV) and 1.88 eV (2.26 eV), respectively.

Owing to the different Brillouin zones it is difficult to directly compare the band structures of LDS and cdSi. Therefore, the effects of the layered character on the electronic structure are sought in the normalized densities of states (DOS). Figure~\ref{multibands}(b) shows the normalized DOSs of eLDS, sLDS and cdSi. Except for some peak shifts, the DOSs of silicites are similar. Owing to the fourfold coordination of Si atoms in all structures, the overall features of DOSs of LDS structures appear to be reminiscent of that of cdSi. This confirms the fact that the overall features of the bands of cdSi can be obtained within the first nearest neighbor coupling \citep{Harrison1}. The total charge density, $|\Psi_T|^2$ presented by inset, depicts that electrons are mainly confined to TDS layers. This is another clear manifestation of the layered character of eLDS and sLDS phases. On the other hand, significant differences are distinguished in the details of the electronic energy structures due to deviations from tetrahedral coordination: (i) Indirect band gaps relatively larger than that of cdSi can offer promising applications in micro and nanoelectronics. (ii) Sharp peaks $E_3$ and $E_4$ near the edges of the valence and conduction bands, originate from the states, which are confined to TDS layers and can add critical functionalities in optoelectronic properties. (iii) A gap opens near the bottom of the valence band at $\sim$ -11 eV; its edge states are also confined to TDS layers.

The in-plane and out of plane static dielectric responses also reflect the layered nature of silicite. As a matter of fact, the calculated in-plane dielectric constant of eLDS (sLDS) is $\epsilon_{\parallel}$=12.52 (12.85), while its out of plane dielectric constant is $\epsilon_{\perp}$=11.69 (11.56). Those values are contrasted with the uniform dielectric constant, $\epsilon$=12.19 of cdSi. In Fig.~\ref{absorbtion} we present the optical absorption spectra of eLDS and cdSi calculated at the RPA level using the Kohn-Sham wave functions and G$_0$W$_0$ corrected eigenvalues. The frequency dependent dielectric matrix takes different values in the in-plane and out of the plane directions of eLDS while for cdSi it is isotropic. One can see that the optical absorption of eLDS is significantly enhanced in the visible range compared to cdSi which makes it a potential candidate material for photovoltaic applications. This enhancement is still present when we rigidly shift the absorption spectra by the amount we get from G$_0$W$_0$ corrections \citep{Onida1}.
\chapter{Germanene, Stanene and Other 2D Materials} \label{Germanene}

Germanene and stanene (also sometimes written stannene or called tinene) are 2D materials composed of germanium and tin atoms respectively arranged in a honeycomb structure similarly to graphene and silicene.  The atomic structure of freestanding germanene and stanene is buckled like in the case of silicene (see Fig.~\ref{relaxation}). DFT calculations \citep{Kresse1} performed by projector augmented wave (PAW) method \citep{Blochl1} and adopting PBE functional \citep{Perdew1} result in a lattice constants 4.06~\AA~and 4.67~\AA~and buckling heights of 0.69~\AA~ and 0.85~\AA~for germanene and stanene respectively. The structure of germanene was first theoretically proposed together with that of silicene \citep{Takeda1} while stanene was explored later \citep{Cahangirov1,Sahin1,Xu1}.

\begin{figure}[t]
\includegraphics[width=11.5cm]{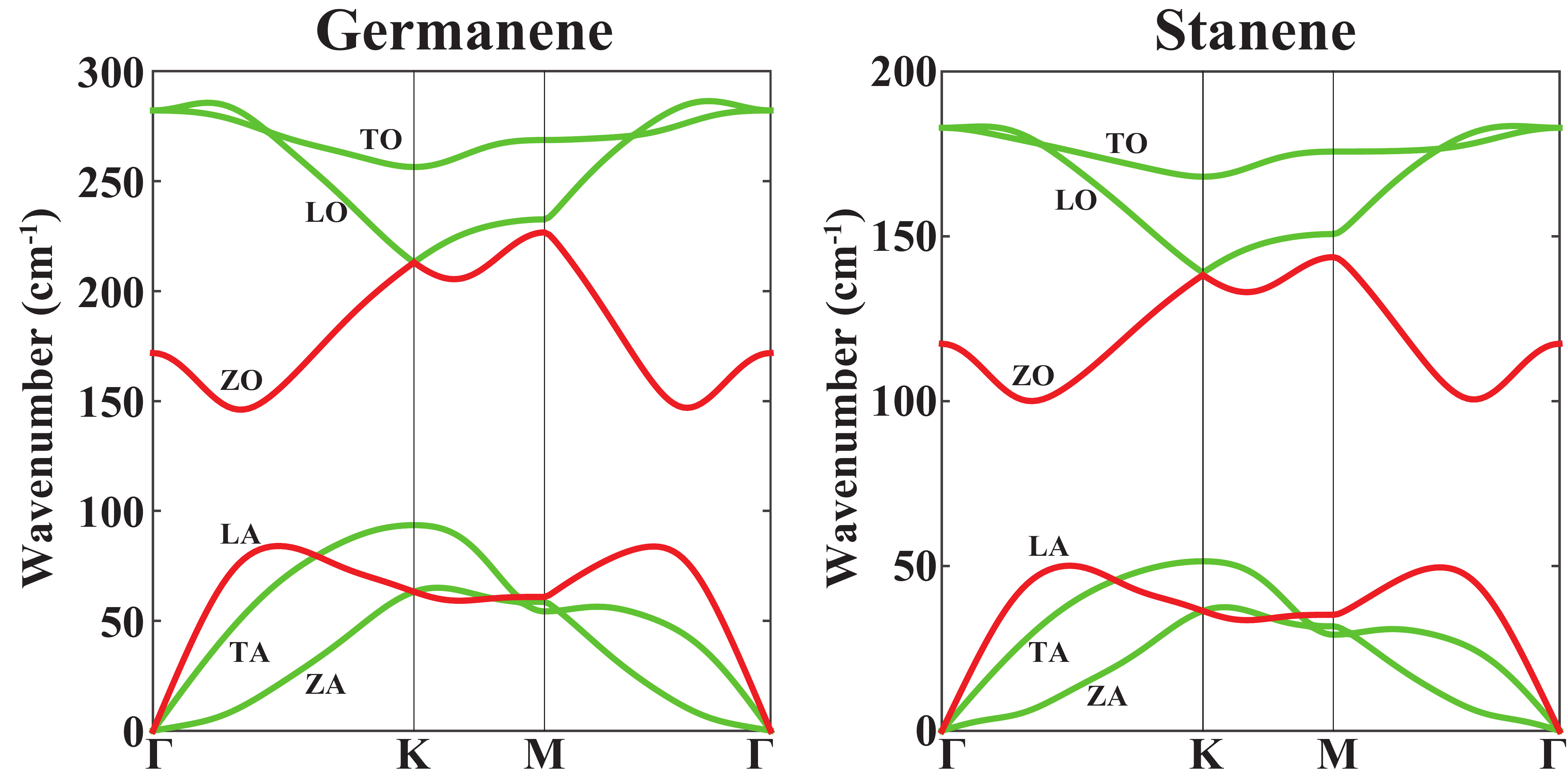}
\caption{The phonon modes of buckled germanene and stanene. The out of plane, transverse and longitudinal acoustic and optical modes are denoted by ZA, TA, LA, ZO, TO and LO, respectively. The LA and ZO modes of germanene and stanene (shown by red lines) change significantly upon transformation from a planar (not shown here) to the buckled structure. A similar change is observed in silicene as shown in Fig~\ref{phonons}.}
\label{gphonons}
\end{figure}

The vibrational modes of germanene and stanene are much softer than those of silicene and graphene. In fact, the maximum of the longitudinal optical (LO) mode is 183 cm$^{-1}$, 286 cm$^{-1}$,  555 cm$^{-1}$ and 1605 cm$^{-1}$ for stanene, germanene, silicene and graphene, respectively. Some studies have found imaginary frequencies in the out-of-plane acoustic (ZO) modes of germanene and stanene near the Brillouin zone center \citep{Cahangirov1,Sahin1,Tang1}. However, these imaginary frequencies indicating instability are just artifacts caused by the numerical accuracy that is not enough to evaluate the modes of germanene and stanene that are too soft. One way to get rid of them is to use a finer mesh grid in the DFT calculation. The phonon modes of germanene and stanene resulting from this kind of calculation are presented in Fig.~\ref{gphonons}. Here, one can see the similarities between the modes of germanene and stanene with those of silicene, shown in Fig~\ref{phonons}.

The fact that all phonon modes are positive indicate the stability of germanene and stanene. However, the larger ionic radius of germanium and tin atoms make them even more in favor of $sp^3$ hybridization compared to silicon. This is mirrored in the bond angles of germanene and stanene that are $\theta$=112.4$\degree$  and 111.4$\degree$ which can be translated into the hybridization $sp^D$ through $D=-1/cos(\theta)$ yielding 2.63 and 2.74, respectively. The corresponding value for silicene is 2.27, as shown in Chapter~\ref{Freestanding}. This means that the 2D structure of germanene and stanene are even more delicate and vulnerable to forming 3D structures by clustering. However, stabilization of germanene and stanene could be possible on certain substrates as it is the case for silicene.

\begin{figure}[t]
\includegraphics[width=11.5cm]{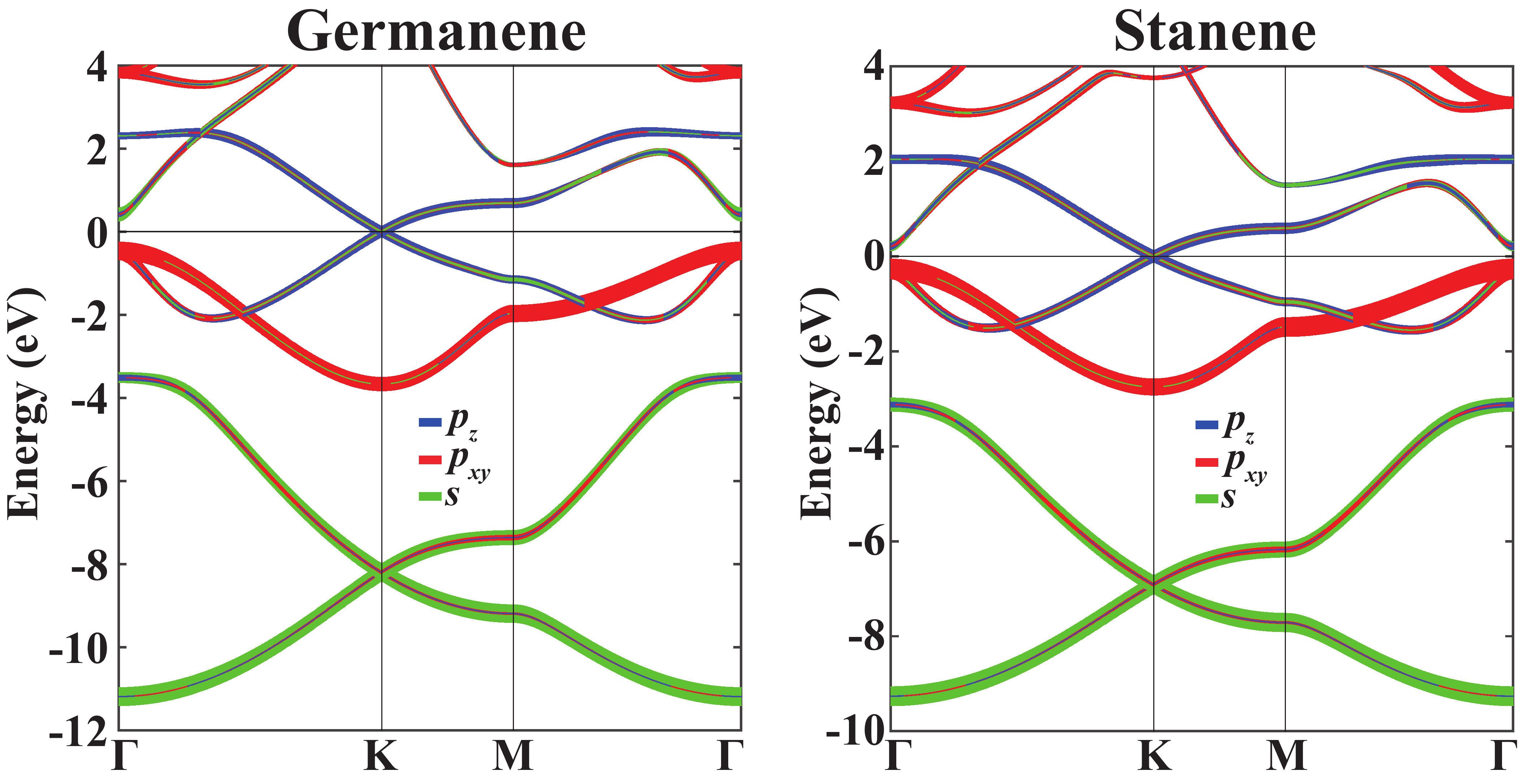}
\caption{The electronic structures of buckled germanene and stanene. The orbital character of each state is shown by colors. The width of each line is proportional to the contribution from the orbital denoted by the color of that line. The orbital with lower contribution is plotted on top of the one that has higher contribution, which makes all contributions visible.}
\label{gbands}
\end{figure}

The electronic energy bands of germanene and stanene are presented in Fig.~\ref{gbands}. The linearly crossing bands are preserved both in buckled germanene and stanene. The contribution from $s$ and $p_{xy}$ states to these bands is enhanced due to higher buckling compared to silicene. The Fermi velocity was calculated (using PAW-PBE) to be 0.52$\times$10$^6$ m/s for germanene which is almost equal to that of silicene (see Chapter~\ref{Freestanding}) while for stanene it is 0.47$\times$10$^6$ m/s. The most dramatic change, compared to silicene, happens near the Brillouin zone center ($\Gamma$-point) where the gap shrinks from 3.13~eV for silicene to 0.88~eV and 0.48~eV for germanene and stanene, respectively. This gap shrinks even further, down to zero, when stanene is fluorinated. However, it opens back if spin-orbit coupling is included in the calculation \citep{Xu1}. This results in an intriguing electronic structure discussed below.

The exotic electronic properties of silicene, like being a topological insulator, are further pronounced in germanene and stanene due to an even larger spin-orbit coupling in germanium and tin atoms compared to silicon. Including spin-orbit coupling opens a 1.55~meV, 23.9~meV and 73.5~meV gap between the linearly crossing bands of silicene, germanene and stanene \citep{Liu1,Liu2}. This is already enough to observe the quantum spin Hall effect at temperatures much higher than that of the liquid nitrogen. Remarkably, it is possible to go even further by functionalizing stanene \citep{Xu1}. Fluorinating stanene saturates its $\pi$-orbitals opening a 2~eV band gap at the K-point. However, it also closes the gap at the $\Gamma$-point, if spin-orbit is not included. Including spin-orbit coupling induces a 0.3~eV gap making fluorinated stanene a large-gap quantum spin Hall insulator that can operate at room temperature. Similar to the well known case of HgTe quantum wells \citep{Bernevig1,Konig1}, the electronic states of fluorinated stanene become topologically nontrivial through $s$-$p$ type band inversion at the $\Gamma$-point \citep{Xu1}. No such inversion occurs in hydrogenated stanene, also called stanane. Accordingly, the nanoribbons of stanene and fluorinated stanene have helical edge states linearly crossing at the Fermi level while stanane nanoribbons don't have such states \citep{Xu1}.

Another way to functionalize stanene is to pattern it with the dumbbell structures \citep{Tang1}. Similar to the silicene case (see Chapter~\ref{Freestanding}) the most energetically favorable structure is large honeycomb dumbbell stanene (LHDT). The LHDT structure is 0.18~eV/atom more favorable than buckled stanene. The band structure calculation including spin-orbit coupling results in a 40~meV non-trivial gap at the $\Gamma$-point of LDHT. In this case, the spin-orbit coupling lifts the degeneracy of two $p_{xy}$ states at the valence band edge and the band inversion occurs between the $p_z$ state at the conduction band edge and one of the  $p_{xy}$ states. Similar to the case of fluorinated stanene, zigzag nanoribbons of LDHT were shown to have helical edge states. 

\begin{figure}[t]
\includegraphics[width=11.5cm]{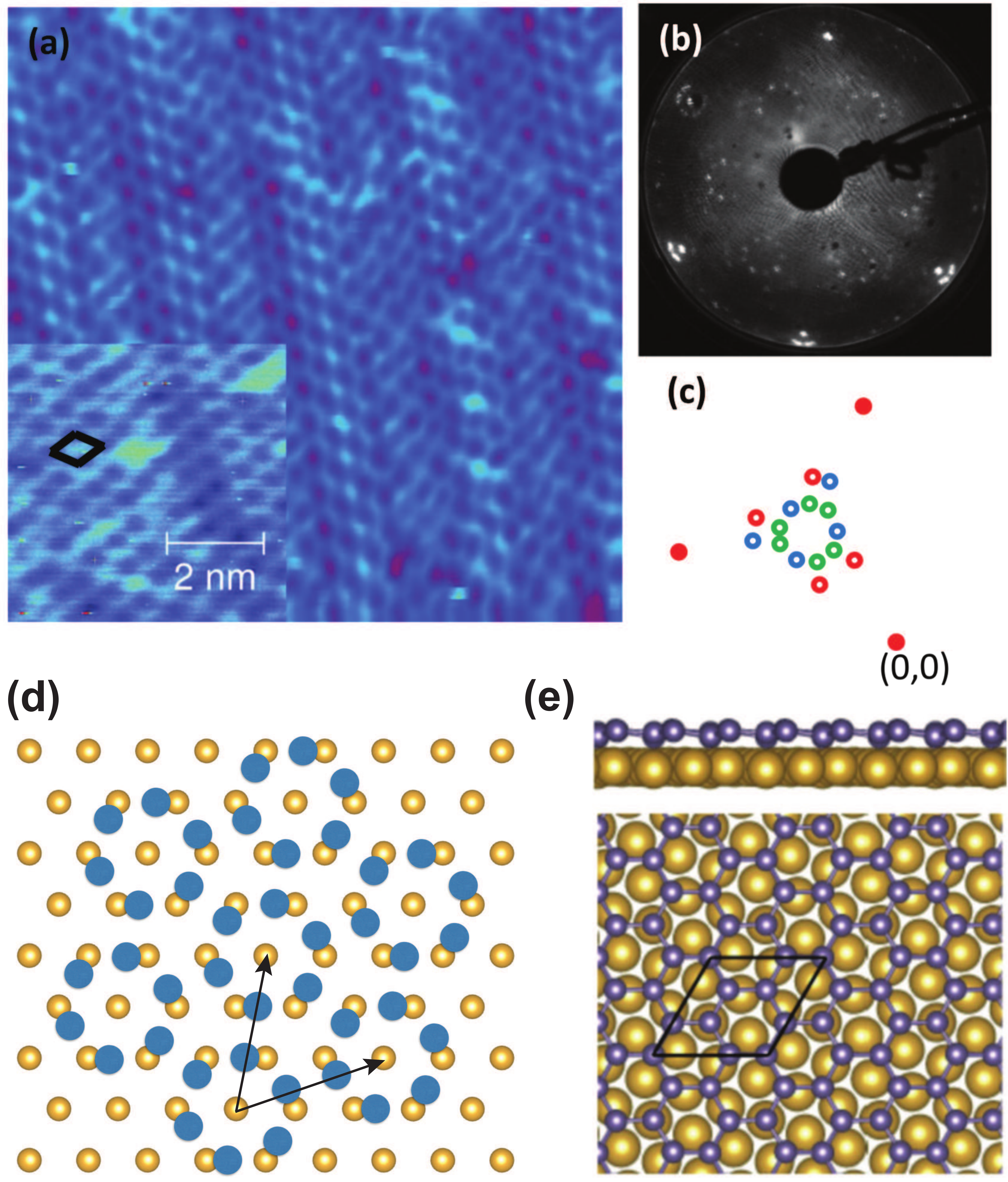}
\caption{(a) 16.2 nm$\times$16.2 nm STM image of the modulated honeycomb $\sqrt{7}\times\sqrt{7}$ superstructure with a close-up in the bottom left corner (sample bias: −1.12 V, 1.58 nA; the $\sqrt{7}\times\sqrt{7}$ unit cell is drawn in black); (b) associated LEED pattern taken at 59 V; (c) schematic illustration of one sixth of the pattern, filled dots: hidden (0,0) spot and integer order spots, open circles: spots corresponding to the $\sqrt{7}\times\sqrt{7}$ superstructure (in red), the $\sqrt{19}\times\sqrt{19}$ one (in green) and the 5$\times$5 (in blue). (d) Initial position of Ge atoms (blue) on the top layer of Au(111) (gold). In this configuration, each Ge atom has the same environment. The arrows correspond to both $\sqrt{3}\times\sqrt{3}$ supercell of germanene and $\sqrt{7}\times\sqrt{7}$ supercell of Au(111) surface. (e) Top and side view of atomic structure after optimization. Adapted from \citep{Davila1}.}
\label{geau1}
\end{figure}

Very recently, both germanene and stanene were synthesized on substrates. Germanene was synthesized on Au(111) substrates by molecular beam epitaxy \citep{Davila1}. In this respect, synthesis of germanene is very similar to that of silicene \citep{Vogt1}, only here silver sample was exchanged for a gold (111) one and the silicon source was replaced by a germanium evaporator to deposit Ge atoms onto a clean Au(111) surface prepared in a standard fashion by Ar+ ion bombardment and annealing. Low energy electron diffraction (LEED) and STM observations were performed at room temperature (RT) at different stages of the growth, carried out at several substrate temperatures to determine potential candidates for germanene in an overall multiphase diagram, as was already the case for silicon deposition onto Ag(111). Germanene phases were obtained at 200 $\degree$C growth temperature at about 1 ML coverage, as estimated from the 32 \% attenuation of the Au 4$f_{7/2}$ core level intensity. It covers extended regions, larger than 50$\times$50 nm$^2$ in size, with a honeycomb appearance and a very small corrugation of just 0.01~nm, as well as with a weak long range modulation in STM imaging, as displayed in Fig.~\ref{geau1}. Core-level spectroscopy revealed that Ge 3d states can be fitted with a very narrow, asymmetric single component essentially signaling a unique environment of the germanium atoms at the surface while spots corresponding to $\sqrt{7}\times\sqrt{7}$, $\sqrt{19}\times\sqrt{19}$ and 5$\times$5 superstructure of Au(111) were observed in the LEED pattern. STM image also supports a $\sqrt{7}\times\sqrt{7}$ superstructure.

Actually, it is possible to place Ge atoms on top of the Au(111) surface in such a way that each of them feels the same environment. This arrangement is shown in Fig.~\ref{geau1}(d). It turns out that this particular configuration corresponds to $\sqrt{3}\times\sqrt{3}$ germanene matched by $\sqrt{7}\times\sqrt{7}$ Au(111). This explains the $\sqrt{7}\times\sqrt{7}$ superstructure observed in the LEED and the STM measurements. Furthermore, this matching is made possible by stretching germanene layer which would make it rather flat. This is also in accordance with STM findings. DFT calculations starting from such a configuration results in optimized structures that reproduces the experimental STM profile \citep{Davila1}.

Even more uniform phase of germanene was synthesized at only 80$\degree$C on the Al(111) surface \citep{Derivaz1}. In this case, 2$\times$2 supercell of germanene is matched by 3$\times$3 supercell of the Al(111) surface. Stanene, on the other hand, was synthesized on Bi$_2$Te$_3$(111) substrate \citep{Zhu1}. Due to the lattice match stanene has no reconstruction. STM and ARPES measurements have confirmed the atomic and electronic structures of stanene on Bi$_2$Te$_3$(111) obtained by DFT calculations.

\begin{figure}[t]
\begin{center}
\includegraphics[width=9.5cm]{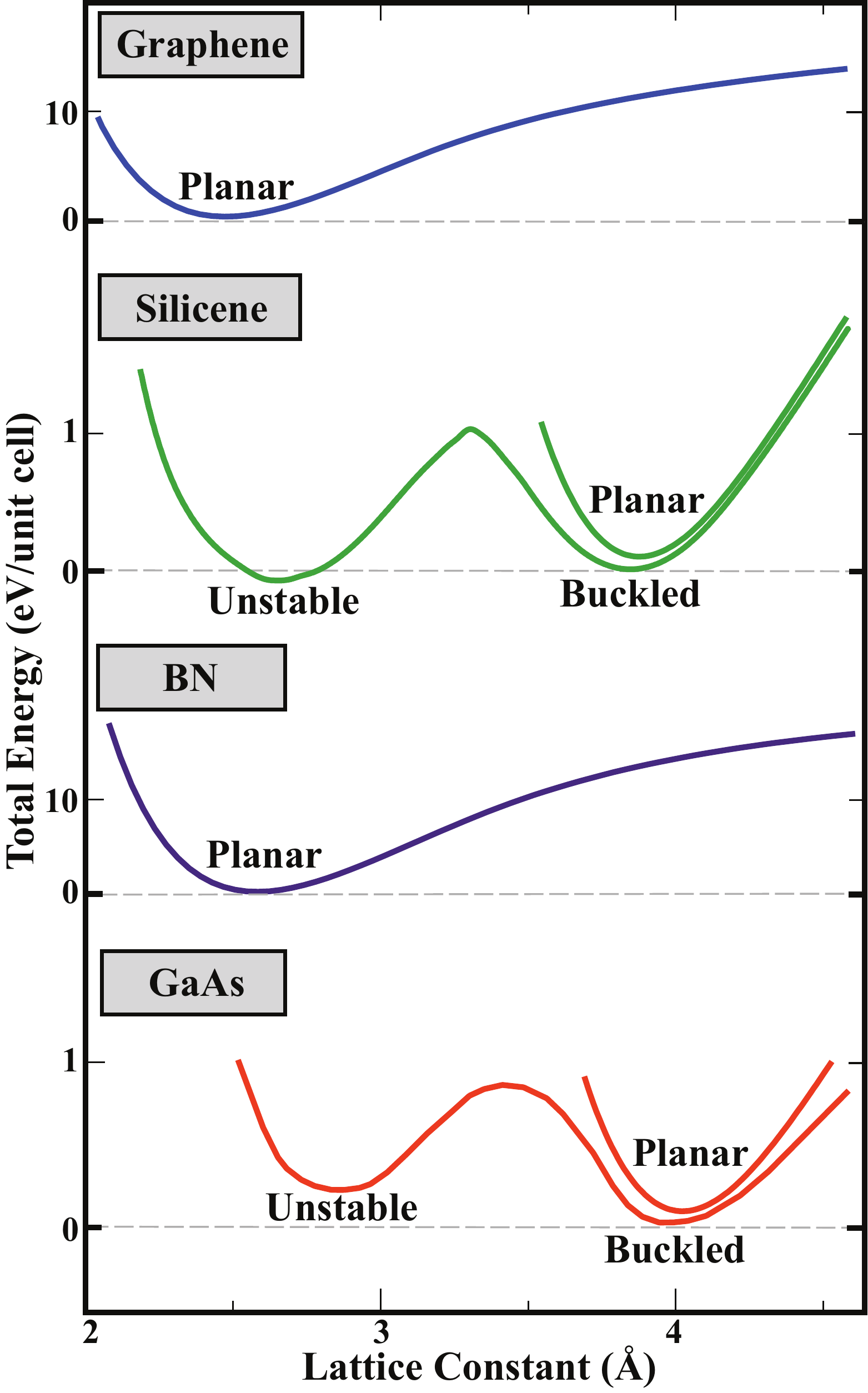}
\end{center}
\caption{Comparison of the total energy per unit cell versus the lattice constant for graphene, silicene, hexagonal BN sheet and GaAs sheet. For each structure, zero of energy is set to the minimum energy of a stable configuration. Adapted from \citep{Sahin1}.}
\label{energetics}
\end{figure}

\section{Binary Compounds of Group IV and Group III-V Elements}

The elemental 2D materials like silicene, germanene and stanene are natural extensions of graphene. One can extend the search for novel 2D materials further by taking inspiration from hexagonal boron nitride (BN) that has been studied in detail \citep{Rubio1,Chuanhong1}. This approach leads to consideration of the binary compounds of Group IV elements like SiC, GeSn and etc. and those of Group III-V elements like GaAs, BP and others \citep{Sahin1}.

In the Fig.~\ref{energetics} we present a general trend for the energy per unit cell versus the lattice constant. Some structures like graphene and hexagonal boron nitride have only one energy minimum corresponding to a planar geometry while others like silicene and GaAs have three energy minima. However, only the slightly buckled structure that has a lattice constant close to that of the planar structure is stable. The other two minima correspond to unstable structures. In fact, by analysing all possible binary compounds of Group IV and Group III-V, it was shown that structures having at least one element from the first row (that is B, C or N) have stable planar minimum while others are buckled \citep{Sahin1}.

\begin{figure}[t]
\begin{center}
\includegraphics[width=11.5cm]{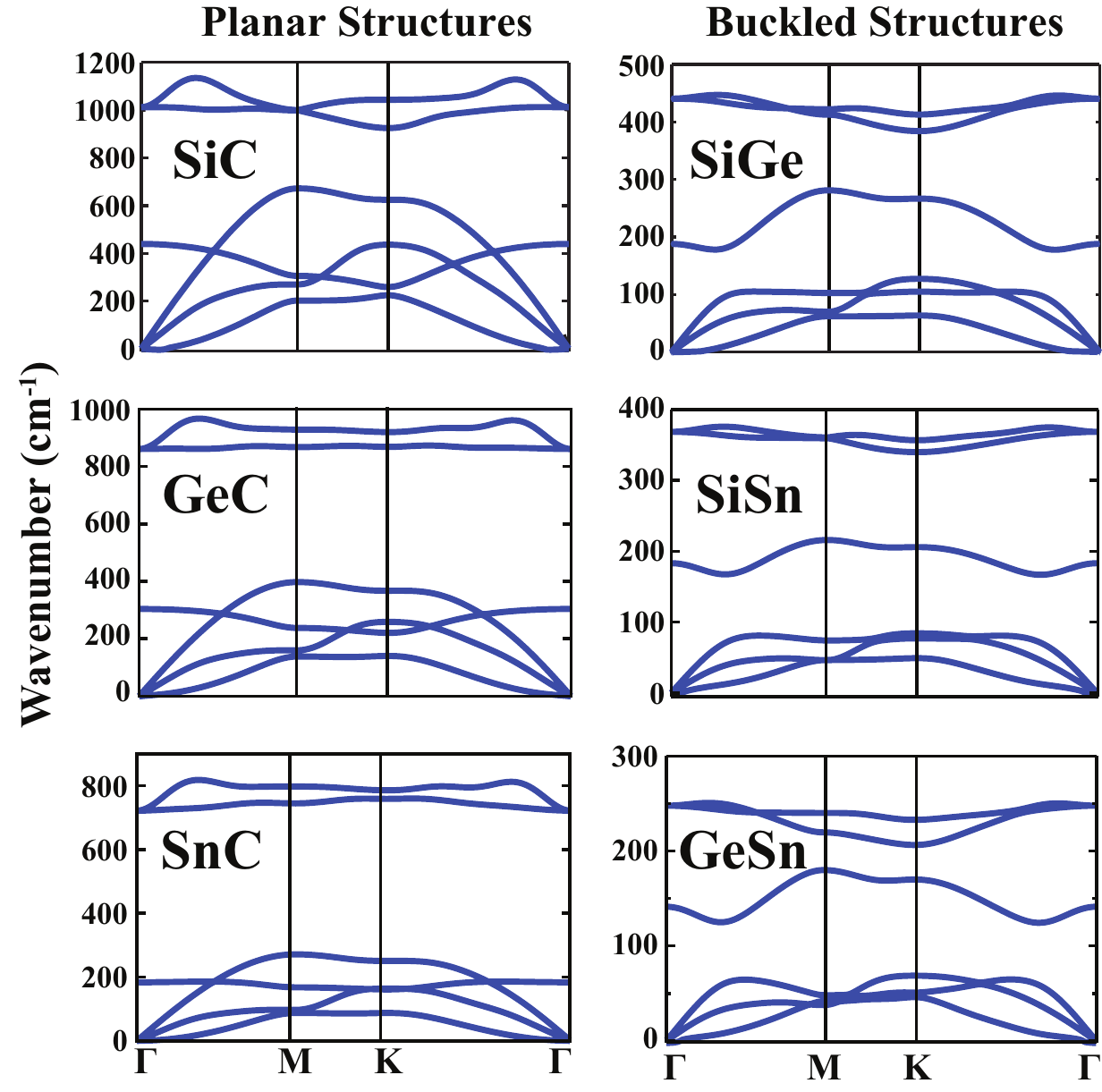}
\end{center}
\caption{Phonon dispersions of planar and buckled structures of 2D materials constituted by two different Group IV elements. Adapted from \citep{Sahin1}.}
\label{ivphonon}
\end{figure}

Phonon dispersions of 2D binary compounds of Group IV elements are presented in the Fig.~\ref{ivphonon}. The longitudinal and transverse optical modes are separated from other modes. This separation is larger when the constituting elements are further apart in the periodic table. In planar structures, the out-of-plane optical mode mixes with acoustical modes while in the buckled structures this mode is in between acoustical and other optical modes. Note that, all modes are positive for any combination of Group IV elements, which means that they are all stable.

\begin{figure}[t]
\begin{center}
\includegraphics[width=11.5cm]{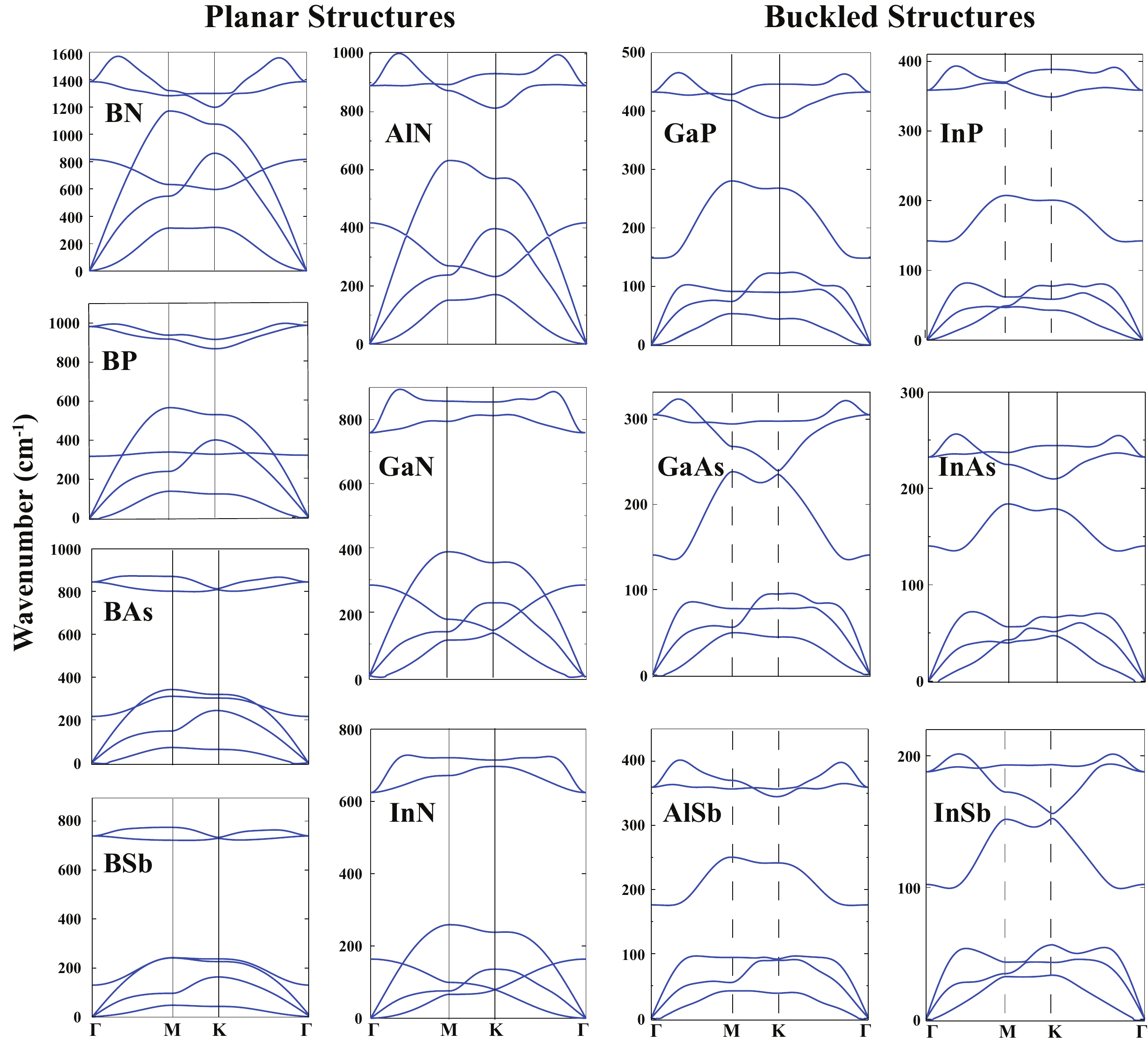}
\end{center}
\caption{Phonon dispersions of planar and buckled structures of 2D materials constituted by Group III-V elements. Adapted from \citep{Sahin1}.}
\label{iiivphonon}
\end{figure}

\begin{figure}[t]
\begin{center}
\includegraphics[width=8.5cm]{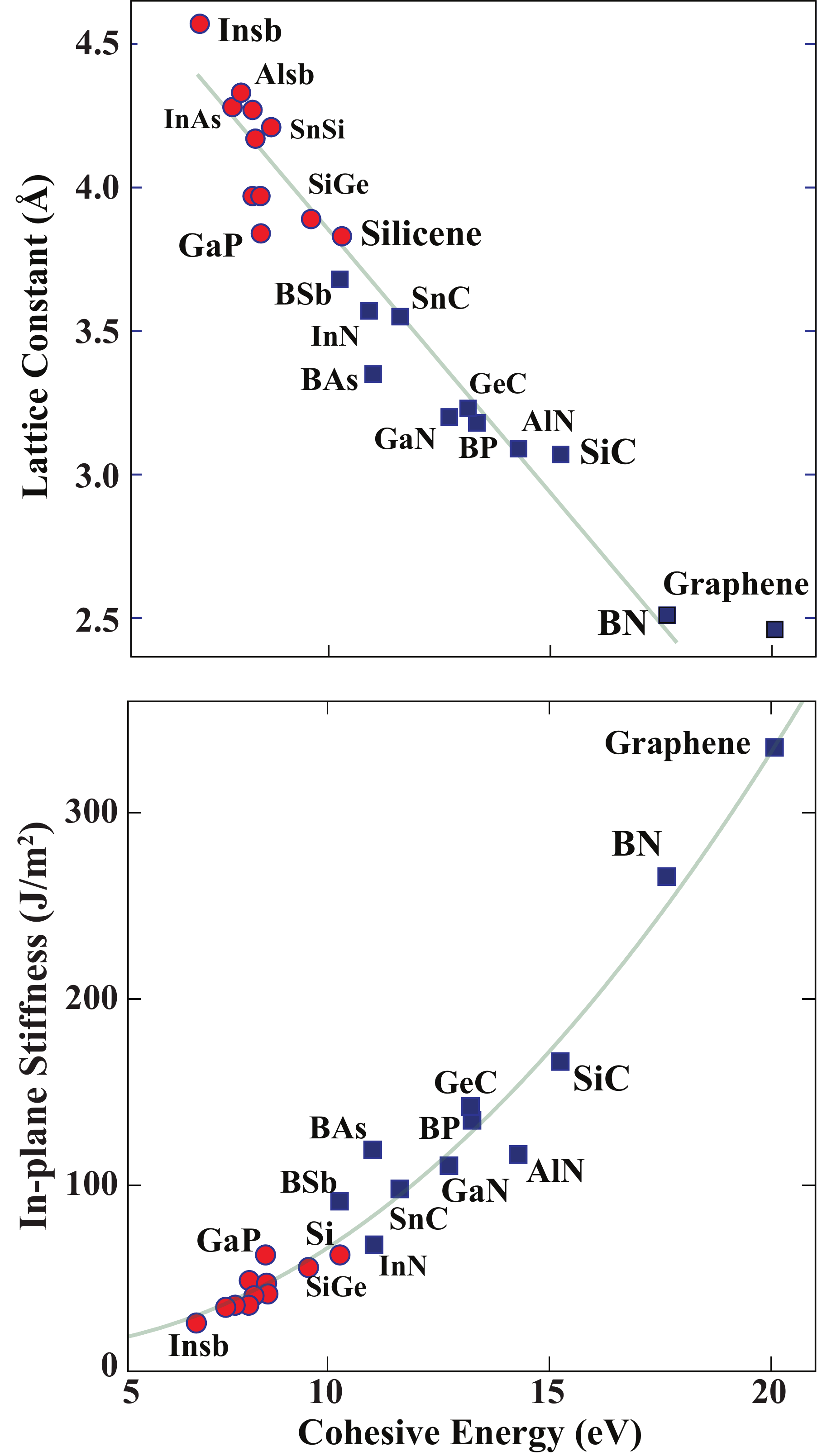}
\end{center}
\caption{Cohesive energy versus lattice constant and in-plane stiffness versus cohesive energy trends for 2D materials constituted by Group IV and Group III-V elements. Planar and buckled structures are denoted by purple squares and red circles, respectively. Adapted from \citep{Sahin1}.}
\label{trends}
\end{figure}

In the Fig.~\ref{iiivphonon}, we present the phonon dispersions of stable 2D materials constituted by Group III-V elements. One can immediately notice the similarity between the phonon dispersions of graphene, germanene and stanene with that of the Group III-V compounds at the same row, namely, BN, GaAs and InSb. Interestingly, 2D AlP that is constituted by elements from the silicon row is unstable, all other combination of elements in the neighbourhood of Al and P are stable. This exception needs further elaboration. The out of plane optical mode comes very close to the other optical modes at the K point of GaAs and InSb structures that are formed from the same row elements. BN is also formed by the same row elements but it is planar and thus has an out-of-plane optical mode that is mixed with acoustical modes, similar to aforementioned planar structures.

There are several clear trends in properties of 2D materials composed of Group IV and Group III-V elements. In Fig.~\ref{trends} we present the lattice constant and the in-plane stiffness versus the cohesive energy per atomic pair. Note that, planar and buckled structures are clearly separated from each other while sharing the same linear dependence on the cohesive energy versus lattice constant plot. In-plane stiffnesses of graphene and hexagonal BN are significantly higher than that of the rest. Here the cohesive energy, $E_c$, is calculated using the expression $E_c = E_T[AB]-E_T[A]-E_T[B]$, where $E_T[AB]$ is the total energy per A-B pair of the optimized honeycomb structure while $E_T[A]$ and $E_T[B]$ are the total energies of free A and B atoms corresponding to nonmagnetic state. In-plane stiffness is calculated by focusing on the harmonic range of the elastic deformation, where  the structure respondes to strain $\epsilon$ linearly. The rectangular supercell is pulled in $x$- and $y$-directions over a regular mesh of strains $\epsilon_x$ and $\epsilon_y$ while the strain energy, $E_{S}=E_{T}(\epsilon)-E_{T}(\epsilon=0)$ is determined. The data is then fitted to a two-dimensional quadratic polynomial expressed by

\begin{equation}\label{equ:stiffness}
E_{S}(\epsilon_{x},\epsilon_{y})=a_{1}\epsilon_{x}^{2}+a_{2}\epsilon_{y}^{2}+a_{3}\epsilon_{x}\epsilon_{y}
\end{equation}

where $\epsilon_{x}$ and $\epsilon_{y}$ are the small strains along $x$- and $y$-directions in the harmonic region. Due to the isotropy of the honeycomb structure $a_{1}=a_{2}$. These parameters can be related to the elastic stiffness constants by $a_{1} = a_{2} = (h\cdot A_{0}/2) \cdot C_{11}$, $a_{3}=(h \cdot A_{0}) \cdot C_{12}$, where $h$ and $A_{0}$ are the effective thickness and equilibrium unitcell area of the system, respectively. Hence one obtains Poisson's ratio $\nu$ = -$\epsilon_{trans}$/$\epsilon_{axial}$, which is equal to $C_{12}/C_{11}=a_{3}/2a_{1}$. Similarly, the in-plane stiffness,  $C=h\cdot C_{11}\cdot(1-(C{}_{11}/C_{12})^{2}$) = (2a$_{1}$-(a$_{3}$)$^{2}$/2a$_{1}$)/(A$_{0}$). The calculated value of the in-plane stiffness of graphene is in agreement with the experimental value of 340 (N/m) \citep{Lee1}.

Band structures of these materials were also investigated both by LDA and $G_0W_0$ methods. Interested reader is referred to \citep{Sahin1}.

\begin{figure}[t]
\includegraphics[width=11.5cm]{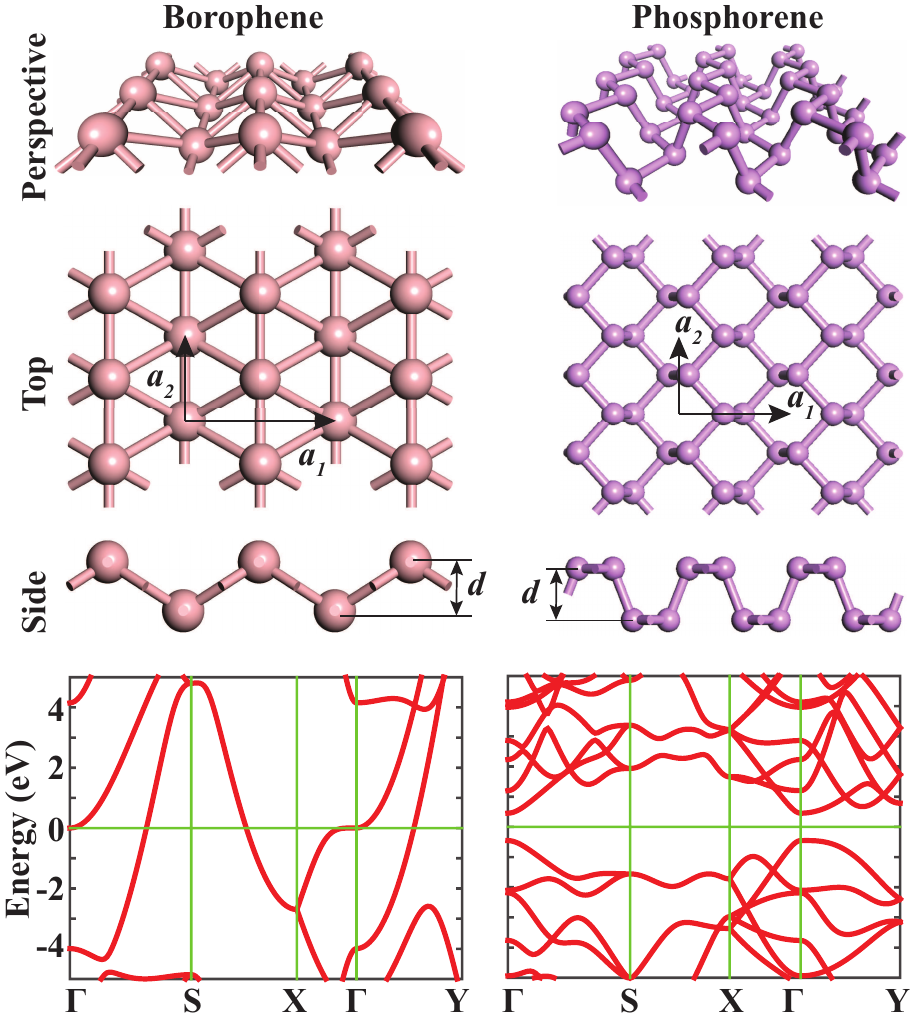}
\caption{Atomic and electronic structure of borophene and phosphorene.}
\label{phosphor}
\end{figure}

\section{Borophene and Phosphorene} \label{phosphorene}

Isolated hexagonal boron sheet is theoretically predicted to be unstable \citep{Boustani1}. Boron has been shown to have metallic layered structure composed of buckled triangular motifs in a partially filled honeycomb structure at a specific hole density \citep{Boustani1, Tang2}. This configuration allows the kind of chemical bonds that boron forms best: bonds among three atoms rather than between two. However, as for silicene, germenene and stanane, a hexagonal boron layer could be stabilized by the interaction with a substrate. Indeed crystalline 2D boron sheets (i.e., borophene) on silver surfaces have been recently synthesized \citep{Mannix2}.

Phosphorene is a 2D material derived from the layered allotrope of phosphorus called ``black phosphorus''. Black phosphorus is the most stable allotrope of phosphorus and was synthesized a century ago \citep{Bridgman1}. It has black color, as the name suggests, and in general is very similar to graphite in terms of both appearance and atomic structure. Despite this similarity, the single layer of black phosphorus, named phosphorene, was unexplored, even for a decade after the synthesis of graphene, the single layer of graphite. Phosphorene became a center of attention after two groups managed to exfoliate single and few layers of phosphorene from black phosphorus \citep{Reich1,Liu3,Li1}. The field-effect transistor made from few layers of phosphorene was shown to reach the charge carrier mobility of 1000 cm$^2$ V$^{-1}$s$^{-1}$ \citep{Liu3}.

The atomic structures of black phosphorus and phosphorene are presented in Fig.~\ref{phosphor}. The structural parameters of the freestanding phosphorene shown in the right panel of Fig.~\ref{phosphor} are obtained by DFT calculations using the PAW-PBE approach \citep{Kresse1,Blochl1,Perdew1}. The thickness of phosphorene which is the height difference between the top and the bottom phosphorus atoms is 2.10~\AA. Although the threefold coordination of atoms in phosphorene reminds the honeycomb structure of silicene, there are significant differences. The structure of phosphorene is reminiscent of the (110) plane of the diamond structure while silicene is related to the (111) plane. However, the bond angles of phosphorene are significantly smaller than the tetrahedral angle 109.5$\degree$ due to the fact that phosphorus is a group V element.

Experiments have shown that black phosphorus has a direct band gap of 0.31-0.36~eV at the $\Gamma$-point \citep{Warschauer1,Narita1,Maruyama1} which is very well reproduced by the DFT calculations \citep{Liu3}. The DFT calculations also show that, as one goes from bulk black phosphorus to single layer phosphorene, the band gap increases linearly with respect to the inverse of the number of phosphorene layers reaching 1.0~eV at the single layer. The band gap of phosphorene becomes 2.31~eV \citep{Cakir1} when the self-energy corrections are included at the $G_0W_0$ level \citep{Hedin1,Shishkin1}.  On the other hand, the excitonic binding energy calculated by solving the Bethe-Salpeter equation \citep{Salpeter1,Onida1} was found to be 0.70~eV  \citep{Cakir1}. Hence, the optical gap is predicted to be 1.61~eV. This is in agreement with the measured photoluminescence spectrum of single layer phosphorene that has a peak at 1.45~eV \citep{Liu3}.

The atomic structure of phosphorene is highly anisotropic compared to other 2D materials that have honeycomb symmetry. This anisotropy is reflected in the optical absorption spectra of phosphorene \citep{Tran1}. Monolayer phosphorene absorbs light polarized along the armchair direction ($a_1$ in Fig.~\ref{phosphor}) with energies 1.1~eV and above. This lower bound in the energy of absorbed armchair-polarized light goes down to 300~meV when the number of layers are increased from monolayer phosphorene to the bulk black phosphorus. On the other hand, light polarized along the zigzag direction ($a_2$ in Fig.~\ref{phosphor}) is absorbed if its energy exceeds 2.5~eV. In this case, the absorption spectrum has minor changes when the number of layers are varied. This means that armchair-polarized light having energy in the range from 1.1~eV to 2.5~eV will be absorbed by phosphorene while zigzag-polarized light with the same energy will pass through. This makes phosphorene a natural linear polarizer with an energy window that can be tuned by the number of layers.

\section{Transition metal dichalcogenides} \label{tmd}

The remarkable properties of monolayer graphene have triggered interest in ultra-thin materials with similar crystal structure. Among these, Transition Metal Dichalcogenides (TMDs) are the most promising. Here we note that especially this field is evolving very fast and we can only mention a few points. Bulk forms of TMDs have been used in industrial applications such as lubricant materials, coating technology, water splitting and petroleum refining. Although there were many studies before, the research on TMDs got an extra boost after the synthesis of graphene.

Although TMDs in bulk form have been investigated for long time, recent advances in experimental techniques revealed many unique properties of single layers of TMDs:
\begin{itemize}
\item while graphene is semimetallic, single layer TMDS are semiconductors,
\item TMDs become much more resistant against mechanical deformations when they are synthesized in the form of a single layer,
\item the band gaps of TMDs strongly depend on the number of layers,
\item even the direct/indirect nature of the allowed optical transitions can be controlled via the number of layers,
\item TMDs exhibit strain-tunable vibrational spectra and electronic properties,
\item in contrast to graphene, the surfaces of TMDs possess tunable chemical properties,
\item TMDs have localized plasmons with a confinement factor that can be an order of magnitude larger than that of graphene
\item chiral light emission from K and K’ valleys is possible in TMDs
\item for different combinations of metal and chalcogen atoms TMDs can form various phases that have diverse electronic properties ranging from semiconducting, superconducting to ferromagnetic.
\end{itemize}
Because of these reasons TMDs are very promising materials for nanoscale electronic and opto-electronic device applications. 

\begin{figure}[t]
\includegraphics[width=11.5cm]{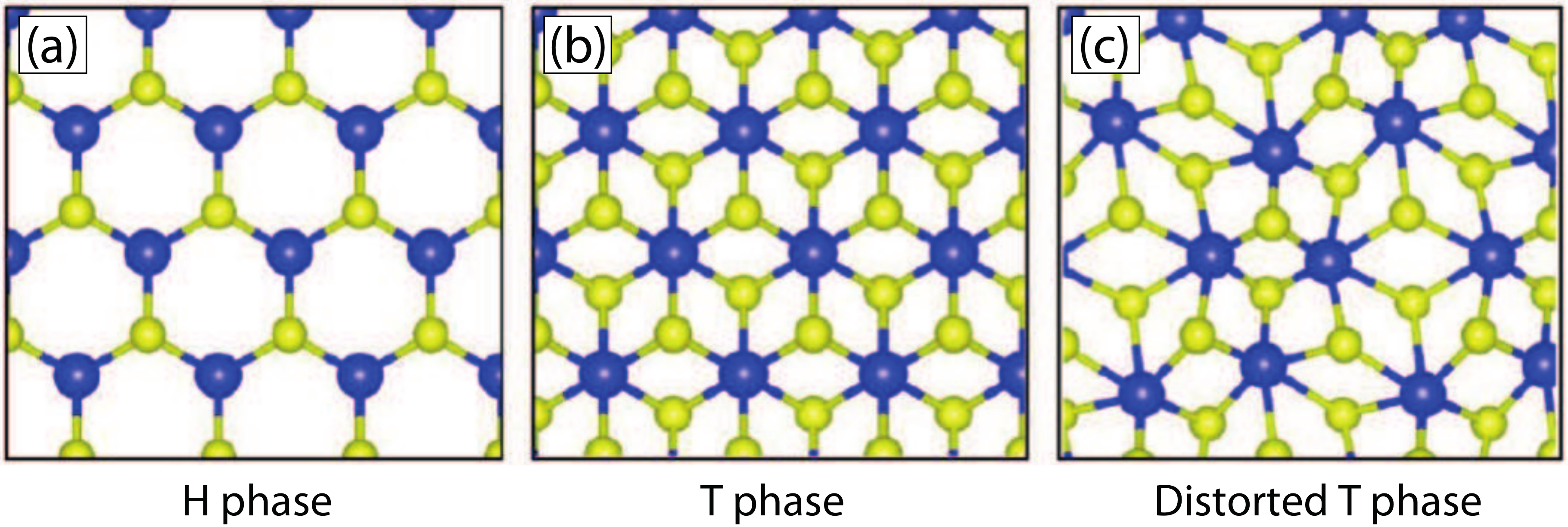}
\caption{\label{httd} Possible phases of TMD crystal structures are visualized from a top view of their single layers.}
\end{figure}

TMDs are a class of materials with the general formula MX$_{2}$, where M is a transition metal element and X is a chalcogen (S, Se or Te).  In each individual layer of their lamellar crystal structure, a monolayer lattice of metal atoms is sandwiched between two chalcogen layers, and, hence, each metal atom is surrounded by six chalcogen atoms. While weak layer-layer interactions in TMDs enable easy exfoliation of individual layers, strong covalent nature of in-plane bonds makes these structures strong and flexible. As shown in Figure \ref{httd}, depending on their constituent atoms, the TMDs’ crystal structures can be found in three different phases; H, T and distorted-T.  In the following we survey these phases in details.

\begin{figure}[t]
\includegraphics[width=11.5cm]{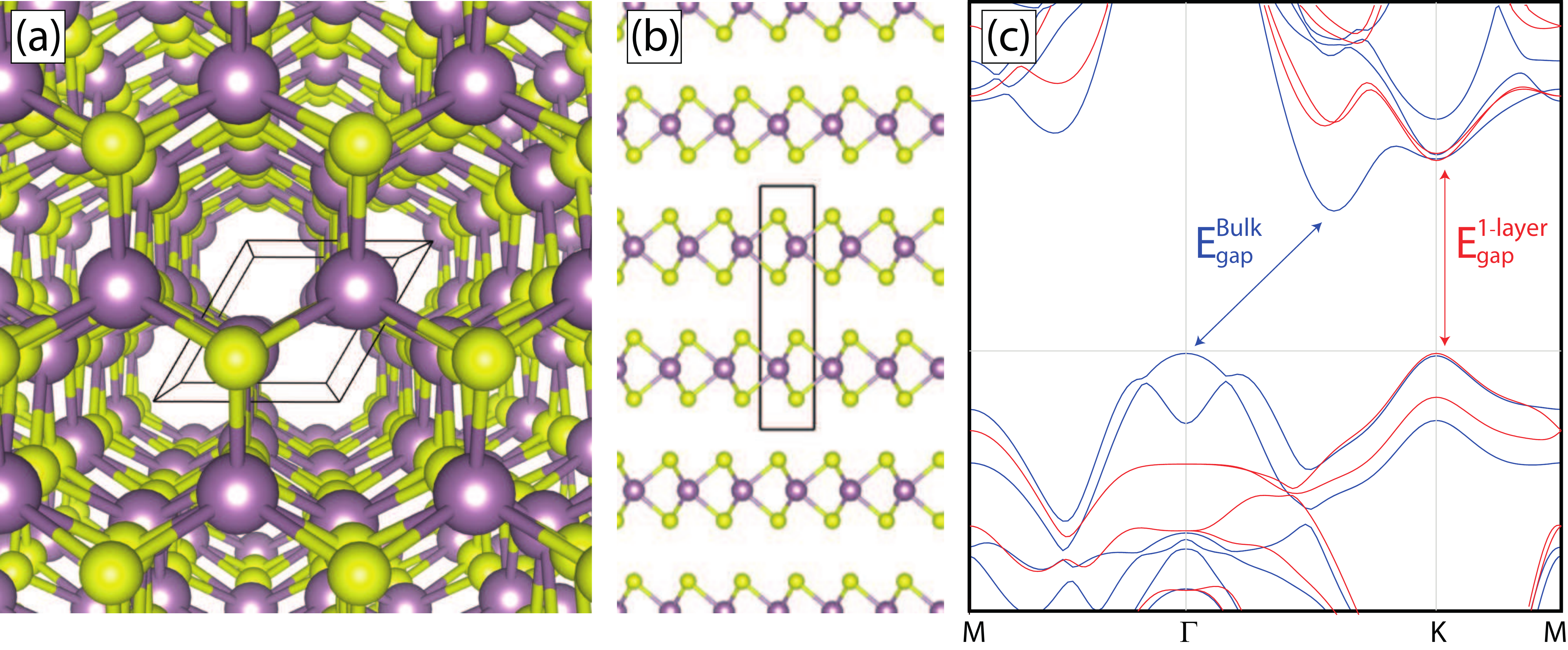}
\caption{\label{hmx2} (a) Perspective top view and (b) side view of MX2s in H phase. The primitive unit cell containing two layers is represented by the black rectangle. (c) Characteristic band dispersions of H-MX2 type bulk (in blue) and single layer (in red) structures.}
\end{figure}

Many of the TMDs such as MoS$_{2}$, MoSe$_{2}$, MoTe$_{2}$, WS$_{2}$ and WSe$_{2}$ crystallize in a hexagonal graphene-like structure. As shown in Figure \ref{httd}(a) and Figure \ref{hmx2}(a), each metal atom binds six chalcogen atoms by forming a prismatic configuration. This trigonal prismatic configuration has D$_{3h}$ lattice symmetry.  Because of large atomic distances between metal atoms the interaction of d-orbital states are negligible. So far, experiments have revealed that single layers of dichalcogenides of molybdenum and tungsten have ground state structures with D$_{3h}$ symmetry. This phase of single layer TMDs is called 1H. In bulk structures, stacking of individual layers results in an alternatingly rotated sequence which is called hexagonal symmetric 2H phase that belongs to the inversion-symmetric D$_{6h}$ point group (see Figure \ref{hmx2}(b)). In contrast to weak inter-layer interactions that allows easy mechanical exfoliation, metal and chalcogen atoms have strong intra-plane bonds that have covalent character.  

Another unique characteristic of the TMDs’ H phase is thickness dependent electronic properties. Starting from the early experiments \citep{mak10,spl10} revealing the strongly enhanced photoluminescence intensity of ultra-thin MoS$_{2}$, the thickness dependent characteristics of this family has been the focus of interest. The main reason of such a dramatic change in the optical spectrum stems from band gap crossover. The characteristic electronic band dispersion of dichalcogenides of Mo and W is presented in Figure  \ref{hmx2}(c). It is seen that the bulk structure is an indirect band gap semiconductor with band edges located at the $\Gamma$ and $\Gamma-K$  points, while the monolayer crystal structure is a semiconductor with a direct band gap at the $K$ high symmetry point. While the top of the valence band is mainly composed of Mo-d orbitals, the conduction band minimum is formed by hybridization of Mo-$d$ and Se-$p$ orbitals. Another interesting point in the electronic dispersion is the large spin-orbit splitting which is always found around the $K$ symmetry point. Such a large spin-orbit splitting has never been reported for other graphene-like materials.

Furthermore, the 1H phase is not the only possible phase for single layer TMDs.  Some of the single layer TMDs such as TaS$_{2}$, TiSe$_{2}$, SnS$_{2}$ and HfS$_{2}$ form trigonal anti-prismatic configurations as shown in Figure \ref{httd}(b). Differing from the H phase, bulk structures with T phase prefer AA type stacking in which same type of atoms are located on top of each other. However, the trigonal anti-prismatic configuration of monolayer T structure belongs to the D$_{3d}$ point group. Although, 1H and 1T configurations result in slightly different hexagonal crystal structures, their electronic spectrum turns out to be very different. Most of the experimental data and theoretical predictions show that 1T structures have metallic character \citep{ta, ti, ti2, hf}. One notes that the spatial overlap of the d-state electrons with neighboring metal and chalcogen atoms plays a more important role in the T phase. 

\begin{figure}[t]
\includegraphics[width=11.5cm]{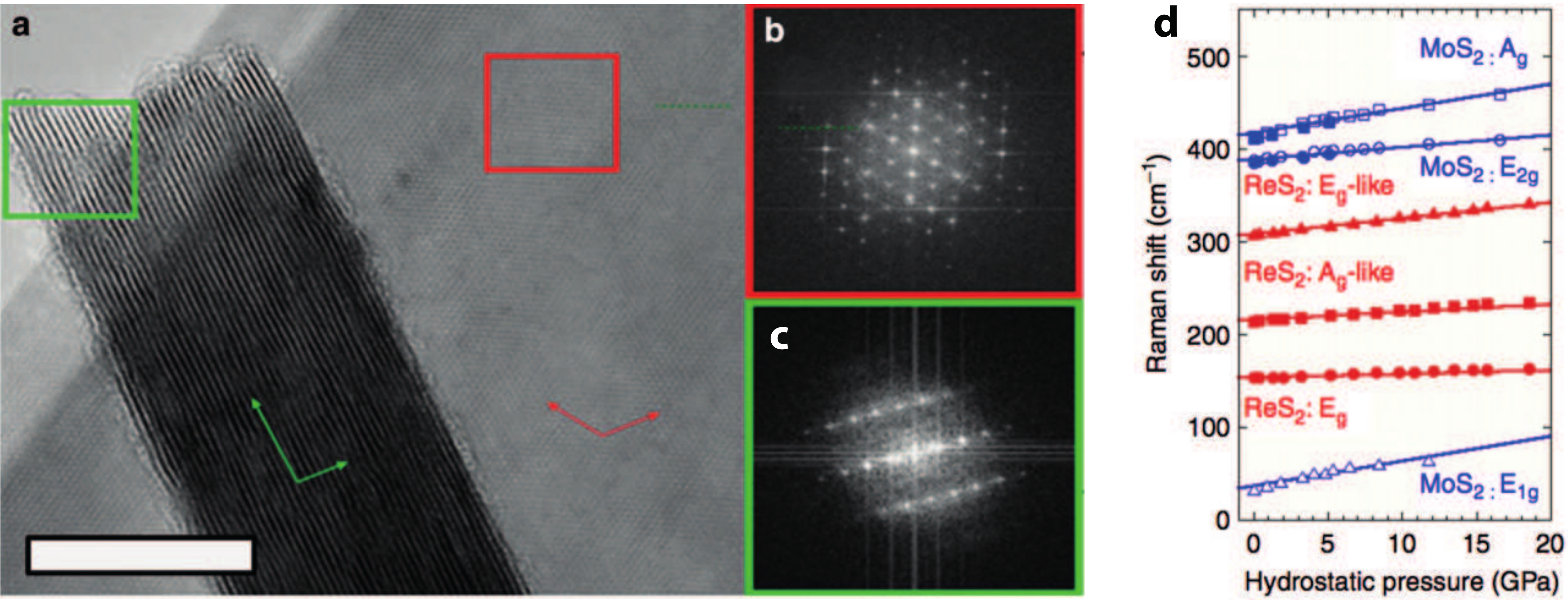}
\caption{\label{tdmx2} (a) HRTEM images of ReS$_{2}$ taken from in- and out-of-plane directions;  scale bar: 20 nm. (b) the flake's basal plane is perpendicular to the electron beam and (c) the other basal plane of the flake is oriented parallel to the beam. (d) pressure dependent shift in Raman-active phonon modes of ReS$_{2}$ and MoS$_{2}$.}
\end{figure}

In contrast to TMDs in the H phase, TMDs forming T phase have robust metallic behavior which is independent from their thickness. In addition, recent studies have also revealed that apart from TMDs some other dihalogen and dichalcogen compounds may form stable monolayer structures that have 1T symmetry. For instance, while FeCl$_{2}$ is a ferromagnetic semiconductor with half-metallic behavior \citep{eng15} SnS$_{2}$ is a direct bandgap semiconductor \citep{ahn15}.

More interestingly, recent studies have also revealed that single layer TMDs may form distorted lattice structures. For instance, it was noticed that mechanically exfoliated ReS$_{2}$ and ReSe$_{2}$ single layers, which are expect to be in the 1T phase, are neither in the 1H nor the 1T phase \citep{res,rese}. As firstly demonstrated for ReS$_{2}$, 1H and 1T phases can only be obtained through simple total energy calculations, but, however, they are dynamically unstable. Since each Re atom has seven valence electrons, there is one dangling electron on each Re atom when the six-coordinated H or T phase is formed. However, when the distorted-T phase is constructed through the formation of Re-Re dimers the total energy of the system is 1.1 eV (per ReS$_{2}$ unit) lowered and the resulting structure was found to be dynamically stable as well.

Due to the increased Re-Re interaction in ReS$_{2}$ the hexagonal lattice is narrowed along one of the lattice vectors and the symmetry of the structure obeys neither D$_{3d}$ nor D$_{3h}$ symmetry. This disturbed lattice structure with its broken symmetry is referred as distorted-1T structure (see Figure \ref{httd}(c)). Here the phase transition is directly driven by the Re atoms. It appears that the use of different metal atoms during the synthesis of MX$_{2}$ structures may result in new TMDs that have various phases and novel electronic properties. Formation of such dimers was also confirmed by HRTEM measurements \citep{res}. Moreover, such a reconstruction also leads to a decrease in intraplanar polarization and to a vanishing interlayer coupling. As a result of these combined unusual effects the ReS$_{2}$ and ReSe$_{2}$ crystal structures become electronically and vibrationally decoupled. As presented in Figure \ref{tdmx2}(d), as a consequence of Re-induced inter planar decoupling, the eigenfrequencies of the Raman-active vibrational modes of ReS$_{2}$ change very slightly even under very high hydrostatic pressures. However, mode hardening in the phonon frequencies of MoS$_{2}$ is much larger than in ReS$_{2}$.

More interestingly, recent studies also revealed that distorted phases of TMDs can also be induced by simple adsorption processes. Although the T phase of MoS$_2$ has never been experimentally observed, it is possible to stabilize a T phase over the H phase through adsorption of Li atoms on the MoS$_2$ surface \citep{Esfahani1}. Moreover, the dynamical stability of the resulting 1T-Li$_2$MoS$_2$ structure was investigated by performing phonon and molecular dynamics calculations at finite temperatures. They showed that the perfect T structure is unstable toward clustering of Mo and Li atoms even at low temperatures, while the distorted structure remains stable up to about 500 K. Therefore, as supported by phonon calculations, fully lithium covered MoS$_2$, Li$_2$MoS$_2$, has a distorted 1T structure that resembles the diamond-like dimerized crystal structure of ReX$_2$ (X=S, Se).

\begin{figure}[t]
\includegraphics[width=11.5cm]{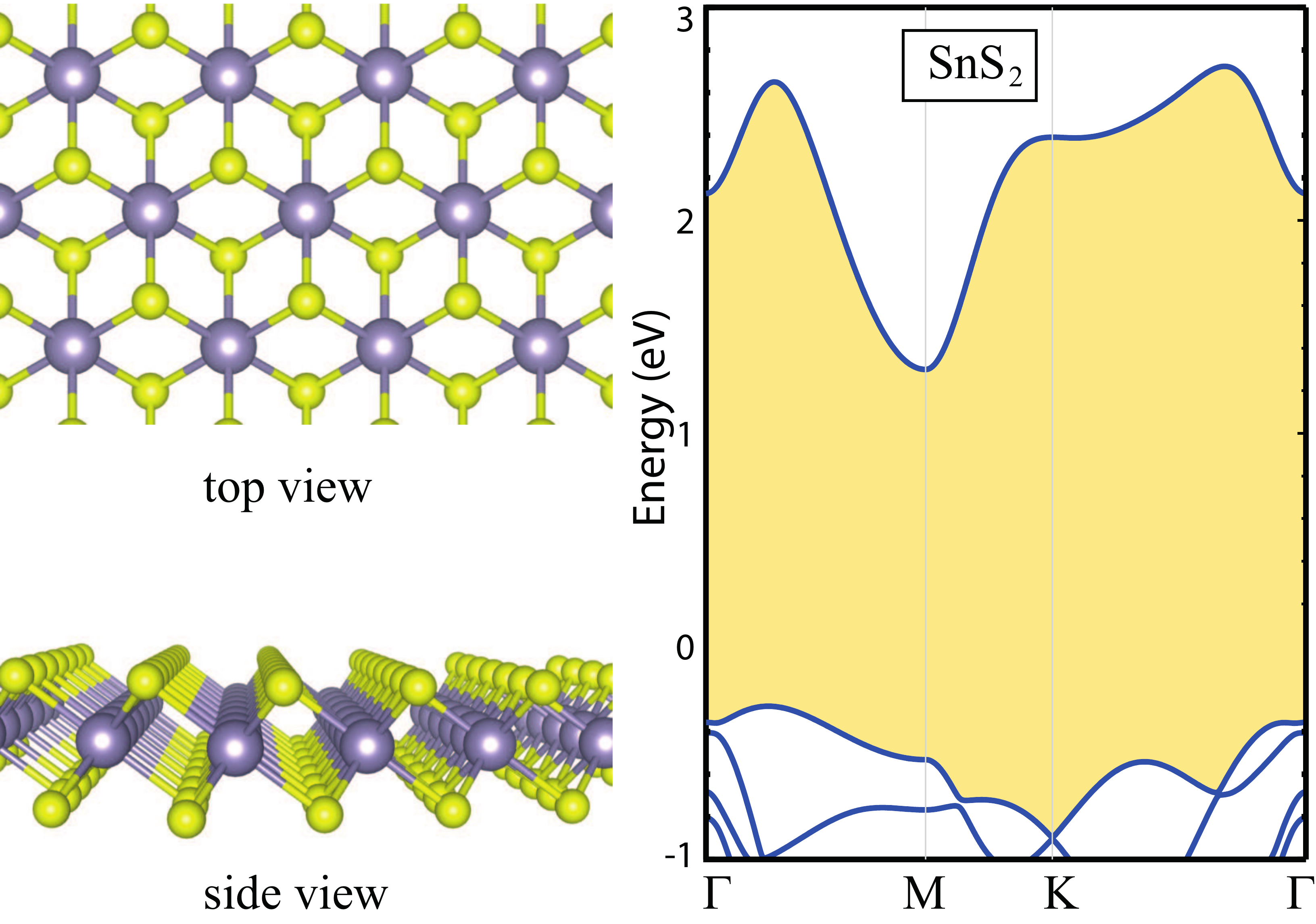}
\includegraphics[width=11.5cm]{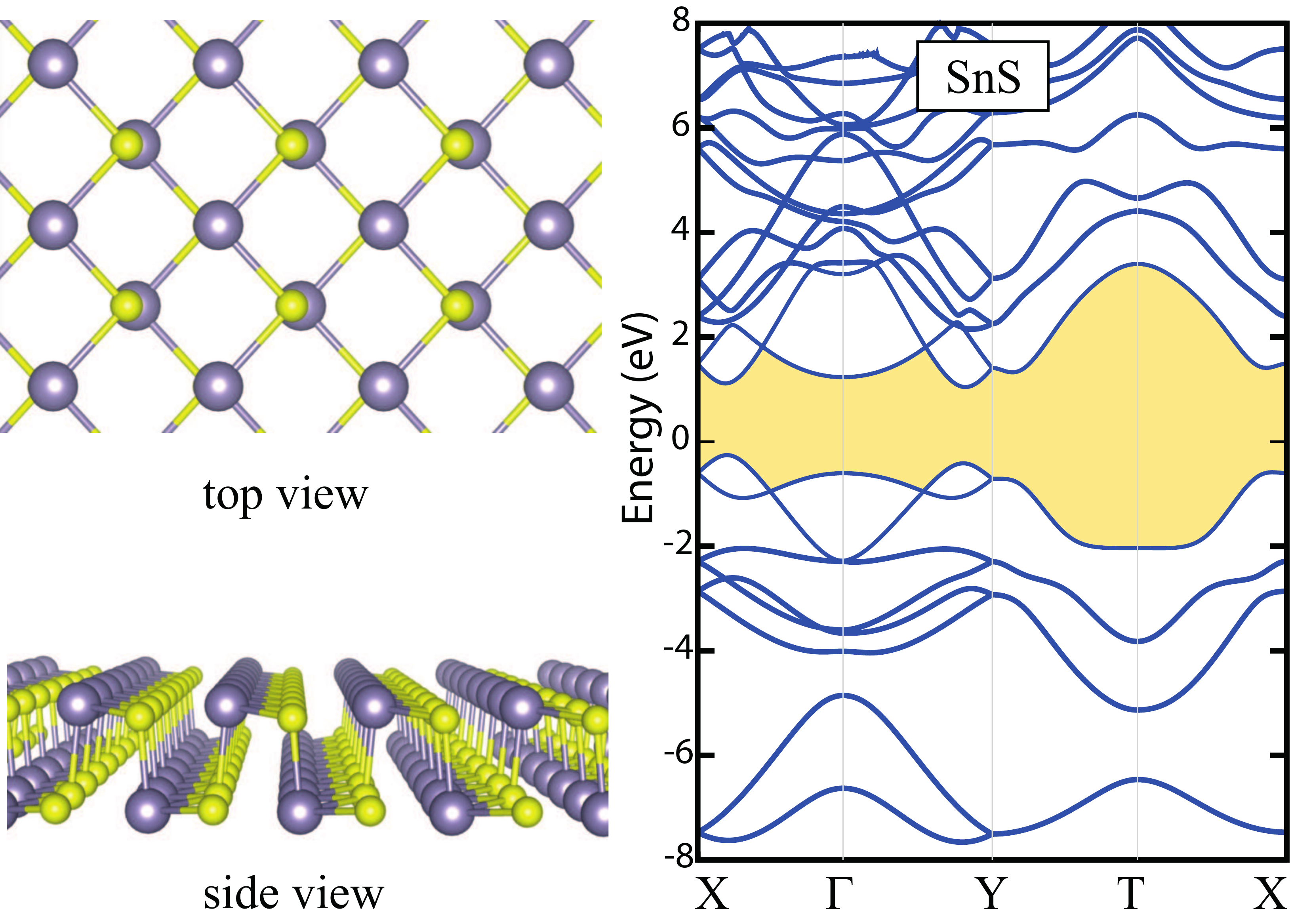}
\caption{\label{sns} Atomic structure and electronic band dispersion of SnS$_2$ (top) and SnS (bottom).}
\end{figure}

\section{Chalcogenides of Sn}

Very recent experiments have also shown that apart from the elemental monolayer crystal structure of tin, called stanene, various tin sulphide structures can be stabilized in ultra-thin single layer form.

Similar to TMDs MoS$_{2}$ and WS$_{2}$  monolayer SnS$_{2}$ possesses a three-atomic hexagonal primitive unit cell. However, differing from TMDs that have mostly a 1H phase, SnS$_{2}$ crystallizes in a 1T phase. As shown in  Figure ~\ref{sns}, the trigonally arranged subplane of Sn atoms is sandwiched by two S-subplanes.  While the 1H phase belongs to the $P\overline{6}m2$ space group where subplanes are $ABA$ stacked, the 1T phase is a member of the $P\overline{3}m2$ space group where subplanes are $ABC$ stacked. Recently, Bacaksiz \textit{et al.} found that the 1T phase of the SnS$_{2}$ is 875 meV/cell more favorable over the 1H phase, by performing ab initio DFT calculations. The primitive lattice vectors of 1T-SnS$_{2}$ can be given as  $\textbf{v}_{1}=a(\frac{1}{2},\frac{\sqrt{3}}{2},0)$, $\textbf{v}_{2}=a(\frac{1}{2},-\frac{\sqrt{3}}{2},0)$ where $|\textbf{v}_{1}|=|\textbf{v}_{2}|$ and $a$ is the lattice constant ($a=b$). In this cell the fractional coordinates of Sn, top S and bottom S atoms are given by $(\frac{|v_{1}|}{2},\frac{|v_{1}|}{2},0)$, $(\frac{|v_{1}|}{6},\frac{|v_{1}|}{6},\frac{c}{2})$, and  $(\frac{5|v_{1}|}{6},\frac{5|v_{1}|}{6},-\frac{c}{2})$, respectively (where $c$ is the distance between the subplanes of S atoms). In this configuration the lattice parameter and Sn-S inter-atomic distance are calculated to be 2.59 and 3.68 \AA, respectively.

While TMDs such as MoS$_{2}$ and WS$_{2}$ are direct band gap semiconductors whose valence band and conduction band edges are located at the K symmetry point, the 1T-SnS$_{2}$ crystal is a semiconductor with an indirect bandgap of 1.58 eV. The HSE06 corrected bandgap is found to be 2.40 eV. It was also shown that 1T-SnS$_{2}$ has a nonmagnetic ground state. In addition, a possibility of easy tuning of the stacking sequence of SnS$_{2}$ bilayers by two different ways, charging and application of loading pressure, was described.

Very recently, Ahn \textit{et al.} reported the successful synthesis of monolayer SnS$_{2}$ on SiO$_{2}$/Si substrates by a vapor transport method \citep{ahn15}. In the growth procedure while the precursor molecules are SnO$_{2}$ and S$_{2}$, the products are determined by the added H$_{2}$. It was also measured that SnS$_{2}$ exhibits $n$-type electronic character with a 2.77 eV band gap, which implies presence of defects or impurities at its surface of SnS$_{2}$. The interest on monolayer orthorhombic crystals was triggered by the synthesis of phosphorene that has easy tunable and highly anisotropic electronic properties. In their study, Ahn \textit{et al.} also achieved the synthesis of orthorhombic SnS crystals by tuning the amount of H$_{2}$ added during gas-phase synthesis. 

Experiments also revealed that  $n$-type doping occurs in synthesized orthorhombic SnS which has an energy band gap of 1.26 eV. As shown in Figure \ref{sns}, differing from hexagonal 1T phase of SnS$_{2}$, the orthorhombic phase possesses a direct band gap at the vicinity of the zone center which makes it more suitable for optoelectronic device applications.

\section{Portlandite and Brucite}

\begin{figure}[t]
\includegraphics[width=11.5cm]{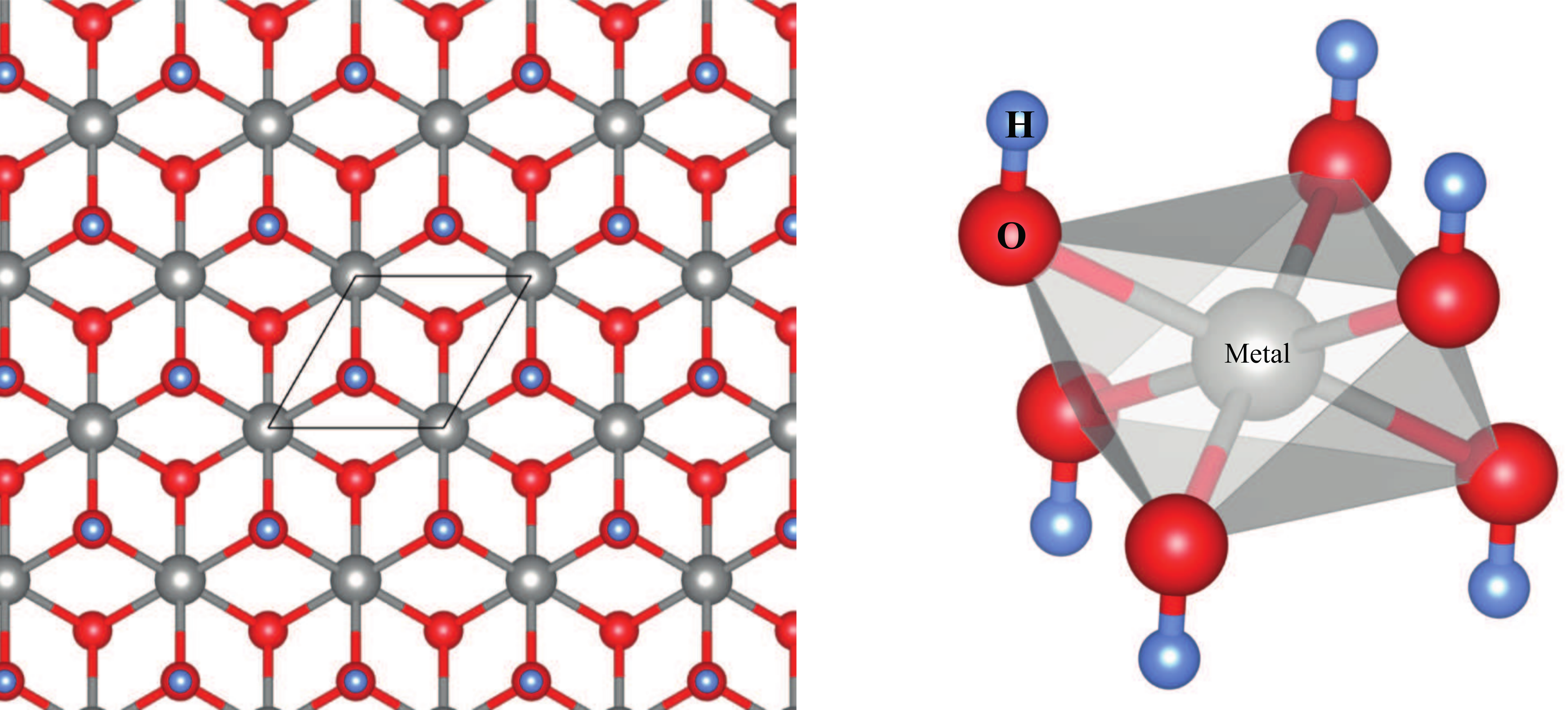}
\caption{\label{moh} Top and tilted side view of M(OH)$_{2}$. The primitive unitcell is shown by the black lozenge. The six-coordinated bonding nature of the metal atom is shown by polyhedra. Grey, red and blue balls represent metal atom, oxygen and hydrogen, respectively.}
\end{figure}

The synthesis of monolayer graphene from bulk graphite not only gave researchers, for the first time, access to a quasi-two dimensional (2D) crystal, but it also drew significant attention to similar layered bulk structures such as Alkaline Earth Metal Hydroxides (AEMHs). AEMHs can be represented by the general chemical formula M(OH)$_{2}$ where M is the alkaline-earth metal. Bulk single crystals of AEMHs are layered structures that belong to the P3m1 space group. In this layered crystal structure shown in Figure \ref{moh} the metal atoms fill all of the octahedral sites in alternating layers, while the OH pairs form a hexagonal close packed arrangement. In AEMHs, while the bonding has mainly ionic character, it has also partial covalent character.

\begin{figure}[t]
\includegraphics[width=11.5cm]{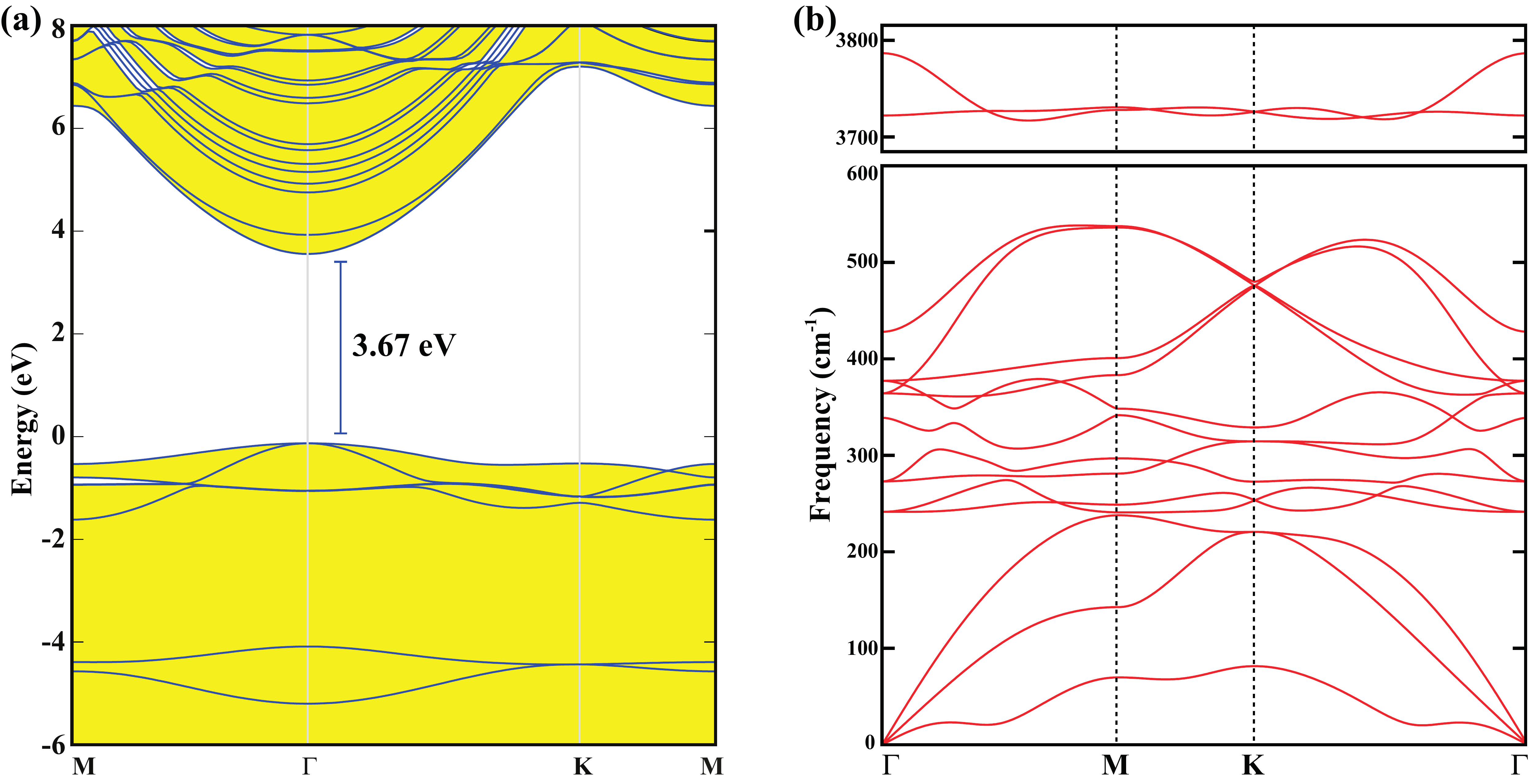}
\caption{\label{caoh} (a) Electronic band dispersion and (b) vibrational spectrum of single layer Ca(OH)$_2$. }
\end{figure}

\begin{figure}[t]
\includegraphics[width=11.5cm]{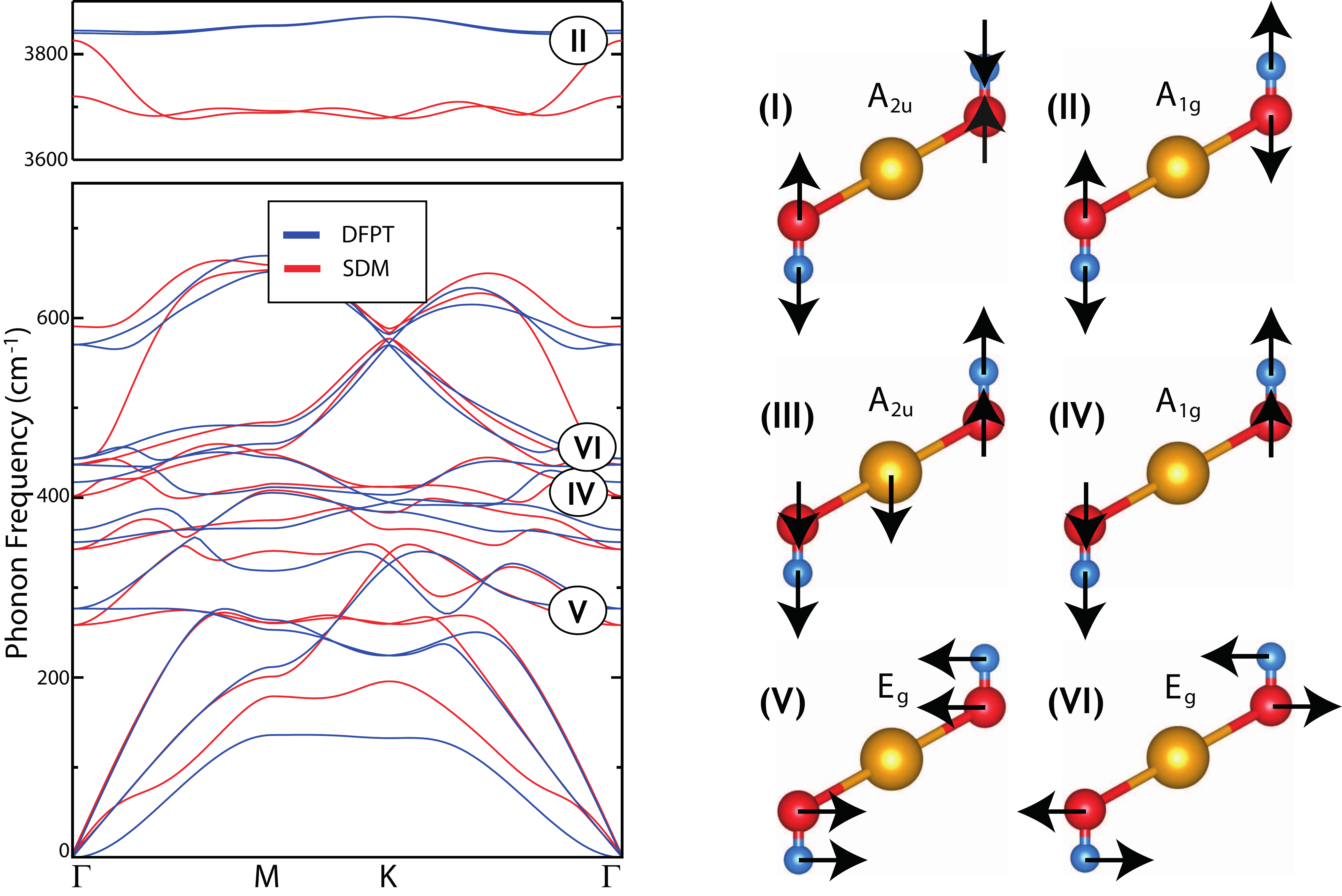}
\caption{\label{mgoh} Left panel: phonon dispersion of monolayer brucite calculated by using two different methodologies. Right panel: characteristic eigenmotions. Orange, red and blue balls represent Mg, O and H atoms, respectively.}
\end{figure}

One of the most well-known AEMH is portlandite, presented by the chemical formula Ca(OH)$_{2}$, which is the main product of hydration of portland cement.  Although portlandite has been known for a long time, synthesis and understanding the characteristic properties of its monolayer structures have not been studied until 2015. In our study we, experimentally and theoretically investigated bulk, bilayer, and monolayer portlandites \citep{erp15}. Firstly we showed that single crystal Ca(OH)$_{2}$ monolayers can be exfoliated from their bulk form onto different substrates. By performing ab initio total energy calculations, the lattice parameter, the Ca-O distance and the O-H distance are calculated as 3.62, 2.38 and 0.97 \AA, respectively. The electronic band dispersion shown in Figure \ref{caoh}(a) shows that single layer portlandite is a direct semiconductor with a large bandgap. While GGA-PBE  approximated bandgap is 3.67 eV, HSE06 approach gives bandgap of 5.16 eV which is usually close to the optical bandgap of the materials.

Analysis of the phonon dispersion of the monolayer structure revealed that it is a dynamically stable crystal structure and the decomposition of the vibration representation of optical modes at the zone center is  $\Gamma$ = 4E$_{u}$ + 2A$_{2u}$ + 4E$_{g}$ + 2A$_{1g}$. As shown in the right panel of Figure \ref{caoh}(b), there are four Raman-active phonon branches around 240, 350, 390, and 3700-3800 cm$^{-1}$. It is also worth noting that as opposed to other TMDCs structures that have the 1T phase, the presence of H atoms results in the existence of two different E$_{g}$ and A$_{1g}$ modes. It was also found that, similar to the Raman shift measurements observed from the bulk crystal structure, the monolayer material also has high-frequency OH stretching modes at 3700-3800 cm$^{-1}$.

Another well-known AEMH is brucite, with the chemical formula Mg(OH)$_{2}$. The bulk brucite mineral has been used widely in industrial applications as magnesium source and flame retardant. Our experimental measurements and calculations revealed that single layer crystal structures of AEMHs of Ca and Mg share many of the same electronic and vibrational characteristics. While bulk brucite has a direct bandgap of 6.17 eV at the $\Gamma$ point, (HSE06 corrected bandgap), the direct bandgap of monolayer brucite is calculated to be 4.80 eV. Regarding the indirect-to-direct bandgap crossover in well-known TMDs, AEMHs that have a robust, thickness-independent band character, have an entirely different behavior.  Moreover, our experiments demonstrated that when MoS$_{2}$ is supported by Mg(OH)$_{2}$, its photoluminescence intensity is enhanced significantly.  It is seen that due to the structural stability, thickness-independent robust electronic behavior, good dielectric property and unique optical performance in heterostructure applications AEMHs hold promise for future nanoscale device applications.

\section{Concluding Remarks}

Layered materials offer a new perspective in the development of low dimensional materials with a wealth of novel electronic and mechanical  properties. An exciting aspect of 2D materials is stacking them into structures that are still very thin, such that we take advantage of the vastly different properties of different layered materials, creating previously unimagined devices  made of atomically thick components. Each 2D material known or to be discovered is like a Lego brick that we can use to build new materials with tailored properties.

We have been living with those layered materials for very long time: graphite, hex-BN, the family of transition metal dichalgogenides (TMD), layered oxides, etc. All of them are well characterized in a bulk 3D stacked configuration, however the boom of the field of 2D materials started with the rise of graphene and its derivatives. Exfoliation of already existing bulk materials offer a direct way to get layered single (and few layers) of hex-BN and TMDs. The impressive rise of graphene, due to its exceptional electronic and mechanical properties, has spurred many scientists to searching for alternative 2D materials with heavy elements such as  Si, Ge and Sn. This race has led to the synthesis and/or prediction of many other atomically-thin materials.  These materials promise to improve nano electronics and open up entirely new ways of carrying information, for example by exploiting quantum properties such as the electron spin (“spintronics”),  or the presence of two electron’s valley states in many 2D materials that offers a direct way to digital switches (“valleytronics”). Since no net flow of charge is required in spintronics or valleytronics the problem of heating and power dissipation might disappear.

In the last years we have witnessed an important incorporation in that family of layered material based on new single element layered structures made of group III (B), group IV (Si,Ge and Sn) and group V (P)  elements have been proposed and few of them already synthesized not in isolated form but supported on metallic substrates.  Particularly the 2D group IV materials that include graphene, silicene, germanene and stanene and group V phosphorene have attracted enormous interests due to their exotic electronic properties and potential  applications in nanoelectroics, spintronics, optical materials as well as an ideal testbed for the understanding of novel phenomena as topological superconductivity, near-room-temperature quantum (“half-integer”) anomalous Hall (QAH) effect enhanced thermoelectric performance and dissipationless electric conduction at room temperature. The QSHE is of broad interest because of its scientific importance as a novel quantum state of matter and its potential for technological applications in the fields of spintronics, valleytronics and quantum computation.

As the material research community turns its attention to 2D materials beyond graphene, new 2D materials are discovered (or re-discovered) on a weekly basis. There are estimated to be a few hundred more that are waiting to be explored, and the research on 2D materials will stay active in the foreseeable future. Among the vast number of 2D materials, elemental 2D materials occupies a special place due to their diverse properties and their mono-elemental nature. It’s conceivable that eventually only few 2D materials will find their way into industry, if at all. We believe that elemental 2D materials will be among those selected few.
\chapter{Strain Engineering of 2D Materials} \label{Strain}

When bulk structures are thinned down to their monolayers, degree of orbital interactions, mechanical properties and electronic band dispersion of the crystal structure become highly sensitive to the amount of applied strain. The source of strain on the ultra-thin lattice structure can be (i) an external device or a flexible substrate that can stretch or compress the structure, (ii) the lattice mismatch between the layer and neighboring layers or (iii) stress induced by STM or AFM tip.

\section{ MoS$_{2}$}

\begin{figure}[t]
\begin{center}
\includegraphics[width=8.8cm]{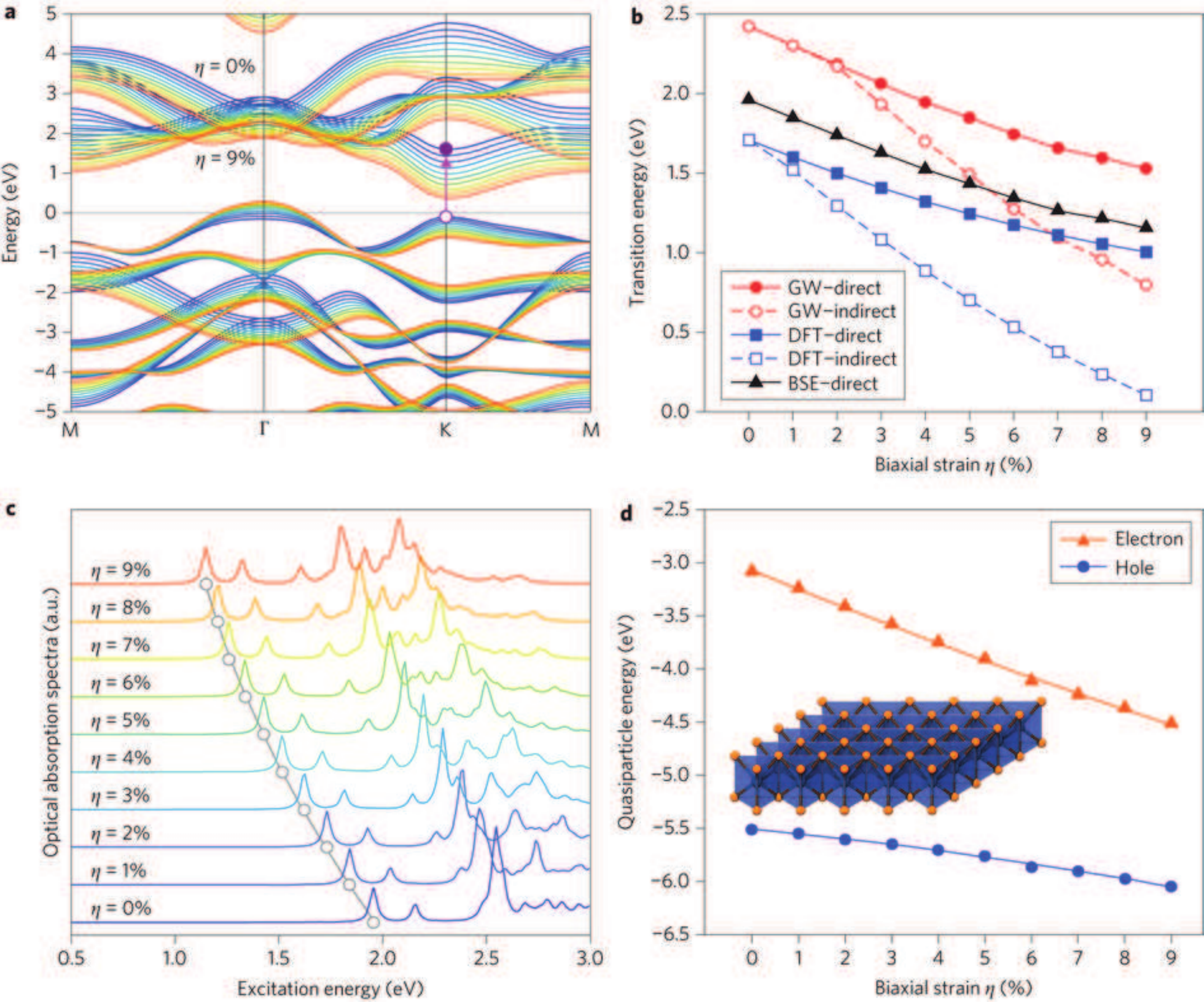}
\end{center}
\caption{\label{mos1} (a) Strain dependent electronic band structure of monolayer MoS$_{2}$ under different biaxial strains from 0 (blue line) to 9 percent (red line). (b) Direct and indirect bandgaps under different biaxial strains calculated using DFT and GW. Strain-dependent (c)  absorption spectra and (d) quasi-particle energy of electron and hole. Adapted from \citep{Feng2}.}
\end{figure}

\begin{figure}[b]
\begin{center}
\includegraphics[width=8.8cm]{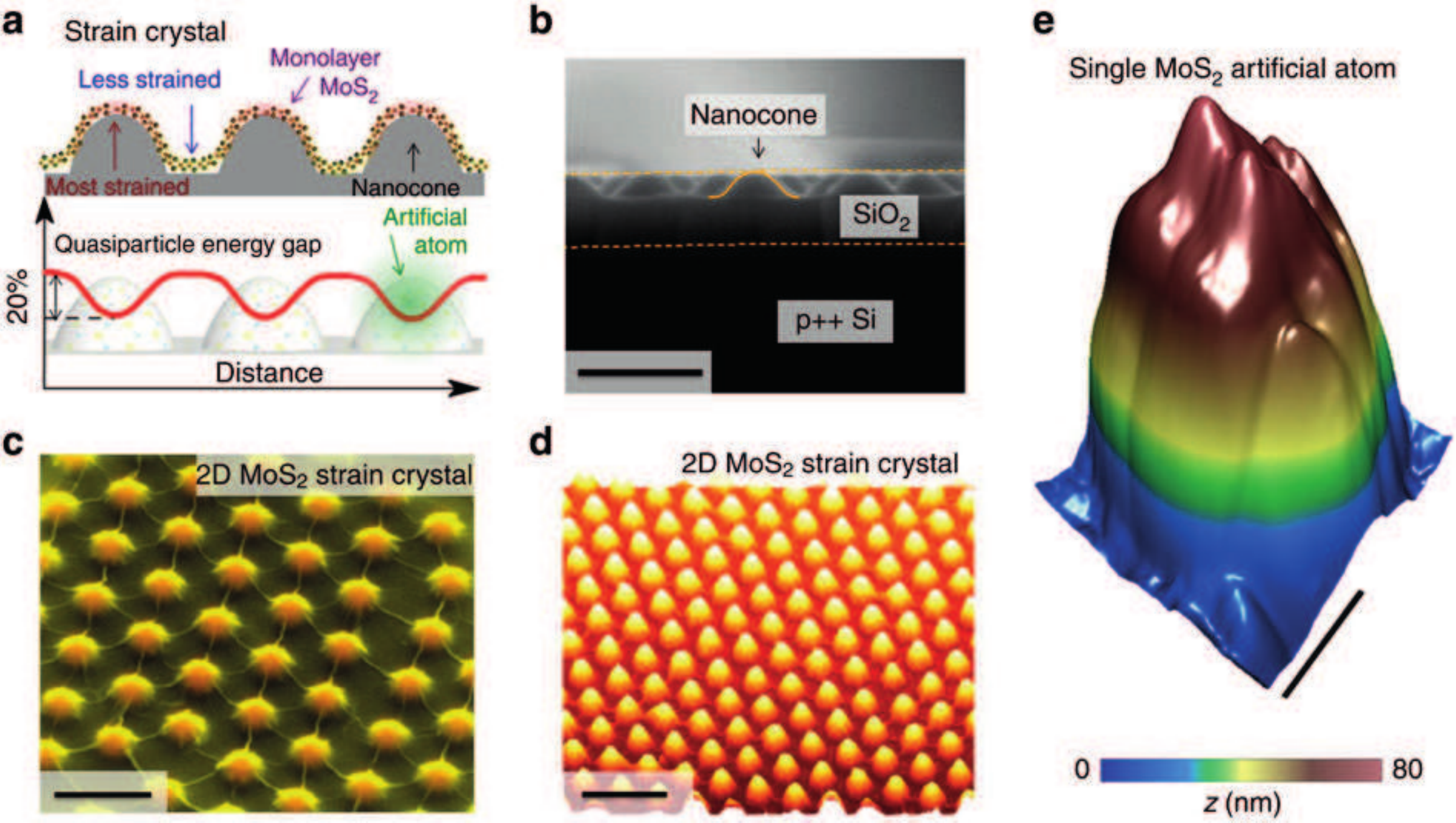}
\end{center}
\caption{\label{mos2} (a) Schematic of strained MoS$_{2}$ indented by SiO$_{2}$ nanocones, where the regions on the tips of nanocones exhibit highest tensile strain while the areas between nanocones are less strained. (b) Cross-sectional scanning electron microscopy (SEM) image of the nanocone substrate. Solid curve delineates the nanocone shape. Scale bar is 500 nm. (c) Tilted false-color SEM image of the 2D strained MoS$_{2}$ crystal defined by the nanocone array. Scale bar is 500 nm. (d) AFM topography of the 2D MoS$_{2}$ strain crystal. Scale bar is 1 μm. (e) STM topography of a single ‘artificial atom’ building block within the crystal. Scale bar is 100 nm. Adapted from \citep{Li2}}
\end{figure}

Among the ultra-thin TMDs MoS$_{2}$ is the most studied one due to its easy synthesis, semiconducting nature, atomically perfect surface structure and abundance. It has also been shown experimentally and computationally that elastic strain is a viable agent for creating a continuously varying bandgap profile in an initially homogeneous, atomically thin single layer MoS$_{2}$ which is highly desirable for its use in photovoltaics, photocatalysis and photodetection \citep{Feng2}. As shown by Feng \textit{et al.} the electronic bandgap of the single layer structure can be tuned significantly and reversibly upon biaxial strain (see Fig.~\ref{mos1}). Therefore, it was also shown that single layer MoS$_{2}$ based photovoltaic devices can capture broad range of the solar spectrum upon application of strain. Moreover, these devices can concentrate excitons and charge carriers in desired domains of the crystal structure.

Another device example is optoelectronic crystal of artificial atoms in strain-textured MoS$_{2}$ \citep{Li2}. This was achieved by transferring as-grown MoS$_{2}$ sheet onto assembled SiO$_{2}$ nanocones and then soaking it in ethylene glycol to remove the trapped air bubbles while optimizing the strain (see Fig.~\ref{mos2}). The evaporation of ethylene glycol from the interface generates capillary force that applies local strain to the MoS$_{2}$ sheet. It was shown that a synthetic superlattice of capillary-pressure-induced nanoindentation of monolayer molybdenum disulphide forms an optoelectronic crystal capable of broadband light absorption which covers the entire visible wavelength and most intensive wavelengths of the solar spectrum. It was also discussed by Li \textit{et al.} that such two-dimensional semiconductors with spatially textured band gaps represent a new class of materials, which may find applications in next-generation optoelectronics or photovoltaics.

\begin{figure}[t]
\begin{center}
\includegraphics[width=10.5cm]{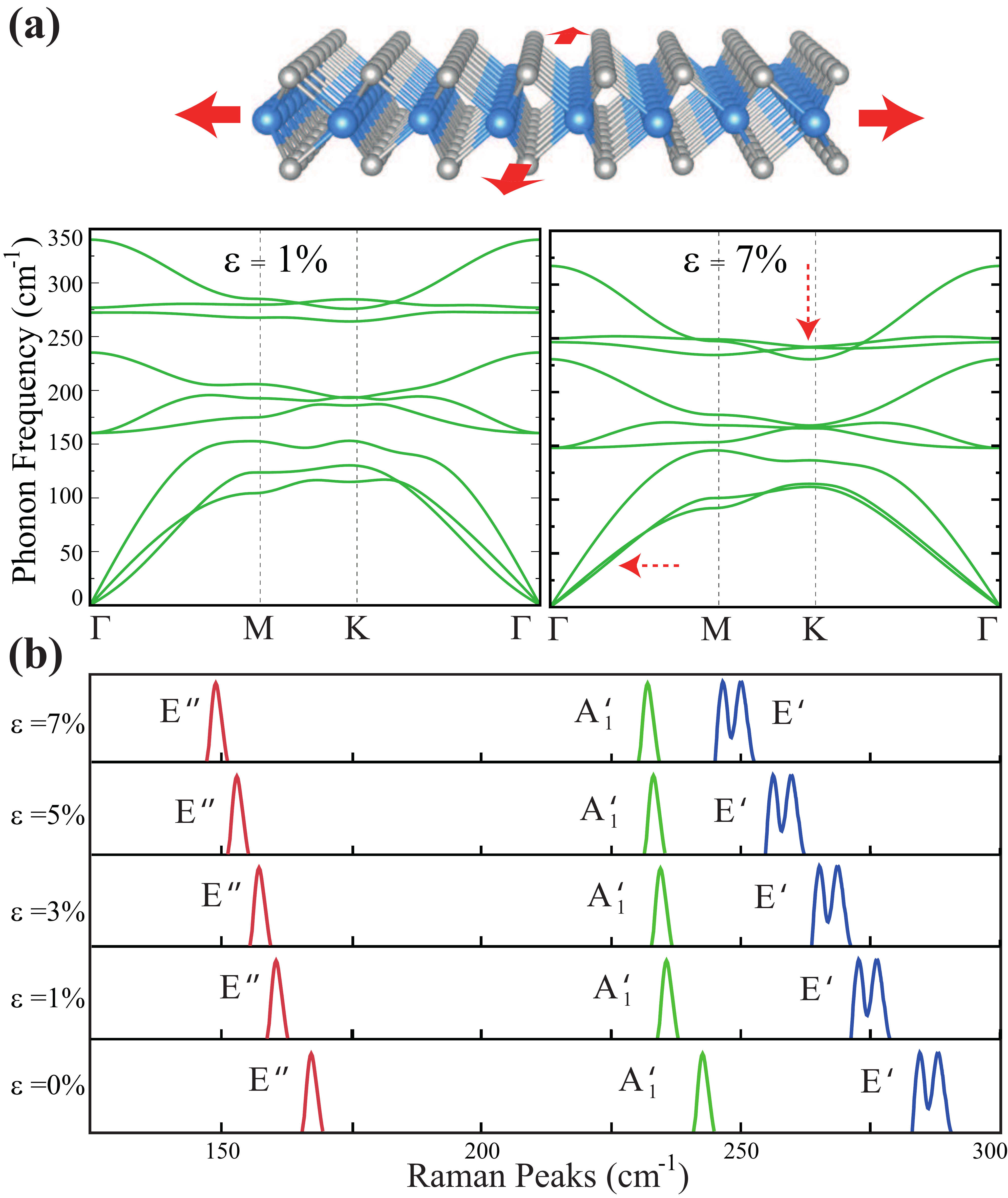}
\end{center}
\caption{\label{mose1} Strain-dependent phonon dispersion and frequency shift in single layer MoSe$_{2}$. Adapted from \citep{Horzum1}.}
\end{figure}

\begin{figure}[t]
\begin{center}
\includegraphics[width=10.5cm]{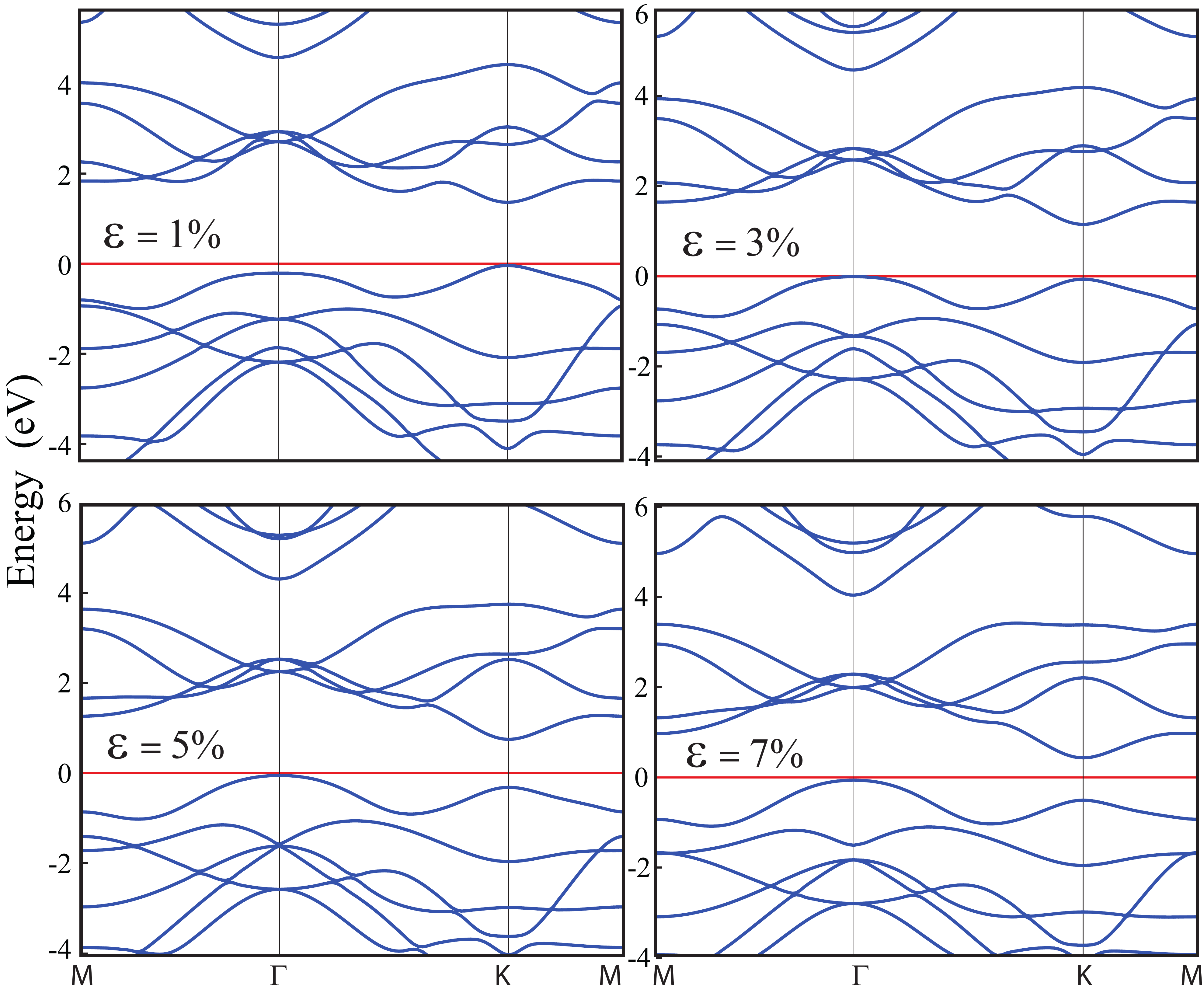}
\end{center}
\caption{\label{mose2} Evolution of the electronic band structure of single layer MoSe$_{2}$ upon biaxial strain. Valence band edge is set to zero. Adapted from \citep{Horzum1}.}
\end{figure}

\begin{figure}[t]
\begin{center}
\includegraphics[width=9.0cm]{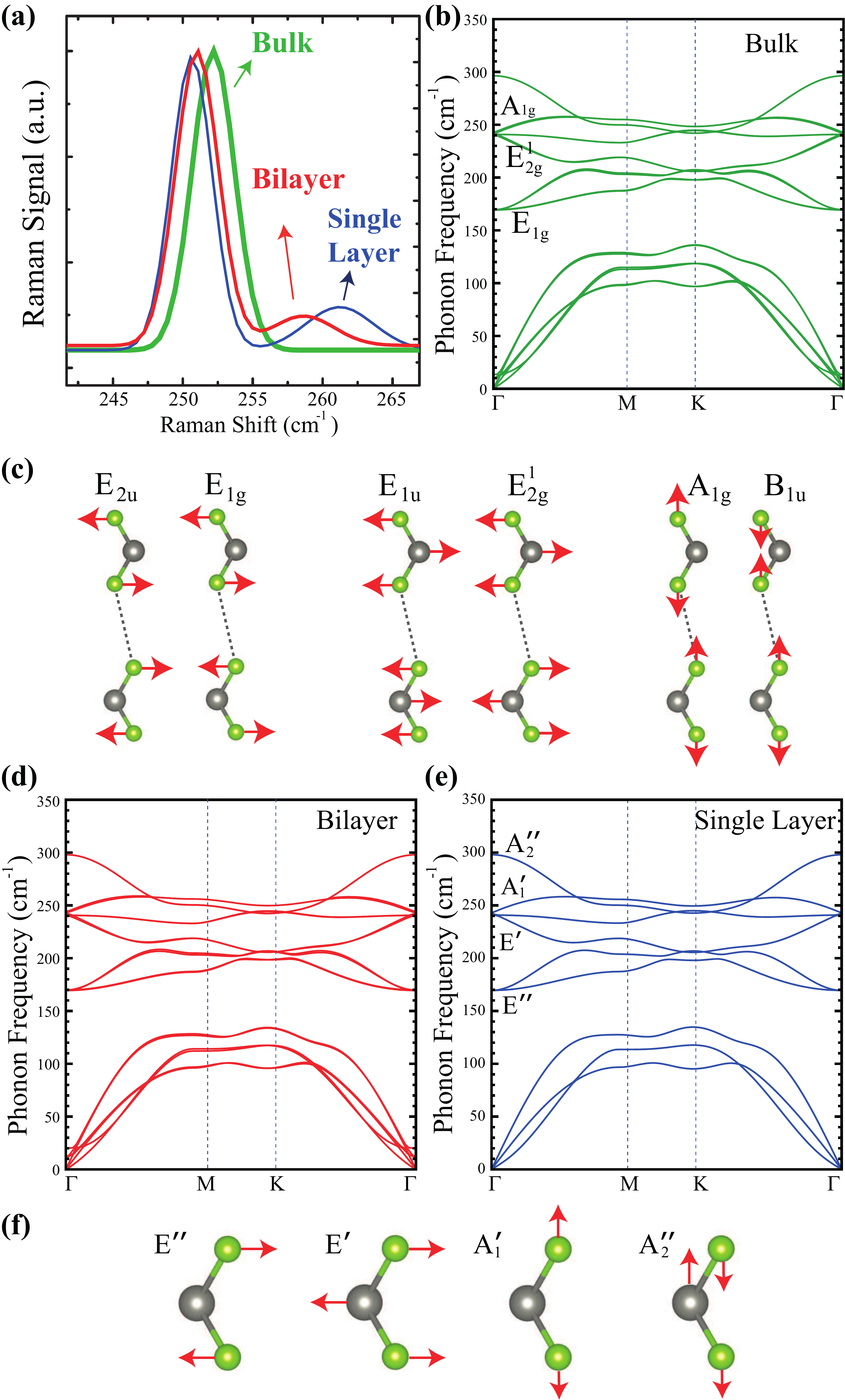}
\end{center}
\caption{\label{wse1} (a) Experimentally observed Raman spectrum bulk,bilayer and monolayer WSe$_{2}$. (b) Calculated phonon dispersion and (c) atomic displacements of the bulk structure. Phonon dispersion of (d) bilayer and (e) monolayer structure and (f) atomic displacements in monolayer structure. Adapted from \citep{Sahin4}.}
\end{figure}

\section{ MoSe$_{2}$}

Lattice vibrations of single layer MoSe$_{2}$ crystal that has the D$_{3h}$ point group symmetry can be characterized by nine phonon branches including three acoustic and six optical branches. While LA and TA acoustic branches have linear dispersion, the frequency of the out of plane flexural (ZA) mode has a quadratic dispersion in the vicinity of q = 0.  Decomposition of optical vibrations at the zone center reveals that there are 2E$^{''}$, 2E$^{'}$, A$_{1}$$^{'}$ and A$_{2}$$^{''}$ modes. While atoms have in-plane motion in E modes they move in out-of-plane direction in A modes. Among these 6 optical branches only 2E$^{''}$, 2E$^{'}$ and A$_{1}^{'}$ modes are Raman-active and measurable in Raman intensity measurements. Ab initio calculations show that the lattice dynamics of single layer MoSe$_{2}$ is quite sensitive to the applied strain \citep{Horzum1}. As shown in Figure \ref{mose1} while the phonon branches having in-plane motion symmetry are softened significantly, out-of-plane Raman-active phonon branch are almost insensitive to applied biaxial strain.

Horzum \textit{et al.} also showed that not only vibrational properties but also electronic characteristics of the monolayer MoSe$_{2}$ can be tuned by applied biaxial strain. As presented in Figure \ref{mose2}, when the tensile strain is increased, electronic states around the $K$ symmetry point rapidly shift down to lower energies and the shift in other regions of the Brillouin Zone is quite slower. Therefore, after a certain strain value (3 percent) monolayer MoSe$_{2}$ which is a direct bandgap semiconductor in its ground state becomes an indirect semiconductor.

\begin{figure}[t]
\begin{center}
\includegraphics[width=11.5cm]{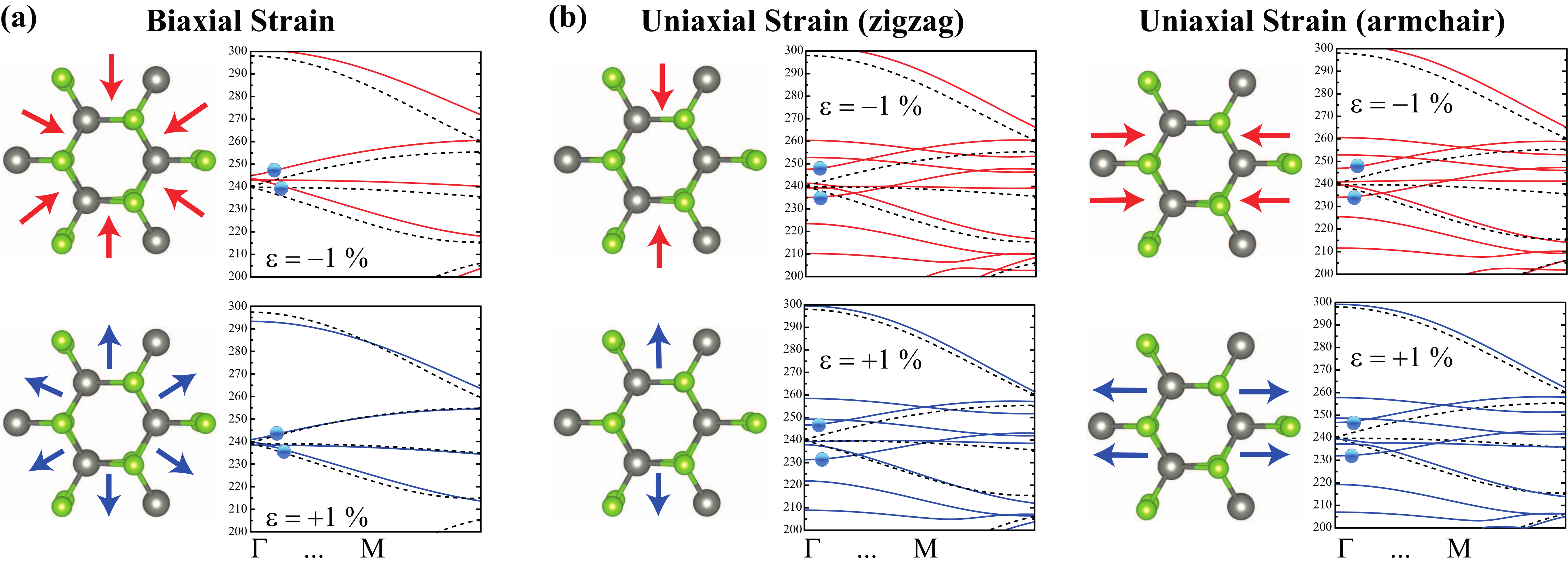}
\end{center}
\caption{\label{wse2} Strain-dependent phonon dispersion and frequency shift in single layer WSe$_{2}$. Adapted from \citep{Sahin4}.}
\end{figure}

\section{ WSe$_{2}$}

Following theoretical findings on single layer MoSe$_{2}$, one can expect similar behavior in electronic and vibrational characteristics of single layer WSe$_{2}$. Both structures have the same crystal symmetry and both Mo and W atoms have the same chemical characteristics. However, Sahin \textit{et al.} reported some unique features, that do not exist in other TMDs, in the vibrational spectrum of WSe$_{2}$ \citep{Sahin4}. As shown in Fig.~\ref{wse1}(a), experimentally observed most prominent Raman peak is located at 252 cm$^{-1}$. While other semiconducting TMDs are characterized by well-separated E$_{g}$ and A$_{g}$ Raman peaks, bulk WSe$_{2}$ has just one prominent peak in its spectrum. Experiments revealed that there is an additional Raman-active mode which is only visible when the structure becomes ultra-thin. Deeper understanding of this anomaly was provided by theoretical calculations performed \citep{Sahin4}. As shown in Figure~\ref{wse1}(b), in theoretically observed phonon dispersion of the bulk WSe$_{2}$ well-separated E$_{g}$ and A$_{g}$ Raman-active modes have an "accidental degeneracy". Furthermore, it was also seen that bilayer and monolayer crystal structures of WSe$_{2}$ possess the same unique feature (see Figures \ref{wse1}(d)-(e)). 

Then, the question is why the accidental degeneracy, which is a characteristics of WSe$_{2}$, is broken in experiments. Since the secondary Raman-active peak (shoulder of the most prominent peak) in spectrum is visible only in ultra-thin crystals, an explanation based on the substrate-induced strain is quite reasonable. As shown in Figure \ref{wse2}, while the biaxial strain on monolayer crystal structure has no influence on the accidentally degenerate modes, these phonon branches become well-separated upon application of mild uniaxial strain. The response of the crystal structure against tensile strain is found to be the same in both zigzag and armchair directions. Therefore, one can conclude the reason of broken accidentally degeneracy as the substrate-induced uniaxial strain. 

\begin{figure}[t]
\begin{center}
\includegraphics[width=11.5cm]{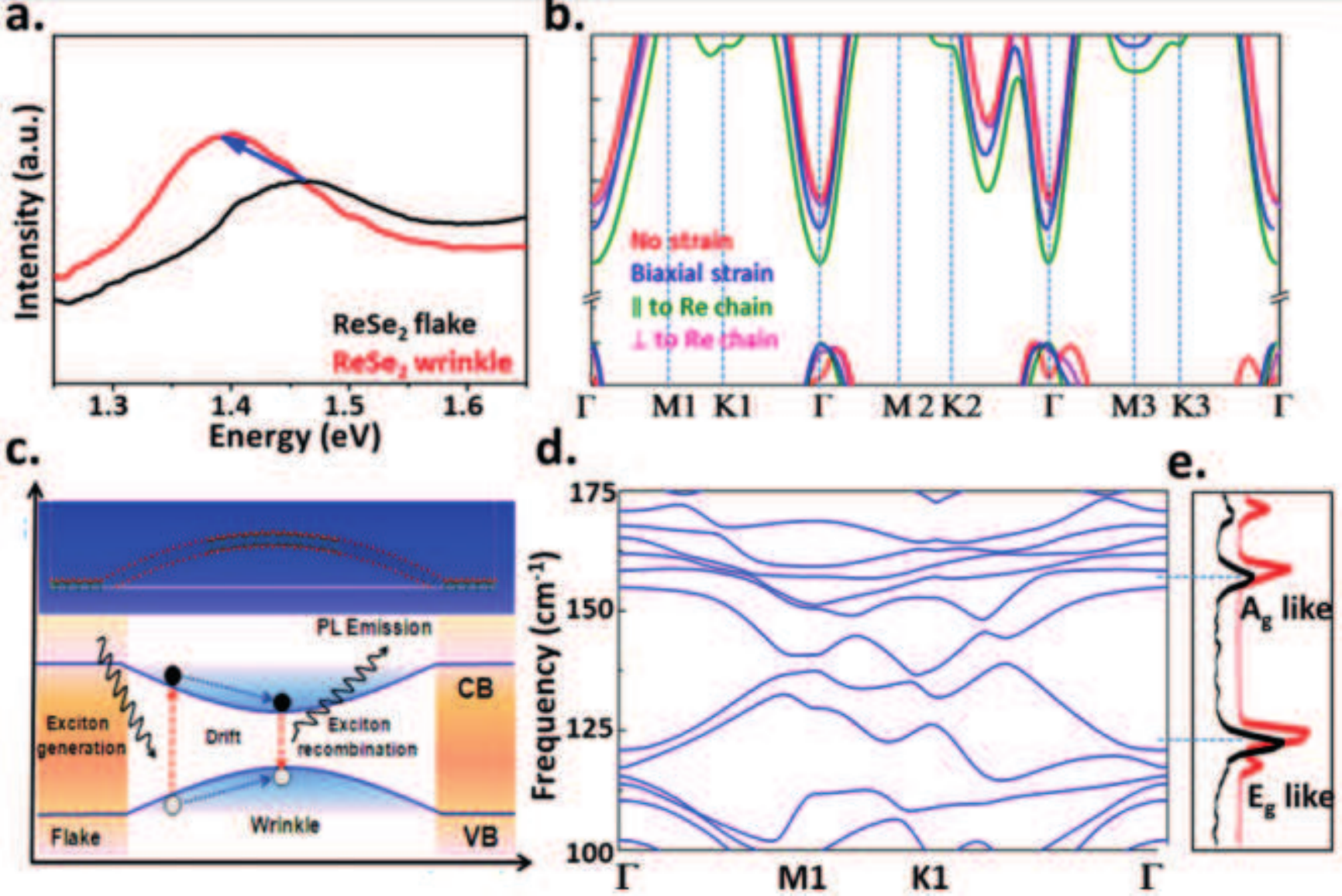}
\end{center}
\caption{\label{rese1} Strain-dependent phonon dispersion and frequency shift in single layer ReSe$_{2}$. Adapted from \citep{Yang2}.}
\end{figure}

\begin{figure}[b]
\begin{center}
\includegraphics[width=11.5cm]{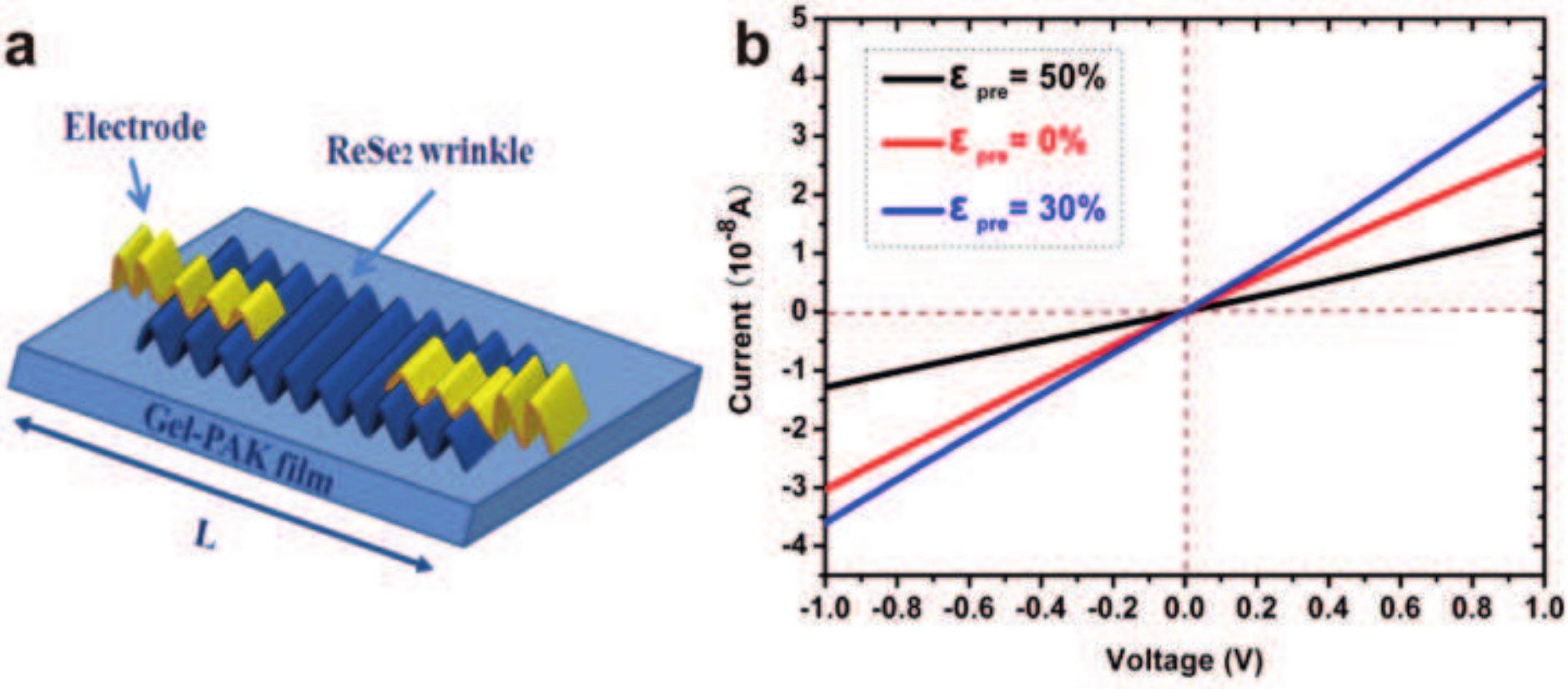}
\end{center}
\caption{\label{rese2} Electrical characterization on strained ReSe$_{2}$. (a) Schematic drawing of a two-terminal ReSe$_{2}$ wrinkles device. (b) I-V curves of a twoterminal ReSe$_{2}$ wrinkles device under different prestrain. Adapted from \citep{Yang2}.}
\end{figure}

\section{ ReSe$_{2}$}

In addition to large amount of research effort on dichalcogenides of molybdenum and tungsten (group-VI TMDs), recent studies have also focused on crystal structures of group-VII TMDs such as ReS$_{2}$ and ReSe$_{2}$ \citep{res,Yang2}. Differing from other TMDs extra electron in the d-orbitals of Rhenium atom leads to large differences. For example, ReS$_{2}$ and ReSe$_{2}$ lack dimensional-crossover-driven indirect-to-direct gap transition but possess strong in-plane anisotropy and a rich Raman spectrum.

In a recent study, Yang \textit{et al.} showed that local strain induced by generation of wrinkles reduces the optical gap as evidenced by red-shifted photoluminescence peak (Fig.~\ref{rese1}(a)-(c)). It was also shown that locally-strained domain is not only provides a enhanced PL intensity but also a suitable playground for exciton recombination \citep{Yang2}. In addition, theoretical calculations, shown in Fig.~\ref{rese1}(b), revealed that while the effect of local strain along the direction perpendicular to Re-Re dimers is ignorable, strain applied along the direction parallel to the dimers significantly alters the band gap. It was also experimentally demonstrated that formation of wrinkles may also leads to the presence of local magnetism. Furthermore, locally wrinkled domains in in ReSe$_{2}$ have also strain-tunable I-V characteristics as shown in Fig.~\ref{rese2}.

\begin{figure}[t]
\begin{center}
\includegraphics[width=7.5cm]{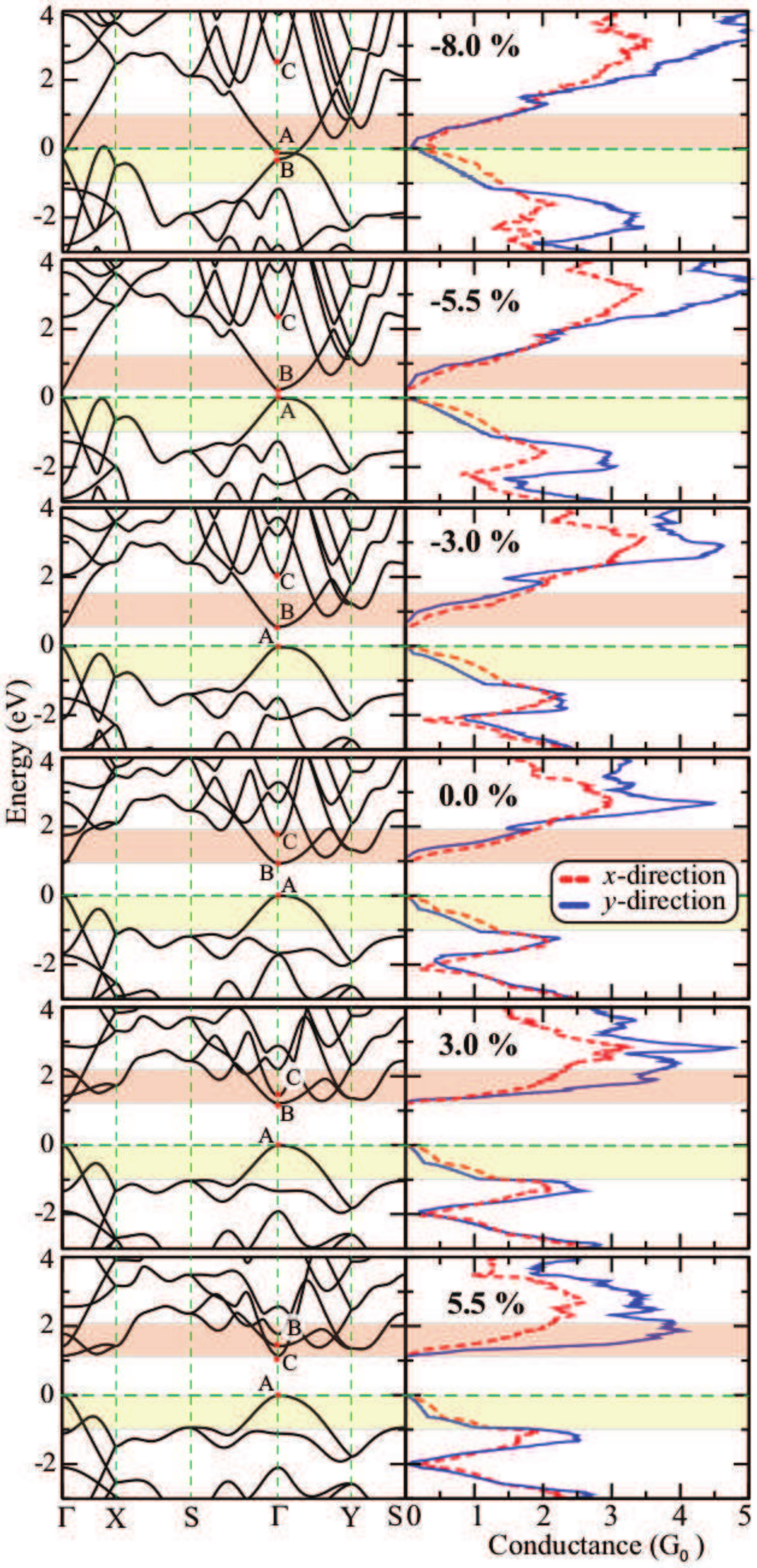}
\end{center}
\caption{\label{bp1} Evolution of electronic band dispersion and conductance of single-layer phosphorene under biaxial strain. A, B, and C mark the top of the valence band, bottom of the conduction band, and the second lowest conduction band at the $\Gamma$ point, respectively. Shaded regions depict the upper part of the valence and lower part of the conduction bands.}
\end{figure}

\section{Black Phosphorus}

Another promising material in the family of ultra-thin crystal structures is phosphorene (black phosphorus). Structural stability, anisotropic structural properties, a high on-current, a high hole field-effect mobility, and a high on/off ratio in few layer phosphorene FETs make black phosphorus suitable for various optoelectronic nanoscale device applications. Buscema \textit{et al.} demonstrated that black phosphorus is an appealing candidate for tunable photodetection applications due its unique properties such as (i) FETs allowing for ambipolar operation in the dark state
and (ii) broadband (from the visible region up to 940 nm) with fast detection (rise time of about 1 ms) when illuminated \citep{Buscema1}.

In addition, strain dependent electronic, quantum transport and optical properties of black phosphorus were investigated by a recent theoretical study \citep{Horzum2}. As shown in Fig.~\ref{bp1}, single layer phosphorus has different responses against compressive and tensile strain. While tensile strain has no significant effect on the electronic structure, energy bandgap is quite sensitive to the compressive strain. It is also seen that while the states at the valence band edge are not affected by the strain, conduction band edges, especially around the $\Gamma$ point, rapidly shifts down in energy space. Therefore, such behavior results in metallicity after 6 percent compressive strain. However, rapid shift in C band edge upon tensile strain may also lead to metallicity in monolayer phosphorus. Furthermore, stemming from the structural anisotropy, black phosphorus shows different quantum transport characteristics in armchair and zigzag directions.

To conclude, graphene, silicene and similar ultra-thin crystals are quite promising building blocks for a wide range of strain-based device applications such as strain sensors, stretchable electrodes, flexible field-effect transistors, electromechanical, piezoelectric devices.
\bibliography{References}

\begin{thebibliography}{202}
\providecommand{\natexlab}[1]{#1}
\providecommand{\url}[1]{\texttt{#1}}
\providecommand{\urlprefix}{}

\bibitem[{Abersfelder et~al.(2010)Abersfelder, White, Rzepa, and
  Scheschkewitz}]{Abersfelder1}
Abersfelder, K., White, A.J.P., Rzepa, H.S., Scheschkewitz, D.: A tricyclic
  aromatic isomer of hexasilabenzene.
\newblock Science {\bf 327}, 564--566 (2010)

\bibitem[{Acun et~al.(2013)Acun, Poelsema, Zandvliet, and van Gastel}]{Acun1}
Acun, A., Poelsema, B., Zandvliet, H.J.W., van Gastel, R.: The instability of
  silicene on {A}g(111).
\newblock Applied Physics Letters {\bf 103}, 263119 (2013)

\bibitem[{Ahn et~al.(2015)Ahn, Lee, Heo, Sung, Kim, Hwang, and Jo}]{ahn15}
Ahn, J.H., Lee, M.J., Heo, H., Sung, J.H., Kim, K., Hwang, H., Jo, M.H.:
  Deterministic two-dimensional polymorphism growth of hexagonal n-type
  {SnS$_2$} and orthorhombic p-type {SnS} crystals.
\newblock Nano Letters {\bf 15}, 3703--3708 (2015)

\bibitem[{Aierken et~al.(2015)Aierken, Sahin, Iyikanat, Horzum, Suslu, Chen,
  Senger, Tongay, and Peeters}]{erp15}
Aierken, Y., Sahin, H., Iyikanat, F., Horzum, S., Suslu, A., Chen, B., Senger,
  R.T., Tongay, S., Peeters, F.M.: Portlandite crystal: Bulk, bilayer, and
  monolayer structures.
\newblock Phys. Rev. B {\bf 91}, 245413 (2015)

\bibitem[{Allen et~al.(2013)Allen, Berlijn, Casavant, and Soler}]{Allen1}
Allen, P.B., Berlijn, T., Casavant, D.A., Soler, J.M.: Recovering hidden
  {B}loch character: Unfolding electrons, phonons, and slabs.
\newblock Phys. Rev. B {\bf 87}, 085322 (2013)

\bibitem[{An et~al.(2014)An, Wang, Vasilopoulos, Liu, Chen, Dong, and
  Zhai}]{An1}
An, R.L., Wang, X.F., Vasilopoulos, P., Liu, Y.S., Chen, A.B., Dong, Y.J.,
  Zhai, M.X.: Vacancy effects on electric and thermoelectric properties of
  zigzag silicene nanoribbons.
\newblock The Journal of Physical Chemistry C {\bf 118}, 21339--21346 (2014)

\bibitem[{Arafune et~al.(2013)Arafune, Lin, Nagao, Kawai, and
  Takagi}]{Arafune1}
Arafune, R., Lin, C.L., Nagao, R., Kawai, M., Takagi, N.: Comment on
  {``E}vidence for {D}irac fermions in a honeycomb lattice based on
  silicon{''}.
\newblock Phys. Rev. Lett. {\bf 110}, 229701 (2013)

\bibitem[{Aufray et~al.(2010)Aufray, Kara, Vizzini, Oughaddou, L\'eandri,
  Ealet, and {Le Lay}}]{Aufray1}
Aufray, B., Kara, A., Vizzini, S., Oughaddou, H., L\'eandri, C., Ealet, B., {Le
  Lay}, G.: Graphene-like silicon nanoribbons on {A}g(110): A possible
  formation of silicene.
\newblock Applied Physics Letters {\bf 96}, 183102 (2010)

\bibitem[{Barton and Burns(1978)}]{Barton1}
Barton, T.J., Burns, G.T.: Unambiguous generation and trapping of a
  silabenzene.
\newblock Journal of the American Chemical Society {\bf 100}, 5246--5246 (1978)

\bibitem[{Bernevig et~al.(2006)Bernevig, Hughes, and Zhang}]{Bernevig1}
Bernevig, B.A., Hughes, T.L., Zhang, S.C.: Quantum spin hall effect and
  topological phase transition in {HgTe} quantum wells.
\newblock Science {\bf 314}, 1757--1761 (2006)

\bibitem[{Bianco et~al.(2013)Bianco, Butler, Jiang, Restrepo, Windl, and
  Goldberger}]{bia13}
Bianco, E., Butler, S., Jiang, S., Restrepo, O.D., Windl, W., Goldberger, J.E.:
  Stability and exfoliation of germanane: A germanium graphane analogue.
\newblock ACS Nano {\bf 7}, 4414--4421 (2013)

\bibitem[{Bl\"ochl(1994)}]{Blochl1}
Bl\"ochl, P.E.: Projector augmented-wave method.
\newblock Phys. Rev. B {\bf 50}, 17953--17979 (1994)

\bibitem[{Boustani et~al.(1999)Boustani, Quandt, Hernández, and
  Rubio}]{Boustani1}
Boustani, I., Quandt, A., Hernández, E., Rubio, A.: New boron based
  nanostructured materials.
\newblock The Journal of Chemical Physics {\bf 110} (1999)

\bibitem[{Bridgman(1914)}]{Bridgman1}
Bridgman, P.W.: Two new modifications of phosphorus.
\newblock Journal of the American Chemical Society {\bf 36}, 1344--1363 (1914)

\bibitem[{Brumfiel(2013)}]{Brumfiel1}
Brumfiel, G.: Sticky problem snares wonder material.
\newblock Nature {\bf 495}, 152--153 (2013)

\bibitem[{Buscema et~al.(2014)Buscema, Groenendijk, Steele, van~der Zant, and
  Castellanos-Gomez}]{Buscema1}
Buscema, M., Groenendijk, D.J., Steele, G.A., van~der Zant, H.S.J.,
  Castellanos-Gomez, A.: {Photovoltaic effect in few-layer black phosphorus PN
  junctions defined by local electrostatic gating} {\bf 5} (2014)

\bibitem[{Cahangirov et~al.(2009)Cahangirov, Topsakal, Akt\"urk,
  \ifmmode~\mbox{\c{S}}\else \c{S}\fi{}ahin, and Ciraci}]{Cahangirov1}
Cahangirov, S., Topsakal, M., Akt\"urk, E., \ifmmode~\mbox{\c{S}}\else
  \c{S}\fi{}ahin, H., Ciraci, S.: Two- and one-dimensional honeycomb structures
  of silicon and germanium.
\newblock Phys. Rev. Lett. {\bf 102}, 236804 (2009)

\bibitem[{Cahangirov et~al.(2010)Cahangirov, Topsakal, and
  Ciraci}]{Cahangirov2}
Cahangirov, S., Topsakal, M., Ciraci, S.: Armchair nanoribbons of silicon and
  germanium honeycomb structures.
\newblock Phys. Rev. B {\bf 81}, 195120 (2010)

\bibitem[{Cahangirov et~al.(2013)Cahangirov, Audiffred, Tang, Iacomino, Duan,
  Merino, and Rubio}]{Cahangirov3}
Cahangirov, S., Audiffred, M., Tang, P., Iacomino, A., Duan, W., Merino, G.,
  Rubio, A.: Electronic structure of silicene on {A}g(111): Strong
  hybridization effects.
\newblock Phys. Rev. B {\bf 88}, 035432 (2013)

\bibitem[{Cahangirov et~al.(2014{\natexlab{a}})Cahangirov,
  \"Oz\ifmmode~\mbox{\c{c}}\else \c{c}\fi{}elik, Rubio, and
  Ciraci}]{Cahangirov5}
Cahangirov, S., \"Oz\ifmmode~\mbox{\c{c}}\else \c{c}\fi{}elik, V.O., Rubio, A.,
  Ciraci, S.: Silicite: The layered allotrope of silicon.
\newblock Phys. Rev. B {\bf 90}, 085426 (2014{\natexlab{a}})

\bibitem[{Cahangirov et~al.(2014{\natexlab{b}})Cahangirov,
  \"Oz\ifmmode~\mbox{\c{c}}\else \c{c}\fi{}elik, Xian, Avila, Cho, Asensio,
  Ciraci, and Rubio}]{Cahangirov4}
Cahangirov, S., \"Oz\ifmmode~\mbox{\c{c}}\else \c{c}\fi{}elik, V.O., Xian, L.,
  Avila, J., Cho, S., Asensio, M.C., Ciraci, S., Rubio, A.: Atomic structure of
  the
  $\sqrt{3}\phantom{\rule{0.16em}{0ex}}\ifmmode\times\else\texttimes\fi{}\phantom{\rule{0.16em}{0ex}}\sqrt{3}$
  phase of silicene on {A}g(111).
\newblock Phys. Rev. B {\bf 90}, 035448 (2014{\natexlab{b}})

\bibitem[{\c{C}ak{\i}r et~al.(2014)\c{C}ak{\i}r, Sahin, and Peeters}]{Cakir1}
\c{C}ak{\i}r, D., Sahin, H., Peeters, F.M.: Tuning of the electronic and
  optical properties of single-layer black phosphorus by strain.
\newblock Phys. Rev. B {\bf 90}, 205421 (2014)

\bibitem[{Chen et~al.(2013{\natexlab{a}})Chen, Feng, and Wu}]{Chen4}
Chen, L., Feng, B., Wu, K.: Observation of a possible superconducting gap in
  silicene on {A}g(111) surface.
\newblock Applied Physics Letters {\bf 102}, 081602 (2013{\natexlab{a}})

\bibitem[{Chen et~al.(2013{\natexlab{b}})Chen, Li, Feng, Ding, Qiu, Cheng, Wu,
  and Meng}]{Chen2}
Chen, L., Li, H., Feng, B., Ding, Z., Qiu, J., Cheng, P., Wu, K., Meng, S.:
  Spontaneous symmetry breaking and dynamic phase transition in monolayer
  silicene.
\newblock Phys. Rev. Lett. {\bf 110}, 085504 (2013{\natexlab{b}})

\bibitem[{Chen et~al.(2012)Chen, Liu, Feng, He, Cheng, Ding, Meng, Yao, and
  Wu}]{Chen1}
Chen, L., Liu, C.C., Feng, B., He, X., Cheng, P., Ding, Z., Meng, S., Yao, Y.,
  Wu, K.: Evidence for {D}irac fermions in a honeycomb lattice based on
  silicon.
\newblock Phys. Rev. Lett. {\bf 109}, 056804 (2012)

\bibitem[{Chen et~al.(2013{\natexlab{c}})Chen, Liu, Feng, He, Cheng, Ding,
  Meng, Yao, and Wu}]{Chen3}
Chen, L., Liu, C.C., Feng, B., He, X., Cheng, P., Ding, Z., Meng, S., Yao, Y.,
  Wu, K.: Reply to the comment by {R. A}rafune et al.
\newblock Phys. Rev. Lett. {\bf 110}, 229702 (2013{\natexlab{c}})

\bibitem[{Chen and Weinert(2014)}]{Chen5}
Chen, M.X., Weinert, M.: {Revealing the Substrate Origin of the Linear
  Dispersion of Silicene/Ag(111)}.
\newblock Nano letters  (2014)

\bibitem[{Chiappe et~al.(2014)Chiappe, Scalise, Cinquanta, Grazianetti, van~den
  Broek, Fanciulli, Houssa, and Molle}]{Chiappe1}
Chiappe, D., Scalise, E., Cinquanta, E., Grazianetti, C., van~den Broek, B.,
  Fanciulli, M., Houssa, M., Molle, A.: Two-dimensional {S}i nanosheets with
  local hexagonal structure on a {MoS$_2$} surface.
\newblock Advanced Materials {\bf 26}, 2096--2101 (2014)

\bibitem[{Cinquanta et~al.(2013)Cinquanta, Scalise, Chiappe, Grazianetti,
  van~den Broek, Houssa, Fanciulli, and Molle}]{Cinquanta1}
Cinquanta, E., Scalise, E., Chiappe, D., Grazianetti, C., van~den Broek, B.,
  Houssa, M., Fanciulli, M., Molle, A.: Getting through the nature of silicene:
  An sp2â“sp3 two-dimensional silicon nanosheet.
\newblock The Journal of Physical Chemistry C {\bf 117}, 16719--16724 (2013)

\bibitem[{Cudazzo et~al.(2010)Cudazzo, Attaccalite, Tokatly, and
  Rubio}]{Cudazzo1}
Cudazzo, P., Attaccalite, C., Tokatly, I.V., Rubio, A.: Strong charge-transfer
  excitonic effects and the {B}ose-{E}instein exciton condensate in graphane.
\newblock Phys. Rev. Lett. {\bf 104}, 226804 (2010)

\bibitem[{D\'avila et~al.(2014)D\'avila, Xian, Cahangirov, Rubio, and {Le
  Lay}}]{Davila1}
D\'avila, M.E., Xian, L., Cahangirov, S., Rubio, A., {Le Lay}, G.: Germanene: a
  novel two-dimensional germanium allotrope akin to graphene and silicene.
\newblock New Journal of Physics {\bf 16}, 095002 (2014)

\bibitem[{De~Padova et~al.(2013{\natexlab{a}})De~Padova, Avila, Resta,
  Razado-Colambo, Quaresima, Ottaviani, Olivieri, Bruhn, Vogt, Asensio, and
  Lay}]{DePadova6}
De~Padova, P., Avila, J., Resta, A., Razado-Colambo, I., Quaresima, C.,
  Ottaviani, C., Olivieri, B., Bruhn, T., Vogt, P., Asensio, M.C., Lay, G.L.:
  The quasiparticle band dispersion in epitaxial multilayer silicene.
\newblock Journal of Physics: Condensed Matter {\bf 25}, 382202
  (2013{\natexlab{a}})

\bibitem[{De~Padova et~al.(2012)De~Padova, Kubo, Olivieri, Quaresima, Nakayama,
  Aono, and Le~Lay}]{DePadova5}
De~Padova, P., Kubo, O., Olivieri, B., Quaresima, C., Nakayama, T., Aono, M.,
  Le~Lay, G.: Multilayer silicene nanoribbons.
\newblock Nano Letters {\bf 12}, 5500--5503 (2012)

\bibitem[{De~Padova et~al.(2008{\natexlab{a}})De~Padova, Leandri, Vizzini,
  Quaresima, Perfetti, Olivieri, Oughaddou, Aufray, and {Le Lay}}]{DePadova2}
De~Padova, P., Leandri, C., Vizzini, S., Quaresima, C., Perfetti, P., Olivieri,
  B., Oughaddou, H., Aufray, B., {Le Lay}, G.: Burning match oxidation process
  of silicon nanowires screened at the atomic scale.
\newblock Nano Letters {\bf 8}, 2299--2304 (2008{\natexlab{a}})

\bibitem[{De~Padova et~al.(2014)De~Padova, Ottaviani, Quaresima, Olivieri,
  Imperatori, Salomon, Angot, Quagliano, Romano, Vona, Muniz-Miranda, Generosi,
  Paci, and Lay}]{DePadova7}
De~Padova, P., Ottaviani, C., Quaresima, C., Olivieri, B., Imperatori, P.,
  Salomon, E., Angot, T., Quagliano, L., Romano, C., Vona, A., Muniz-Miranda,
  M., Generosi, A., Paci, B., Lay, G.L.: 24 h stability of thick multilayer
  silicene in air.
\newblock 2D Materials {\bf 1}, 021003 (2014)

\bibitem[{De~Padova et~al.(2011)De~Padova, Quaresima, Olivieri, Perfetti, and
  {Le Lay}}]{DePadova4}
De~Padova, P., Quaresima, C., Olivieri, B., Perfetti, P., {Le Lay}, G.:
  sp$^2$-like hybridization of silicon valence orbitals in silicene
  nanoribbons.
\newblock Applied Physics Letters {\bf 98}, 081909 (2011)

\bibitem[{De~Padova et~al.(2010)De~Padova, Quaresima, Ottaviani, Sheverdyaeva,
  Moras, Carbone, Topwal, Olivieri, Kara, Oughaddou, Aufray, and {Le
  Lay}}]{DePadova3}
De~Padova, P., Quaresima, C., Ottaviani, C., Sheverdyaeva, P.M., Moras, P.,
  Carbone, C., Topwal, D., Olivieri, B., Kara, A., Oughaddou, H., Aufray, B.,
  {Le Lay}, G.: Evidence of graphene-like electronic signature in silicene
  nanoribbons.
\newblock Applied Physics Letters {\bf 96}, 261905 (2010)

\bibitem[{De~Padova et~al.(2008{\natexlab{b}})De~Padova, Quaresima, Perfetti,
  Olivieri, Leandri, Aufray, Vizzini, and {Le Lay}}]{DePadova1}
De~Padova, P., Quaresima, C., Perfetti, P., Olivieri, B., Leandri, C., Aufray,
  B., Vizzini, S., {Le Lay}, G.: Growth of straight, atomically perfect, highly
  metallic silicon nanowires with chiral asymmetry.
\newblock Nano Letters {\bf 8}, 271--275 (2008{\natexlab{b}})

\bibitem[{De~Padova et~al.(2013{\natexlab{b}})De~Padova, Vogt, Resta, Avila,
  Razado-Colambo, Quaresima, Ottaviani, Olivieri, Bruhn, Hirahara, Shirai,
  Hasegawa, Carmen~Asensio, and Le~Lay}]{DePadova8}
De~Padova, P., Vogt, P., Resta, A., Avila, J., Razado-Colambo, I., Quaresima,
  C., Ottaviani, C., Olivieri, B., Bruhn, T., Hirahara, T., Shirai, T.,
  Hasegawa, S., Carmen~Asensio, M., Le~Lay, G.: Evidence of {D}irac fermions in
  multilayer silicene.
\newblock Applied Physics Letters {\bf 102}, 163106 (2013{\natexlab{b}})

\bibitem[{Derivaz et~al.(2015)Derivaz, Dentel, Stephan, Hanf, Mehdaoui, Sonnet,
  and Pirri}]{Derivaz1}
Derivaz, M., Dentel, D., Stephan, R., Hanf, M.C., Mehdaoui, A., Sonnet, P.,
  Pirri, C.: Continuous germanene layer on {A}l(111).
\newblock Nano Letters {\bf 15}, 2510--2516 (2015)

\bibitem[{Ding et~al.(1991)Ding, Chan, and Ho}]{Ding1}
Ding, Y.G., Chan, C.T., Ho, K.M.: Structure of the ($\surd${}3
  \ifmmode\times\else\texttimes\fi{} $\surd${}3 ) \textit{R}
  30\ifmmode^\circ\else\textdegree\fi{} {A}g/{S}i(111) surface from
  first-principles calculations.
\newblock Phys. Rev. Lett. {\bf 67}, 1454--1457 (1991)

\bibitem[{Drummond et~al.(2012)Drummond, Z\'olyomi, and Fal'ko}]{Drummond1}
Drummond, N.D., Z\'olyomi, V., Fal'ko, V.I.: Electrically tunable band gap in
  silicene.
\newblock Phys. Rev. B {\bf 85}, 075423 (2012)

\bibitem[{Du et~al.(2014)Du, Zhuang, Liu, Xu, Eilers, Wu, Cheng, Zhao, Pi, See,
  Peleckis, Wang, and Dou}]{Du14}
Du, Y., Zhuang, J., Liu, H., Xu, X., Eilers, S., Wu, K., Cheng, P., Zhao, J.,
  Pi, X., See, K.W., Peleckis, G., Wang, X., Dou, S.X.: Tuning the band gap in
  silicene by oxidation.
\newblock ACS Nano {\bf 8}, 10019--10025 (2014)

\bibitem[{Durgun et~al.(2005)Durgun, Tongay, and Ciraci}]{Durgun1}
Durgun, E., Tongay, S., Ciraci, S.: Silicon and {III-V} compound nanotubes:
  Structural and electronic properties.
\newblock Phys. Rev. B {\bf 72}, 075420 (2005)

\bibitem[{Dutta and Wakabayashi(2015)}]{Dutta1}
Dutta, S., Wakabayashi, K.: {Momentum shift of Dirac cones in the
  silicene-intercalated compound CaSi2}.
\newblock Physical Review B {\bf 91}, 201410 (2015)

\bibitem[{Elias et~al.(2009)Elias, Nair, Mohiuddin, Morozov, Blake, Halsall,
  Ferrari, Boukhvalov, Katsnelson, Geim, and Novoselov}]{Elias1}
Elias, D.C., Nair, R.R., Mohiuddin, T.M.G., Morozov, S.V., Blake, P., Halsall,
  M.P., Ferrari, A.C., Boukhvalov, D.W., Katsnelson, M.I., Geim, A.K.,
  Novoselov, K.S.: Control of graphene's properties by reversible
  hydrogenation: Evidence for graphane.
\newblock Science {\bf 323}, 610--613 (2009)

\bibitem[{Enriquez et~al.(2014)Enriquez, Kara, Mayne, Dujardin, Jamgotchian,
  Aufray, and Oughaddou}]{Enriquez1}
Enriquez, H., Kara, A., Mayne, A.J., Dujardin, G., Jamgotchian, H., Aufray, B.,
  Oughaddou, H.: Atomic structure of the
  ($2\sqrt{3}\times2\sqrt{3}r$)30$\degree$ of silicene on {A}g(111) surface.
\newblock Journal of Physics: Conference Series {\bf 491}, 012004 (2014)

\bibitem[{Esfahani et~al.(2015)Esfahani, Leenaerts, Sahin, Partoens, and
  Peeters}]{Esfahani1}
Esfahani, D.N., Leenaerts, O., Sahin, H., Partoens, B., Peeters, F.M.:
  Structural transitions in monolayer {MoS$_2$} by lithium adsorption.
\newblock The Journal of Physical Chemistry C {\bf 119}, 10602--10609 (2015)

\bibitem[{Ezawa(2012)}]{Ezawa1}
Ezawa, M.: Valley-polarized metals and quantum anomalous {H}all effect in
  silicene.
\newblock Phys. Rev. Lett. {\bf 109}, 055502 (2012)

\bibitem[{Fagan et~al.(2000)Fagan, Baierle, Mota, da~Silva, and
  Fazzio}]{Fagan1}
Fagan, S.B., Baierle, R.J., Mota, R., da~Silva, A.J.R., Fazzio, A.: Ab initio
  calculations for a hypothetical material: Silicon nanotubes.
\newblock Phys. Rev. B {\bf 61}, 9994--9996 (2000)

\bibitem[{Feng et~al.(2012{\natexlab{a}})Feng, Ding, Meng, Yao, He, Cheng,
  Chen, and Wu}]{Feng1}
Feng, B., Ding, Z., Meng, S., Yao, Y., He, X., Cheng, P., Chen, L., Wu, K.:
  Evidence of silicene in honeycomb structures of silicon on {A}g(111).
\newblock Nano Letters {\bf 12}, 3507--3511 (2012{\natexlab{a}})

\bibitem[{Feng et~al.(2012{\natexlab{b}})Feng, Qian, Huang, and Li}]{Feng2}
Feng, J., Qian, X., Huang, C.W., Li, J.: {Strain-engineered artificial atom as
  a broad-spectrum solar energy funnel}.
\newblock Nature Photonics {\bf 6}, 866--872 (2012{\natexlab{b}})

\bibitem[{Fleurence et~al.(2012)Fleurence, Friedlein, Ozaki, Kawai, Wang, and
  Yamada-Takamura}]{Fleurence1}
Fleurence, A., Friedlein, R., Ozaki, T., Kawai, H., Wang, Y., Yamada-Takamura,
  Y.: Experimental evidence for epitaxial silicene on diboride thin films.
\newblock Phys. Rev. Lett. {\bf 108}, 245501 (2012)

\bibitem[{Fukaya et~al.(2013)Fukaya, Mochizuki, Maekawa, Wada, Hyodo, Matsuda,
  and Kawasuso}]{Fukaya1}
Fukaya, Y., Mochizuki, I., Maekawa, M., Wada, K., Hyodo, T., Matsuda, I.,
  Kawasuso, A.: Structure of silicene on a {A}g(111) surface studied by
  reflection high-energy positron diffraction.
\newblock Phys. Rev. B {\bf 88}, 205413 (2013)

\bibitem[{Gao et~al.(2012)Gao, Gao, Cannuccia, Taha-Tijerina, Balicas, Mathkar,
  Narayanan, Liu, Gupta, Peng, Yin, Rubio, and Ajayan}]{Gao1}
Gao, G., Gao, W., Cannuccia, E., Taha-Tijerina, J., Balicas, L., Mathkar, A.,
  Narayanan, T.N., Liu, Z., Gupta, B.K., Peng, J., Yin, Y., Rubio, A., Ajayan,
  P.M.: Artificially stacked atomic layers: Toward new van der {W}aals solids.
\newblock Nano Letters {\bf 12}, 3518--3525 (2012)

\bibitem[{Garcia et~al.(2011)Garcia, de~Lima, Assali, and Justo}]{Garcia1}
Garcia, J.C., de~Lima, D.B., Assali, L.V.C., Justo, J.F.: Group iv graphene-
  and graphane-like nanosheets.
\newblock The Journal of Physical Chemistry C {\bf 115}, 13242--13246 (2011)

\bibitem[{Geim and Grigorieva(2013)}]{Geim1}
Geim, A.K., Grigorieva, I.V.: {V}an der {W}aals heterostructures.
\newblock Nature {\bf 499}, 419--425 (2013)

\bibitem[{Gori et~al.(2013)Gori, Pulci, Ronci, Colonna, and Bechstedt}]{Gori1}
Gori, P., Pulci, O., Ronci, F., Colonna, S., Bechstedt, F.: Origin of
  {D}irac-cone-like features in silicon structures on {A}g(111) and {A}g(110).
\newblock Journal of Applied Physics {\bf 114}, 113710 (2013)

\bibitem[{Guo et~al.(2013)Guo, Furuya, Iwata, and Oshiyama}]{Guo1}
Guo, Z.X., Furuya, S., Iwata, J.I., Oshiyama, A.: Absence and presence of
  {D}irac electrons in silicene on substrates.
\newblock Phys. Rev. B {\bf 87}, 235435 (2013)

\bibitem[{Guo and Oshiyama(2014)}]{Guo2}
Guo, Z.X., Oshiyama, A.: Structural tristability and deep {D}irac states in
  bilayer silicene on {A}g(111) surfaces.
\newblock Phys. Rev. B {\bf 89}, 155418 (2014)

\bibitem[{Guzm\'an-Verri and Lew Yan~Voon(2007)}]{Guzman1}
Guzm\'an-Verri, G.G., Lew Yan~Voon, L.C.: Electronic structure of silicon-based
  nanostructures.
\newblock Phys. Rev. B {\bf 76}, 075131 (2007)

\bibitem[{Han et~al.(2007)Han, \"Ozyilmaz, Zhang, and Kim}]{Han1}
Han, M.Y., \"Ozyilmaz, B., Zhang, Y., Kim, P.: Energy band-gap engineering of
  graphene nanoribbons.
\newblock Phys. Rev. Lett. {\bf 98}, 206805 (2007)

\bibitem[{Harrison and Ciraci(1974)}]{Harrison1}
Harrison, W.A., Ciraci, S.: Bond-orbital model {II}.
\newblock Phys. Rev. B {\bf 10}, 1516--1527 (1974)

\bibitem[{Hedin(1965)}]{Hedin1}
Hedin, L.: New method for calculating the one-particle {G}reen's function with
  application to the electron-gas problem.
\newblock Phys. Rev. {\bf 139}, A796--A823 (1965)

\bibitem[{Hoffmann(2013)}]{Hoffmann1}
Hoffmann, R.: Small but strong lessons from chemistry for nanoscience.
\newblock Angewandte Chemie International Edition {\bf 52}, 93--103 (2013)

\bibitem[{Hohenberg and Kohn(1964)}]{Hohenberg1}
Hohenberg, P., Kohn, W.: Inhomogeneous electron gas.
\newblock Phys. Rev. {\bf 136}, B864--B871 (1964)

\bibitem[{Horzum et~al.(2014)Horzum, \ifmmode \mbox{\c{C}}\else
  \c{C}\fi{}ak\ifmmode \imath \else~\i \fi{}r, Suh, Tongay, Huang, Ho, Wu,
  Sahin, and Peeters}]{Horzum2}
Horzum, S., \ifmmode \mbox{\c{C}}\else \c{C}\fi{}ak\ifmmode \imath \else~\i
  \fi{}r, D., Suh, J., Tongay, S., Huang, Y.S., Ho, C.H., Wu, J., Sahin, H.,
  Peeters, F.M.: Formation and stability of point defects in monolayer rhenium
  disulfide.
\newblock Phys. Rev. B {\bf 89}, 155433 (2014)

\bibitem[{Horzum et~al.(2013)Horzum, Sahin, Cahangirov, Cudazzo, Rubio, Serin,
  and Peeters}]{Horzum1}
Horzum, S., Sahin, H., Cahangirov, S., Cudazzo, P., Rubio, A., Serin, T.,
  Peeters, F.M.: Phonon softening and direct to indirect band gap crossover in
  strained single-layer mose${}_{2}$.
\newblock Phys. Rev. B {\bf 87}, 125415 (2013)

\bibitem[{Houssa et~al.(2011)Houssa, Scalise, Sankaran, Pourtois, Afanas'ev,
  and Stesmans}]{Houssa1}
Houssa, M., Scalise, E., Sankaran, K., Pourtois, G., Afanas'ev, V.V., Stesmans,
  A.: Electronic properties of hydrogenated silicene and germanene.
\newblock Applied Physics Letters {\bf 98}, 223107 (2011)

\bibitem[{Hu et~al.(2013)Hu, Zhang, and Poulikakos}]{Hu1}
Hu, M., Zhang, X., Poulikakos, D.: Anomalous thermal response of silicene to
  uniaxial stretching.
\newblock Phys. Rev. B {\bf 87}, 195417 (2013)

\bibitem[{Huang et~al.(2014)Huang, Deng, Lee, Yoon, Sumpter, Liu, Smith, and
  Wei}]{Hua14}
Huang, B., Deng, H.X., Lee, H., Yoon, M., Sumpter, B.G., Liu, F., Smith, S.C.,
  Wei, S.H.: Exceptional optoelectronic properties of hydrogenated bilayer
  silicene.
\newblock Phys. Rev. X {\bf 4}, 021029 (2014)

\bibitem[{Jahn and Teller(1937)}]{jt}
Jahn, H.A., Teller, E.: Stability of polyatomic molecules in degenerate
  electronic states. i. orbital degeneracy.
\newblock Proceedings of the Royal Society of London A: Mathematical, Physical
  and Engineering Sciences {\bf 161}, 220--235 (1937)

\bibitem[{Jamgotchian et~al.(2012)Jamgotchian, Colignon, Hamzaoui, Ealet,
  Hoarau, Aufray, and Bib\'erian}]{Jamgotchian1}
Jamgotchian, H., Colignon, Y., Hamzaoui, N., Ealet, B., Hoarau, J.Y., Aufray,
  B., Bib\'erian, J.P.: Growth of silicene layers on {A}g(111): unexpected
  effect of the substrate temperature.
\newblock Journal of Physics: Condensed Matter {\bf 24}, 172001 (2012)

\bibitem[{Jamgotchian et~al.(2014)Jamgotchian, Colignon, Ealet, Parditka,
  Hoarau, Girardeaux, Aufray, and Bib\'erian}]{Jamgotchian2}
Jamgotchian, H., Colignon, Y., Ealet, B., Parditka, B., Hoarau, J.Y.,
  Girardeaux, C., Aufray, B., Bib\'erian, J.P.: Silicene on {A}g(111) : domains
  and local defects of the observed superstructures.
\newblock Journal of Physics: Conference Series {\bf 491}, 012001 (2014)

\bibitem[{Jin et~al.(2009)Jin, Lin, Suenaga, and Iijima}]{Chuanhong1}
Jin, C., Lin, F., Suenaga, K., Iijima, S.: Fabrication of a freestanding boron
  nitride single layer and its defect assignments.
\newblock Phys. Rev. Lett. {\bf 102}, 195505 (2009)

\bibitem[{Joe et~al.(2014)Joe, Chen, Ghaemi, Finkelstein, de~la Pena, Gan, Lee,
  Yuan, Geck, MacDougall, Chiang, Cooper, Fradkin, and Abbamonte}]{ti2}
Joe, Y.I., Chen, X.M., Ghaemi, P., Finkelstein, K.D., de~la Pena, G.A., Gan,
  Y., Lee, J.C.T., Yuan, S., Geck, J., MacDougall, G.J., Chiang, T.C., Cooper,
  S.L., Fradkin, E., Abbamonte, P.: {Emergence of charge density wave domain
  walls above the superconducting dome in 1T-TiSe$_2$} {\bf 10}, 421--425
  (2014)

\bibitem[{Kaltsas and Tsetseris(2013)}]{Kaltsas1}
Kaltsas, D., Tsetseris, L.: Stability and electronic properties of ultrathin
  films of silicon and germanium.
\newblock Phys. Chem. Chem. Phys. {\bf 15}, 9710--9715 (2013)

\bibitem[{Kane and Mele(2005)}]{Kane1}
Kane, C.L., Mele, E.J.: Quantum spin {H}all effect in graphene.
\newblock Phys. Rev. Lett. {\bf 95}, 226801 (2005)

\bibitem[{Kang et~al.(2015)Kang, Sahin, and Peeters}]{hf}
Kang, J., Sahin, H., Peeters, F.M.: Mechanical properties of monolayer
  sulphides: a comparative study between {MoS$_2$}{,} {HfS$_2$} and {TiS$_3$}.
\newblock Phys. Chem. Chem. Phys. {\bf 17}, 27742--27749 (2015)

\bibitem[{Kara et~al.(2009)Kara, L\'eandri, D\'avila, De~Padova, Ealet,
  Oughaddou, Aufray, and {Le Lay}}]{Kara1}
Kara, A., L\'eandri, C., D\'avila, M., De~Padova, P., Ealet, B., Oughaddou, H.,
  Aufray, B., {Le Lay}, G.: Physics of silicene stripes.
\newblock Journal of Superconductivity and Novel Magnetism {\bf 22}, 259--263
  (2009)

\bibitem[{Kawahara et~al.(2014)Kawahara, Shirasawa, Arafune, Lin, Takahashi,
  Kawai, and Takagi}]{Kawahara1}
Kawahara, K., Shirasawa, T., Arafune, R., Lin, C.L., Takahashi, T., Kawai, M.,
  Takagi, N.: Determination of atomic positions in silicene on {A}g(111) by
  low-energy electron diffraction.
\newblock Surface Science {\bf 623}, 25 -- 28 (2014)

\bibitem[{K\"onig et~al.(2007)K\"onig, Wiedmann, Br\"une, Roth, Buhmann,
  Molenkamp, Qi, and Zhang}]{Konig1}
K\"onig, M., Wiedmann, S., Br\"une, C., Roth, A., Buhmann, H., Molenkamp, L.W.,
  Qi, X.L., Zhang, S.C.: Quantum spin {H}all insulator state in {HgTe} quantum
  wells.
\newblock Science {\bf 318}, 766--770 (2007)

\bibitem[{Kresse and Joubert(1999)}]{Kresse1}
Kresse, G., Joubert, D.: From ultrasoft pseudopotentials to the projector
  augmented-wave method.
\newblock Phys. Rev. B {\bf 59}, 1758--1775 (1999)

\bibitem[{Kr\"uger and Pollmann(1995)}]{kr}
Kr\"uger, P., Pollmann, J.: Dimer reconstruction of diamond, {S}i, and {G}e
  (001) surfaces.
\newblock Phys. Rev. Lett. {\bf 74}, 1155--1158 (1995)

\bibitem[{Lalmi et~al.(2010)Lalmi, Oughaddou, Enriquez, Kara, Vizzini, Ealet,
  and Aufray}]{Lalmi1}
Lalmi, B., Oughaddou, H., Enriquez, H., Kara, A., Vizzini, S., Ealet, B.,
  Aufray, B.: Epitaxial growth of a silicene sheet.
\newblock Applied Physics Letters {\bf 97}, 223109 (2010)

\bibitem[{Landau(1937)}]{Landau1}
Landau, L.D.: Zur theorie der phasenumwandlungen {II}.
\newblock Phys. Z. Sowjetunion {\bf 11}, 26--35 (1937)

\bibitem[{Lander et~al.(1963)Lander, Gobeli, and Morrison}]{land}
Lander, J.J., Gobeli, G.W., Morrison, J.: Structural properties of cleaved
  silicon and germanium surfaces.
\newblock Journal of Applied Physics {\bf 34} (1963)

\bibitem[{{Le Lay}(1983)}]{Lelay1}
{Le Lay}, G.: Physics and electronics of the
  noble-metal/elemental-semiconductor interface formation: A status report.
\newblock Surface Science {\bf 132}, 169 -- 204 (1983)

\bibitem[{{Le Lay} et~al.(2009){Le Lay}, Aufray, L\'eandri, Oughaddou,
  Biberian, Padova, D\'avila, Ealet, and Kara}]{LeLay2}
{Le Lay}, G., Aufray, B., L\'eandri, C., Oughaddou, H., Biberian, J.P., Padova,
  P.D., D\'avila, M., Ealet, B., Kara, A.: Physics and chemistry of silicene
  nano-ribbons.
\newblock Applied Surface Science {\bf 256}, 524 -- 529 (2009)

\bibitem[{{Le Lay} et~al.(2012){Le Lay}, Padova, Resta, Bruhn, and
  Vogt}]{Lelay3}
{Le Lay}, G., Padova, P.D., Resta, A., Bruhn, T., Vogt, P.: Epitaxial silicene:
  can it be strongly strained?
\newblock Journal of Physics D: Applied Physics {\bf 45}, 392001 (2012)

\bibitem[{Le~Lay(2015)}]{Lelay4}
Le~Lay, G.: 2{D} materials: Silicene transistors.
\newblock Nature Nanotechnology {\bf 10}, 202--203 (2015)

\bibitem[{L\'eandri et~al.(2005)L\'eandri, Lay, Aufray, Girardeaux, Avila,
  D\'avila, Asensio, Ottaviani, and Cricenti}]{Leandri1}
L\'eandri, C., Lay, G.L., Aufray, B., Girardeaux, C., Avila, J., D\'avila, M.,
  Asensio, M., Ottaviani, C., Cricenti, A.: Self-aligned silicon quantum wires
  on {A}g(110).
\newblock Surface Science {\bf 574}, L9--L15 (2005)

\bibitem[{Leb\'egue and Eriksson(2009)}]{Lebegue1}
Leb\'egue, S., Eriksson, O.: Electronic structure of two-dimensional crystals
  from ab initio theory.
\newblock Phys. Rev. B {\bf 79}, 115409 (2009)

\bibitem[{Lee et~al.(2008)Lee, Wei, Kysar, and Hone}]{Lee1}
Lee, C., Wei, X., Kysar, J.W., Hone, J.: Measurement of the elastic properties
  and intrinsic strength of monolayer graphene.
\newblock Science {\bf 321}, 385--388 (2008)

\bibitem[{Lelay et~al.(1978)Lelay, Manneville, and Kern}]{Lelay5}
Lelay, G., Manneville, M., Kern, R.: Cohesive energy of the two-dimensional
  {Si(111) 3$\times$1 Ag and Si(111)$\sqrt{3}$-R(30$\degree$) Ag} phases of the
  silver (deposit)-silicon(111) (substrate) system.
\newblock Surface Science {\bf 72}, 405 -- 422 (1978)

\bibitem[{Lew Yan~Voon et~al.(2010)Lew Yan~Voon, Sandberg, Aga, and
  Farajian}]{Voo10}
Lew Yan~Voon, L.C., Sandberg, E., Aga, R.S., Farajian, A.A.: Hydrogen compounds
  of group-iv nanosheets.
\newblock Applied Physics Letters {\bf 97}, 163114 (2010)

\bibitem[{Li et~al.(2011)Li, Zhou, Wu, Peng, Yan, Zhou, and Liu}]{li11}
Li, B., Zhou, L., Wu, D., Peng, H., Yan, K., Zhou, Y., Liu, Z.: Photochemical
  chlorination of graphene.
\newblock ACS Nano {\bf 5}, 5957--5961 (2011)

\bibitem[{Li et~al.(2015)Li, Contryman, Qian, Ardakani, Gong, Wang, Weisse,
  Lee, Zhao, Ajayan, Li, Manoharan, and Zheng}]{Li2}
Li, H., Contryman, A.W., Qian, X., Ardakani, S.M., Gong, Y., Wang, X., Weisse,
  J.M., Lee, C.H., Zhao, J., Ajayan, P.M., Li, J., Manoharan, H.C., Zheng, X.:
  {Optoelectronic crystal of artificial atoms in strain-textured molybdenum
  disulphide} {\bf 6} (2015)

\bibitem[{Li et~al.(2014)Li, Yu, Ye, Ge, Ou, Wu, Feng, Chen, and Zhang}]{Li1}
Li, L., Yu, Y., Ye, G.J., Ge, Q., Ou, X., Wu, H., Feng, D., Chen, X.H., Zhang,
  Y.: Black phosphorus field-effect transistors.
\newblock Nature Nano. {\bf 9}, 372--377 (2014)

\bibitem[{Lin et~al.(2013)Lin, Arafune, Kawahara, Kanno, Tsukahara, Minamitani,
  Kim, Kawai, and Takagi}]{Lin1}
Lin, C.L., Arafune, R., Kawahara, K., Kanno, M., Tsukahara, N., Minamitani, E.,
  Kim, Y., Kawai, M., Takagi, N.: Substrate-induced symmetry breaking in
  silicene.
\newblock Phys. Rev. Lett. {\bf 110}, 076801 (2013)

\bibitem[{Lin et~al.(2012)Lin, Arafune, Kawahara, Tsukahara, Minamitani, Kim,
  Takagi, and Kawai}]{Lin2}
Lin, C.L., Arafune, R., Kawahara, K., Tsukahara, N., Minamitani, E., Kim, Y.,
  Takagi, N., Kawai, M.: Structure of silicene grown on {A}g(111).
\newblock Applied Physics Express {\bf 5}, 045802 (2012)

\bibitem[{Liu et~al.(2014{\natexlab{a}})Liu, Baimova, Reddy, Law, Dmitriev, Wu,
  and Zhou}]{liu14}
Liu, B., Baimova, J.A., Reddy, C.D., Law, A.W.K., Dmitriev, S.V., Wu, H., Zhou,
  K.: Interfacial thermal conductance of a silicene/graphene bilayer
  heterostructure and the effect of hydrogenation.
\newblock ACS Applied Materials \& Interfaces {\bf 6}, 18180--18188
  (2014{\natexlab{a}})

\bibitem[{Liu et~al.(2011{\natexlab{a}})Liu, Feng, and Yao}]{Liu1}
Liu, C.C., Feng, W., Yao, Y.: Quantum spin {H}all effect in silicene and
  two-dimensional germanium.
\newblock Phys. Rev. Lett. {\bf 107}, 076802 (2011{\natexlab{a}})

\bibitem[{Liu et~al.(2011{\natexlab{b}})Liu, Jiang, and Yao}]{Liu2}
Liu, C.C., Jiang, H., Yao, Y.: Low-energy effective hamiltonian involving
  spin-orbit coupling in silicene and two-dimensional germanium and tin.
\newblock Phys. Rev. B {\bf 84}, 195430 (2011{\natexlab{b}})

\bibitem[{Liu et~al.(2014{\natexlab{b}})Liu, Neal, Zhu, Luo, Xu, Tom\'~anek,
  and Ye}]{Liu3}
Liu, H., Neal, A.T., Zhu, Z., Luo, Z., Xu, X., Tom\'~anek, D., Ye, P.D.:
  Phosphorene: An unexplored {2}d semiconductor with a high hole mobility.
\newblock ACS Nano {\bf 8}, 4033--4041 (2014{\natexlab{b}})

\bibitem[{Liu et~al.(2014{\natexlab{c}})Liu, Wang, Liu, Jia, Vogt, Quaresima,
  Ottaviani, Olivieri, De~Padova, and Lay}]{Liu4}
Liu, Z.L., Wang, M.X., Liu, C., Jia, J.F., Vogt, P., Quaresima, C., Ottaviani,
  C., Olivieri, B., De~Padova, P., Lay, G.L.: The fate of the
  2$\sqrt{3}\times$2$\sqrt{3}$r(30$\degree$) silicene phase on {A}g(111).
\newblock APL Materials {\bf 2}, 092513 (2014{\natexlab{c}})

\bibitem[{Liu et~al.(2014{\natexlab{d}})Liu, Wang, Xu, Ge, Lay, Vogt, Qian,
  Gao, Liu, and Jia}]{Liu5}
Liu, Z.L., Wang, M.X., Xu, J.P., Ge, J.F., Lay, G.L., Vogt, P., Qian, D., Gao,
  C.L., Liu, C., Jia, J.F.: Various atomic structures of monolayer silicene
  fabricated on {A}g(111).
\newblock New Journal of Physics {\bf 16}, 075006 (2014{\natexlab{d}})

\bibitem[{van Loenen et~al.(1987)van Loenen, Demuth, Tromp, and
  Hamers}]{Loenen1}
van Loenen, E.J., Demuth, J.E., Tromp, R.M., Hamers, R.J.: Local electron
  states and surface geometry of {S}i(111)-$\sqrt{3}\times\sqrt{3}$ {A}g.
\newblock Phys. Rev. Lett. {\bf 58}, 373--376 (1987)

\bibitem[{Mak et~al.(2010)Mak, Lee, Hone, Shan, and Heinz}]{mak10}
Mak, K.F., Lee, C., Hone, J., Shan, J., Heinz, T.F.: Atomically thin
  ${\mathrm{mos}}_{2}$: A new direct-gap semiconductor.
\newblock Phys. Rev. Lett. {\bf 105}, 136805 (2010)

\bibitem[{Mannix et~al.(2014)Mannix, Kiraly, Fisher, Hersam, and
  Guisinger}]{Mannix1}
Mannix, A.J., Kiraly, B., Fisher, B.L., Hersam, M.C., Guisinger, N.P.: Silicon
  growth at the two-dimensional limit on {A}g(111).
\newblock ACS Nano {\bf 8}, 7538--7547 (2014)

\bibitem[{Mannix et~al.(2015)Mannix, Zhou, Kiraly, Wood, Alducin, Myers, Liu,
  Fisher, Santiago, Guest, Yacaman, Ponce, Oganov, Hersam, and
  Guisinger}]{Mannix2}
Mannix, A.J., Zhou, X.F., Kiraly, B., Wood, J.D., Alducin, D., Myers, B.D.,
  Liu, X., Fisher, B.L., Santiago, U., Guest, J.R., Yacaman, M.J., Ponce, A.,
  Oganov, A.R., Hersam, M.C., Guisinger, N.P.: Synthesis of borophenes:
  Anisotropic, two-dimensional boron polymorphs.
\newblock Science {\bf 350}, 1513--1516 (2015)

\bibitem[{Marutheeswaran et~al.(2014)Marutheeswaran, Pancharatna, and
  Balakrishnarajan}]{Marutheeswaran1}
Marutheeswaran, S., Pancharatna, P.D., Balakrishnarajan, M.M.: Preference for a
  propellane motif in pure silicon nanosheets.
\newblock Phys. Chem. Chem. Phys. {\bf 16}, 11186--11190 (2014)

\bibitem[{Maruyama et~al.(1981)Maruyama, Suzuki, Kobayashi, and
  Tanuma}]{Maruyama1}
Maruyama, Y., Suzuki, S., Kobayashi, K., Tanuma, S.: Synthesis and some
  properties of black phosphorus single crystals.
\newblock Physica {\bf 105B}, 99 -- 102 (1981)

\bibitem[{Matthes et~al.(2013)Matthes, Pulci, and Bechstedt}]{Matthes1}
Matthes, L., Pulci, O., Bechstedt, F.: Massive {D}irac quasiparticles in the
  optical absorbance of graphene, silicene, germanene, and tinene.
\newblock Journal of Physics: Condensed Matter {\bf 25}, 395305 (2013)

\bibitem[{Meng et~al.(2013)Meng, Wang, Zhang, Du, Wu, Li, Zhang, Li, Zhou,
  Hofer, and Gao}]{Meng1}
Meng, L., Wang, Y., Zhang, L., Du, S., Wu, R., Li, L., Zhang, Y., Li, G., Zhou,
  H., Hofer, W.A., Gao, H.J.: Buckled silicene formation on {I}r(111).
\newblock Nano Letters {\bf 13}, 685--690 (2013)

\bibitem[{Mermin(1968)}]{Mermin1}
Mermin, N.D.: Crystalline order in two dimensions.
\newblock Phys. Rev. {\bf 176}, 250--254 (1968)

\bibitem[{Meyer et~al.(2007)Meyer, Geim, Katsnelson, Novoselov, Booth, and
  Roth}]{Meyer1}
Meyer, J.C., Geim, A.K., Katsnelson, M.I., Novoselov, K.S., Booth, T.J., Roth,
  S.: The structure of suspended graphene sheets.
\newblock Nature {\bf 446}, 60--63 (2007)

\bibitem[{Moras et~al.(2014)Moras, Mentes, Sheverdyaeva, Locatelli, and
  Carbone}]{Moras1}
Moras, P., Mentes, T.O., Sheverdyaeva, P.M., Locatelli, A., Carbone, C.:
  Coexistence of multiple silicene phases in silicon grown on {A}g(111).
\newblock Journal of Physics: Condensed Matter {\bf 26}, 185001 (2014)

\bibitem[{Morishita and Spencer(2015)}]{mor15}
Morishita, T., Spencer, M.J.S.: {How silicene on Ag(111) oxidizes: microscopic
  mechanism of the reaction of O$_2$ with silicene}.
\newblock Scientific Reports {\bf 5}, 17570 (2015)

\bibitem[{Nair et~al.(2010)Nair, Ren, Jalil, Riaz, Kravets, Britnell, Blake,
  Schedin, Mayorov, Yuan, Katsnelson, Cheng, Strupinski, Bulusheva, Okotrub,
  Grigorieva, Grigorenko, Novoselov, and Geim}]{nai10}
Nair, R.R., Ren, W., Jalil, R., Riaz, I., Kravets, V.G., Britnell, L., Blake,
  P., Schedin, F., Mayorov, A.S., Yuan, S., Katsnelson, M.I., Cheng, H.M.,
  Strupinski, W., Bulusheva, L.G., Okotrub, A.V., Grigorieva, I.V., Grigorenko,
  A.N., Novoselov, K.S., Geim, A.K.: Fluorographene: A two-dimensional
  counterpart of teflon.
\newblock Small {\bf 6}, 2877--2884 (2010)

\bibitem[{Narita et~al.(1983)Narita, Akahama, Tsukiyama, Muro, Mori, Endo,
  Taniguchi, Seki, Suga, Mikuni, and Kanzaki}]{Narita1}
Narita, S., Akahama, Y., Tsukiyama, Y., Muro, K., Mori, S., Endo, S.,
  Taniguchi, M., Seki, M., Suga, S., Mikuni, A., Kanzaki, H.: Electrical and
  optical properties of black phosphorus single crystals.
\newblock Physica {\bf 117B-118B}, 422 -- 424 (1983)

\bibitem[{{Nature Research Highlights}(2012)}]{Nature1}
{Nature Research Highlights}: Graphene's silicon cousin.
\newblock Nature {\bf 485}, 9 (2012)

\bibitem[{Ni et~al.(2012)Ni, Liu, Tang, Zheng, Zhou, Qin, Gao, Yu, and
  Lu}]{Ni1}
Ni, Z., Liu, Q., Tang, K., Zheng, J., Zhou, J., Qin, R., Gao, Z., Yu, D., Lu,
  J.: Tunable bandgap in silicene and germanene.
\newblock Nano Letters {\bf 12}, 113--118 (2012)

\bibitem[{Noguchi et~al.(2015)Noguchi, Sugawara, Yaokawa, Hitosugi, Nakano, and
  Takahashi}]{Noguchi1}
Noguchi, E., Sugawara, K., Yaokawa, R., Hitosugi, T., Nakano, H., Takahashi,
  T.: {Direct Observation of Dirac Cone in Multilayer Silicene Intercalation
  Compound CaSi2}.
\newblock Advanced Materials {\bf 27}, 856--860 (2015)

\bibitem[{Novoselov et~al.(2005)Novoselov, Geim, Morozov, Jiang, Katsnelson,
  Grigorieva, Dubonos, and Firsov}]{Novoselov2}
Novoselov, K.S., Geim, A.K., Morozov, S.V., Jiang, D., Katsnelson, M.I.,
  Grigorieva, I.V., Dubonos, S.V., Firsov, A.A.: Two-dimensional gas of
  massless {D}irac fermions in graphene.
\newblock Nature {\bf 438}, 197--200 (2005)

\bibitem[{Novoselov et~al.(2004)Novoselov, Geim, Morozov, Jiang, Zhang,
  Dubonos, Grigorieva, and Firsov}]{Novoselov1}
Novoselov, K.S., Geim, A.K., Morozov, S.V., Jiang, D., Zhang, Y., Dubonos,
  S.V., Grigorieva, I.V., Firsov, A.A.: Electric field effect in atomically
  thin carbon films.
\newblock Science {\bf 306}, 666--669 (2004)

\bibitem[{olyomi et~al.(2014)olyomi, Wallbank, and Fal'ko}]{Zolyomi1}
olyomi, V.Z., Wallbank, J.R., Fal'ko, V.I.: Silicane and germanane:
  tight-binding and first-principles studies.
\newblock 2D Materials {\bf 1}, 011005 (2014)

\bibitem[{Onida et~al.(2002)Onida, Reining, and Rubio}]{Onida1}
Onida, G., Reining, L., Rubio, A.: Electronic excitations: density-functional
  versus many-body {G}reen's-function approaches.
\newblock Rev. Mod. Phys. {\bf 74}, 601--659 (2002)

\bibitem[{Osborn et~al.(2011)Osborn, Farajian, Pupysheva, Aga, and
  Voon}]{Osborn1}
Osborn, T.H., Farajian, A.A., Pupysheva, O.V., Aga, R.S., Voon, L.L.Y.: Ab
  initio simulations of silicene hydrogenation.
\newblock Chemical Physics Letters {\bf 511}, 101 -- 105 (2011)

\bibitem[{Ozaydin et~al.(2015)Ozaydin, Sahin, Kang, Peeters, and Senger}]{ti}
Ozaydin, H.D., Sahin, H., Kang, J., Peeters, F.M., Senger, R.T.: Electronic and
  magnetic properties of 1 {T-TiSe$_2$} nanoribbons.
\newblock 2D Materials {\bf 2}, 044002 (2015)

\bibitem[{Ozaydin et~al.(2014)Ozaydin, Sahin, Senger, and Peeters}]{ta}
Ozaydin, H.D., Sahin, H., Senger, R.T., Peeters, F.M.: Formation and diffusion
  characteristics of {P}t clusters on graphene, {1H-MoS$_2$} and {1T-TaS$_2$}.
\newblock Annalen der Physik {\bf 526}, 423--429 (2014)

\bibitem[{\"Oz\c{c}elik et~al.(2014)\"Oz\c{c}elik, Cahangirov, and
  Ciraci}]{Ozcelik3}
\"Oz\c{c}elik, V.O., Cahangirov, S., Ciraci, S.: Stable single-layer
  honeycomblike structure of silica.
\newblock Phys. Rev. Lett. {\bf 112}, 246803 (2014)

\bibitem[{\"Oz\c{c}elik and Ciraci(2013)}]{Ozcelik1}
\"Oz\c{c}elik, V.O., Ciraci, S.: Local reconstructions of silicene induced by
  adatoms.
\newblock The Journal of Physical Chemistry C {\bf 117}, 26305--26315 (2013)

\bibitem[{\"Oz\c{c}elik et~al.(2013)\"Oz\c{c}elik, Gurel, and
  Ciraci}]{Ozcelik2}
\"Oz\c{c}elik, V.O., Gurel, H.H., Ciraci, S.: Self-healing of vacancy defects
  in single-layer graphene and silicene.
\newblock Phys. Rev. B {\bf 88}, 045440 (2013)

\bibitem[{Pandey(1981)}]{Pandley1}
Pandey, K.C.: New $\ensuremath{\pi}$-bonded chain model for
  si(111)-(2\ifmmode\times\else\texttimes\fi{}1) surface.
\newblock Phys. Rev. Lett. {\bf 47}, 1913--1917 (1981)

\bibitem[{Peierls(1934)}]{Peierls1}
Peierls, R.E.: Bemerkungen \"uber umwandlungstemperaturen.
\newblock Helv. Phys. Acta {\bf 7}, 81--83 (1934)

\bibitem[{Peplow(2015)}]{Peplow1}
Peplow, M.: Graphene's cousin silicene makes transistor debut.
\newblock Nature {\bf 518}, 17--18 (2015)

\bibitem[{Perdew et~al.(1996)Perdew, Burke, and Ernzerhof}]{Perdew1}
Perdew, J.P., Burke, K., Ernzerhof, M.: Generalized gradient approximation made
  simple.
\newblock Phys. Rev. Lett. {\bf 77}, 3865--3868 (1996)

\bibitem[{Phillips(1973)}]{ph}
Phillips, J.: Excitonic instabilities, vacancies, and reconstruction of
  covalent surfaces.
\newblock Surface Science {\bf 40}, 459 -- 469 (1973)

\bibitem[{Poppendieck et~al.(1978)Poppendieck, Ngoc, and Webb}]{po}
Poppendieck, T.D., Ngoc, T.C., Webb, M.B.: An electron diffraction study of the
  structure of silicon (100).
\newblock Surface Science {\bf 75}, 287 -- 315 (1978)

\bibitem[{Qiu et~al.(2015)Qiu, Fu, Xu, Oreshkin, Shao, Li, Meng, Chen, and
  Wu}]{Qiu1}
Qiu, J., Fu, H., Xu, Y., Oreshkin, A.I., Shao, T., Li, H., Meng, S., Chen, L.,
  Wu, K.: Ordered and reversible hydrogenation of silicene.
\newblock Phys. Rev. Lett. {\bf 114}, 126101 (2015)

\bibitem[{Rahman et~al.(2015)Rahman, Nakagawa, and Mizuno}]{Rahman1}
Rahman, M.S., Nakagawa, T., Mizuno, S.: Growth of {S}i on {A}g(111) and
  determination of large commensurate unit cell of high-temperature phase.
\newblock Japanese Journal of Applied Physics {\bf 54}, 015502 (2015)

\bibitem[{Reich(2014)}]{Reich1}
Reich, E.S.: {Phosphorene excites materials scientists}.
\newblock Nature {\bf 506}, 19 (2014)

\bibitem[{Resta et~al.(2013)Resta, Leoni, Barth, Ranguis, Becker, Bruhn, Vogt,
  and Le~Lay}]{Resta1}
Resta, A., Leoni, T., Barth, C., Ranguis, A., Becker, C., Bruhn, T., Vogt, P.,
  Le~Lay, G.: Atomic structures of silicene layers grown on {A}g(111): Scanning
  tunneling microscopy and noncontact atomic force microscopy observations.
\newblock Sci. Rep. {\bf 3} (2013)

\bibitem[{Ronci et~al.(2010)Ronci, Colonna, Cricenti, De~Padova, Ottaviani,
  Quaresima, Aufray, and {Le Lay}}]{Ronci1}
Ronci, F., Colonna, S., Cricenti, A., De~Padova, P., Ottaviani, C., Quaresima,
  C., Aufray, B., {Le Lay}, G.: Low temperature {STM/STS} study of silicon
  nanowires grown on the {A}g(110) surface.
\newblock Physica Status Solidi (C) {\bf 7}, 2716--2719 (2010)

\bibitem[{Rubio et~al.(1994)Rubio, Corkill, and Cohen}]{Rubio1}
Rubio, A., Corkill, J.L., Cohen, M.L.: Theory of graphitic boron nitride
  nanotubes.
\newblock Phys. Rev. B {\bf 49}, 5081--5084 (1994)

\bibitem[{Sahin et~al.(2009)Sahin, Cahangirov, Topsakal, Bekaroglu, Akt\"urk,
  Senger, and Ciraci}]{Sahin1}
Sahin, H., Cahangirov, S., Topsakal, M., Bekaroglu, E., Akt\"urk, E., Senger,
  R.T., Ciraci, S.: Monolayer honeycomb structures of group{-IV} elements and
  {III-V} binary compounds: First-principles calculations.
\newblock Phys. Rev. B {\bf 80}, 155453 (2009)

\bibitem[{Sahin et~al.(2015)Sahin, Leenaerts, Singh, and Peeters}]{sah15}
Sahin, H., Leenaerts, O., Singh, S.K., Peeters, F.M.: Graphane.
\newblock Wiley Interdisciplinary Reviews: Computational Molecular Science {\bf
  5}, 255--272 (2015)

\bibitem[{Sahin and Peeters(2013)}]{Sahin2}
Sahin, H., Peeters, F.M.: Adsorption of alkali, alkaline-earth, and 3$d$
  transition metal atoms on silicene.
\newblock Phys. Rev. B {\bf 87}, 085423 (2013)

\bibitem[{Sahin et~al.(2013{\natexlab{a}})Sahin, Sivek, Li, Partoens, and
  Peeters}]{Sahin3}
Sahin, H., Sivek, J., Li, S., Partoens, B., Peeters, F.M.: Stone-{W}ales
  defects in silicene: Formation, stability, and reactivity of defect sites.
\newblock Phys. Rev. B {\bf 88}, 045434 (2013{\natexlab{a}})

\bibitem[{Sahin et~al.(2013{\natexlab{b}})Sahin, Tongay, Horzum, Fan, Zhou, Li,
  Wu, and Peeters}]{Sahin4}
Sahin, H., Tongay, S., Horzum, S., Fan, W., Zhou, J., Li, J., Wu, J., Peeters,
  F.M.: Anomalous raman spectra and thickness-dependent electronic properties
  of {WS}e$_2$.
\newblock Phys. Rev. B {\bf 87}, 165409 (2013{\natexlab{b}})

\bibitem[{Salomon et~al.(2014)Salomon, Ajjouri, Lay, and Angot}]{Salomon1}
Salomon, E., Ajjouri, R.E., Lay, G.L., Angot, T.: Growth and structural
  properties of silicene at multilayer coverage.
\newblock Journal of Physics: Condensed Matter {\bf 26}, 185003 (2014)

\bibitem[{Salpeter and Bethe(1951)}]{Salpeter1}
Salpeter, E.E., Bethe, H.A.: A relativistic equation for bound-state problems.
\newblock Phys. Rev. {\bf 84}, 1232--1242 (1951)

\bibitem[{Sato et~al.(1999)Sato, Nagao, and Hasegawa}]{Sato1}
Sato, N., Nagao, T., Hasegawa, S.: Si(111)-($\sqrt{3}\times\sqrt{3}$)-{A}g
  surface at low temperatures: symmetry breaking and surface twin boundaries.
\newblock Surface Science {\bf 442}, 65 -- 73 (1999)

\bibitem[{Schlier and Farnsworth(1959)}]{sch59}
Schlier, R.E., Farnsworth, H.E.: Structure and adsorption characteristics of
  clean surfaces of germanium and silicon.
\newblock The Journal of Chemical Physics {\bf 30} (1959)

\bibitem[{Schwierz et~al.(2015)Schwierz, Pezoldt, and Granzner}]{Schwierz1}
Schwierz, F., Pezoldt, J., Granzner, R.: Two-dimensional materials and their
  prospects in transistor electronics.
\newblock Nanoscale {\bf 7}, 8261--8283 (2015)

\bibitem[{Shirai et~al.(2014)Shirai, Shirasawa, Hirahara, Fukui, Takahashi, and
  Hasegawa}]{Shirai1}
Shirai, T., Shirasawa, T., Hirahara, T., Fukui, N., Takahashi, T., Hasegawa,
  S.: Structure determination of multilayer silicene grown on {A}g(111) films
  by electron diffraction: Evidence for {A}g segregation at the surface.
\newblock Phys. Rev. B {\bf 89}, 241403 (2014)

\bibitem[{Shishkin and Kresse(2007)}]{Shishkin1}
Shishkin, M., Kresse, G.: Self-consistent {GW} calculations for semiconductors
  and insulators.
\newblock Phys. Rev. B {\bf 75}, 235102 (2007)

\bibitem[{Si et~al.(2014)Si, Liu, Xu, Wu, Gu, and Duan}]{Si14}
Si, C., Liu, J., Xu, Y., Wu, J., Gu, B.L., Duan, W.: Functionalized germanene
  as a prototype of large-gap two-dimensional topological insulators.
\newblock Phys. Rev. B {\bf 89}, 115429 (2014)

\bibitem[{Sivek et~al.(2013)Sivek, Sahin, Partoens, and Peeters}]{Silvek1}
Sivek, J., Sahin, H., Partoens, B., Peeters, F.M.: Adsorption and absorption of
  boron, nitrogen, aluminum, and phosphorus on silicene: Stability and
  electronic and phonon properties.
\newblock Phys. Rev. B {\bf 87}, 085444 (2013)

\bibitem[{Sofo et~al.(2007)Sofo, Chaudhari, and Barber}]{Sofo1}
Sofo, J.O., Chaudhari, A.S., Barber, G.D.: Graphane: {A} two-dimensional
  hydrocarbon.
\newblock Phys. Rev. B {\bf 75}, 153401 (2007)

\bibitem[{Son et~al.(2006{\natexlab{a}})Son, Cohen, and Louie}]{Son2}
Son, Y.W., Cohen, M.L., Louie, S.G.: Energy gaps in graphene nanoribbons.
\newblock Phys. Rev. Lett. {\bf 97}, 216803 (2006{\natexlab{a}})

\bibitem[{Son et~al.(2006{\natexlab{b}})Son, Cohen, and Louie}]{Son1}
Son, Y.W., Cohen, M.L., Louie, S.G.: Half-metallic graphene nanoribbons.
\newblock Nature {\bf 444}, 347--349 (2006{\natexlab{b}})

\bibitem[{Splendiani et~al.(2010)Splendiani, Sun, Zhang, Li, Kim, Chim, Galli,
  and Wang}]{spl10}
Splendiani, A., Sun, L., Zhang, Y., Li, T., Kim, J., Chim, C.Y., Galli, G.,
  Wang, F.: Emerging photoluminescence in monolayer mos2.
\newblock Nano Letters {\bf 10}, 1271--1275 (2010)

\bibitem[{Stephan et~al.(2015)Stephan, Hanf, and Sonnet}]{Stephan1}
Stephan, R., Hanf, M.C., Sonnet, P.: Spatial analysis of interactions at the
  silicene/{A}g interface: first principles study.
\newblock Journal of Physics: Condensed Matter {\bf 27}, 015002 (2015)

\bibitem[{Takayanagi et~al.(1985)Takayanagi, Tanishiro, Takahashi, and
  Takahashi}]{Takayanagi1}
Takayanagi, K., Tanishiro, Y., Takahashi, S., Takahashi, M.: Structure analysis
  of {S}i(111)-7$\times$7 reconstructed surface by transmission electron
  diffraction.
\newblock Surface Science {\bf 164}, 367 -- 392 (1985)

\bibitem[{Takeda and Shiraishi(1994)}]{Takeda1}
Takeda, K., Shiraishi, K.: Theoretical possibility of stage corrugation in {S}i
  and {G}e analogs of graphite.
\newblock Phys. Rev. B {\bf 50}, 14916--14922 (1994)

\bibitem[{Tang and Ismail-Beigi(2007)}]{Tang2}
Tang, H., Ismail-Beigi, S.: Novel precursors for boron nanotubes: The
  competition of two-center and three-center bonding in boron sheets.
\newblock Phys. Rev. Lett. {\bf 99}, 115501 (2007)

\bibitem[{Tang et~al.(2014)Tang, Chen, Cao, Huang, Cahangirov, Xian, Xu, Zhang,
  Duan, and Rubio}]{Tang1}
Tang, P., Chen, P., Cao, W., Huang, H., Cahangirov, S., Xian, L., Xu, Y.,
  Zhang, S.C., Duan, W., Rubio, A.: Stable two-dimensional dumbbell stanene: A
  quantum spin {H}all insulator.
\newblock Phys. Rev. B {\bf 90}, 121408 (2014)

\bibitem[{Tao et~al.(2015)Tao, Cinquanta, Chiappe, Grazianetti, Fanciulli,
  Dubey, Molle, and Akinwande}]{Tao1}
Tao, L., Cinquanta, E., Chiappe, D., Grazianetti, C., Fanciulli, M., Dubey, M.,
  Molle, A., Akinwande, D.: Silicene field-effect transistors operating at room
  temperature.
\newblock Nature Nanotechnology {\bf 10}, 227--231 (2015)

\bibitem[{Tongay et~al.(2014)Tongay, Sahin, Ko, Luce, Fan, Liu, Zhou, Huang,
  Ho, Yan, Ogletree, Aloni, Ji, Li, Li, Peeters, and Wu}]{res}
Tongay, S., Sahin, H., Ko, C., Luce, A., Fan, W., Liu, K., Zhou, J., Huang,
  Y.S., Ho, C.H., Yan, J., Ogletree, D.F., Aloni, S., Ji, J., Li, S., Li, J.,
  Peeters, F.M., Wu, J.: {Monolayer behaviour in bulk ReS2 due to electronic
  and vibrational decoupling} {\bf 5} (2014)

\bibitem[{Torun et~al.(2015)Torun, Sahin, Singh, and Peeters}]{eng15}
Torun, E., Sahin, H., Singh, S.K., Peeters, F.M.: Stable half-metallic
  monolayers of {FeCl$_2$}.
\newblock Applied Physics Letters {\bf 106}, 192404 (2015)

\bibitem[{Tran et~al.(2014)Tran, Soklaski, Liang, and Yang}]{Tran1}
Tran, V., Soklaski, R., Liang, Y., Yang, L.: Layer-controlled band gap and
  aniso-tropic excitons in few-layer black phosphorus.
\newblock Phys. Rev. B {\bf 89}, 235319 (2014)

\bibitem[{Tsoutsou et~al.(2013)Tsoutsou, Xenogiannopoulou, Golias, Tsipas, and
  Dimoulas}]{Tsoutsou1}
Tsoutsou, D., Xenogiannopoulou, E., Golias, E., Tsipas, P., Dimoulas, A.:
  {Evidence for hybrid surface metallic band in (4 {\texttimes} 4) silicene
  on Ag(111)}.
\newblock Applied Physics Letters {\bf 103}, 231604 (2013)

\bibitem[{Vlieg et~al.(1989)Vlieg, Gon, Veen, MacDonald, and Norris}]{Vlieg1}
Vlieg, E., Gon, A.V.D., Veen, J.V.D., MacDonald, J., Norris, C.: The structure
  of {S}i(111)-($\sqrt{3}\times\sqrt{3}$)r30$\degree$-{A}g determined by
  surface {X}-ray diffraction.
\newblock Surface Science {\bf 209}, 100 -- 114 (1989)

\bibitem[{Vogt et~al.(2014)Vogt, Capiod, Berthe, Resta, De~Padova, Bruhn,
  Le~Lay, and Grandidier}]{Vogt2}
Vogt, P., Capiod, P., Berthe, M., Resta, A., De~Padova, P., Bruhn, T., Le~Lay,
  G., Grandidier, B.: Synthesis and electrical conductivity of multilayer
  silicene.
\newblock Applied Physics Letters {\bf 104}, 021602 (2014)

\bibitem[{Vogt et~al.(2012)Vogt, De~Padova, Quaresima, Avila, Frantzeskakis,
  Asensio, Resta, Ealet, and Le~Lay}]{Vogt1}
Vogt, P., De~Padova, P., Quaresima, C., Avila, J., Frantzeskakis, E., Asensio,
  M.C., Resta, A., Ealet, B., Le~Lay, G.: Silicene: Compelling experimental
  evidence for graphenelike two-dimensional silicon.
\newblock Phys. Rev. Lett. {\bf 108}, 155501 (2012)

\bibitem[{Wang et~al.(2013)Wang, Pi, Ni, Liu, Lin, Xu, and Yang}]{Wang2}
Wang, R., Pi, X., Ni, Z., Liu, Y., Lin, S., Xu, M., Yang, D.: Silicene oxides:
  formation, structures and electronic properties.
\newblock Sci. Rep. {\bf 3} (2013)

\bibitem[{Wang et~al.(2015)Wang, Liu, and Tu}]{Wan15}
Wang, X., Liu, H., Tu, S.T.: First-principles study of half-fluorinated
  silicene sheets.
\newblock RSC Adv. {\bf 5}, 6238--6245 (2015)

\bibitem[{Wang and Cheng(2013)}]{Wang1}
Wang, Y.P., Cheng, H.P.: Absence of a {D}irac cone in silicene on {A}g(111):
  First-principles density functional calculations with a modified effective
  band structure technique.
\newblock Phys. Rev. B {\bf 87}, 245430 (2013)

\bibitem[{Warschauer(1963)}]{Warschauer1}
Warschauer, D.: Electrical and optical properties of crystalline black
  phosphorus.
\newblock Journal of Applied Physics {\bf 34}, 1853--1860 (1963)

\bibitem[{Wei and Jacob(2013)}]{Wei13}
Wei, W., Jacob, T.: Strong many-body effects in silicene-based structures.
\newblock Phys. Rev. B {\bf 88}, 045203 (2013)

\bibitem[{Wierzbicki et~al.(2015)Wierzbicki, Barna\ifmmode~\acute{s}\else
  \'{s}\fi{}, and Swirkowicz}]{Wierzbicki1}
Wierzbicki, M., Barna\ifmmode~\acute{s}\else \'{s}\fi{}, J., Swirkowicz, R.:
  Thermoelectric properties of silicene in the topological- and band-insulator
  states.
\newblock Phys. Rev. B {\bf 91}, 165417 (2015)

\bibitem[{Xu et~al.(2013{\natexlab{a}})Xu, Liang, Shi, and Chen}]{Xu2}
Xu, M., Liang, T., Shi, M., Chen, H.: Graphene-like two-dimensional materials.
\newblock Chemical Reviews {\bf 113}, 3766--3798 (2013{\natexlab{a}})

\bibitem[{Xu et~al.(2014)Xu, Zhuang, Du, Feng, Zhang, Liu, Lei, Wang, Spencer,
  Morishita, Wang, and Dou}]{Xu14}
Xu, X., Zhuang, J., Du, Y., Feng, H., Zhang, N., Liu, C., Lei, T., Wang, J.,
  Spencer, M., Morishita, T., Wang, X., Dou, S.X.: {Effects of Oxygen
  Adsorption on the Surface State of Epitaxial Silicene on Ag(111)}.
\newblock Scientific Reports {\bf 4}, 7543 (2014)

\bibitem[{Xu et~al.(2013{\natexlab{b}})Xu, Yan, Zhang, Wang, Xu, Tang, Duan,
  and Zhang}]{Xu1}
Xu, Y., Yan, B., Zhang, H.J., Wang, J., Xu, G., Tang, P., Duan, W., Zhang,
  S.C.: Large-gap quantum spin {H}all insulators in tin films.
\newblock Phys. Rev. Lett. {\bf 111}, 136804 (2013{\natexlab{b}})

\bibitem[{Yang et~al.(2014)Yang, Cahangirov, Cantarero, Rubio, and
  D'Agosta}]{Yang1}
Yang, K., Cahangirov, S., Cantarero, A., Rubio, A., D'Agosta, R.:
  Thermoelectric properties of atomically thin silicene and germanene
  nanostructures.
\newblock Phys. Rev. B {\bf 89}, 125403 (2014)

\bibitem[{Yang et~al.(2015{\natexlab{a}})Yang, Wang, Sahin, Chen, Li, Li,
  Suslu, Peeters, Liu, Li, and Tongay}]{rese}
Yang, S., Wang, C., Sahin, H., Chen, H., Li, Y., Li, S.S., Suslu, A., Peeters,
  F.M., Liu, Q., Li, J., Tongay, S.: Tuning the optical, magnetic, and
  electrical properties of {ReSe$_2$} by nanoscale strain engineering.
\newblock Nano Letters {\bf 15}, 1660--1666 (2015{\natexlab{a}})

\bibitem[{Yang et~al.(2015{\natexlab{b}})Yang, Wang, Sahin, Chen, Li, Li,
  Suslu, Peeters, Liu, Li, and Tongay}]{Yang2}
Yang, S., Wang, C., Sahin, H., Chen, H., Li, Y., Li, S.S., Suslu, A., Peeters,
  F.M., Liu, Q., Li, J., Tongay, S.: Tuning the optical, magnetic, and
  electrical properties of rese2 by nanoscale strain engineering.
\newblock Nano Letters {\bf 15}, 1660--1666 (2015{\natexlab{b}})

\bibitem[{Yao et~al.(2007)Yao, Ye, Qi, Zhang, and Fang}]{Yao1}
Yao, Y., Ye, F., Qi, X.L., Zhang, S.C., Fang, Z.: Spin-orbit gap of graphene:
  First-principles calculations.
\newblock Phys. Rev. B {\bf 75}, 041401 (2007)

\bibitem[{Ye et~al.(2014)Ye, Shao, Zhao, Yang, and Wang}]{Ye1}
Ye, X.S., Shao, Z.G., Zhao, H., Yang, L., Wang, C.L.: Intrinsic carrier
  mobility of germanene is larger than graphene{'}s: first-principle
  calculations.
\newblock RSC Adv. {\bf 4}, 21216--21220 (2014)

\bibitem[{Zberecki et~al.(2014{\natexlab{a}})Zberecki, Swirkowicz, and
  Barna\ifmmode~\acute{s}\else \'{s}\fi{}}]{Zberecki3}
Zberecki, K., Swirkowicz, R., Barna\ifmmode~\acute{s}\else \'{s}\fi{}, J.: Spin
  effects in thermoelectric properties of {A}l- and {P}-doped zigzag silicene
  nanoribbons.
\newblock Phys. Rev. B {\bf 89}, 165419 (2014{\natexlab{a}})

\bibitem[{Zberecki et~al.(2014{\natexlab{b}})Zberecki, Swirkowicz, Wierzbicki,
  and Barnas}]{Zberecki1}
Zberecki, K., Swirkowicz, R., Wierzbicki, M., Barnas, J.: Enhanced
  thermoelectric efficiency in ferromagnetic silicene nanoribbons terminated
  with hydrogen atoms.
\newblock Phys. Chem. Chem. Phys. {\bf 16}, 12900--12908 (2014{\natexlab{b}})

\bibitem[{Zberecki et~al.(2013)Zberecki, Wierzbicki,
  Barna\ifmmode~\acute{s}\else \'{s}\fi{}, and Swirkowicz}]{Zberecki2}
Zberecki, K., Wierzbicki, M., Barna\ifmmode~\acute{s}\else \'{s}\fi{}, J.,
  Swirkowicz, R.: Thermoelectric effects in silicene nanoribbons.
\newblock Phys. Rev. B {\bf 88}, 115404 (2013)

\bibitem[{Zhang and Yan(2012)}]{Zha12b}
Zhang, C.W., Yan, S.S.: First-principles study of ferromagnetism in
  two-dimensional silicene with hydrogenation.
\newblock The Journal of Physical Chemistry C {\bf 116}, 4163--4166 (2012)

\bibitem[{Zhang et~al.(2015{\natexlab{a}})Zhang, Yang, and Yao}]{Zhang1}
Zhang, L.D., Yang, F., Yao, Y.: Possible electric-field-induced superconducting
  states in doped silicene.
\newblock Sci. Rep. {\bf 5} (2015{\natexlab{a}})

\bibitem[{Zhang et~al.(2012)Zhang, Li, Hu, Wu, and Zhu}]{Zha12}
Zhang, P., Li, X., Hu, C., Wu, S., Zhu, Z.: First-principles studies of the
  hydrogenation effects in silicene sheets.
\newblock Physics Letters A {\bf 376}, 1230 -- 1233 (2012)

\bibitem[{Zhang et~al.(2001)Zhang, Chu, Cheung, Wang, and Lee}]{Zhang2}
Zhang, R.Q., Chu, T.S., Cheung, H.F., Wang, N., Lee, S.T.: High reactivity of
  silicon suboxide clusters.
\newblock Phys. Rev. B {\bf 64}, 113304 (2001)

\bibitem[{Zhang et~al.(2015{\natexlab{b}})Zhang, Song, and Dou}]{Zha15b}
Zhang, W.B., Song, Z.B., Dou, L.M.: The tunable electronic structure and
  mechanical properties of halogenated silicene: a first-principles study.
\newblock J. Mater. Chem. C {\bf 3}, 3087--3094 (2015{\natexlab{b}})

\bibitem[{Zheng and Zhang(2012)}]{Zheng2}
Zheng, F.B., Zhang, C.w.: {The electronic and magnetic properties of
  functionalized silicene: a first-principles study}.
\newblock Nanoscale Research Letters {\bf 7}, 422 (2012)

\bibitem[{Zheng et~al.(2013)Zheng, Zhang, Yan, and Li}]{Zheng1}
Zheng, F.B., Zhang, C.W., Yan, S.S., Li, F.: Novel electronic and magnetic
  properties in {N} or {B} doped silicene nanoribbons.
\newblock J. Mater. Chem. C {\bf 1}, 2735--2743 (2013)

\bibitem[{Zhu et~al.(2015)Zhu, Chen, Xu, Gao, Guan, Liu, Qian, Zhang, and
  Jia}]{Zhu1}
Zhu, F.F., Chen, W.J., Xu, Y., Gao, C.L., Guan, D.D., Liu, C.H., Qian, D.,
  Zhang, S.C., Jia, J.F.: {Epitaxial growth of two-dimensional stanene.}
\newblock Nature materials  (2015)

\end{thebibliography}
\bibliographystyle{springer}



\end{document}